\documentclass[longtitle,preprint,5p,times,twocolumn]{elsarticle}

\usepackage{amssymb}
\usepackage{amsmath}

\usepackage{gensymb}
\usepackage{caption}
\usepackage{subcaption} %RM: i added this for sib figure
\usepackage{xspace}
\usepackage{longtable}
\usepackage{pdflscape}
\usepackage{multirow}
\usepackage{ragged2e}
\usepackage{booktabs}
\usepackage{hyperref}
\usepackage{float}
\usepackage{adjustbox}%marginboxes
\usepackage{lineno}

\newcommand{\geant}

\begin{document}
\hypersetup{pdflinkmargin=2.8pt}%fixing red boxes - test
\begin{frontmatter}

%%%%%%%%%%%%%%%%%%%  header  (packages, 
%\input{header}
\title{Detector Requirements and Simulation Results for the EIC Exclusive, Diffractive and Tagging Physics Program using the ECCE Detector Concept}
\def\theaffn{\arabic{affn}} % added to use numbers instead of letters with a long author list

%
% Full list of ECCE Collaboratotions
%
% Click Refresh to get the latest version
%
% Goto the following to edit:
% https://www.overleaf.com/1234763137gmqgnzsrpsdj
%
%===============================
% Student and postdocs

\author[UKansas]{A.~Bylinkin}
\author[LANL]{C.~T.~Dean}
\author[York]{S.~Fegan}
\author[UH]{D.~Gangadharan}
\author[Glasgow]{K.~Gates}
\author[Regina]{S.~J.~D.~Kay}
\author[MIT]{I.~Korover}
\author[CFNS,StonyBrook,WandM]{W.B.~Li}
\author[USTC]{X.~Li}
\author[Glasgow]{R.~Montgomery}
% ...

%
% Full ECCE author list
% Click Refresh to get the latest version
%
% Goto the following to edit:
% https://www.overleaf.com/1234763137gmqgnzsrpsdj

%
% Please fill in any missing information
%
%
% Example of the format:
%
% \affiliation[]{organization={},
%	 addressline={},
%	 city={},
%	 postcode={},
%	 state={},
%	 country={}} 
%

% ...
\affiliation[Glasgow]{organization={University of Glasgow},
	 city={Glasgow},
	 postcode={},
	 country={UK}} 
 
% ...
\affiliation[LANL]{organization={Los Alamos National Laboratory},
	 city={},
	 postcode={},
	 state={NM},
	 country={USA}} 
 
% ...
\affiliation[MIT]{organization={Massachusetts Institute of Technology},
	 addressline={77 Massachusetts Ave.},
	 city={Cambridge},
	 postcode={02139},
	 state={MA},
	 country={USA}} 
 
% ...
\affiliation[CFNS]{organization={Center for Frontiers in Nuclear Science},
	 city={Stony Brook},
	 postcode={11794},
	 state={NY},
	 country={USA}} 
 
\affiliation[StonyBrook]{organization={Stony Brook University},
	 addressline={100 Nicolls Rd.},
	 city={Stony Brook},
	 postcode={11794},
	 state={NY},
	 country={USA}} 
 
% ...
\affiliation[USTC]{organization={University of Science and Technology of China},
     city={Hefei},
     country={China}}

% ...
\affiliation[UH]{organization={University of Houston},
	 city={Houston},
	 postcode={},
	 state={TX},
	 country={USA}} 
 
% ...
\affiliation[UKansas]{organization={Unviersity of Kansas},
	 addressline={1450 Jayhawk Blvd.},
	 city={Lawrence},
	 postcode={66045},
	 state={KS},
	 country={USA}} 
 
% ...
\affiliation[Regina]{organization={ University of Regina},
	 city={Regina},
	 postcode={S4S 0A2},
	 state={SK},
	 country={Canada}} 
 
% ...
\affiliation[WandM]{organization={The College of William and Mary},
	 city={Williamsburg},
	 state={VA},
	 country={USA}} 
 
% ...
\affiliation[York]{organization={University of York},
	 city={York},
	 postcode={YO10 5DD},
	 country={UK}} 
 
% ...

\begin{abstract}

This article presents a collection of simulation studies using the ECCE detector concept in the context of the EIC's exclusive, diffractive, and tagging physics program, which aims to further explore the rich quark-gluon structure of nucleons and nuclei. To successfully execute the program, ECCE proposed to utilize the detector system close to the beamline to ensure exclusivity and tag ion beam/fragments for a particular reaction of interest. Preliminary studies confirm the proposed technology and design satisfy the requirements. The projected physics impact results are based on the projected detector performance from the simulation at 10 or 100~fb$^{-1}$ of integrated luminosity. Additionally, insights related to a potential second EIC detector are documented, which could serve as a guidepost for future development. 

\end{abstract}


\end{frontmatter}

\setcounter{tocdepth}{1}
\tableofcontents 

%% 
%% Input Sections of the Document
%%

%-----------------------------------------
\section {Introduction}

\label{sec:overview}
The planned Electron-Ion Collider (EIC), to be constructed at Brookhaven National Laboratory (BNL) in partnership with the Thomas Jefferson National Accelerator Facility (JLab), is considered the next generation ``dream machine'' to further explore the quark and gluon substructure of hadrons and nuclei, and provide scientific opportunities for the upcoming decades.

The scientific mission at the EIC was summarized in a 2018 report by the National Academies of Science (NAS)~\cite{NAP25171}: 
\begin{itemize}

\item While the longitudinal momenta of quarks and gluons in nucleons and nuclei have been measured with great precision at previous facilities -- most notably CEBAF at JLab and the HERA collider at DESY -- the \textbf{full three-dimensional momentum and spatial structure} of nucleons are not fully elucidated, particularly including spin, which requires the separation of the intrinsic spin of the constituent particles from their orbital motion.  

\item These studies will also provide insight into how the mutual interactions of quarks and gluons generate the \textbf{nucleon mass and the masses of other hadrons}.  The nucleon mass is one of the single most important scales in all of physics, as it is the basis for nuclear masses and thus the mass of essentially all of the visible matter.

\item The \textbf{density of gluons and sea quarks} which carry the smallest $x_B$, the fraction of the nuclear momentum (or that of its constituent nucleons), can grow so large that their mutual interactions enter a non-linear regime where elegant, universal features emerge in what may be a new, distinct state of matter characterized by a “saturation momentum scale”.  Probing this state requires high energy beams and large nuclear size, and will answer longstanding questions raised by the heavy ion programs at RHIC and the LHC. 
\end{itemize}

To accomplish the physics program, the EIC requires an accelerator capable of delivering: 1) Highly polarized electron ($\sim$70\%) and proton ($\sim$70\%) beams; 2) Ion beams from deuterons to heavy nuclei such as gold, lead, or uranium; 3) Variable $ep$ center-of-mass energies from 20-140 GeV at high collision luminosity of $10^{33}$--$10^{34}$~cm$^{-2}$~s$^{-1}$. Additionally, the EIC requires a comprehensive and hermetic detector to record final-state particles produced by the $ep$ and $e$A scattering.

The EIC Comprehensive Chromodynamics Experiment (ECCE) is a detector proposal that was designed to address the full scope of the EIC physics program as presented in the EIC White Paper~\cite{Accardi:2016etf} and the NAS report. In parallel, the EIC community developed two additional detector proposals: ATHENA (A Totally Hermetic Electron Nucleus Apparatus)~\cite{ATHENA} and CORE (COmpact detectoR for the EIC)~\cite{CORE}. All three proposals were submitted to the EIC Detector Proposal Advisory Panel (DPAP) and thoroughly accessed. Following the recommendations of the DPAP (in March 2022), all three proposals joined efforts to form the ePIC (Electron Proton-and Ion Collider Experiment) Collaboration\footnote{https://www.jlab.org/conference/EPIC} to complete the design of the first detector of the EIC project. 

The specific requirements on each of the ECCE detector systems follow from the more general detector requirements described in the EIC Yellow Report (YR)~\cite{khalek2021science}.  Through the judicious use of existing equipment, ECCE can be built within the budget envelope set out by the EIC project while also managing schedule risks~\cite{ecce-paper-det-2022-01}.

The YR also identified a set of detector performance requirements that flow down from the physics requirements of the EIC science program articulated in the NAS report:
\begin{itemize}
\item The outgoing electron must be distinguished from other produced particles in the event, with a pion rejection of $10^3$--$10^4$ even at large angles, in order to characterize the kinematic properties of the initial scattering process.  These include $x_B$ and the squared momentum transfer ($Q^{2}$).  

\item A large-acceptance magnetic spectrometer is needed to measure the scattered electron momentum, as well as those of the other charged hadrons and leptons. The magnet dimensions and field strength should be matched to the scientific program and the medium-energy scale of the EIC. This requires a nearly 4$\pi$ angular aperture, and the ability to precisely make measurements of the sagitta of its curved trajectory, to measure its momentum down to low $p_T$ (transverse momentum), and to determine its point of origin, in order to distinguish particles from the charm and bottom hadron decays. 

\item A high-purity hadron particle identification (PID) system, able to provide continuous $e/\pi$ and $K/\pi$ discrimination to the highest momentum (60 GeV), is important for identifying particles containing different light-quark flavors.  

\item A hermetic electromagnetic calorimeter system %including both electromagnetic and hadronic sections, 
-- with matching hadronic sections -- is required to measure neutral particles (particularly photons and neutrons) and, in tandem with the spectrometer, to reconstruct hadronic jets that carry kinematic information of the struck quark or gluon, as well as its radiative properties via its substructure.  

\item Far-Forward detector systems, in the direction of the outgoing hadron beam, are needed in order to perform measurements of deeply-virtual Compton scattering (DVCS) through exclusive production as well as diffractive processes, e.g. by measuring the small deflections of the incoming proton and suppressing incoherent interactions with nuclei.  

\item Far-Backward detectors, in the direction of the outgoing electron beam, are needed to reach the very lowest values of $Q^2$, and to measure luminosity to extract the absolute cross-section and spin-dependent asymmetries (with high precision). 
\end{itemize}

The ECCE concept reuses the BaBar~\cite{AUBERT02, AUBERT13} superconducting solenoid (which will be operated at 1.4~T) as well as the sPHENIX~\cite{sPHENIX} barrel flux return and hadronic calorimeter.
These two pieces of equipment are currently being installed in RHIC Interaction Point 8 (IR8) as part of the sPHENIX detector.
Engineering studies have confirmed that these critical components can be relocated to IP6, where the EIC project plans to site the on-project detector.
Additional details concerning ECCE subsystems, performance and selected physics objectives are provided in separate articles within this same
collection.\cite{
ecce-paper-det-2022-01,
ecce-paper-det-2022-02,
ecce-paper-det-2022-03,
ecce-note-comp-2021-01-nim,
ecce-paper-comp-2022-01,
ecce-paper-comp-2022-02,
ecce-paper-phys-2022-01,
ecce-paper-phys-2022-02,
ecce-paper-phys-2022-03}

Among different types of $ep$ interactions studied by ECCE, in the \textbf{exclusive} processes: all final state particles are detected and reconstructed, and in the \textbf{diffractive} processes: no exchange of color-charge between the initial and final state nucleon. These physics processes share a commonality of requiring detection (\textbf{tagging}) of the interacted (recoiled) nucleons and electrons close to the outgoing beamlines. Specialized detector systems are required to perform such measurements to high precision. The unifying theme of this paper is to introduce the design and technology used by these specialized detector systems, and summarize the physics simulation studied based on the expected detector performance. The conclusion of these studies signifies feasibility based on the realistic detector acceptance based on the current base knowledge, however, the process-specific energy-dependent efficiencies and the resolution of the reconstructed kinematics variables will be further studied in future publications.

This paper is structured as follows: Sec.~\ref{sec:FF_detector} provides a short overview of the ECCE detector and a detailed description of the far-forward region (FFR); Sec.~\ref{sec:sim_framework} provides a brief description of the structure and workflow of the ECCE simulation and analysis framework; Sec.~\ref{sec:physics} presents and discusses the physics impact related to the Exclusive, Diffractive and Tagging sections; Sec.~\ref{sec:IP8} discusses some improvements and complementary information associated with the unique second beam focused in IP8; and, finally a summary is presented in Sec.~\ref{sec:summary}.

%-----------------------------------------
\section {ECCE central detector and far-farword components}

\label{sec:FF_detector}
\label{sec:ECCEDet}

\begin{figure*}[th]
    \centering
    \includegraphics[width=0.8\linewidth]{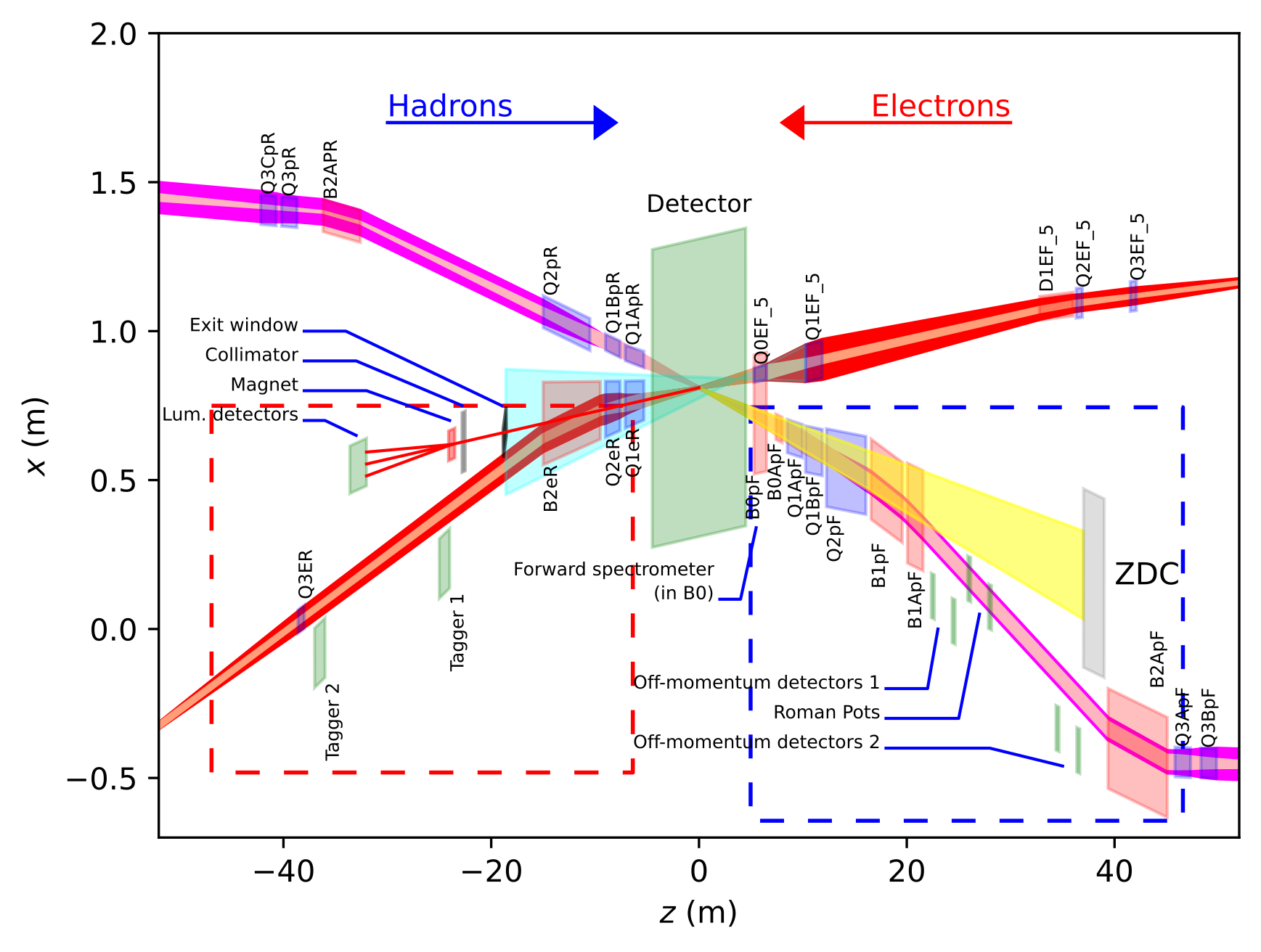}
    \caption{IP6 interaction region. The far-forward region is encircled by the blue dashed box, and the far-backward region is encircled by the red dashed box. It is important to point out that the layout of the ECCE central detector (green box) reflects the asymmetrical nature of the $ep$ collision laboratory frame at the EIC. The incoming electron and ion beam form a crossing angle of 25~mrad. Note that the central detector is parallel to the electron beam line.
    }
    \label{fig:IP6}
\end{figure*}

The ECCE detector consists of three major components: the central detector, the far-forward, and the far-backward systems. The ECCE central detector has a cylindrical geometry based on the BaBar/sPHENIX superconducting solenoid (nominally operated at 1.4~T), and has three primary subdivisions: the barrel (pseudorapidity coverage $|\eta| < 1$), the forward endcap ($-3.5 < \eta < -1$), and the backward endcap ($1 < \eta < 3.5$). The ``forward'' region is defined as the hadron/nuclear beam direction and ``backward'' refers to the electron beam direction. These are illustrated in the beam-crossing schematic of Fig.~\ref{fig:IP6}. It is important to note that the electron and ion beams cross at a 25~mrad  angle and that the electron beam passes down the axis of the central detector, parallel to the magnetic field lines. 

The purpose of the far-forward and far-backward detectors is to measure the reaction kinematics of the colliding systems. This information is vital for the interpretation of the data from the central detectors. The goal of the far-backward system is to determine the luminosity, and measure the momentum of the scattered electron, while the far-forward detectors are designed around detecting the forward (close to the hadron beamline) charged hadrons, neutrons, photons, and light nuclei or nuclear fragment photons over the maximum possible acceptance with high position and momentum resolution. 
 
ECCE's barrel, far-forward, and far-backward detector systems were implemented and studied using a Geant4 simulation~\cite{Agostinelli:2002hh} within the Fun4all framework~\cite{fun4all} (see Sec.~\ref{sec:sim_framework} for further detail).

\subsection{A brief description of central detector}

The layout of the ECCE central detector is intended to be asymmetrical. In the laboratory frame of the EIC, the collisions are asymmetric as the ion beam will carry higher momentum and interact will the electron beam at 25~mrad angle (the crossing angle). The incoming electron beam is defined as $\eta\rightarrow - \infty$. Note that the central detector is parallel to the electron beam line, therefore, the spectating or recoiled nucleon could be tagged by the integrated detector systems along the beam momentum. 

The ECCE central barrel detector features a hybrid-tracking detector design using three state-of-the-art technologies to determine vertex positions (for both primary and decay vertices), track momenta, and distance of closest approach with high precision over the $|\eta| \le 3.5$ region with full azimuthal coverage. This tracking detector consists of the Monolithic Active Pixel Sensor (MAPS) based silicon vertex/tracking subsystem, the $\mu$RWELL tracking subsystem, and the AC-LGAD outer tracker, which also serves as the ToF detector. 
 
The PID system in the barrel, forward, and backward endcaps consists of high-performance DIRC (hpDIRC), dual-radiator Ring Imaging Cherenkov (dRICH), and modular RICH (mRICH), respectively. Their key features are:

\begin{description}
\item[hpDIRC] with coverage of $-1< \eta < 1$, provides PID separation with 3$\sigma$ (standard deviations) or more for $\pi/K$ up to 6 GeV/c, $e/\pi$ up to 1.2 GeV/c, and $K/p$ up to 12 GeV/c.

\item[dRICH] with coverage of $1 < \eta<3.5$ (hadron direction), is designed to provide hadron identification in the forward endcap with 3$\sigma$ or more for $\pi/K$ from 0.7 GeV to 50 GeV, and for $e/\pi$ from $\sim$100 MeV up to 15 GeV/c.

\item[mRICH] with coverage of $-3.5 < \eta < -1$ (electron direction), is to achieve 3$\sigma$ $K/p$ separation in the momentum range from 3 to 10 GeV/c, within the physical constraints of the ECCE detector. It also provides excellent $e/p$ separation for momenta below 2 GeV. In addition, the RICH detectors contribute to $e/\pi$ identification. e.g., when combined with an EM calorimeter, the mRICH and hpDIRC will provide excellent suppression of the low-momentum $\pi^{\pm}$ backgrounds, which can limit the ability to measure the scattered electron in kinematics where it loses most of its
energy.

\end{description}

The ECCE electromagnetic calorimeter system consists of three major components, it allows high-precision electron/hadron detection and suppression in the backward, barrel, and forward directions. Hadronic calorimetry is essential for the barrel and forward endcap regions for hadron and jet reconstruction. Jet yields in the backward region were found to be sufficiently infrequent that hadronic calorimetry would provide little to no scientific benefit. 

\begin{description}

\item[EEMC] The Electron Endcap EM Calorimeter is a high-resolution electromagnetic calorimeter designed for precise measurement of scattered  electrons and final-state photons towards the electron endcap. The design of the EEMC is based on an array of 3000 lead tungstate (PbWO$_4$) crystals of size 2~cm $\times$ 2~cm $\times$ 20~cm and readout by SiPMs yielding an expected energy resolution of $2\%/\sqrt{E} \oplus 1\%$. 

\item[oHCAL and iHCAL] The energy resolution of reconstructed jets in the central barrel will be dominated by the track momentum resolution, as the jets in this region have relatively low momentum, and the measurement of the energy in the hadronic calorimeter does not improve knowledge of the track momentum. The primary use for a hadronic calorimeter in the central barrel will be to collect neutral hadronic energy. The sPHENIX Outer Hadronic Calorimeter (oHCAL) will be reused, which instruments the barrel flux return steel of the BaBar solenoid to provide hadronic calorimetry with an energy resolution of $75\%/ \sqrt{E} \oplus 14.5\%$. There is also a plan to instrument the support for the barrel electromagnetic calorimeter to provide an additional longitudinal segment of hadronic calorimetry. This will provide an Inner Hadronic Calorimeter (iHCAL) layer very similar in design to the sPHENIX inner HCAL. The primary inner HCAL is useful to monitor shower leakage from the barrel electromagnetic calorimeter as well as improve the calibration of the combined calorimeter system.

\item[BEMC] The barrel electromagnetic calorimeter (BEMC) is a projective homogeneous calorimeter based on an inorganic scintillator material that produces shower due to high $Z$ components. Scintillating Glass (SciGlass) blocks of size 4~cm $\times$ 4~cm $\times$ 45.5~cm, plus an additional 10cm of radial readout space. SciGlass has an expected energy resolution of $2.5\%/\sqrt{E} \oplus 1.6\%$~\cite{ecce-note-det-2021-01}, comparable to PbWO$_4$ for a significantly lower cost. The BEMC's optimal acceptance region is ($-1.4 < \eta < 1.1$).

\item[FEMC and LFHCAL] The forward ECal (FEMC) will be a Pb-Scintillator shashlik calorimeter (the scintillator layers consists of polystyrene panels). It is placed after the tracking and PID detectors and made up of two half disks with a radius of $\approx$1.83~m. It employs modern techniques for the readout as well as scintillation tile separation. The towers were designed to be smaller than the Moli\`ere radius in order to allow for further shower separation at high rapidity. The longitudinally segmented forward HCal (LFHCAL) is a Steel-Tungsten-Scintillator calorimeter. It is made up of two half disks with a radius of $\approx$2.6~m. The LFHCAL towers have an active depth of 1.4~m with additional space for the readout of $\approx$20–30~cm depending on their radial position. Each tower consists of 70 layers of 1.6~cm absorber and 0.4~cm scintillator material. For the first 60 layers, the absorber material is steel, while the last 10 layers serve as the tail catcher and are thus made out of tungsten to maximize the interaction length within the available space. The front face of the tower is 5 $\times$ 5~cm$^2$.

\end{description}

Further details of the central barrel detector stack are described in Ref.~\cite{ecce-paper-det-2022-01}. 

\begin{figure*}[h]
    \centering
    \includegraphics[width=0.9\linewidth]{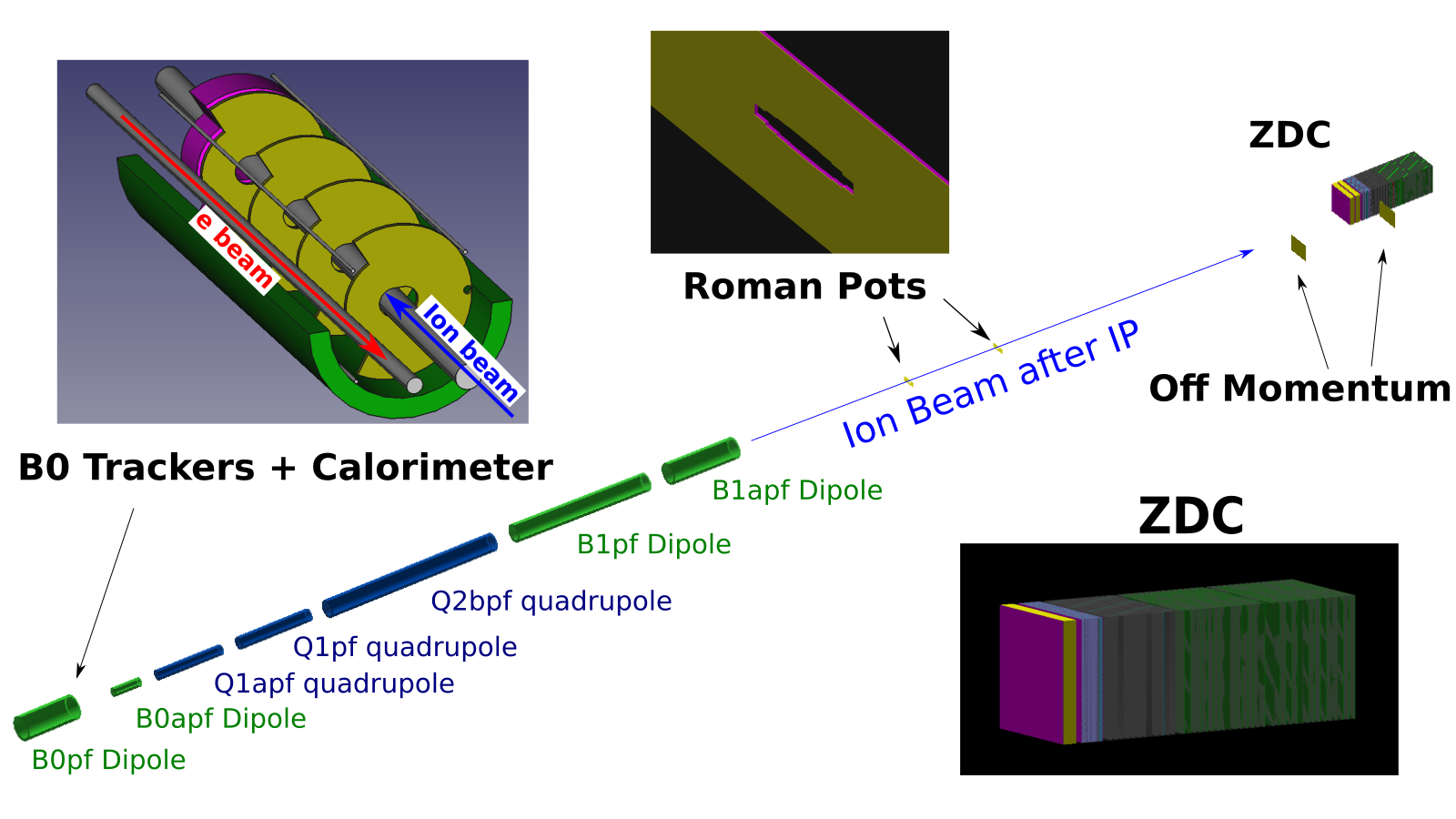}
    \caption{The layout of the EIC far-forward region.}
    \label{fig:far-forward}
\end{figure*}

 \begin{table*}[ht]
   \caption{Summary of far-forward detector locations and angular acceptances for charged hadrons, neutrons, photons, and light nuclei or nuclear fragments. In some cases, the angular acceptance is not uniform in $\phi$, as noted. For the three silicon detectors (Roman Pots, Off-Momentum Detectors, and B0 spectrometer), the 2D size of the silicon planes are given and the thickness is not specified. For the Roman Pots and Off-Momentum Detectors, the simulations have two silicon planes spaced 2~m apart, while the B0 detectors have four silicon planes evenly spaced along the first 1~m length of the B0pf dipole magnet bore. The planes have a hole for the passage of the hadron beam pipe that has a radius of 3.2~cm. }
    \label{tab:FF_accep}
    \renewcommand{\arraystretch}{1.4}
    \centering
    \resizebox{\textwidth}{!}{
    \begin{tabular}{lcccc}
    \toprule
        \textbf{Detector}\ & \textbf{(x,z) Position [m]} & \textbf{Dimensions}& \textbf{$\theta$ [mrad]} & \textbf{Notes} \\
         \midrule
         ZDC & (-0.96, 37.5) & (60 cm, 60 cm, 1.62 m) & $\theta < $ 5.5 & $\sim$4.0 mrad at $\phi = \pi$ \\
         Roman Pots (2 stations) & (-0.83, 26.0), (-0.92, 28.0) & (30 cm, 10 cm) & $0.0 < \theta$ $< 5.5$ & 10$\,\sigma$ cut. \\
         Off-Momentum Detector & (-1.62, 34.5), (-1.71, 36.5) & (50 cm, 35 cm) & $0.0 < \theta < 5.0$ & $0.4 < x_{L} < 0.6$ \\
         B0 Trackers and Calorimeter & (x = -0.15, $5.8<$\,z\,$< 7.0$) & (32 cm, 38 m) & $6.0 < \theta < 22.5$ & $\sim$20\,mrad at $\phi$=0 \\
      \bottomrule
    \end{tabular}
    }
\end{table*}

\subsection{Schematics of the far-forward}

Operating forward detectors at colliders will be a challenge since space is very limited and radiation loads and backgrounds are high. To simplify the operation of such a complex system of detectors, a uniform, and common technology (such as the central barrel) for electromagnetic calorimetry (PbWO$_4$) and tracking (AC-LGAD) is explored and proposed. Such uniformity also allows for the implementation of common monitoring and calibration systems. The luminosity will be determined using complementary approaches following what was learned from HERA, as described in the YR.  

A schematic of the far-forward detectors is shown in Fig.~\ref{fig:far-forward}. They include the B0 spectrometer, off-momentum trackers, Roman Pots, and ZDC (see Table~\ref{tab:FF_accep} for position and dimensions). 

\subsection{Zero Degree Calorimeter (ZDC)}

The Zero-Degree Calorimeter (ZDC) plays an important role in many physics topics.  The production of exclusive vector mesons in diffraction processes from electron-nucleus  collisions is one of the important measurements. For the coherent processes, where the nucleus remains intact, the momentum-transfer ($t$) dependent cross section can be related to the transverse spatial distribution of gluons in the nucleus, which is sensitive to gluon saturation. In this case, however, the coherence of the reaction needs to be determined precisely.  Incoherent events can be isolated by identifying the break-up of the excited nucleus. The evaporated neutrons produced by the break-up in the diffraction process can be used in most cases (about 90\%) to separate coherent processes~\cite{Chang:2021jnu}. In addition, photons from the de-excitation of the excited nuclei can help identify incoherent processes even in the absence of evaporated neutrons. Therefore, in order to identify coherent events over a wide $t$ range, neutrons and photons must be accurately measured near zero degrees.

The geometry of the collision is important to understand the characteristics of each event in electron-nucleus collisions. It has been proposed that collision geometry can be studied by tagging it with  the multiplicity of forward neutrons emitted near zero degrees~(see for instance~\cite{Chang:2022hkt}). Determining the geometry of the collision, such as the ``travel length" of the struck partons in the nucleus, which correlates with the impact parameters of the collision, is very useful in the study of nuclear matter effects. Determining the geometry of the collision will allow us to understand the nuclear structure  with greater accuracy.

The physics requirements of the ZDC are summarised in Table~\ref{tab:zdc_phys_req}.

\begin{table*}[]
    \caption{Physics requirement for ZDC.}
    \label{tab:zdc_phys_req}
    \centering
    \begin{tabular}{|p{16mm}|p{25mm}|p{30mm}|p{20mm}|p{45mm}|}
    \hline
         & Energy range & Energy  & Position  & Others \\
         &              & resolution & resolution & \\
    \hline
    \multirow{4}{*}{Neutrons}  & up to the beam energy &$\dfrac{50\%}{\sqrt{E}} \oplus 5\%$, ideally $\dfrac{35\%}{\sqrt{E}} \oplus 2\%$ & $\dfrac{3\mathrm{mrad}}{\sqrt{E}}$ & Acceptance: 60~cm $\times$ 60~cm \\
    \cline{3-5}
    & & \multicolumn{3}{p{90mm}|}{\small Note:}\\
    & & \multicolumn{3}{p{90mm}|}{\small The acceptance is required for meson structure measurements.}\\
    & & \multicolumn{3}{p{90mm}|}{\small Pion structure measurements may require a position resolution of 1 mm. }\\
    \hline
    \multirow{7}{*}{Photons}  & $0.1-1$~GeV &  $20-30\%$ & & Efficiency: $90-99\%$ \\
\cline{3-5}
& & \multicolumn{3}{p{90mm}|}{\small Note:}\\
& & \multicolumn{3}{p{90mm}|}{\small Used as a veto in $e$Pb exclusive J/$\psi$ production} \\
\cline{2-5}
        & $20-40$~GeV & $\dfrac{35\%}{\sqrt{E}}$ & 0.5--1~mm& \\
        \cline{3-5}
        & & \multicolumn{3}{p{90mm}|}{\small Note:}\\
        & & \multicolumn{3}{p{90mm}|}{\small u-channel exclusive electromagnetic $\pi^0$ production has a milder requirement of $\frac{45\%}{\sqrt{E}} \oplus 7\%$ and 2 cm, respectively. Events will have two photons, but single-photon tagging is also useful.}\\
         & & \multicolumn{3}{p{90mm}|}{\small  Kaon structure measurement requires tagging a neutron and 2 or 3 photons, as decay products of $\Lambda$ or $\Sigma$. }\\ 
    \hline
    \end{tabular}
\end{table*}

\begin{figure}[h!]
    \centering
    \includegraphics[width=0.9\linewidth]{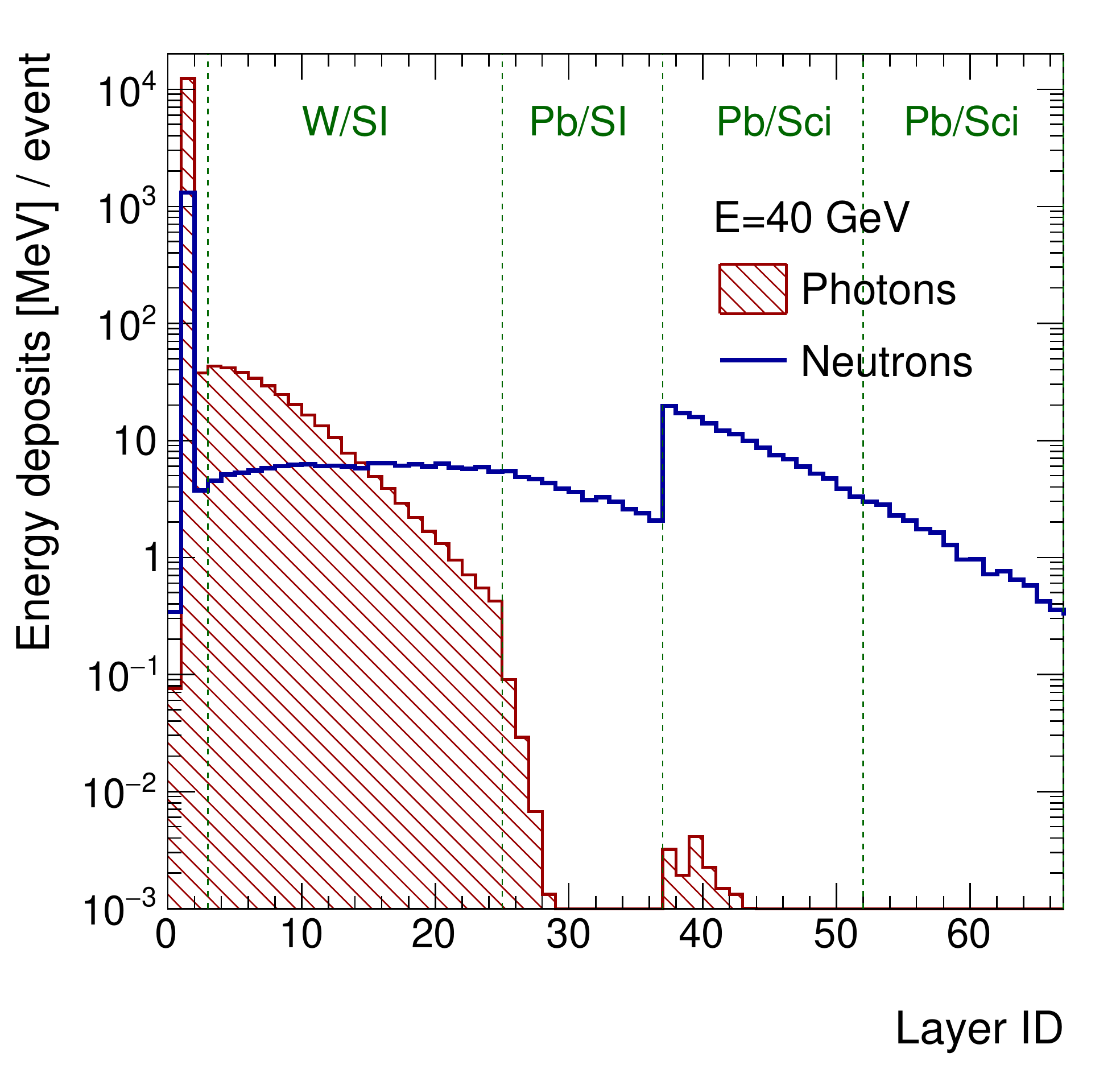}
    \caption{Simulated energy deposits in each layer of active materials of the ZDC, namely a silicon, crystal, or scintillator layer, shown for 40 GeV single photons and single neutrons. The first silicon layer has the layer ID = 0 and the next crystal layer has ID = 1. Other detector parts are indicated in the figure. The shown energy deposits are averaged values for an event, where a single photon or a neutron hits the ZDC by a particle gun.}
    \label{fig:zdc_energy_layer}
\end{figure}

\subsubsection{ZDC design}

The ZDC design is shown in Fig.~\ref{fig:far-forward} (bottom right). It consists of four different calorimeters. Particles come in from the left side of the figure. The detector consists of a 7 cm crystal layer (yellow) with a silicon pixel layer attached (magenta), 22 layers of Tungsten/Silicon planes (light purple) with an additional silicon pixel layer attached in front, 12 layers of Lead/Silicon planes (gray), and 30 layers of Lead/Scintillator planes (green), corresponding to the thickness of $8X_0$, $22X_0$, $2\lambda_I$, and $5 \lambda_I$, respectively. The energy deposition in each layer of active material in shown in Fig.~\ref{fig:zdc_energy_layer}. The total size is 60~cm$\times$60~cm$\times$162~cm and the weight is greater than 6 tons.

\textbf{Crystal calorimeter: } 
For good measurement of low-energy photons, the first part of ZDC is designed to use a layer of crystal calorimeter towers which is 7~cm in thickness. The layer consists of  $3 \times 3$~cm$^2$ crystals in an array of $20 \times 20$. $\mathrm{PbWO}_4$ is considered as the material choice for the crystal, but LYSO is another candidate as the radiation hardness of $\mathrm{PbWO}_4$ could be an issue. In front of the crystal layer, a silicon pixel layer, which has the same design as in the W/SI calorimeter, is attached. 

\begin{figure*}
    \centering
    \includegraphics[width=0.45\linewidth]{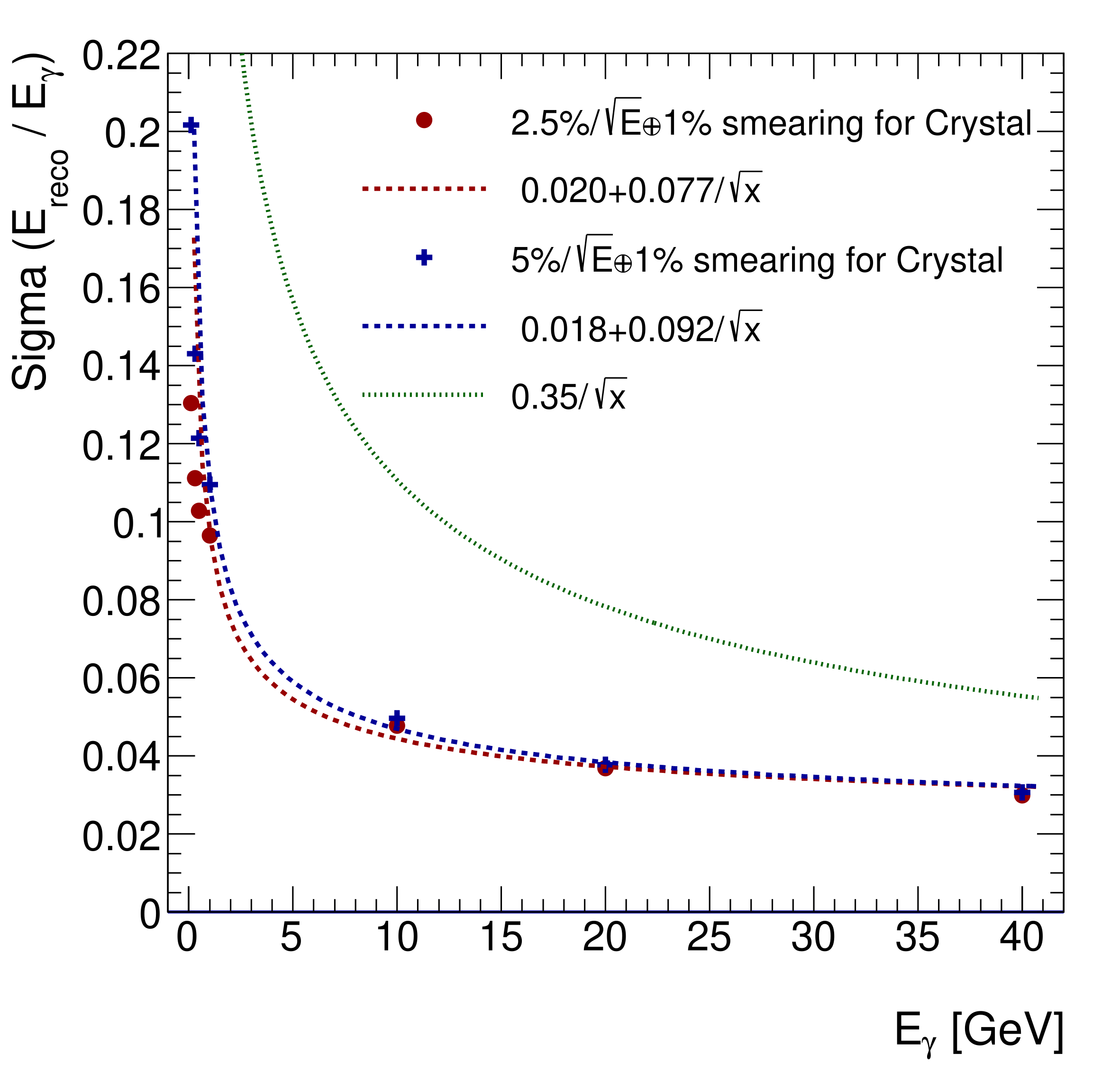} ~
    \includegraphics[width=0.45\linewidth]{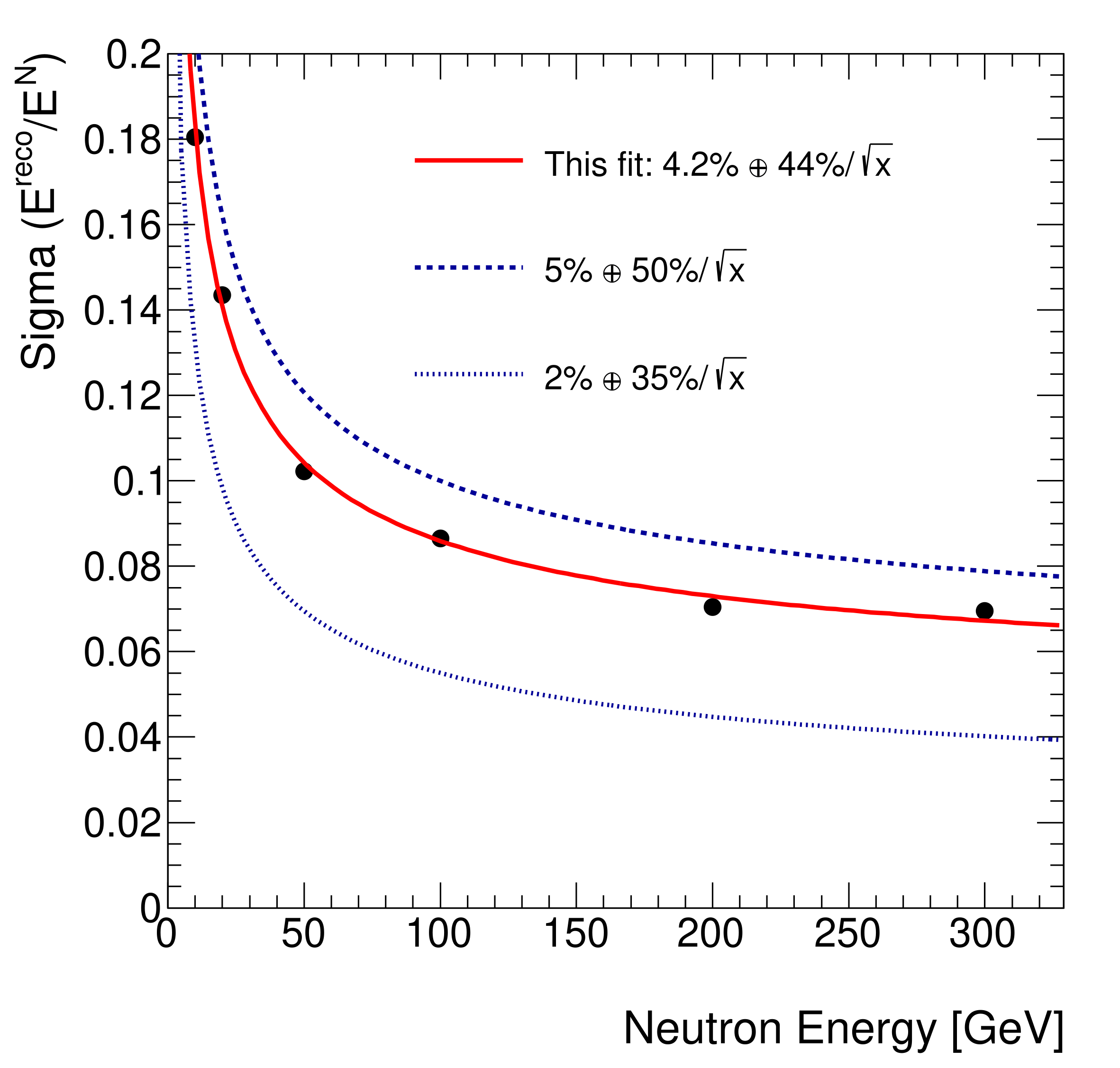}
    \caption{Estimation of the energy resolution for single photons (left) and single neutrons (right). The photon or neutron energy is reconstructed from the deposited energy in each active material. The readout system was not included in the simulation. The energy of the crystal layer is smeared by $2.5\%/\sqrt{E}\oplus1\%$ and is compared to $5\%/\sqrt{E}\oplus1\%$ smearing for the photon case. The estimated resolution is fitted as a function of the induced energy and compared to the physics requirements, $35\%/\sqrt{E}$ for photons and $50\%/\sqrt{E}\oplus5\%$ for neutrons.}
    \label{fig:zdc_eres_neutron}
\end{figure*}

\textbf{W/SI sampling calorimeter: }
This is an ALICE FoCal~\cite{ALICE-PUBLIC-2019-005} style calorimeter and consists of tungsten plates and silicon sensor planes placed one after the other. 
It will measure the rest of the photon energy and extract the shower development of photons and neutrons. 
The tungsten plates have 3.5~mm thickness ($\sim1X_0$) and the silicon sensor planes have a thickness of $300-320~\mu\mathrm{m}$. 
Two types of silicon sensors are considered. 
Pad sensors have $1 \times 1$~cm$^2$ segmentation, while pixel sensors have $3 \times 3$~mm$^2$. 
There are 22 tungsten layers and each of these layers is followed by a silicon pad layer except for the $11^{th}$ and $22^{nd}$ tungsten layers. 
For those tungsten layers, a silicon pixel layer is inserted instead of a pad layer. 
Another silicon pixel layer is attached in front of the first tungsten layer, for the photon position measurement. 
The W/SI calorimeter has 22 tungsten layers, 20 silicon pad layers, and 3 silicon pixel layers in total.

\textbf{Pb/SI sampling calorimeter: }
This is a calorimeter with 3~cm-thick lead planes as absorbers and silicon pad layers as active material, where the pad-layer design is as in the W/SI calorimeter.
The silicon layers (with good radiation hardness) are used for the measurement of the neutron shower development. It consists of 12 lead layers and 12 silicon pad layers.

\textbf{Pb/Sci sampling calorimeter: } 
This is to measure hadron shower energy and uses 3-cm-thick lead planes as absorbers with 2-mm-thick scintillator planes as active material. The calorimeter is segmented as $10 \times 10$~cm$^2$ on a plane and 15 layers of scintillator planes will be read out together, comprising a tower. The length of a tower is 48~cm. The Pb/Sci calorimeter has 6$\times$6 towers in the transverse and two towers in the longitudinal direction. In total, it consists of 30 layers of lead planes and 30 layers of scintillator planes. 

\subsubsection{Simulated performance study}
The performance of the designed ZDC was studied using the Geant4 simulation~\cite{Agostinelli:2002hh}. 
In the simulation, a single photon or a neutron is shot at the center of the ZDC plane. The readout system is not implemented in the simulation but the deposited energy in the active materials is studied. The materials for the readout system were not fully implemented for the crystals and the scintillator layers.\footnote{For each layer of silicon plane, a readout board with chips is inserted.} Empty spaces were used to represent the readout planes, thus, the study provides an optimistic estimation. 

Fig.~\ref{fig:zdc_energy_layer} shows the deposited energy in each layer of ZDC active materials for photons and neutrons with an energy of 40~GeV. It shows a clear difference in the ZDC response against photons and neutrons. Photons deposit more energy in the crystal layer and early layers in the W/Si calorimeter, while neutrons continuously deposit their energy to the scintillator layers, owing to the difference in their shower development.

The photon energy is reconstructed from the crystal layer and the W/SI calorimeter.  In the crystal, a tower with $E>15$~MeV is taken as a seed and $3\times3$ towers build a cluster.  The crystal energy is smeared by $2.5\%/\sqrt{E} \oplus 1\%$ (note that $5\%/\sqrt{E} \oplus 1\%$ was also studied). In the resolution that follows (and throughout this paper), $\sqrt{E}$ is taken to be in units of GeV. The energy in the W/SI calorimeter is reconstructed from a $9 \times 9$~cm$^2$ region of interest (RoI), with a scale factor corresponding to the sampling fraction.  The neutron energy is reconstructed from all the crystal, W/SI, Pb/SI, and Pb/Sci calorimeters.  The W/SI, Pb/SI, and Pb/Sci calorimeters need scale factors in order to convert the energy deposits in the active material to the reconstructed energy, as corrections for the sampling fraction and the $e/h$ compensation.  For extraction of the factors, the crystal calorimeter is taken out from the simulation, and neutrons are shot directly on the sampling calorimeters.  In this setup, the factors are determined by fitting the following function:  \begin{displaymath}
E_N= a\cdot E_\text{SI in W/SI} + b\cdot E_\text{SI in Pb/SI} + c\cdot E_\text{Sci. in Pb/Sci}, 
\end{displaymath} 
where $a$, $b$, and $c$ are the scale factors, performed for $E_N=$20, 40, 60, 80, and 100 GeV. 

The estimated energy resolution is shown in Fig.~\ref{fig:zdc_eres_neutron}. For high-energy photons, the resolution is well below the requirement stated in the YR. For the low energy photons, the estimated resolution for 100~MeV photons using 5\% smearing reaches 20\% but is still acceptable. The neutron energy resolution is larger than the ideal value of $35\%/\sqrt{E}\oplus 2\%$, but is smaller than the required value of $50\%/\sqrt{E}\oplus5\%$.

Position reconstruction is accomplished using the first silicon pixel layer after the crystal calorimeter. 
For 40~GeV and 20~GeV photons, the position resolution is estimated as 1.1~mm and 1.5~mm, respectively.
On the crystal layer, the cluster finding efficiency is $>95\%$ for both 20 GeV photons and 100 MeV photons, with a seed energy requirement of 15~MeV for the clustering. 

Although the simulation results are optimistic without the readout system's geometry and materials, the results show a reasonable performance of the ZDC, which practically fulfills the physics requirements listed in Table~\ref{tab:zdc_phys_req}.

\subsection{Roman Pots}
%\subsection{Roman Pots}

The LHC forward-proton detectors have shown the capability of thin silicon detectors to deliver both excellent precisions in position and timing with pixelated detectors~\cite{Royon:2014wkw,Albrow:2015ois}. 
 
\begin{figure}[h]
      \centering
      \includegraphics[width=0.8\linewidth]{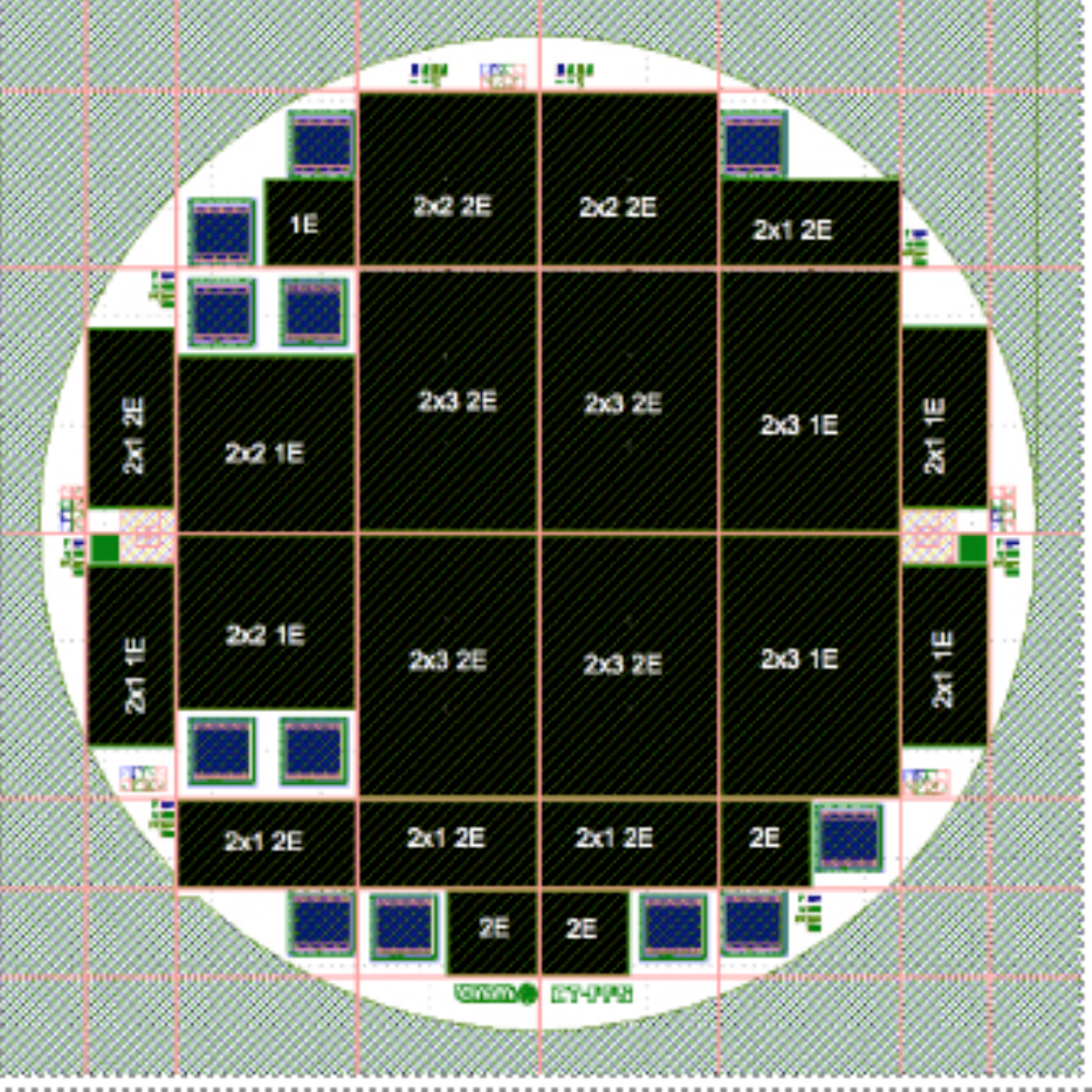}
      \caption{Layout of the CMS Roman Pot silicon sensors \cite{Ravera:2016odg}.}
      \label{fig:CTPPS_Sensors}
  \end{figure}

 The Roman Pots (RP) envisioned for ECCE largely follow the concept outlined in the YR, namely the use of AC-LGADs to provide both precise timing and excellent position resolution. The sensor will be laid out in a grid pattern.  Fig.~\ref{fig:CTPPS_Sensors} shows an example of such a layout from CMS. 
 
\begin{table*}[t]
    \centering
    \begin{tabular}{l|c|c}
    \textbf{Parameter} & \multicolumn{2}{c}{\textbf{Interaction Point/Region}} \\ 
    & IP6 & IP8 \\\hline
    Beam crossing angle & 25 mrad & 35 mrad \\
    Outer radius of B0 detector & 19 cm & 23.5 cm \\
    Spanning angle Packman & 240 deg & 240 deg \\
    Detector cut off for hadron beam pipe, tracker & 3.5 $\times$ 9.5 cm & 3.5 $\times$ 10.5 cm \\
    Detector cut off for hadron beam pipe, calorimeter & 3.5 $\times$ 10.0 cm & 3.5 $\times$ 11.2 cm \\
    Pipe hole offset in x-axis w.r.t. the center of the B0 magnet& -1.0 cm & -1.4 cm \\
    `PAC-man' cut off for electron beam pipe, radius difference & 7 cm & 7 cm\\
    Si layer thickness & 0.1 cm & 0.1 cm\\
    Dead material (Cu) thickness & 0.2 cm & 0.2 cm \\
    B0 EM section ($\rm PbW0_4$) thickness & 10 cm & 10 cm\\
    B0 EM section z-position (relative to the B0-magnet) & 48 cm & 48 cm\\
    \end{tabular}
    \caption{Shape parameters of the B0 detector. These are derived using specifications provided in Ref.~\cite{IP6design}.}
    \label{tab:B0}
\end{table*}

It is essential that such detectors be temperature stabilized. This can be accomplished by using a cooled heat sink to pull heat off the detector via a copper bus. We propose using a foam metal heat sink that will be cooled via compressed air. Such systems have already been deployed at the LHC by a group from  the  Technical University of Prague. The timing and resolution of the RP layers are similar to the expected values of the B0 tracker (identical in technology).

\subsection{B0 magnet detector stack}

\begin{figure*}[t!]
    \centering
    \includegraphics[width=0.4\linewidth]{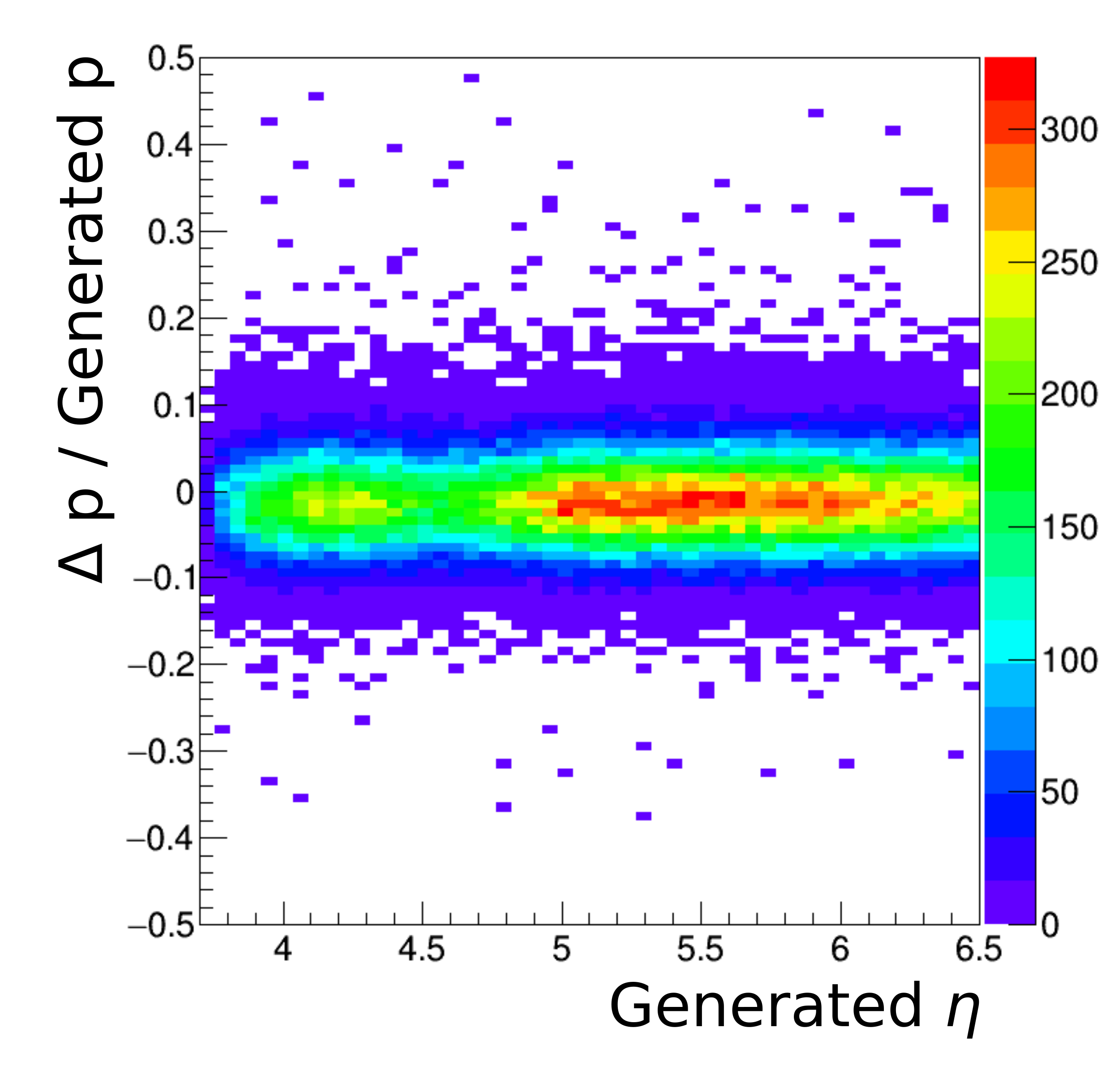} ~~
    \includegraphics[width=0.4\linewidth]{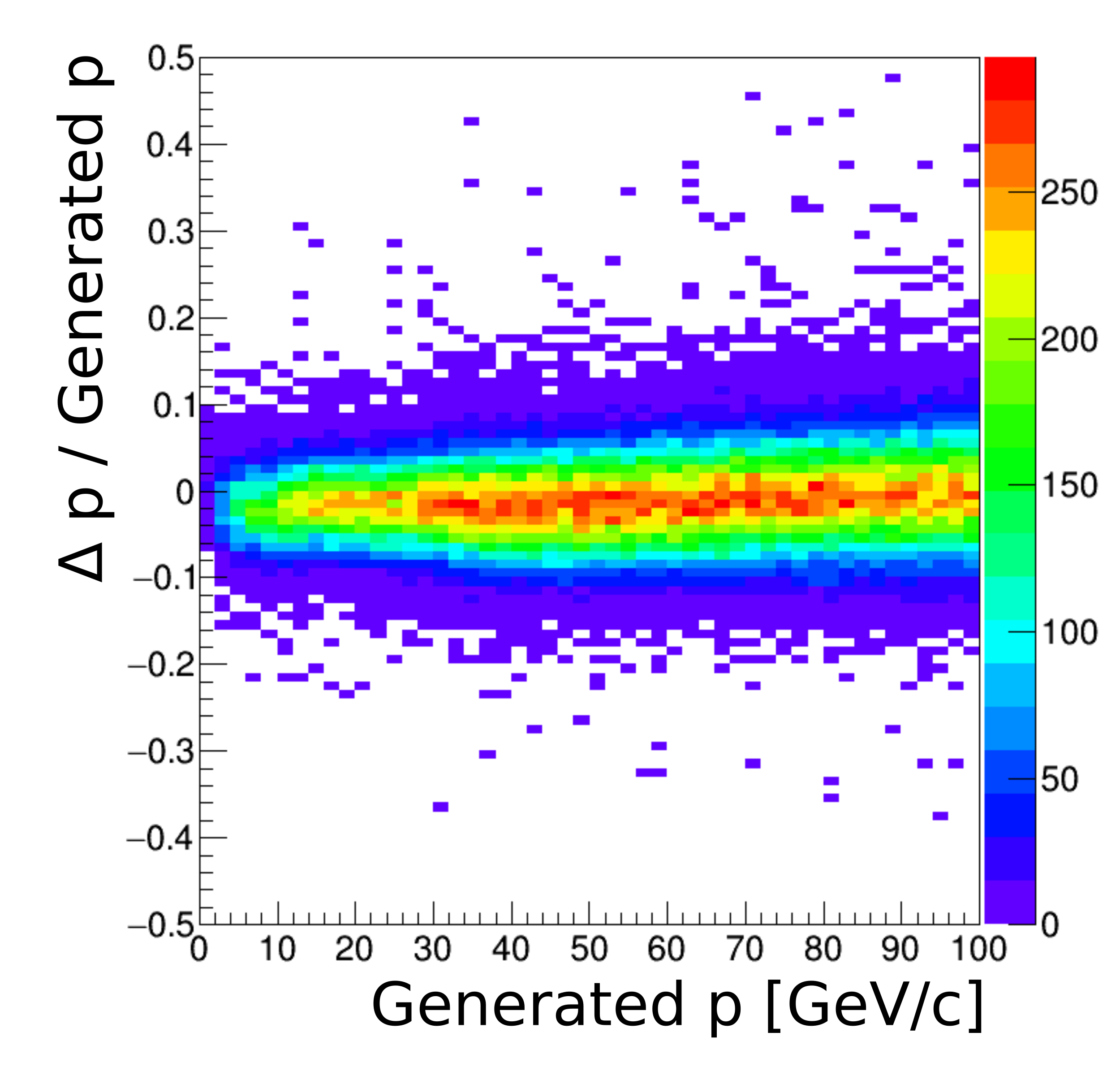}
    \caption{B0 tracking resolution. Difference between reconstructed and true momentum scaled by the true momentum as a function of $\eta$ (top) and generated momentum (bottom).
    }
    \label{fig:B0_scaled_momentum_difference}
%\end{figure*}
%\begin{figure*}[t]
    \centering
    \includegraphics[width=0.385\linewidth]{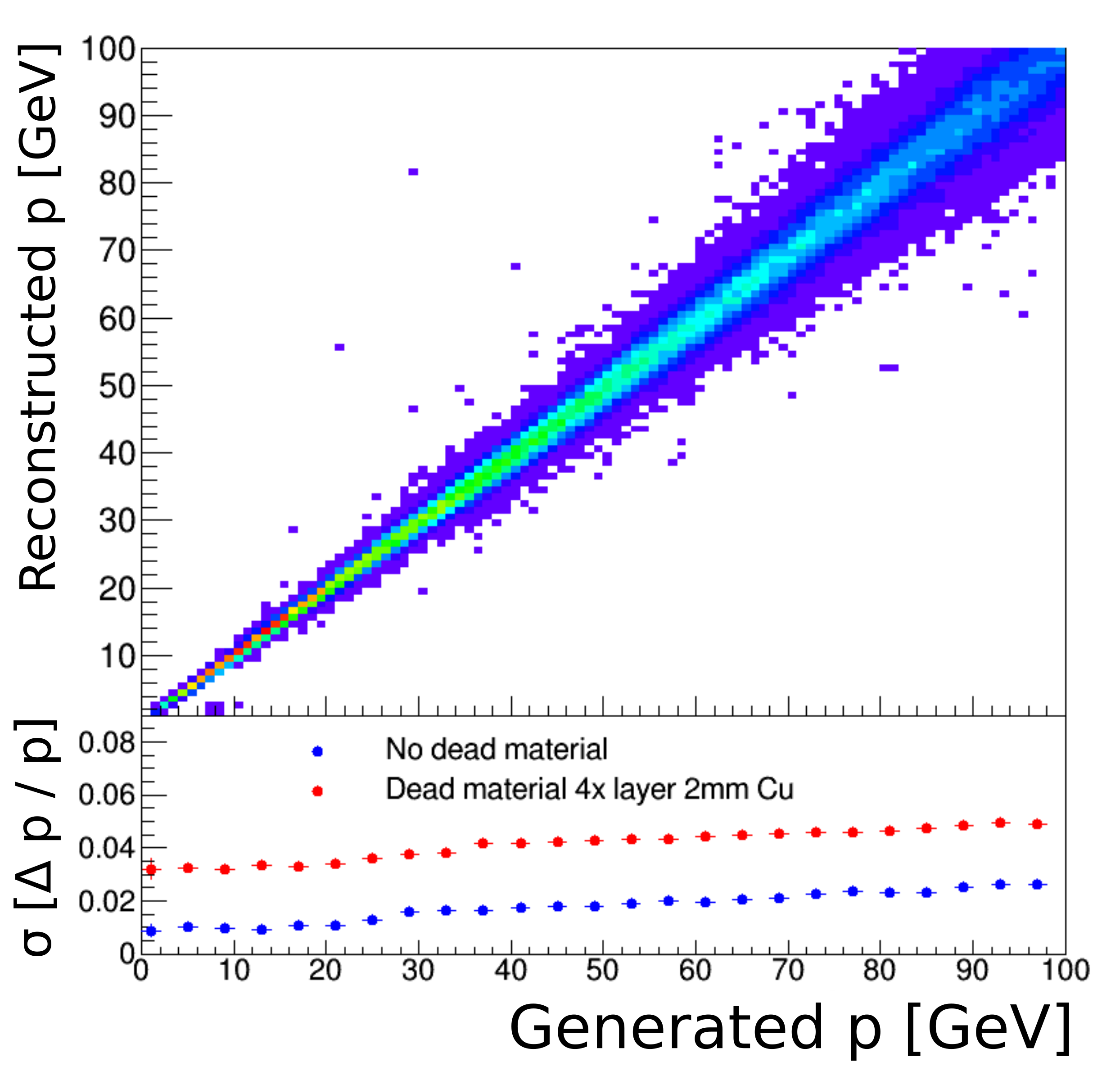} ~~~
    \includegraphics[width=0.4\linewidth]{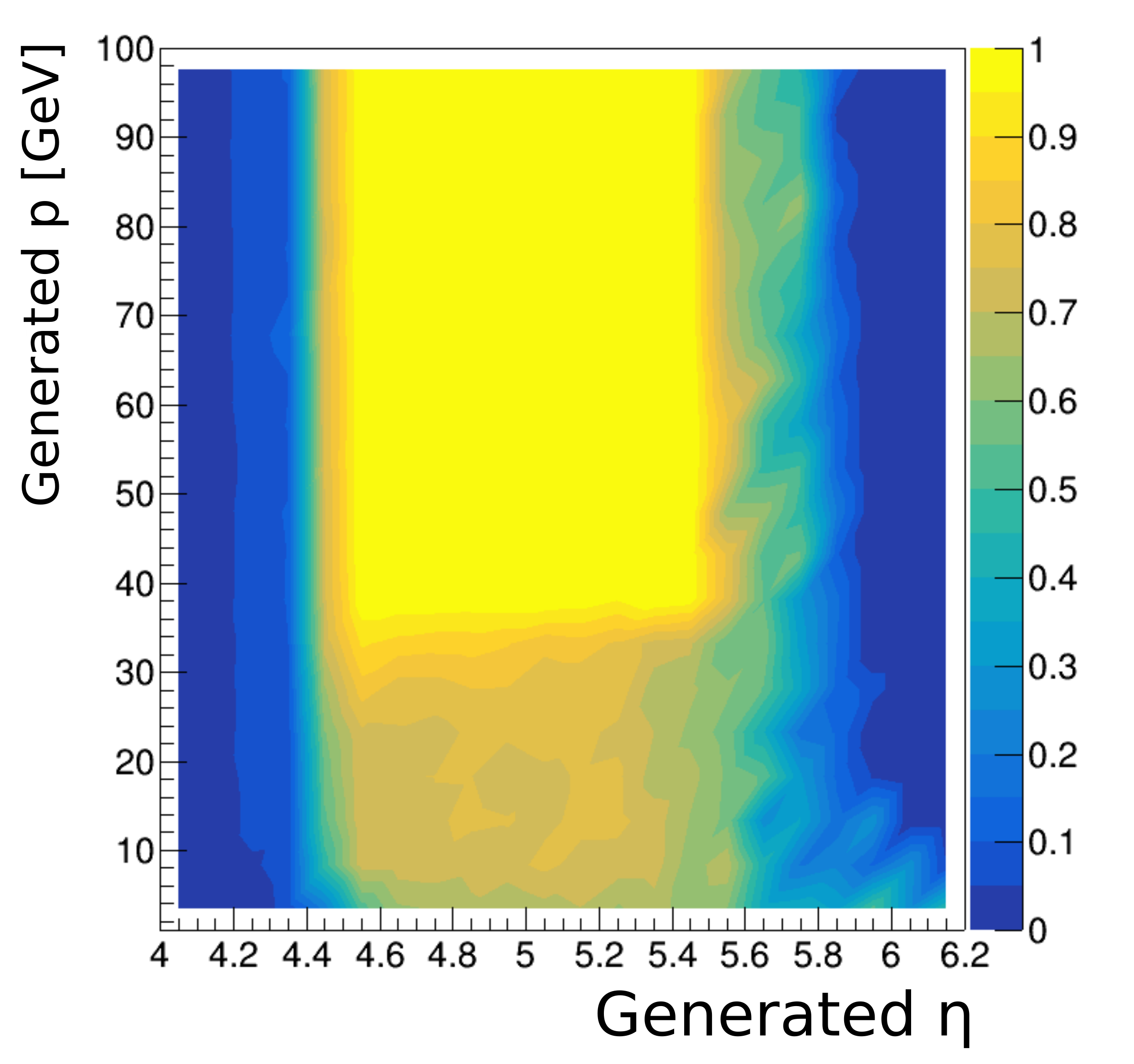}
    \caption{Reconstructed momentum and its resolution for $\mu^{-}$ tracks found in the B0 tracker (top). Acceptance of the B0 tracker (bottom). A particle gun was used for this study.} 
    \label{fig:B0_sigma}
\end{figure*}

The tracker and calorimeter stack inside of the B0 magnet provide detection capability for far-forward charged tracks and photons. Such capability is important for forward ($\eta > 3$) particle measurements as well as event characterization and separation. 

The B0 spectrometer is located inside the B0pf dipole magnet. Its main use is to measure forward-going hadrons and photons to identify exclusive reactions. The B0 acceptance is defined by the B0pf magnet. Its design is challenging due to the two beam pipes (electron and hadron) that must be accommodated and the fact that these pipes are not parallel to each other, due to the 25~mrad IP6 crossing angle. Moreover, service access to the detectors inside of the dipole is only possible from the IP side, where the distance between beam pipes is the narrowest. To satisfy these constraints, the B0 detector design requires the use of compact and efficient detection technologies. 

The B0 detector stack design uses four AC-LGAD tracker layers with 30 cm spacing between each layer (top left Fig.~\ref{fig:far-forward} in yellow). These will provide charged particle detection for $6<\theta<22.5$~mrad. The use of such sensors will provide good position and timing resolutions.  The AC-LGAD sensors will have a $3.2 \times 3.2$~cm$^2$ area, with four dedicated ASIC units on each sensor. In addition, a PbWO$_4$ calorimeter (Fig.~\ref{fig:far-forward} top left in magenta)  will be positioned behind the fourth tracking layer 683~cm away from the IP. The calorimeter is constructed from 10 cm long $3 \times 2$~cm$^2$ PbWO$_4$ crystals positioned to leave 7~cm for the detector and readout system (before the B0 magnet exit). In order to consume less space inside the magnet, the processing of the signals from the detector will be performed outside the magnet volume. Both trackers and the calorimeter have oval holes in the center to accommodate the hadron beam pipe, and a cutaway on the side to accommodate the electron beam and allow installation and service of the detector system. An additional circular cutoff (with 2~cm radius) on the side opposite the electron beam pipe is assumed for cabling in each detector plane.

The parameters of the B0 detector are summarized in Table~\ref{tab:B0} for the two IPs.
%for IP6 as an example.
To help visualize the trackers and calorimeter layout within the compact B0 magnet, CAD drawings (in realistic dimensions) are documented in \ref{app:B0}.

\begin{figure}[h!]
    \centering 
    \includegraphics[width=0.495\linewidth]{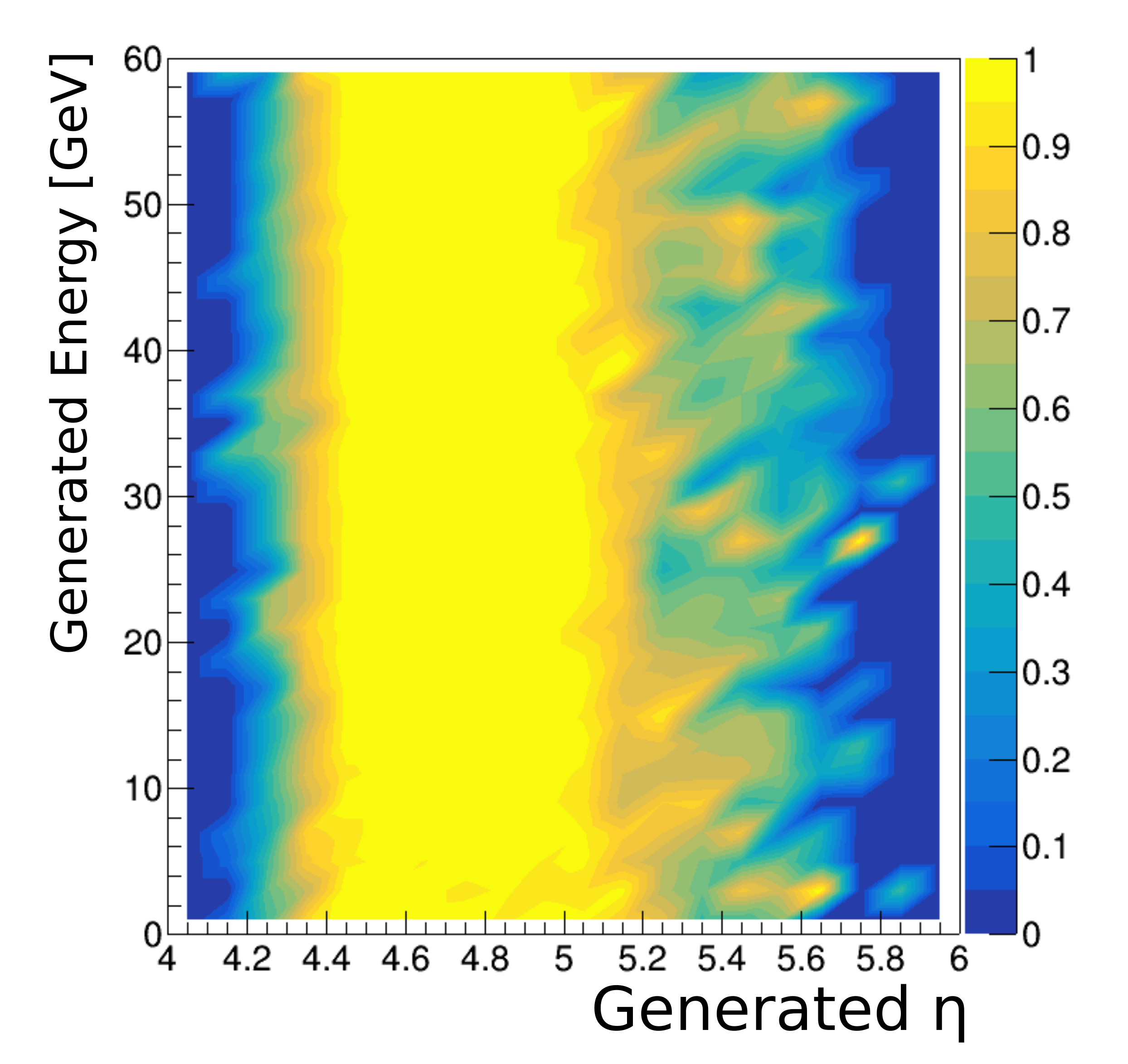}
    \includegraphics[width=0.495\linewidth]{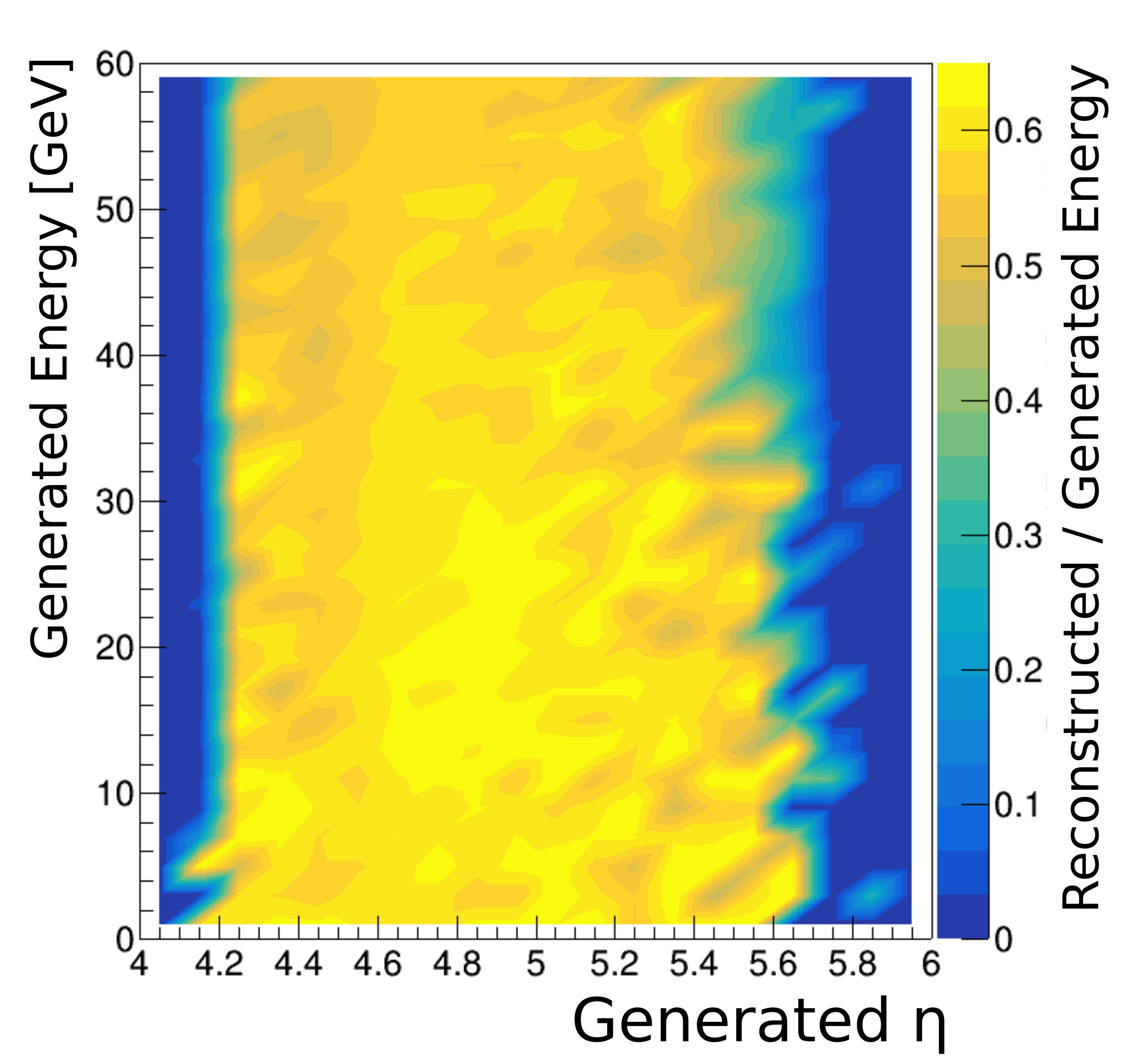}
    \caption{The left plot shows the photon detection acceptance in the B0 EM calorimeter; the right plot is the ratio between the reconstructed and generated photon energies.}
    \label{fig:B0_ECAL}
\end{figure}

\subsubsection{Track Reconstruction in the B0 Calorimeter}

Reconstructing tracks requires an accurate understanding of the magnetic field in the B0 magnet. The field map implemented in the simulation combined the field map of the central detector (1.4~T) with that of the B0 dipole magnet (1.18~T). A Kalman filter was used to reconstruct the track momentum of generated $\mu^{-}$ in the momentum range $1 < p < 100$~GeV, using the reconstructed hits in the tracking layers and this field. 
Fig.~\ref{fig:B0_scaled_momentum_difference} shows the
difference between the reconstructed and true momentum of the track, scaled by its true momentum as a function of $\eta$ (top) and generated momentum (bottom). 
This difference was found to be uniform as a function of pseudorapidity and increasing slightly with the momentum of the generated particle, and staying below 2\% for the studied kinematic region.

The simulated momentum and its resolution $\sigma[\Delta p/ p]$ are shown in Fig.~\ref{fig:B0_sigma} (top), as a function of the truth momentum; the momentum resolution is less than 5\% for the studied kinematic region. The effect of the presence of dead material (2~mm of Cu after each Si plane) on the momentum resolution is also shown and estimated to degrade the resolution by 2\% uniformly as a function of $p$. Fig.~\ref{fig:B0_sigma} (bottom) also shows the acceptance of the B0 tracker in the pseudorapidity-momentum plane.

\subsubsection{Photon Reconstruction in the B0 Calorimeter}
The studies of the efficiency of photon detection with the B0 electromagnetic calorimeter have been performed for photons going from the interaction vertex in the forward direction in the pseudorapidity range $4 < \eta < 6$ and having energy $0 < E_{\gamma} < 60$~GeV. The granularity of the crystals of the B0 EM section was assumed to be $2 \times 2$~cm$^2$. 

The photon reconstruction algorithm search is based on a matrix of $2 \times 2$ crystals. Other algorithms, for example, based on a Swiss-cross pattern, are being considered and require further study.

The acceptance of the calorimeter in the $\eta-E_{\gamma}$ plane and the average ratio of the reconstructed to generated energy are shown in Fig.~\ref{fig:B0_ECAL} (left) and Fig.~\ref{fig:B0_ECAL} (right), respectively. In general, about 60\% of the energy is reconstructed within a 2x2 crystal grid.

\begin{figure}[h]
\centering
\includegraphics[width=0.78\columnwidth]{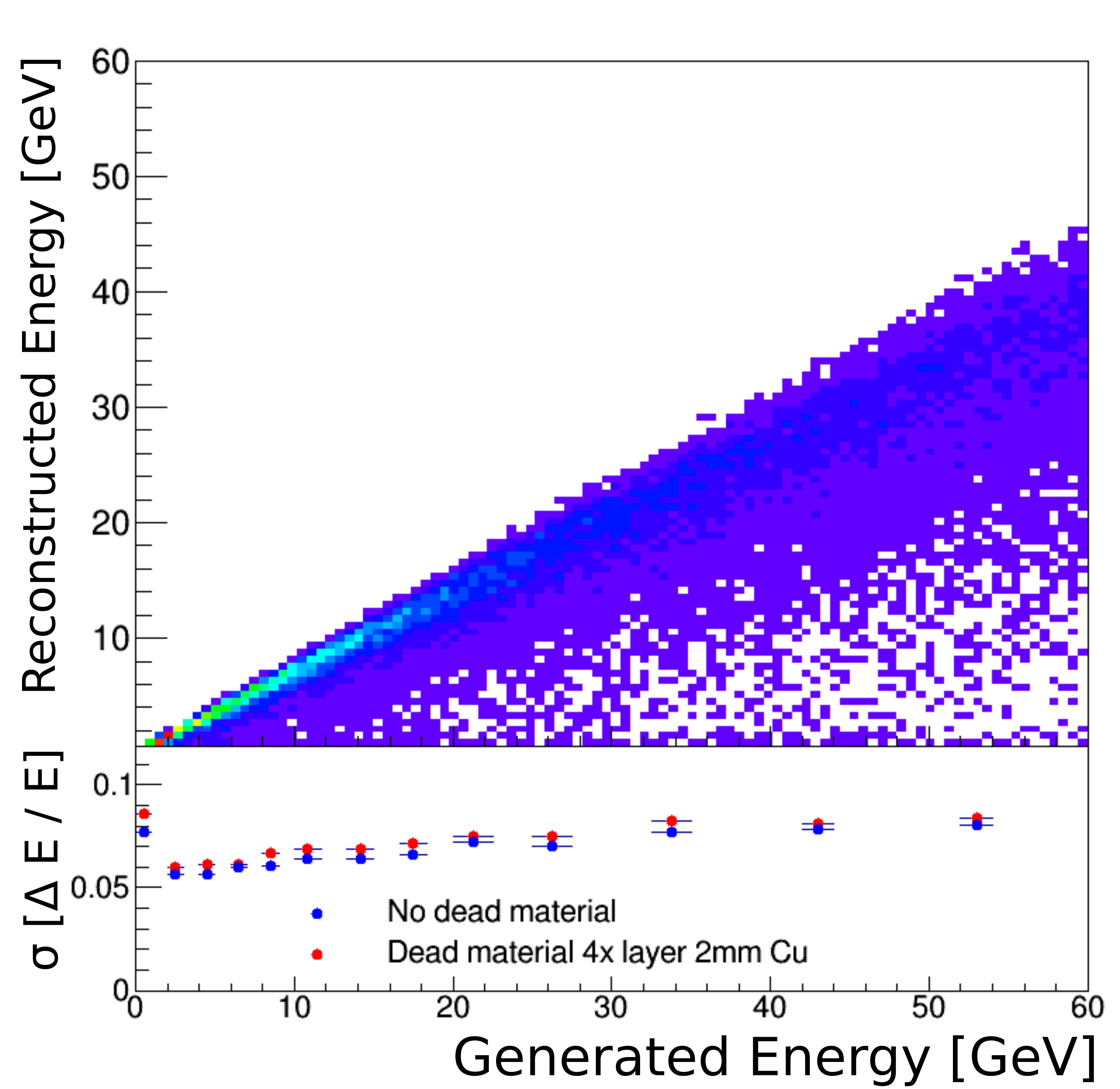}\\
\includegraphics[width=0.78\columnwidth]{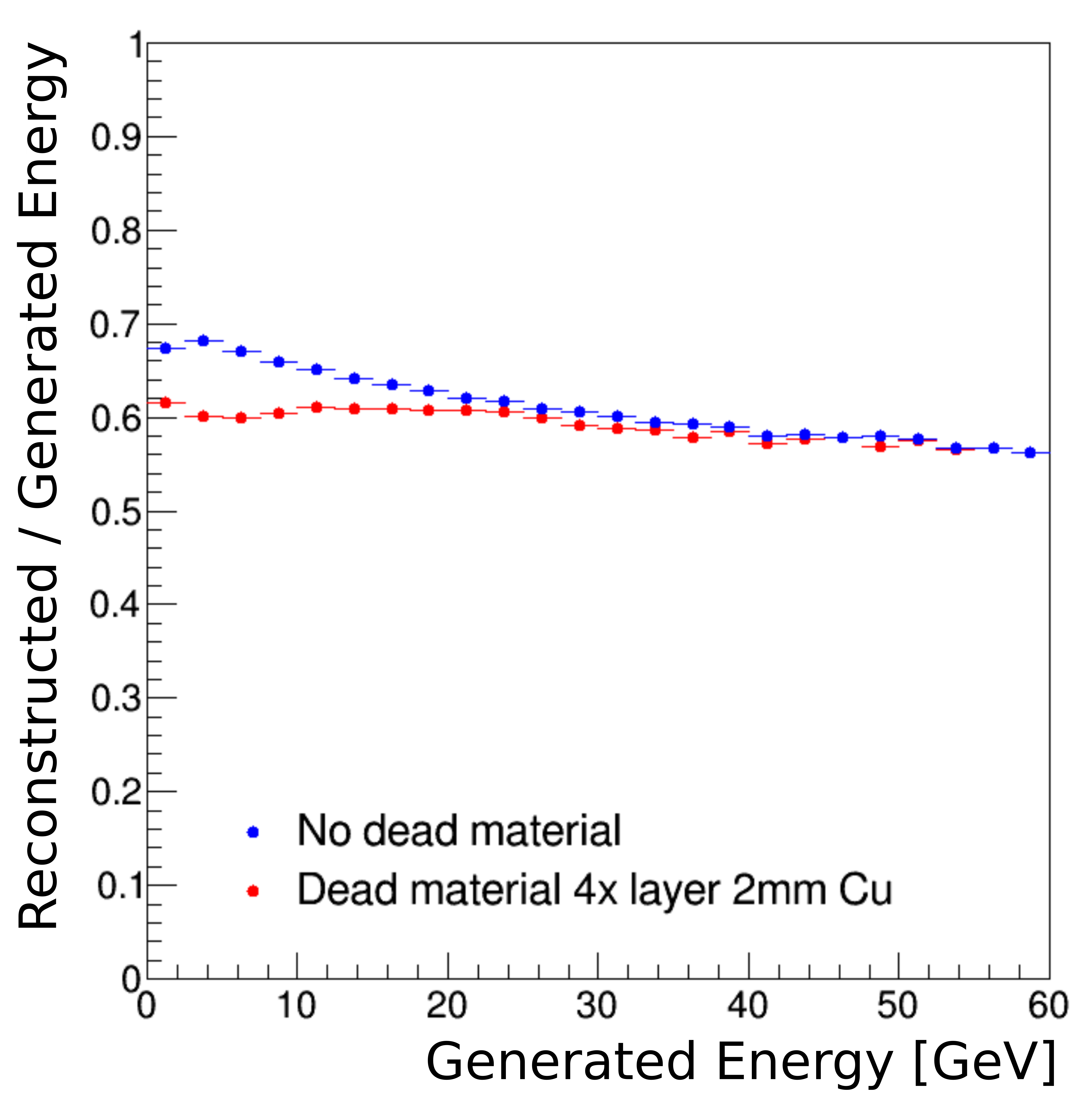}
\caption{(Top) Reconstructed energy of photons and its resolution in the B0 calorimeter; (bottom) effect of the presence of dead material layers in the B0 tracker on the efficiency of photon reconstruction with the B0 calorimeter.}
\label{fig:B0_ECAL_sigma}
\end{figure}

A scatter plot of the reconstructed versus generated photon energy together with the energy resolution is shown in Fig.~\ref{fig:B0_ECAL_sigma} (top). The resolution is found to be below 7\% for the studied kinematic region.
The fraction  of photon energy that is  reconstructed within the B0 calorimeter as a function of  photon energy $E_{\gamma}$ is portrayed in Fig.~\ref{fig:B0_ECAL_sigma} (bottom). The effect of dead material layers (the 2~mm of Cu after each silicon tracking plane) on the efficiency of photon reconstruction with the B0 calorimeter is also shown and does not exceed 10\%.

%-----------------------------------------
\section {Simulation, reconstruction and analysis framework}
\label{sec:sim_framework}

\begin{figure}
    \centering
    \includegraphics[width=0.85\linewidth]{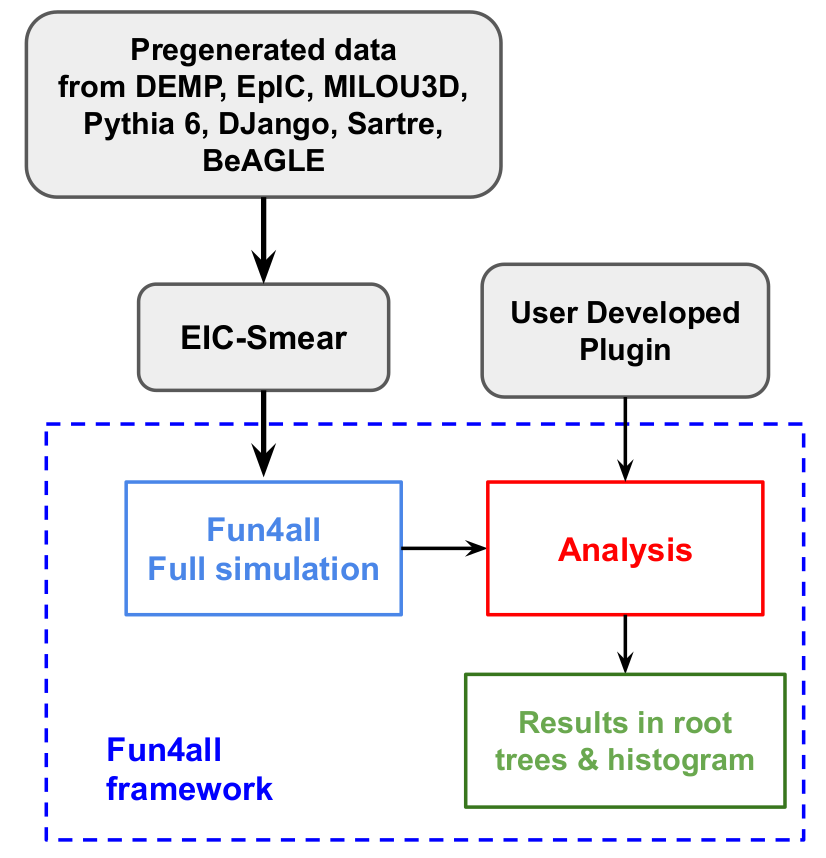}
    \caption{Simulation workflow, from the generated input from a variety of physics event generators to the data output in the format of root trees and histograms.}
    \label{fig:sim_flow}
\end{figure}

The ECCE proto-collaboration made a conservative decision to utilize developed, supported, and established software tools to support the proposal writing process in 2021. The primary consideration was the condensed proposal writing timeline, as several data production campaigns would be necessary to allow the physics and detector working groups to analyze data as well as exercise the full simulation production system. Under such context, the Fun4All software framework was chosen to perform Geant4 simulations~\cite{fun4all}.  

Fun4All is an integrated simulation, reconstruction, and analysis framework. Fun4All is an actively developed event processing framework that was originally written for the PHENIX experiment~\cite{sPHENIX_software}. In 2015, the framework was moved to an open-source project and is now used by the sPHENIX and SpinQuest~\cite{SpinQuest} experiments. As the EIC-related activities increased towards the proposal, a significant amount of software infrastructure was created to support EIC-related studies prior to proto-collaboration formation, such as the various Fun4All related repositories in Ref.~\cite{EIC_fun4all}. This, and ongoing Fun4All software development, was the basis for the studies that were performed to develop the ECCE proposal.

A workflow diagram for using the Fun4All is shown in Fig.~\ref{fig:sim_flow}. As the input to the simulation framework, the users need to generate physics event samples with the generators (a few example generators are shown in the top grey boxes). The fast simulation tool: eic-smear, is used to convert the generated event data into ROOT trees or HepMC2 format, without modifying the underlying event data. The users are also required to write their individual analysis modules to interpret the simulation output, which takes the form of analysis plugins within the Fun4All framework. The beam effects are handled within the Fun4All framework and are further explained in Sec.~\ref{sec:beam_effect}.

The Fun4All framework (enclosed in the blue rectangle) is based upon the Fun4AllServer, which can handle a variety of inputs, reconstruction modules, and outputs. The modularity of the framework allowed users in the detector and physics working groups to develop the relevant code asynchronously, while the computing and simulation teams were then responsible for quality assurance and code integration for deployment in large-scale productions. In this design, various calibration and analysis modules were developed as part of the \textit{coresoftware}\footnotemark[1], \textit{fun4all eicdetectors}\footnotemark[2], \textit{ecce-detectors}\footnotemark[3], and \textit{calibration}\footnotemark[4] repositories. These modules were then aggregated in a series of ROOT macros that were steered by one top macro. The top-most macro defined the event generation, the geometry of the detector, input or output, and anything else that might be relevant for the job. This ran as a standalone ROOT macro to produce the data summary tapes (DSTs) and eventual micro DST data that the physics and detector working groups analyzed as a part of the larger simulation campaigns.
\footnotetext[1]{\url{https://github.com/eic/fun4all_coresoftware}}
\footnotetext[2]{\url{https://github.com/eic/fun4all_eicdetectors}}
\footnotetext[3]{\url{https://github.com/ECCE-EIC/ecce-detectors}}
\footnotetext[4]{\url{https://github.com/ECCE-EIC/calibrations}}

\subsection{Beam parameters}
\label{sec:beam_parameter}

\begin{figure}[t!]
    \centering
    \includegraphics[width=0.92\linewidth]{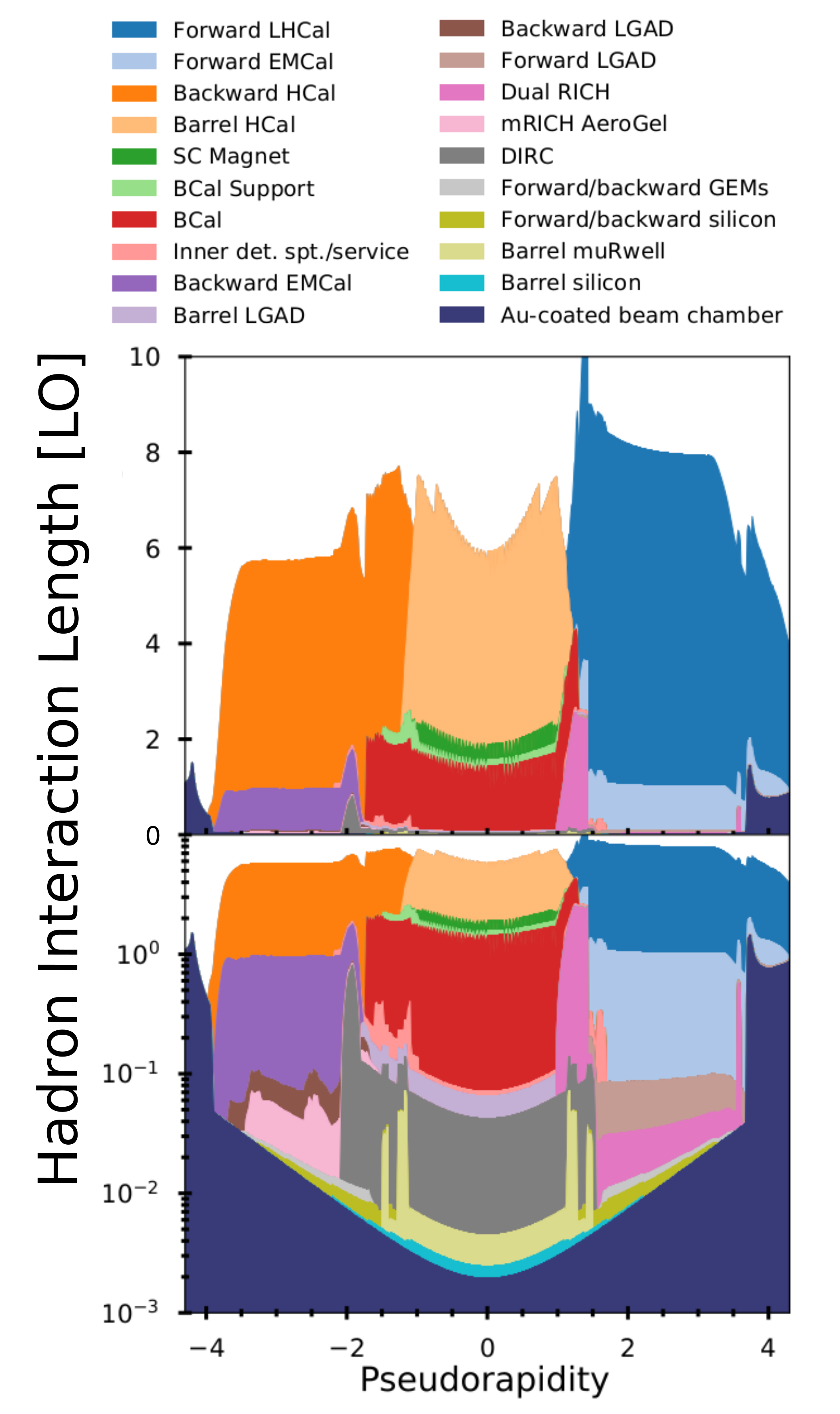}
    \caption{Material scans of the July-concept detector with build tag prop.4. (Top) The radiation length, (bottom) the hadronic interaction length. Note: this setup/study is made prior to the final ECCE detector configuration in the proposal.}
    \label{fig:material}
\end{figure}

To fulfill the physics requirements (see Sec.~\ref{sec:physics}), the EIC accelerator and detector design must enable the detection of scattered protons with a minimum transverse momentum of $p_T = 200$~MeV, which at a hadron beam energy of 275 GeV corresponds to a scattering angle of 730 $\mu$rad in the horizontal plane. The RMS divergence of the proton beam at the IP must not exceed one-tenth of this minimum scattering angle: $\sigma_0 \le 73$~$\mu$rad. This requirement may be violated in the vertical plane, provided the beam divergence in the horizontal plane meets the requirement. A smaller horizontal RMS beam divergence of 56 $\mu$rad allows the detection of 50\% of all scattered protons with a transverse momentum of 200 MeV. For $\sim 10\%$ of the operation time, the EIC will run with a large horizontal beta function at IP, $\beta^*_x$ (related to the transverse beam size at the IP),  that results in this low divergence and thus provides high acceptance at the expense of reduced luminosity; this beam configuration is referred to as the \textbf{high-acceptance} configuration. Because of the large cross-section for small $p_T$, a large amount of data can be collected in a short amount of time. For about 90\% of the time, the EIC will operate at small $\beta^*_x$ for high-luminosity but with a divergence angle exceeding 73 $\mu$rad and this is referred to as the \textbf{high-divergence} configuration. Combining the high-acceptance configuration running (with higher cross-section at lower $p_T$) for a shorter time, with the high-acceptance configuration running (smaller cross-section at higher $p_T$) for a longer run time, a comparable amount of data (between both settings) can be collected at all $p_T$ values from 200 MeV to 1.3 GeV. This scenario substantially increases the effective luminosity of the facility~\cite{EIC_CDR}.

The beam parameters for electron-proton collisions (including the resulting luminosities) at different center-of-mass energies ($\sqrt{s}$) for high-divergence and high-acceptance are listed in Tables 3.3 and 3.4 of Ref.~\cite{EIC_CDR}; the beam parameters for $e$Au collisions (fully stripped gold ions with A=197) are listed in Table 3.5 of Ref.~\cite{EIC_CDR}. All three sets of beam parameters/configurations were implemented and used for the full simulation during physics studies.

\subsection{Applying beam effects on physics data}
\label{sec:beam_effect}

The beam effects are introduced via a generator-agnostic after-burner, which has been integrated into the ECCE software setup since early 2021.  Being the standard procedure to take beam effects into account in the ECCE software, the after-burner implements the beam effects on final-state particles on an event-by-event basis based on the choice of beam configuration, such as high-divergence, high-acceptance, or $e$A scattering.

The beam-parameter after-burner first boosts the generated physics events horizontally, from the head-on frame, towards the beam crossing direction. The amplitude of the boost is $\tan (\theta_{CA}/2)$, ignoring the beam divergence and crab-cavity kick. Here, $\theta_{CA}=25$~mrad, which is the crossing angle at IP6.  In the presence of these variations, the final boost direction and amplitude are chosen according to the final angle between the two beams in the lab frame. In the last step, a simple rotation of $\theta_{CA}/2$ around the vertical axis in the lab coordinate system aligns the electron beam back to the $-z$ axis, which leaves the proton beam with the intended crossing angle of $\theta_{CA}$. A more detailed discussion on the beam effects in EIC simulation is summarized in a technical note~\cite{burner}.

\begin{table*}[t]
    \centering
    \caption{Summary ECCE Exclusive, Diffractive and Tagging physics studies with full simulations. Associated physics event generators and physics objectives of individual topics are given.}
    \label{tab:physics}
    \begin{tabular}{l|c|c|l}
    \textbf{Physics Impact Study Topics} & \textbf{Subsection}          & \textbf{Physics Objective} & \textbf{Event Generator}             \\\hline   
    Pion Form Factor                    & Sec.~\ref{subsec:piff}       & \#4                    & DEMPGen~\cite{url:DEMPGen}          \\
    $\pi$ Structure Function            & Sec.~\ref{subsec:pi_sf}      & \#4                    & EIC\_mesonMC~\cite{url:EIC_mesonMC} \\
    Double Tagged e-He3                 & Sec.~\ref{subsec:double_tag} & \#1                    & DJANGOH ~\cite{CHARCHULA199438}     \\
    ep DVCS                             & Sec.~\ref{subsec:dvcs}       & \#2                    & MILOU3D~\cite{url:MILOU3D, PEREZ04} \\
    eA DVCS via e-He4                   & Sec.~\ref{subsec:dvcsea}     & \#3                    & TOPEG ~\cite{TOPEG}                 \\    
    ep DEMP $J/\psi$                    & Sec.~\ref{sec:DVMPep}        & \#4                    & LAGER ~\cite{lager}                 \\
    TCS                                 & Sec.~\ref{subsec:tcs}        & \#2                    & EpIC ~\cite{EPIC}                   \\
    XYZ Spectroscopy                    & Sec.~\ref{sec:spec_overview} & \#5                    & elSpectro ~\cite{elSpectro}         \\
    \end{tabular}
\end{table*}

\subsection{Simulation campaign status}

Four detector concepts were assembled in the ECCE simulation, one for each simulation campaign. The information and overall simulation status are documented in the wiki database\footnote{\url{https://wiki.bnl.gov/eicug/index.php/ECCE_Simulations_Working_Group}}. The corresponding software branch name for the simulation campaigns are given below.
\begin{enumerate} 
    \item First simulation campaign: June-Concept (2021), which is tagged with proposal software build \textbf{prop.2}.
    \item Second simulation campaign: July-Concept (2021), which is tagged with proposal software build \textbf{prop.4}.
    \item Third simulation campaign: October-Concept (2021) and a variation with an AI-optimized inner tracker, which is tagged with proposal software build \textbf{prop.5}.
    \item Fourth simulation campaign: January-Concept (2022) with the full beam configuration set, which is tagged with proposal software build \textbf{prop.7.1}.
\end{enumerate}
Each software build is developed under a branch at the GitHub repository\footnote{\url{https://github.com/ECCE-EIC/macros}}. The prop.4 simulation is the baseline for the ECCE detector proposal; the material profile as a function of $\eta$ is shown in Fig.~\ref{fig:material}.

\section {Physics impact studies}
\label{sec:physics}

The physics objectives derived from the National Academy of Sciences (NAS) questions (Sec.~\ref{sec:overview}) to the EIC project can be expressed as follows:
\begin{enumerate}
    \item Origin of nucleon spin.
    \item Three-Dimensional structure of nucleons and nuclei.
    \item Gluon structure of nuclei.
    \item Origin of hadron mass.
    \item Science beyond the NAS Report.
\end{enumerate}
See Table~\ref{tab:physics} for the full list of physics topics covered in this paper.

To achieve these objectives, ECCE conducted a variety of studies with \textbf{Exclusive}, \textbf{Diffractive} and \textbf{Tagging} processes utilizing the Fun4All simulation (Sec.~\ref{sec:sim_framework}). 
The said processes were categorized as exclusive electro- and photoproduction of mesons and photons, as well as $ep$ and $e$A vector meson production through a diffractive process.  One commonality among these processes is the requirement that a nucleon (or nucleus) be tagged by the far-forward instrumentation (see Sec.~\ref{sec:FF_detector}). It is important to note that fully reconstructing all final-state particles is experimentally challenging.  Detailed background studies are required in the future to better gauge the sensitivity required to complete the relevant studies under realistic experimental conditions.

\subsection{Pion form factor - $F_{\pi}$}
\label{subsec:piff}
The elastic electromagnetic form factor of the charged pion, $F_{\pi}(Q^2)$, is a rich source of insights into basic features of hadron structure, such as the roles played by confinement and Dynamical Chiral Symmetry Breaking (DCSB) in determining the size and mass of hadrons and defining the transition from the strong- to perturbative-QCD domains.  Studies during the last decade, based on JLab 6-GeV measurements, have generated confidence in the reliability of $\pi^+$ electroproduction as a tool for pion form factor extractions. Forthcoming measurements at the 12-GeV JLab will deliver pion form factor data that are anticipated to bridge the region where QCD transitions from the strong (color confinement, long-distance) to perturbative (asymptotic freedom, short-distance) domains.

The experimental determination of $F_{\pi}$ is challenging.   The theoretically ideal method for determining $F_{\pi}$ would be electron-pion elastic scattering. However, the lifetime of the $\pi^+$ is only 26.0~ns. Since $\pi^+$ targets are not possible, and $\pi^+$ beams with the required properties are not yet available, one must employ  high-energy exclusive electroproduction, $p(e,e^\prime \pi^+n)$.  This is best described as quasi-elastic ($t$-channel) scattering of the electron from the virtual $\pi^+$ cloud of the proton, where $t$ is the Mandelstam momentum transfer $t=(p_{p}-p_{n})^2$ to the target nucleon. Scattering from the $\pi^+$ cloud dominates the longitudinal photon cross section ($d\sigma_L/dt$),  when $|t|\ll m_p^2$ \cite{Carlson:1990zn}.  To reduce background contributions, normally one separates the components of the cross-section due to longitudinal (L) and transverse (T) virtual photons (and the LT, TT interference contributions), via a Rosenbluth separation.  The value of $F_{\pi}(Q^2)$ is determined by comparing the measured $d\sigma_L/dt$ values at small $-t$ to the best available electroproduction model.  The obtained $F_{\pi}$ values are in principle dependent upon the model used, but one anticipates this dependence to be reduced at sufficiently small $-t$.  JLab 6 GeV experiments were instrumental in establishing the reliability of this technique up to $Q^2=2.45$~GeV$^2$ \cite{Huber:2008id}, and extensive further tests are planned as part of JLab E12-19-006 \cite{E12-19-006}.

At the EIC, pion form factor measurements can be extended to still larger $Q^2$, by measuring the unseparated electroproduction cross section ($\sigma_{uns}$) of the Deep Exclusive Meson Production (DEMP) reaction $p(e,e'\pi^+n)$. The value of $F_{\pi}(Q^{2})$ can be determined from these measurements by comparing the measured $\sigma_{uns}$ at low $-t$ to the best available electroproduction model, incorporating pion pole and non-pole contributions. The form factor extraction model would be validated by $\pi^{-}/\pi^{+}$ ratios from deuterium data ($D(e,e'\pi^{-}p)p_{sp}$ and $D(e,e'\pi^{+}n)n_{sp}$) in the same kinematics as the $p(e,e'\pi^+n)$ measurements on the proton.  The measurements would be made over a range of small values of $-t=-(p_p-p_n)^2$, and gauged with theoretical and phenomenological expectations to verify the reliability of the pion form factor extraction in EIC kinematics.

\begin{figure}[t]
%\centering
\begin{center}
\includegraphics[width=0.8\linewidth]{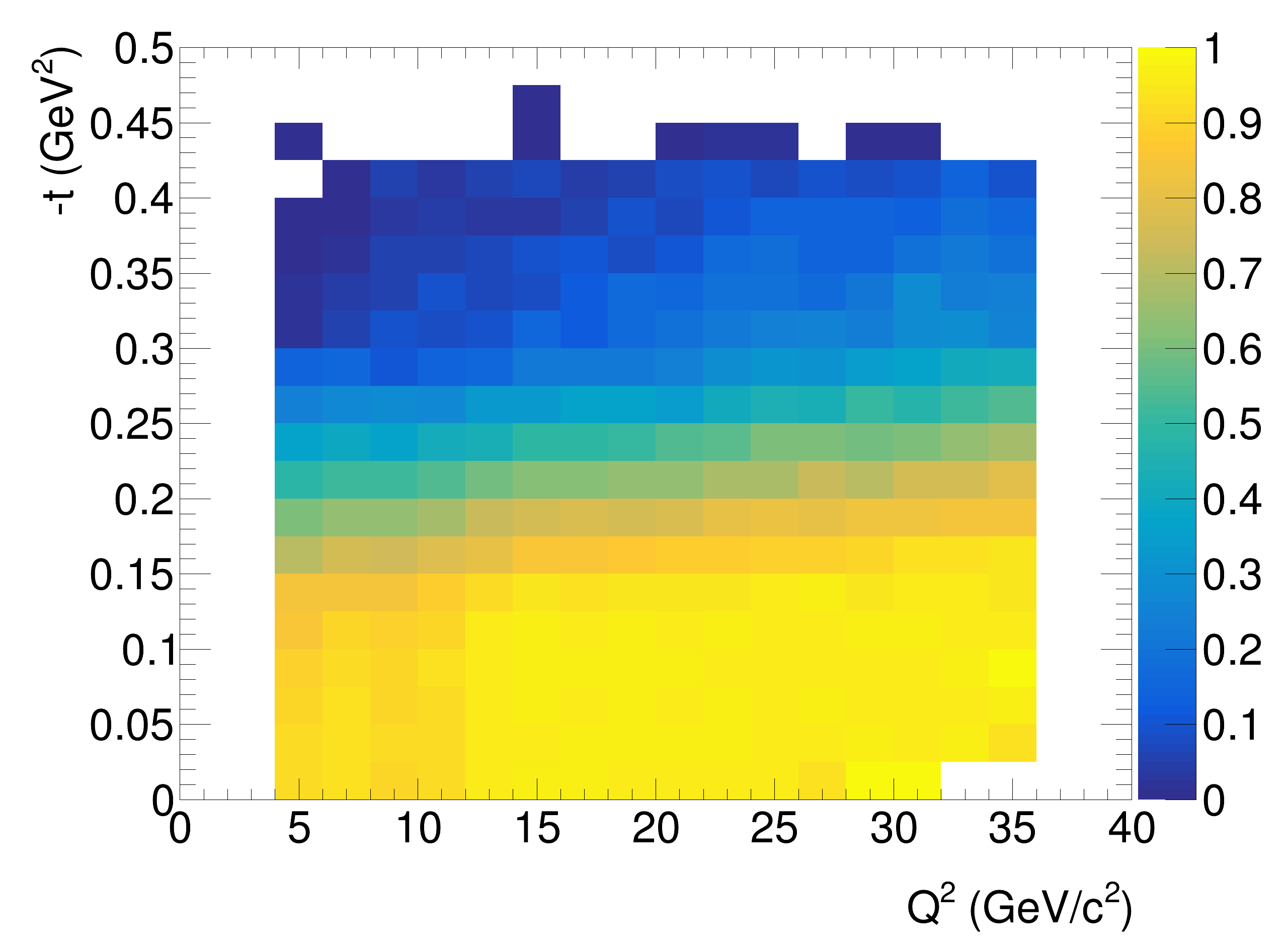}\\
\includegraphics[width=0.8\linewidth]{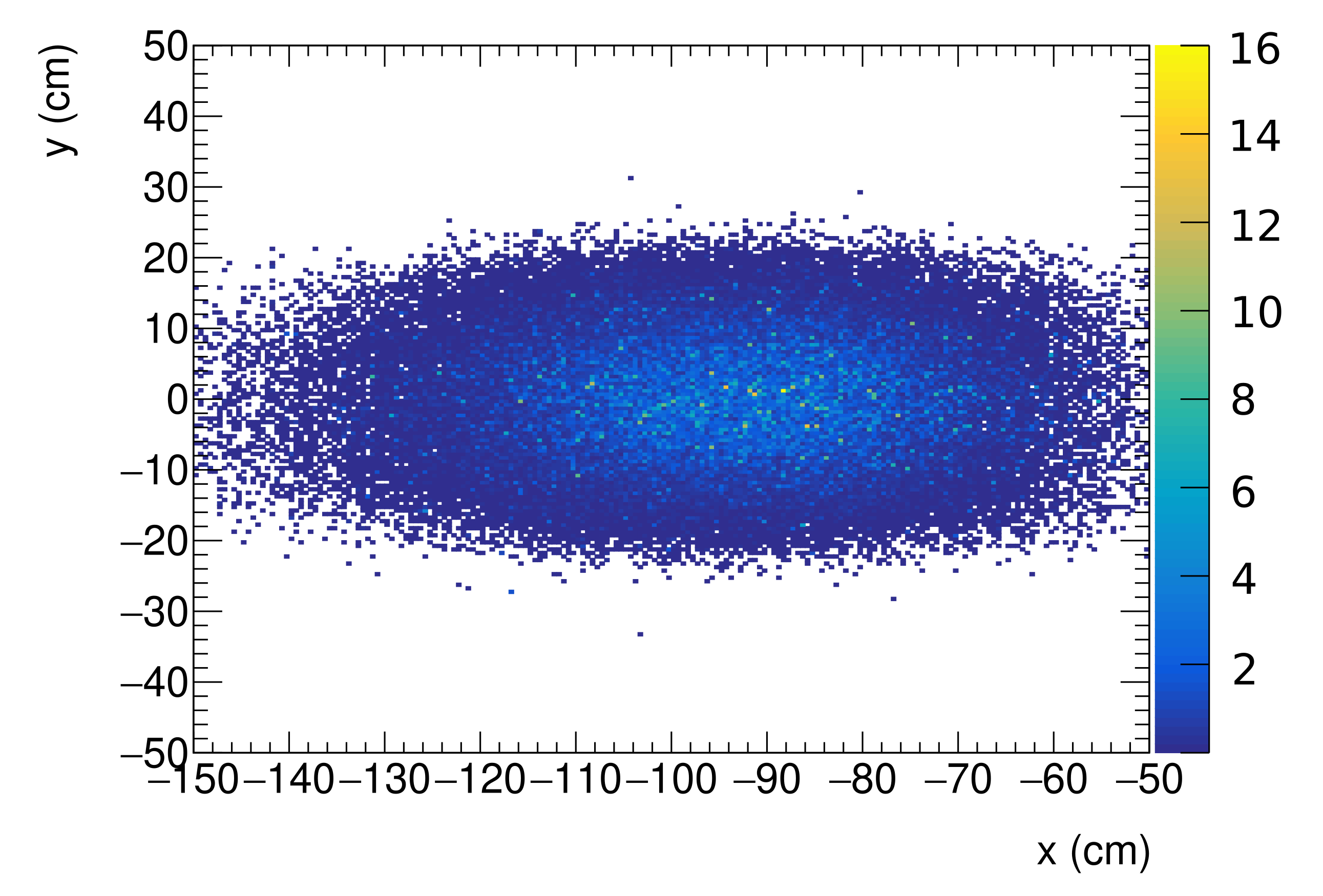}
\caption{[Top] Detection efficiency for $e' \pi^+ n$ triple coincidences in ECCE versus $Q^2$ and $-t$.  [Bottom] Predicted distribution of neutron hits from the DEMP process in the ZDC.}
\label{fig:fpi_acceptance}
\end{center}
\end{figure}

\subsubsection{Kinematics, acceptance and reconstruction resolution}

A DEMP $p(e,e'\pi^+n)$ event generator \cite{url:DEMPGen} was written and used to perform simulations demonstrating the feasibility of pion electric form factor measurements at the EIC. A sample of 0.3M simulated events from the DEMP generator in EIC-Smear format was passed through Fun4All including the ZDC plug-in. The neutrons from the DEMP reactions of interest take 80-98\% of the proton beam momentum and are detected at very forward angles (0--2$^{\circ}$) in the ZDC. The scattered electrons and pions have similar momenta, except that the electrons are distributed over a wider range of angles. For $5 \times 100$ beam energies, the 5--6~GeV/c electrons are primarily scattered 25--45$^{\circ}$ from the electron beam into the lepton end cap and the central barrel detector. The 5--12~GeV/c $\pi^+$ are 7--30$^{\circ}$ from the proton beam and enter the hadron end cap and central barrel detector.

$e'\pi^+n$ triple coincidence events were identified in the simulated data by utilizing a series of conditional selection cuts:
\begin{itemize}
    \item at least one hit in the ZDC, with an associated energy deposit above $40$~GeV.
    \item exactly two charged tracks: a positively charged track going in the $+z$ direction ($\pi^{+}$) and a negatively charged track going in the $-z$ direction ($e'$).
\end{itemize}
Both conditions had to be satisfied for a given event for it to be considered a $e'\pi^+n$ triple coincidence event.

\begin{figure}[t]
    \centering
    \includegraphics[width=0.98\linewidth]{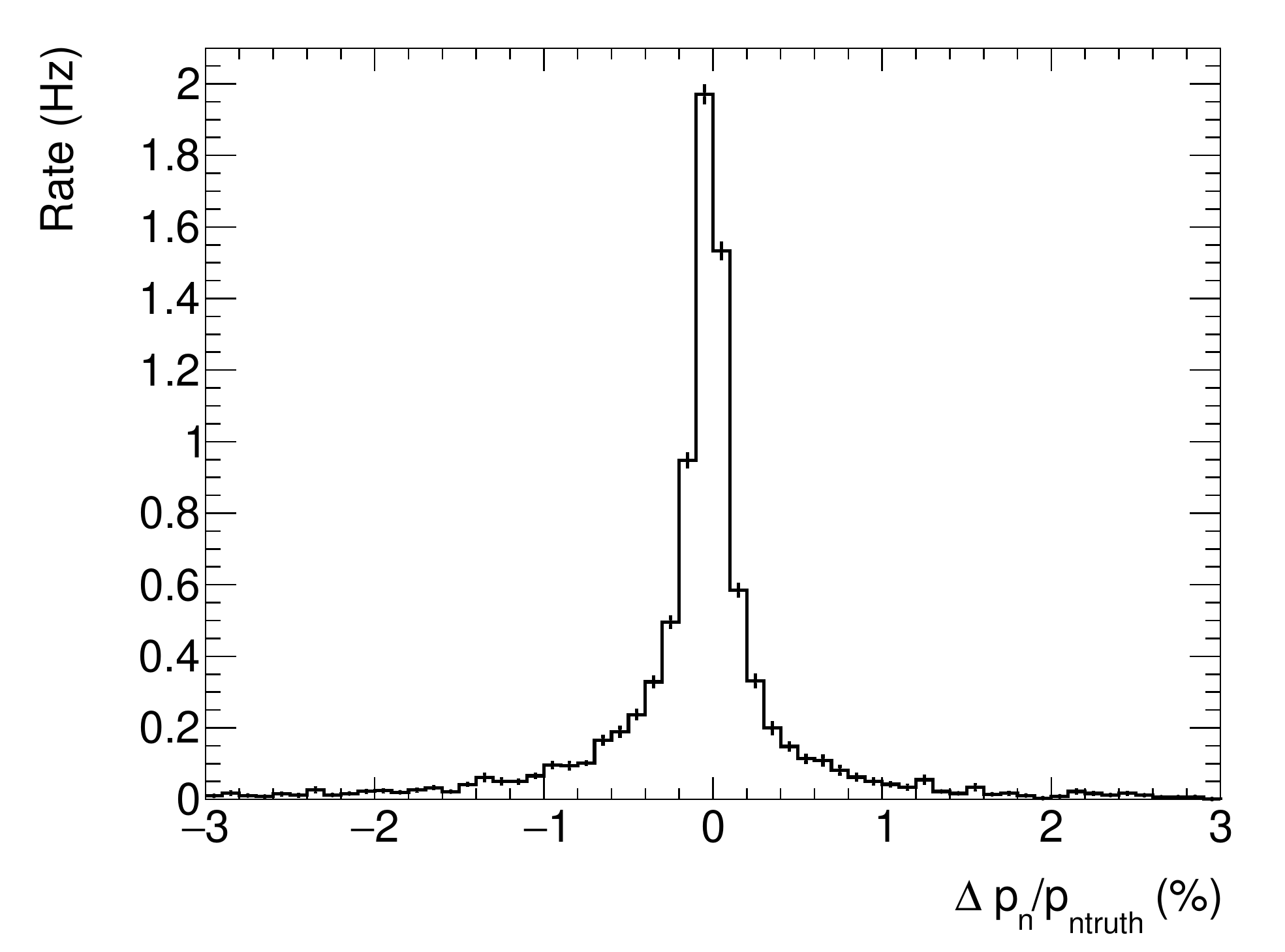}
    \caption{Deviation of the reconstructed neutron track momentum from the neutron ``truth'' track, expressed as a percentage, $\Delta p_{n} = (p_{ntrack} - p_{ntruth})/p_{ntruth}$ for  $e' \pi^+ n$ triple coincidence events.}
    \label{fig:fpi_nresolution}
\end{figure}

The ECCE detection efficiency for these triple coincidence events is fortunately quite high, $\sim80\%$, and nearly independent of $Q^2$.  A density plot of detection efficiency versus $-t$ ($y$-axis) and $Q^2$ ($x$-axis) is shown in the top panel of Fig.~\ref{fig:fpi_acceptance}.  The detection efficiency is highest for the small $-t<0.15$~GeV$^2$ events needed for the pion form factor measurement, decreasing rapidly with $-t$ thereafter.  The $ t$ range of optimal acceptance is dictated by the size of the ZDC, as the energetic neutrons from high $-t$ events are emitted at an angle larger than the ZDC acceptance.  The distribution of neutron hits on the ZDC for $5\times 100$ beam energy up to $-t=0.4$~GeV$^2$ is given in the bottom panel of Fig.~\ref{fig:fpi_acceptance}.

\begin{figure}[h!]
    \centering
    \includegraphics[width=0.8\linewidth]{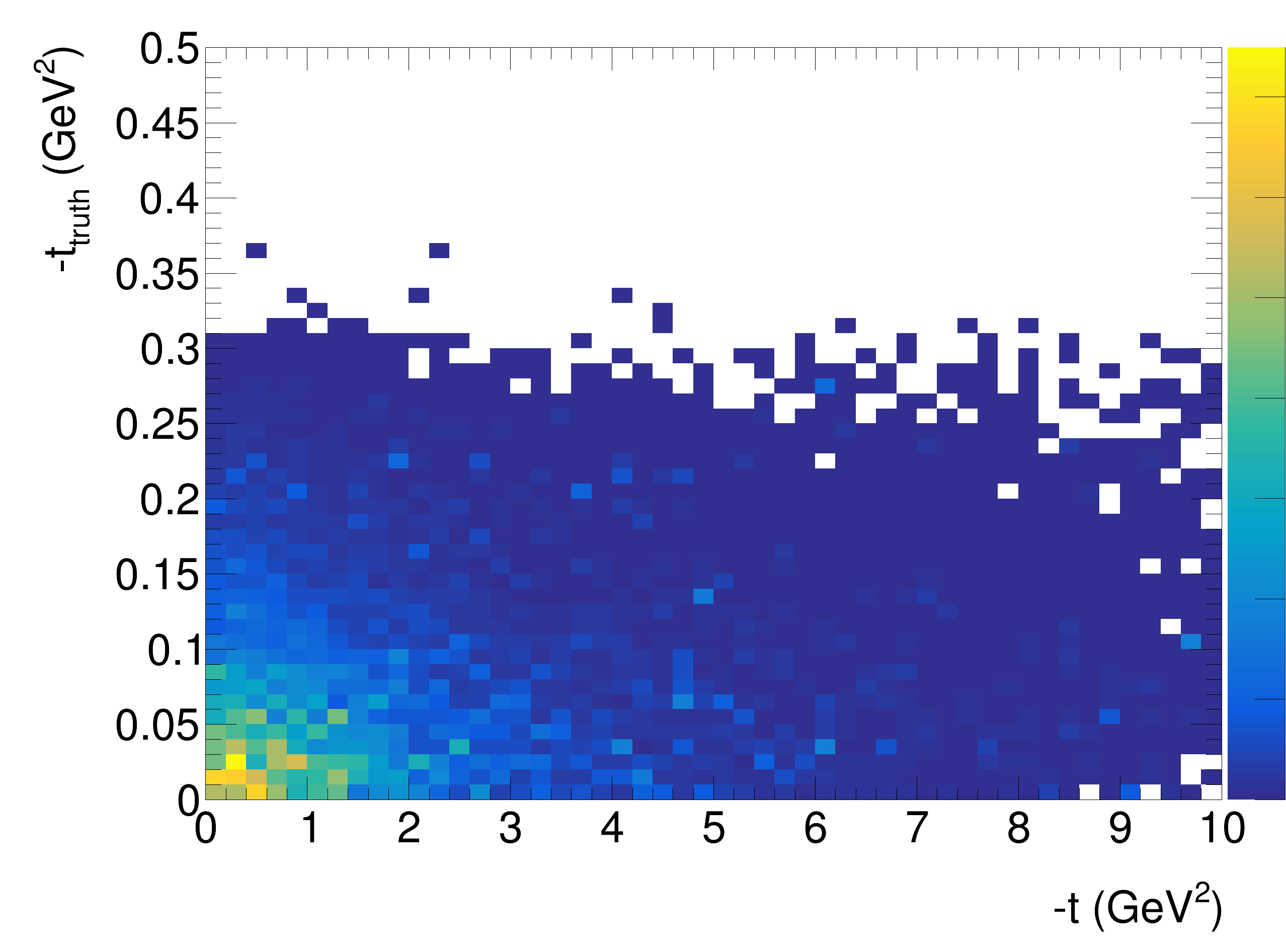}\\
    \includegraphics[width=0.8\linewidth]{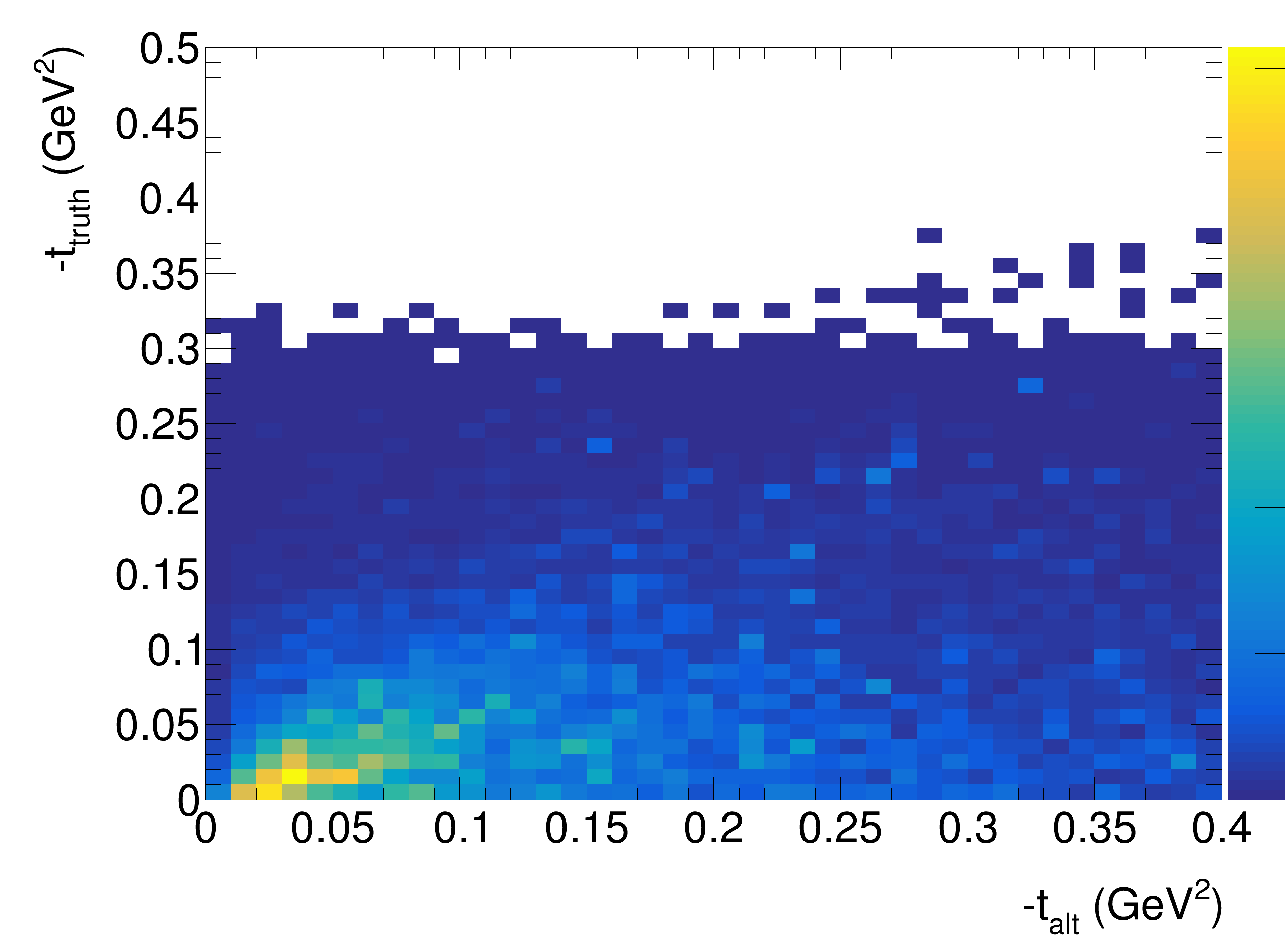}
    \caption{Reconstructed $t$ versus true $t$ for simulated  $e' \pi^+ n$ triple coincidence events with $15 < Q^{2} < 20$~GeV$^{2}$, where $t$ is reconstructed as $t=(p_e-p_{e'}-p_{\pi})^2$ (top) and as $t_{alt}=(p_p-p_n)^2$ (bottom). $p_{n}$ here is the reconstructed neutron track that combines the missing momentum with the ZDC position information. $t$ reconstruction using the lepton and meson information alone shows little correlation with the true value (top), while the reconstruction from the charged tracks and the ZDC position information is more reliable.  Note the vastly different horizontal scales of the two plots.}
    \label{fig:fpi_tresolution}
\end{figure}

The simulation successfully detected and reconstructed the $\pi^{+}$ and $e'$ tracks. The momentum of the detected tracks was reconstructed to within a few percent of the ``true'' momentum for these particles. The two charged tracks were utilized to determine the missing momentum from the reaction, $\vec{p}_{miss}=\vec{p}_e+\vec{p}_p-\vec{p}_{e'}-\vec{p}_{\pi^+}.$ As there is already a requirement for a high-energy hit in the ZDC as a veto, this missing momentum track is treated as being the exclusive neutron track. As discussed in Section \ref{sec:PiFFCuts}, additional cuts were utilized to remove potential contamination from SIDIS or other background reactions.
However, since the hit positions of the neutron track in the ZDC were known to have a high degree of accuracy, they were utilized to ``correct'' the missing momentum track and form a new ``reconstructed neutron track''. The angles, $\theta_{Miss}$ and $\phi_{Miss}$ from the missing momentum track were switched to the values determined from the ZDC hit position $\theta_{ZDC}$ and $\phi_{ZDC}$. The mass of the particle for this track was also fixed to be that of the neutron.
Following these adjustments, the subsequent reconstruction of the neutron track proved to be sufficiently accurate. The resulting reconstructed neutron track momentum was within $1\%$ of the ``true'' momentum, as seen in Fig.~\ref{fig:fpi_nresolution}. 

\begin{figure}[t]
    \centering
    \includegraphics[width=0.98\linewidth]{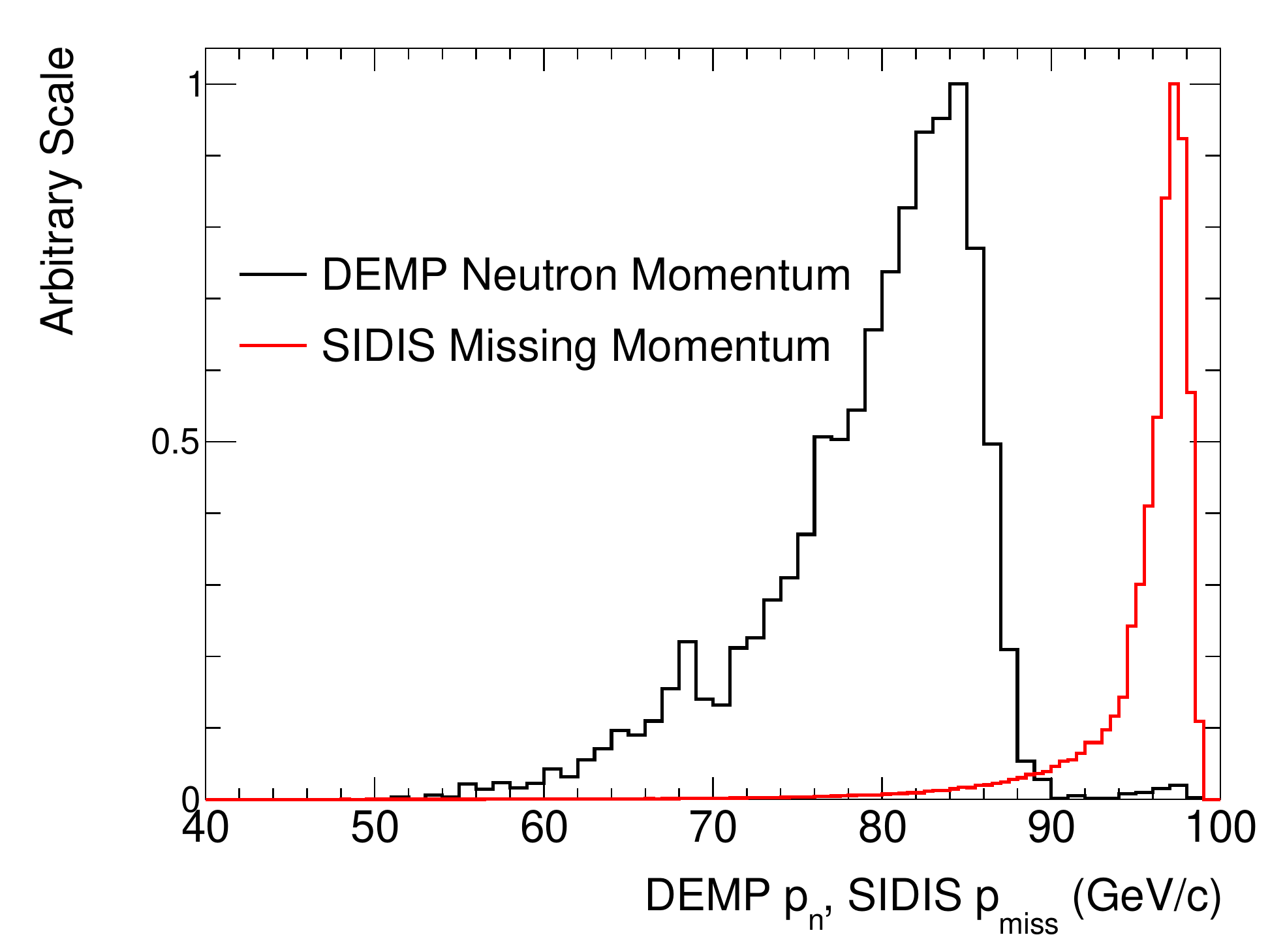}
    \caption{The reconstructed neutron track momentum for DEMP $e' \pi^+ n$ triple coincidence events compared to $\vec{p}_{miss}$ for simulated SIDIS background events ($y$-axis scaled arbitrarily, $\vec{p}_{miss}=\vec{p}_e+\vec{p}_p-\vec{p}_{e'}-\vec{p}_{\pi^+}$). The SIDIS events can be cleanly separated from the DEMP events of interest. Note that both plots display events with $15 < Q^{2} < 20$~GeV$^2$.}
    \label{fig:DEMP_SIDIS_Comp}
\end{figure}

\begin{figure}[h!]
    \centering
    \includegraphics[width=0.98\linewidth]{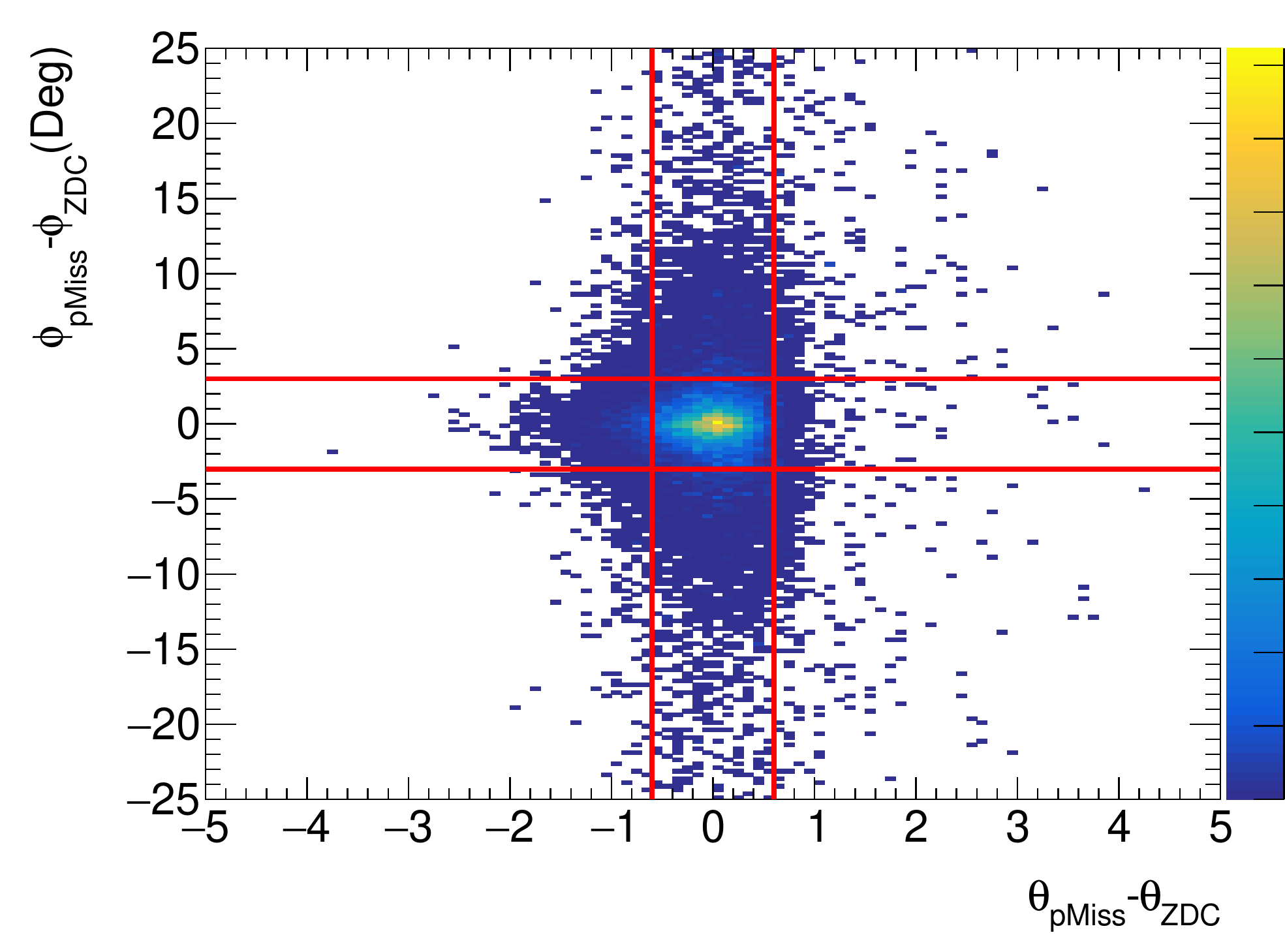}
    \caption{The difference between the reconstructed ($\theta_{pMiss}$, $\phi_{pMiss}$) and detected ($\theta_{ZDC}$, $\phi_{ZDC}$) simulated angles for the neutron in $e' \pi^+ n$ triple  coincidence events. The indicative cut range is shown by the area enclosed within the four red lines,  $-0.6\degree < \theta_{pMiss} - \theta_{ZDC} < 0.6\degree$ and $-3\degree < \phi_{pMiss} - \phi_{ZDC} < 3\degree$.}
    \label{fig:DEMP_AngleDiff}
\end{figure}

Reconstruction of $t=(p_e-p_{e'}-p_{\pi})^2$ from the detected $\pi^{+}$ and $e'$ tracks proved to be highly unreliable, as can be seen in the top panel of Fig.~\ref{fig:fpi_tresolution}. Fortuitously, due to the exclusive nature of the reaction, $t$ can also be calculated from the proton beam and the reconstructed neutron via $t=(p_p-p_n)^2$. With this information, $t$ could be reconstructed from the neutron track in a manner that reproduced the ``true'' value closely (see bottom panel of Fig.~\ref{fig:fpi_tresolution}). This also demonstrates the importance of combining the ZDC hit information with the charged particle tracks to determine the neutron four-momentum. Reliable reconstruction of $t$ is essential for the extraction of the pion form factor from the $p(e,e'\pi^+)n$ data. The cross-section falls rapidly with $-t$ as the distance from the pion pole $(t-m_{\pi}^2)$ is increased. This steep decrease in the cross-section needs to be measured to confirm the dominance of the Sullivan mechanism.

\begin{figure*}[t]
    \centering
    \includegraphics[width=0.33\linewidth]{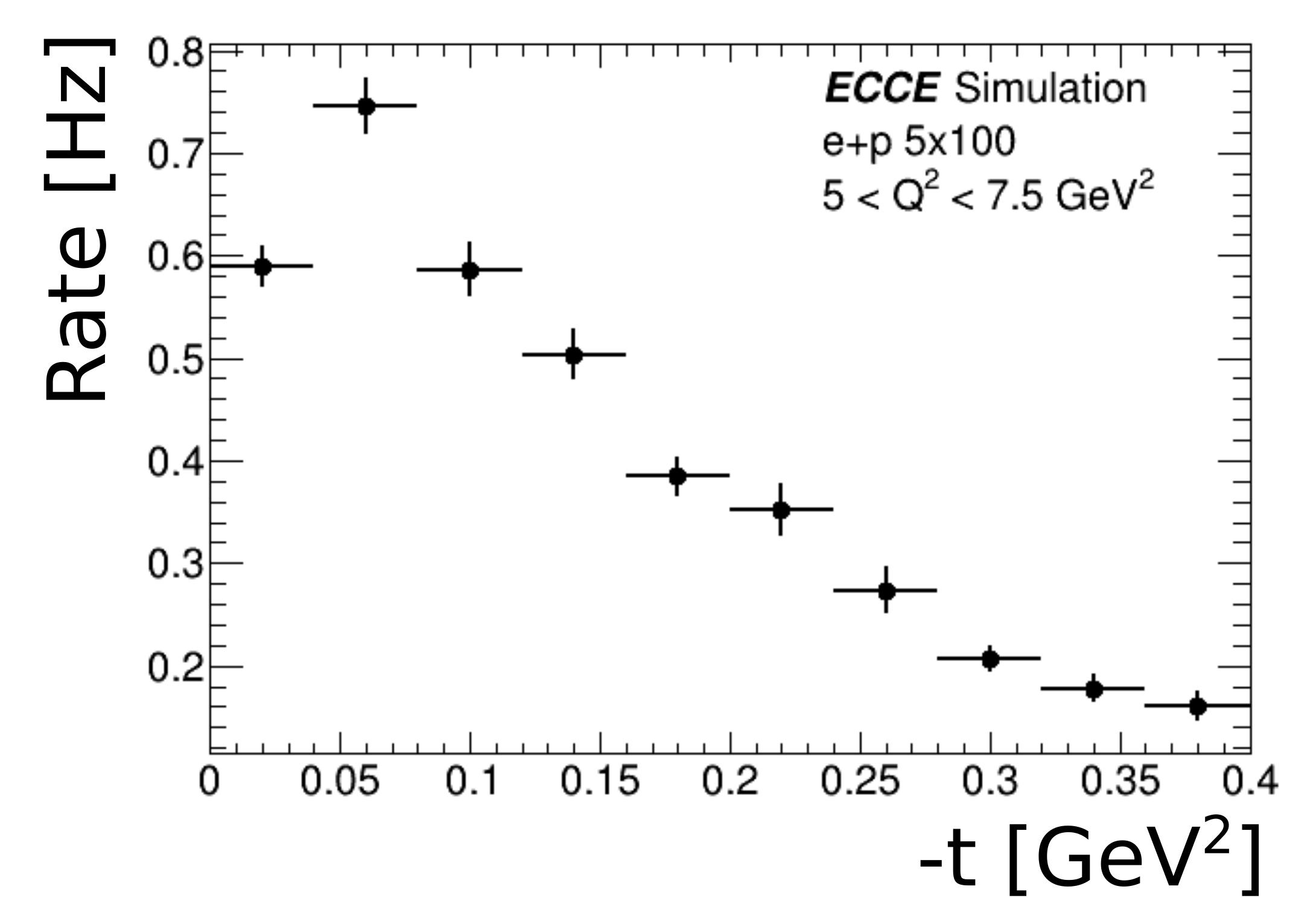}
    \includegraphics[width=0.33\linewidth]{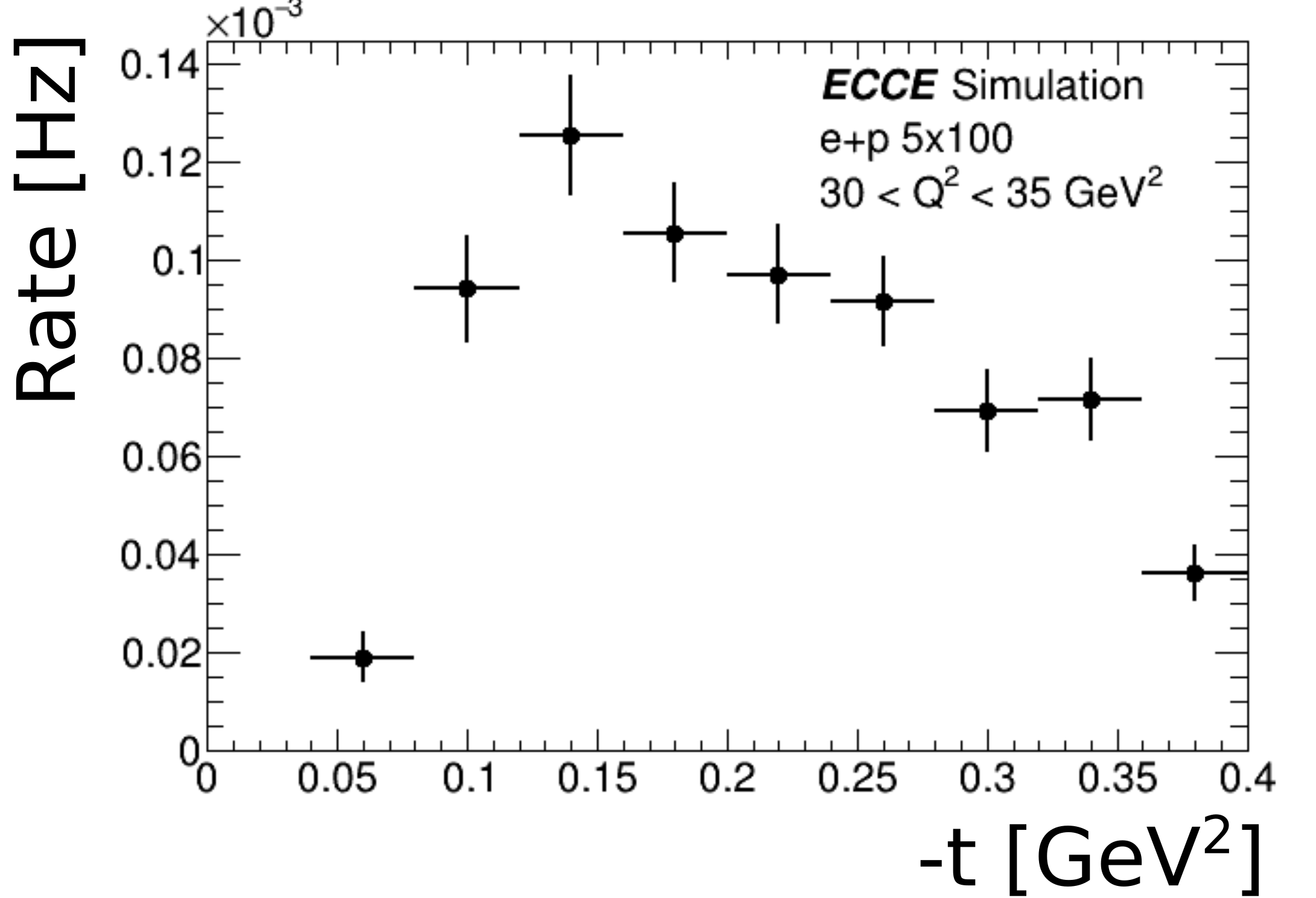}
    \includegraphics[width=0.33\linewidth]{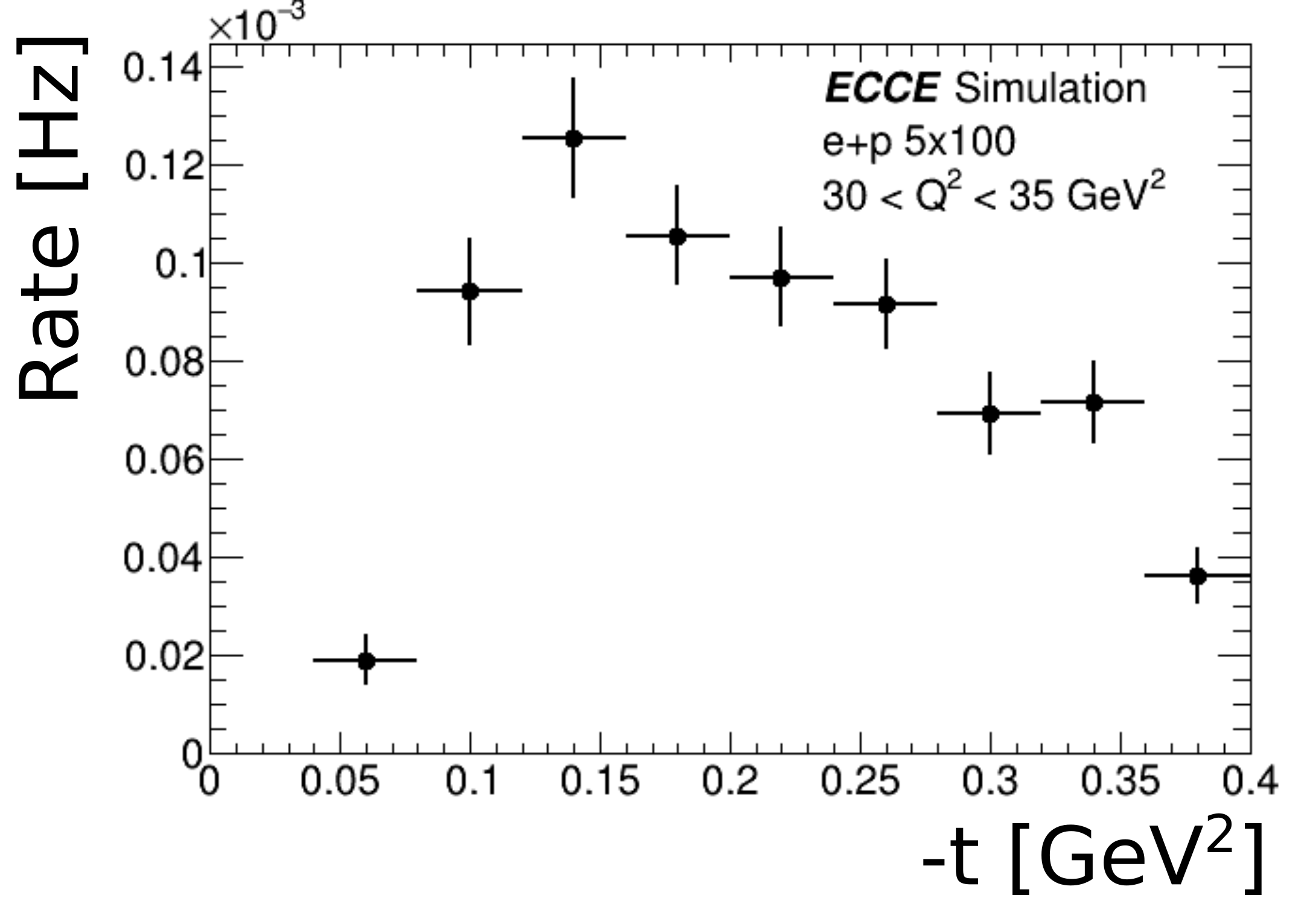}
    \caption{Predicted $e \pi^+ n$ triple coincidence rates for different $Q^{2}$ bins after application of the $p_{miss}$ and $\theta_n$ cuts described in the text. Each $-t$ bin is $0.04$~GeV$^2$ wide. The luminosity assumed in these rate calculations: L$=10^{34}$ cm$^{-2}$s$^{-1}$.}
    \label{fig:fpi_rates}
\end{figure*}

\begin{figure}[h]
\begin{center}
%\vspace{-0.5cm}
\includegraphics[width=0.95\linewidth]{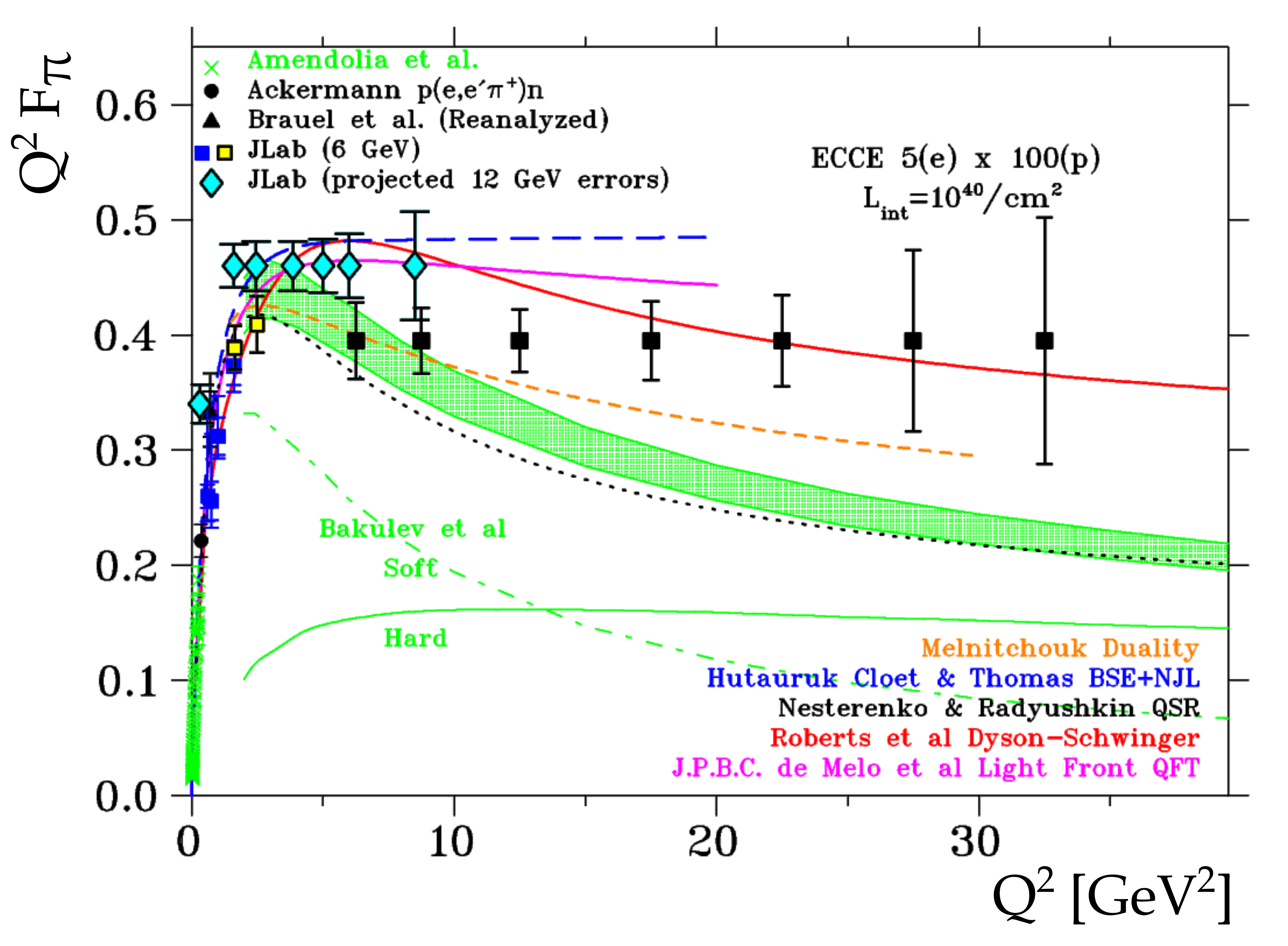}
\caption{\label{fig:ecce_fpi}
Existing data (blue, black, yellow, green) and projected uncertainties for  future data on the pion form factor from JLab (cyan, red) and EIC (black), in comparison to a variety of hadronic structure models.  The ECCE projections  clearly cover a much larger $Q^2$ range than the JLab measurements, providing access to the emergent mass scale in QCD.}
\end{center}
\end{figure}

Our finding that $t$ reconstructed from the baryon information is significantly more reliable than the version reconstructed from the lepton and meson is similar to the studies of $t$ resolution reported as part of exclusive vector meson production studies in the YR (Sec.~8.4.6) and as observed in the TCS study detailed in Sec.~\ref{subsec:tcs}.
The high-quality ZDC proposed by ECCE is clearly of paramount importance to the feasibility of this measurement.

\subsubsection{Other event selection cuts}
\label{sec:PiFFCuts}
Guided by previous work \cite{Aguilar:2019teb}, cuts are applied on the detected neutron angle ($\pm 0.5^{\circ}$ from the outgoing proton beam) and on the missing momentum, computed as $\vec{p}_{miss}=\vec{p}_e+\vec{p}_p-\vec{p}_{e'}-\vec{p}_{\pi^+}$.  The missing momentum cut corresponds to the momentum of the tagged forward-going neutron and is $Q^2$-bin dependent, varying from $p_{miss}<$96~GeV/c at $Q^2$=6.25~GeV$^2$ to $<$77.5~GeV/c at $Q^2$=32.5~GeV$^2$. In earlier studies, these cuts were highly effective in separating DEMP events from background SIDIS ($p(e,e'\pi^{+})X$) events, as can be seen in Fig.~\ref{fig:DEMP_SIDIS_Comp}. After the application of these cuts, the exclusive $p(e,e'\pi^+n)$ events were found to be cleanly separated from the simulated SIDIS events. Due to the compressed ECCE proposal timeline, we did not have time to repeat this study and used the same cuts as our earlier study shown in the YR.

Due to the method used to reconstruct the neutron four-momentum, an additional set of cuts was also implemented. A cut was applied on the difference between the angle reconstructed from the missing momentum of the charged track pair ($\theta_{pMiss}$) and the angle of the neutral particle detected in the ZDC ($\theta_{ZDC}$). A cut was also applied based on the difference in $\phi$. This pair of cuts is likely to be needed to distinguish DEMP events from SIDIS events and will need further study. For now, a conservative, but indicative, cut a range of $-0.6^{\circ} < \theta_{pMiss} - \theta_{ZDC} < 0.6^{\circ}$ and $-3^{\circ} < \phi_{pMiss} - \phi_{ZDC} < 3^{\circ}$ was applied, as shown in Fig.~\ref{fig:DEMP_AngleDiff}.

After application of these cuts, the predicted $p(e,e'\pi^+n)$ event rates at an instantaneous luminosity of $10^{34}$~cm$^{-2}$s$^{-1}$ for $Q^2$ bins over the full simulated kinematic range are shown in Fig.~\ref{fig:fpi_rates}.

\subsubsection{Results}

After the exclusive $\pi^{+}n$ event sample is identified, the next step is to separate the longitudinal cross-section $d\sigma_L/dt$ from $d\sigma_T/dt$, needed for the extraction of the pion form factor.  However, a conventional Rosenbluth separation is impractical at the EIC. Fortunately, at the high $Q^2$ and $W$ accessible at the EIC, phenomenological models predict $\sigma_L\gg\sigma_T$ at small $-t$.  This is expected since in the hard scattering regime, QCD scaling predicts $\sigma_L\propto Q^{-6}$ and $\sigma_T\propto Q^{-8}$, hence $\sigma_L$ is expected to dominate at sufficiently high $Q^2$.  For example, the Vrancx and Ryckebusch Regge-based model \cite{Vrancx:2013fra} predicts $R=\sigma_L/\sigma_T >10$ for $Q^2>10$~GeV$^2$ and $-t<0.06$ GeV$^2$, and $R>25$ for $Q^2>25$~GeV$^2$ and $-t<0.10$~GeV$^2$. Thus, the transverse cross-section contributions are expected in these cases to be only 4-10\%.  The most practical choice appears to be to use a model to isolate the dominant $d\sigma_L/dt$ from the measured $d\sigma_{uns}/dt$.

The value of $F_{\pi}(Q^2)$ is then determined by comparing the measured $d\sigma_{uns}/dt$ at small $-t$ to the best available electroproduction model, incorporating pion pole and non-pole contributions and validated with $\pi^-/\pi^+$ data. The model should have the pion form factor as an adjustable parameter, so that the best-fit value and its uncertainty at fixed $(Q^2, W)$ are obtained by comparison of the magnitude and $t$-dependencies of the model and data. If several models are available, the form factor values obtained with each one can be compared to better understand the model dependence. The importance of additional $p(e,e'\pi^+n)$ model development to improve knowledge of pion form factors cannot be overestimated, and additional activity in this area is strongly encouraged.

Using this technique, ECCE can enable a pion form factor measurement up to $Q^2=32.5\,$GeV$^2$, as shown in Fig.~\ref{fig:ecce_fpi}. Note that the $y$-axis positions of the projected data points in the figure are arbitrary. However, the error bars represent the real projected errors for these points. The errors in the yields are based on the following assumptions: cross sections parameterized from the Regge model in \cite{Choi:2015yia}, integrated luminosity of 10\,fb$^{-1}$ for 5$\times$100 measurement, clean identification of exclusive $p(e,e^\prime\pi^+n)$ events by tagging the forward neutron, and a cross-section systematic uncertainty of 2.5\% point-to-point and 12\% scale (similar to the HERA-H1 pion structure function measurement \cite{Aaron:2010ab}). One should then apply an additional uncertainty, since the form factor will be determined from unseparated, rather than L/T-separated data: $\delta R=R$ systematic uncertainty in the model subtraction to isolate $\sigma_L$, where $R=\sigma_L/\sigma_T=0.013-0.14$ at $-t_{\rm min}$. Finally, the model fitting procedure is used to extract $F_{\pi}(Q^2)$ from the $\sigma_{uns}$ data, where one assumes the applied model is validated at small $-t$ by comparison to data. Additional model uncertainties in the form factor extraction are not estimated here, but the EIC should provide data over a sufficiently large kinematic range to allow the model dependence to be quantified in a detailed analysis.

Regarding the projected uncertainties in Fig.~\ref{fig:ecce_fpi}, for the lowest $Q^2$ bins ($Q^2<10$~GeV$^2$) the uncertainty in $R$ is among the largest systematic uncertainties, arising from the inability to perform an L/T-separation, and the relatively less favorable T/L ratio.  At intermediate $Q^2$ ($10<Q^2<25$~GeV$^2$), the T/L ratio is more favorable and the experimental systematic uncertainties dominate.  The statistical uncertainties dominate the highest $Q^2$ bins ($Q^2>$25~GeV$^2$), as the rates in these regions are very low (see Fig.~\ref{fig:fpi_rates}).

To conclude, the extraction of the pion form factor to high $Q^2$ with ECCE depends on very good ZDC angular resolution for two reasons: 1) the necessity to separate the small exclusive $\pi^+$ cross-section from dominant inclusive backgrounds via $p_{miss}$ and $\theta_n$ cuts, 2) the need to reconstruct $t$ to better than $\sim$0.02~GeV$^2$, such resolution is only possible when reconstructed from the initial proton and final neutron momenta.  The ZDC is thus of crucial importance to the feasibility of a pion form factor measurement.

\subsection{$\pi$ structure function}
\label{subsec:pi_sf}
Studies of the meson structure functions were identified as a key science topic in the YR. The far-forward detection region is particularly important, as the recoiling baryon and its decay products have to be detected with sufficient precision to achieve the desired resolution for meson structure studies. This region provides a broad acceptance for charged and neutral particles for a variety of interactions. For meson structure experiments, it maximizes the kinematic coverage for a range of beam energies. 

\begin{figure}[t!]
\centering
    \centering
    \includegraphics[trim={0 0 0 5cm}, clip, width=0.98\linewidth] {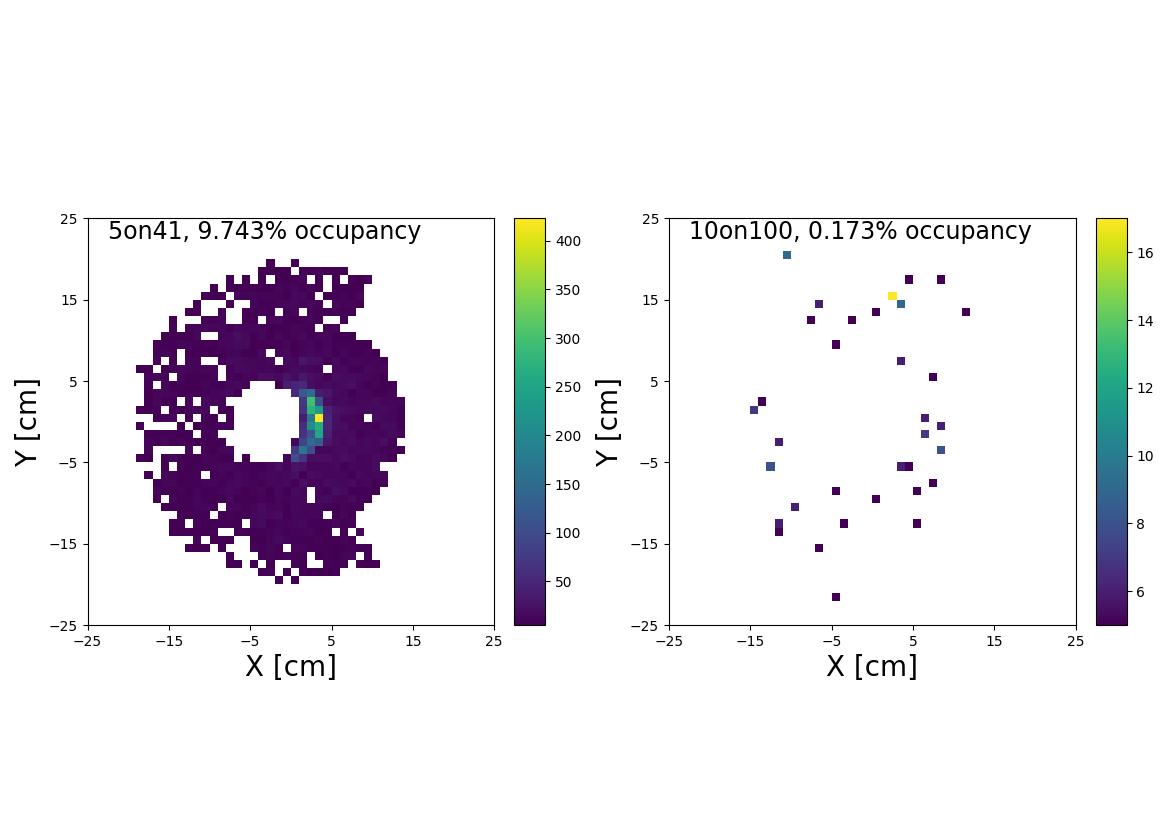}\\[-1cm]
    \centering
    \includegraphics[trim={0 3.8cm 0 5cm}, clip, width=0.98\linewidth] {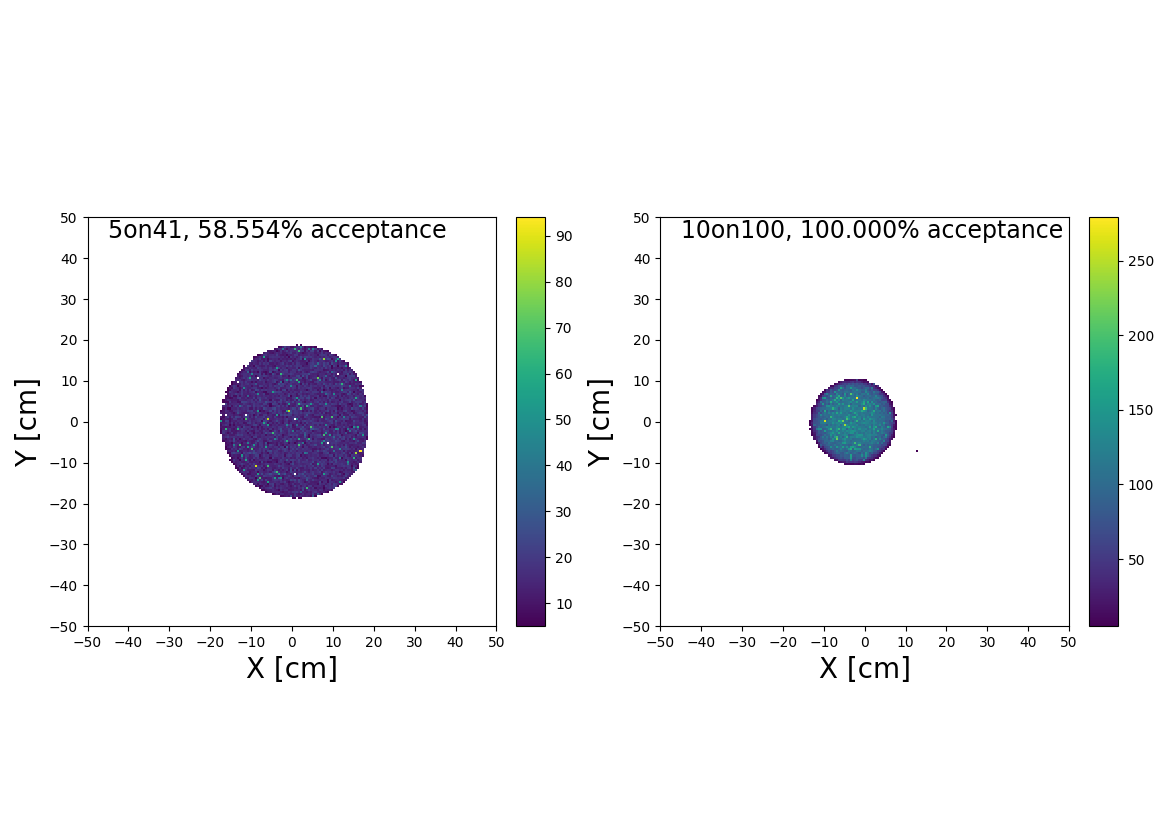}
    \caption{Top plots: B0 occupancy of the simulated leading neutron for $p(e, e'\pi^+n)$ meson structure study at 5$\times$41 (left) and 10$\times$100 (right). Bottom plots: ZDC acceptance of the simulated leading neutron for a range of energies 5$\times$41 (left) and 10$\times$100 (right). }
    \label{fig:b0_zdc_all}
\end{figure}

Similar to the inclusive $ep$ structure-function, the neutron-tagged structure function rises at low $x_B$. As shown by HERA, by determining the neutron-tagged cross-section relative to the inclusive $ep$ cross-section it is possible to precisely determine the leading neutron production~\cite{Aaron:2010ab}. Tagged deep-inelastic scattering (TDIS) can then be used to probe the meson content of the neutron structure function, thus extracting the pion structure function using the Sullivan process.

There are limited data on the pion structure function with (HERA TDIS data~\cite{Andreev2014}) which looked at the low $x_B$ region using the Sullivan process, and the pionic Drell-Yan data~\cite{chang13} from nucleons at large $x_B$. The one-pion exchange seems to be the dominant mechanism that makes it possible to extract the pion structure function through the use of an in-depth model and kinematic studies, which include effects such as re-scattering and absorption. These projected capabilities of EIC (10$^4$ higher luminosity compares to HERA) will directly balance the ratio of Sullivan processed tagged pion structure function measurements in various bins of $t$ to the proton structure measurements. At high $x_{\pi}$, it is possible for a direct comparison to fixed target experiments and Drell-Yan. Upcoming experiments, like  COMPASS++/AMBER Drell-Yan~\cite{Adams2019} and the JLab 12 GeV TDIS experiment~\cite{Adikaram2015}, will be vital consistency checks of pion structure information obtained at the EIC. 

\subsubsection{Kinematics, acceptance, and reconstruction resolution}

The $\pi$ structure studies were conducted over a range of beam energies with the EIC\_mesonMC generator~\cite{url:EIC_mesonMC}. The highest energy of 18$\times$275  was used to maximize the kinematics coverage. However, to improve access to the high $x_{\pi}$ region, alternate lower beam energies 10$\times$100, 5$\times$100, and 5$\times$41 were also utilized. These lower beam energies allow access to this high $x_{\pi}$ regime over a wider range of $Q^2$. 
 
The leading neutrons for these two energy settings are at a very small forward angle while carrying nearly all of the proton beam momentum. These leading neutrons will be detected by the ZDC.

The ZDC must reconstruct the energy and position well enough to constrain both the scattering kinematics and the four-momentum of the pion. Constraining the neutron energy around 35$\%$/$\sqrt{E}$ will assure an achievable resolution in $x_B$. Fig.~\ref{fig:b0_zdc_all} bottom row shows the acceptance plots for neutrons in the ZDC for the two energy settings. As one can see, the spatial resolution of the ZDC plays an important role in the highest energy setting, since it is directly related to the measurements of $p_T$ or $t$. The $t$-reconstruction was produced from the proton beam and the reconstructed neutron via $t=(p_p-p_n)^2$ as outlined in Sec.~\ref{subsec:piff}. For the lowest energy setting, the total acceptance coverage of the ZDC is important. This sets a requirement for the total size of the ZDC to be a minimum of 60$\times$60~cm$^2$. 

The B0 occupancy in Fig.~\ref{fig:b0_zdc_all} top row plots show a significant amount of leading neutrons hitting the detector for the lowest energy settings (i.e. 5$\times$41). The ZDC acceptance in Fig.~\ref{fig:b0_zdc_all} bottom row plots for the leading neutron also show a significant drop in neutron detection for the lowest energy setting (i.e. 5$\times$41). This corresponds to the increased occupancy in the B0. 

\begin{figure}[h!]
\centering
    \includegraphics[width=0.95\linewidth]{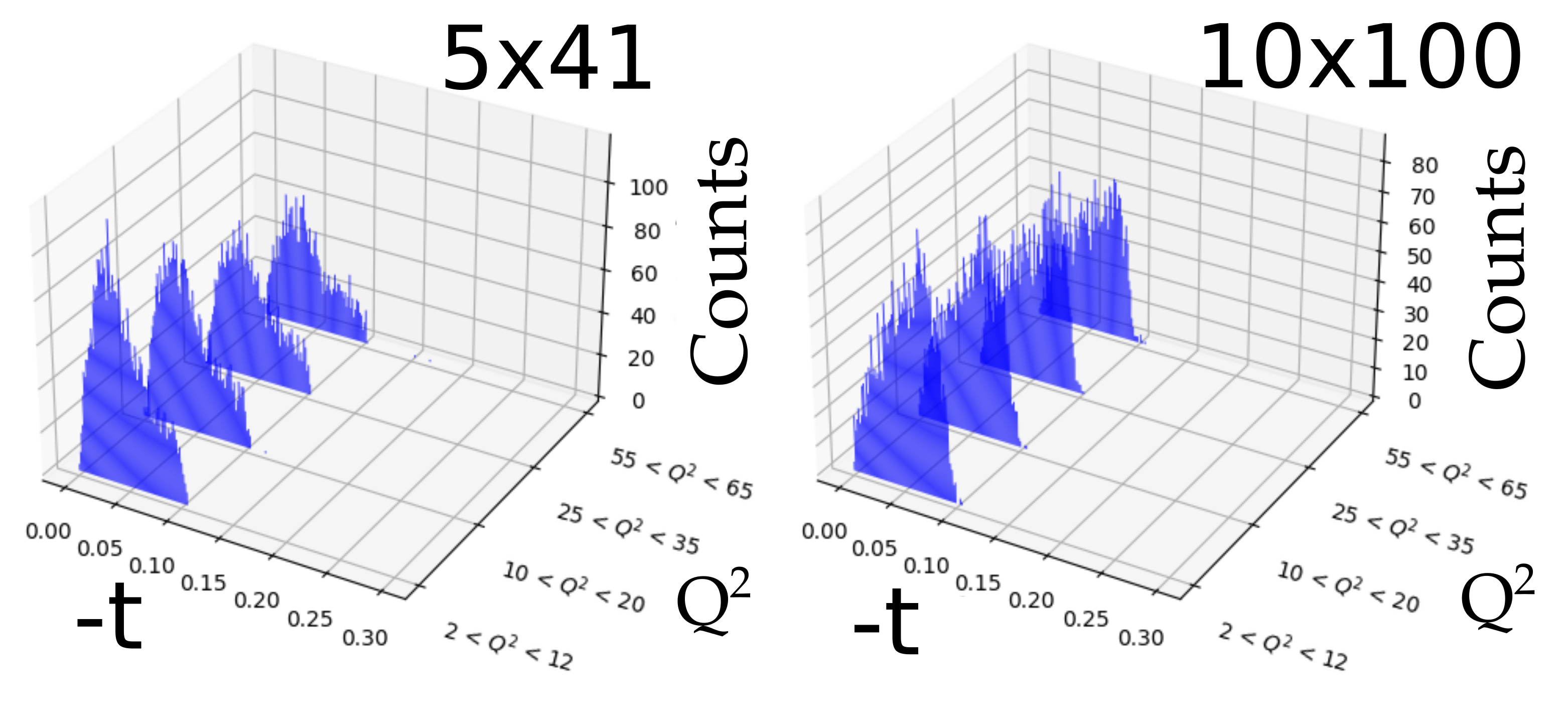}
    \caption{The $-t$ distributions of $p(e, e'\pi^+n)$ meson structure study at 5$\times$41 (left) and 10$\times$100 (right). There are four $Q^2$ bins presented (7, 15, 30, 60~GeV$^2$) of bin width $\pm$5 GeV$^2$. 
    }
    \label{fig:t_Q2_3d}
\end{figure}

\begin{figure}[t!]
\centering
    \centering
    \includegraphics[trim={0 0 0 0.5cm}, clip, width=0.98\linewidth]{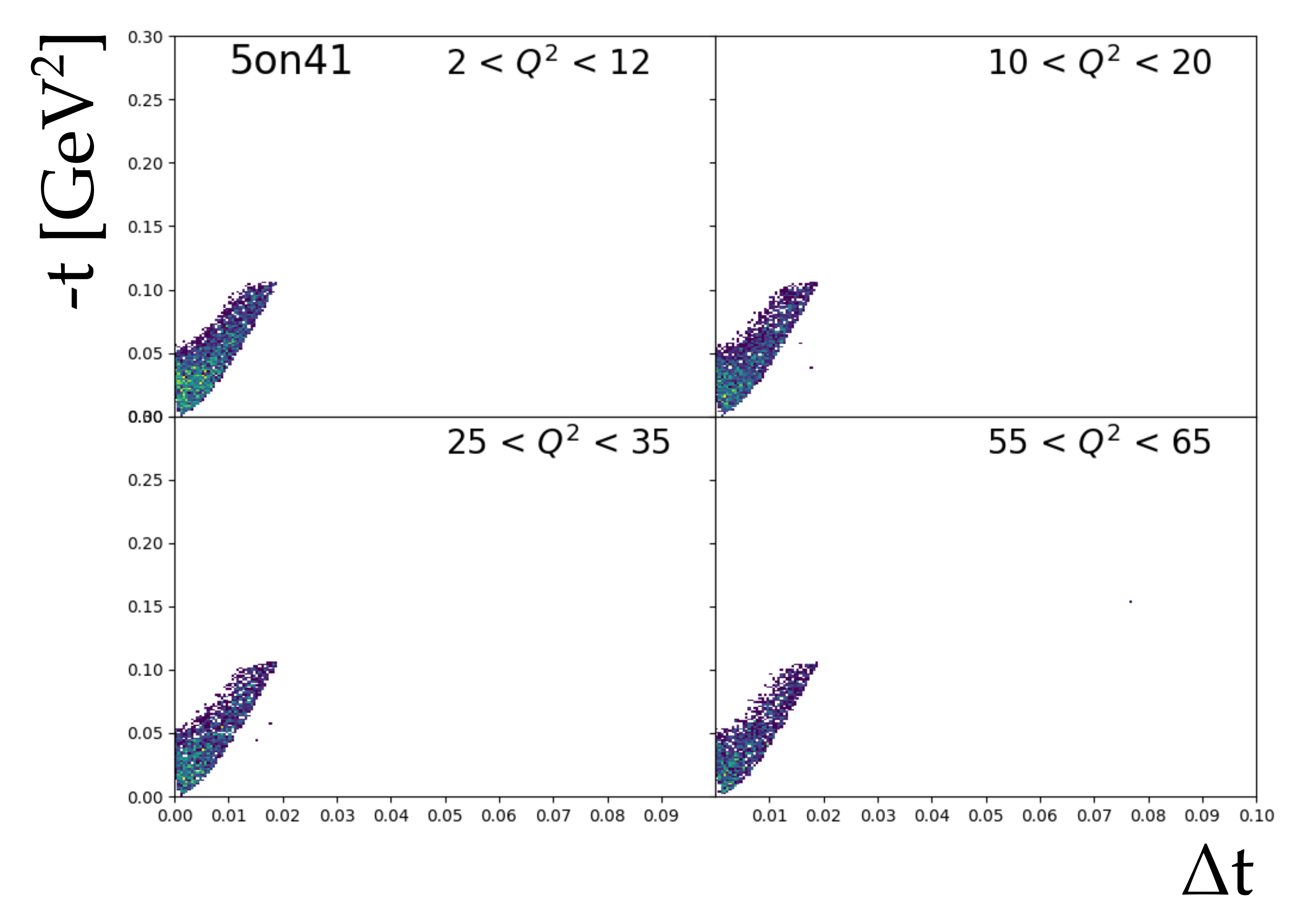}\\[0 cm]
    \centering  
    \includegraphics[trim={0 0 0 0.5cm}, clip, width=0.98\linewidth]{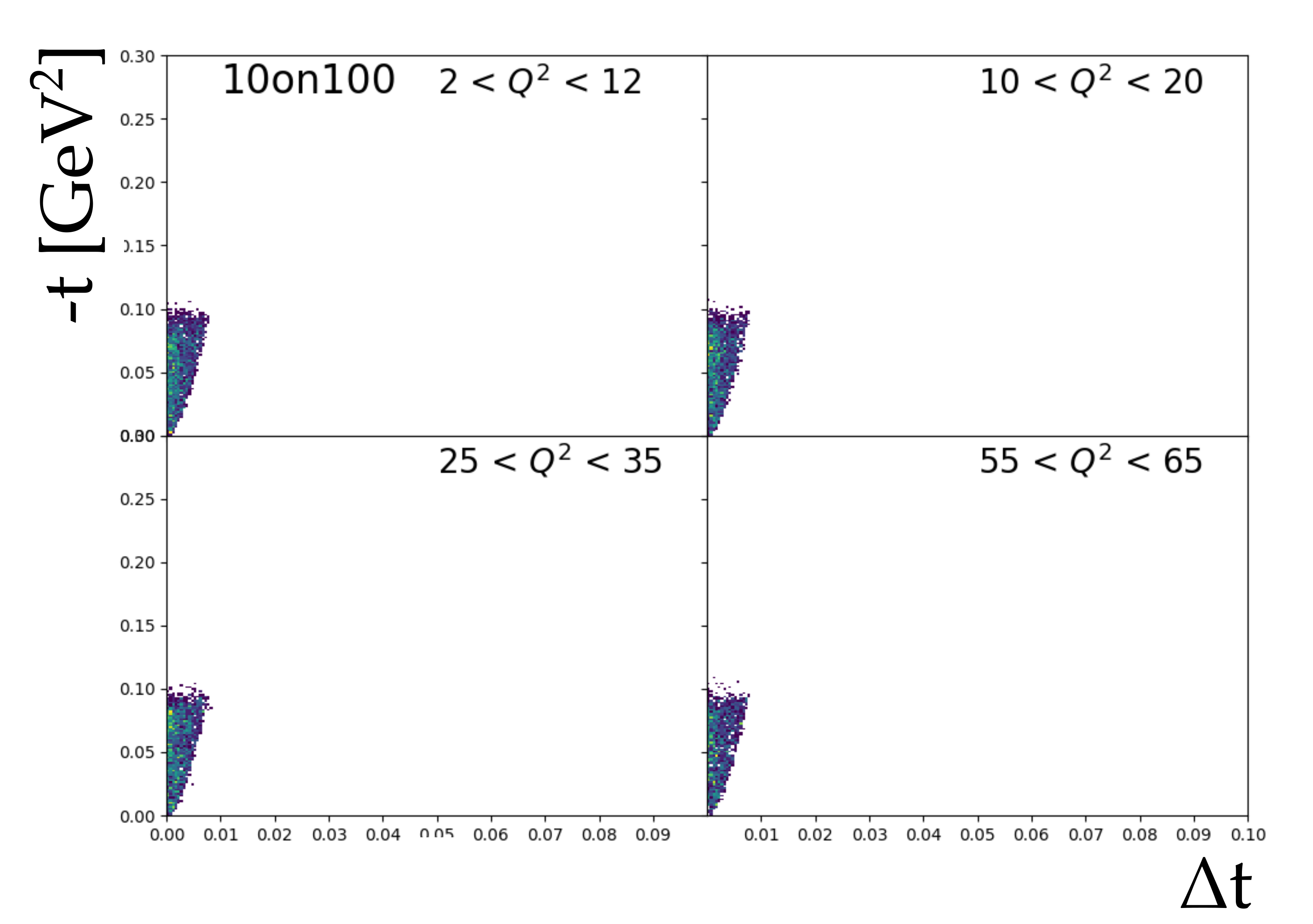}    
    \caption{
    The deviation of generated $-t$ from the detected $t_{truth}$ value for $p(e, e'\pi^+n)$ meson structure study, $\Delta{t}=t-t_{truth}$, for a range of energies (5$\times$41, 10$\times$100) at IP6. There are four $Q^2$ bins presented (7, 15, 30, 60 GeV$^2$) of bin width $\pm$5~GeV$^2$. The lowest energy (5$\times$41) sees a strong deviation. 5$\times$41 is the same energy that sees the drop in ZDC acceptance.
    }
    \label{fig:delta_t_t_Q2_IP6}
\end{figure}

As mentioned earlier, the spatial resolution of the ZDC plays a crucial role in determining measurements of $t$. Fig.~\ref{fig:t_Q2_3d} breaks down the $t$-distribution for the two energies for a range of $Q^2$ bins. The drop in events at the higher $Q^2$ bins is expected for the lower energies. Fig.~\ref{fig:delta_t_t_Q2_IP6} shows the deviation of $t$ from its detected value. The deviation, $\Delta{t}=t-t_{truth}$, is clearly much greater for the lowest energy (5$\times$41), providing a consistent picture of the energy ranges.

\subsubsection{Results}

\begin{figure*}
    \includegraphics[width=1.0\textwidth]{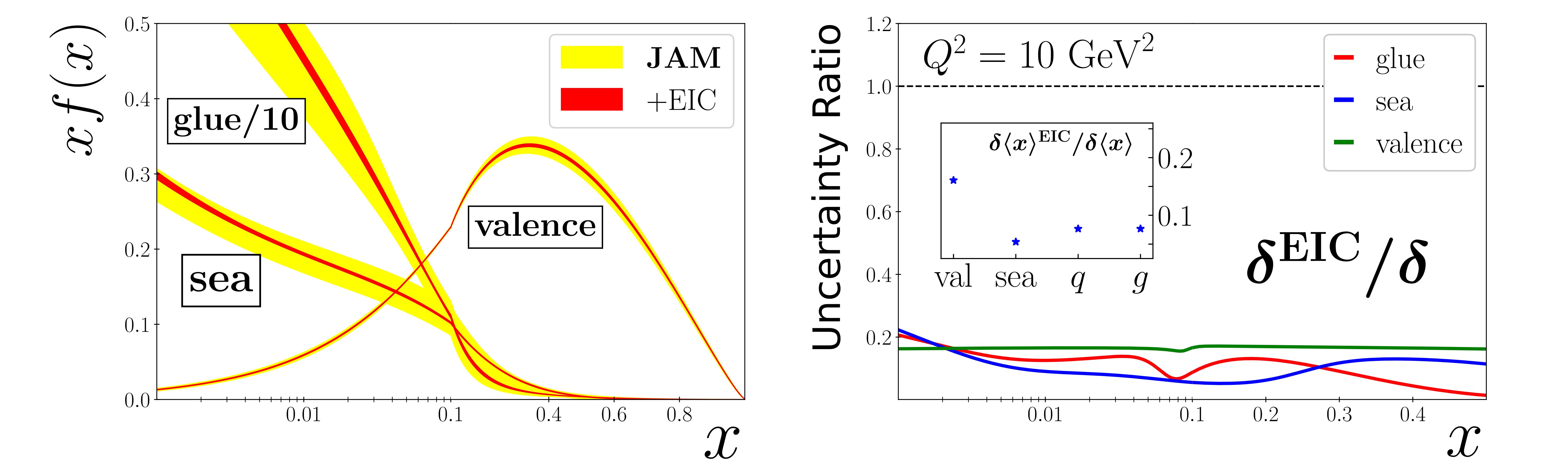}
    \caption{\label{fig:JAM_impact_lesspoints}
    \emph{Left:} Comparison of uncertainties on the pion's valence, sea quark, and gluon PDFs before (yellow bands) and after (red bands) inclusion of EIC data.
    \emph{Right:} Ratio of uncertainties with EIC data to without, $\delta^{\rm EIC}/\delta$, for the valence (green line), sea quark (blue), and gluon (red) PDFs,
    assuming 1.2\% experimental systematic uncertainty but no model systematic uncertainty,
    and (inset) the corresponding ratios of the momentum fraction uncertainties, $\delta\langle x \rangle^{\rm EIC}/\delta\langle x \rangle$, for valence, sea, total quark and gluon PDFs~\cite{Barry-resummation-20}, at a scale $Q^2=10$~GeV$^2$.
    }
\end{figure*}

Statistical uncertainties, with the addition of the leading neutron detection fraction, were incorporated into the overall uncertainty for an integrated luminosity of $\mathcal{L}~=100\,\mathrm{fb}^{-1}$. For 10$\times$100 energy, the coverage in $x_B$ extends down to 10$^{-2}$, with reasonable uncertainties in the mid-to-high $x_B$ region, increasing rapidly as $x\rightarrow1$. Even with these restrictions, the coverage in mid-to-high $x_B$ is unprecedented.

In Fig.~\ref{fig:JAM_impact_lesspoints}, we show the impact of EIC data on the pion PDFs themselves
and their uncertainties, folding in the estimated systematic uncertainty and the projected statistical uncertainties from the simulations. The resulting access to a significant range of $Q^2$ and $x_B$, for appropriately small $-t$, will allow for much-improved insights in the gluonic content of the pion.

\begin{figure}
    \includegraphics[width=0.45\textwidth]{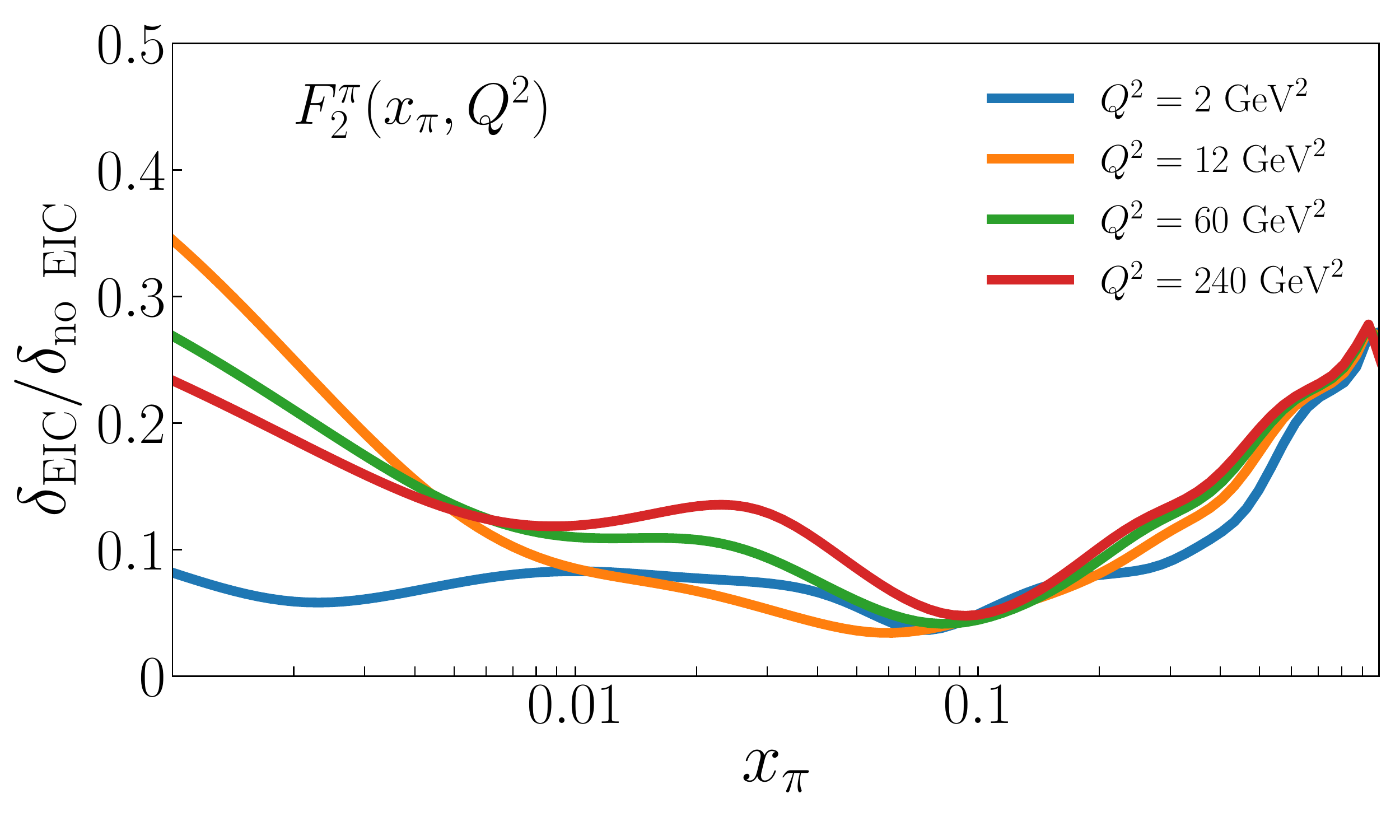}
    \caption{\label{fig:F2pi}
   The ratio of the uncertainty of the $F_2^\pi$ structure function from the global fit with and without including EIC projected data to the uncertainty of the $F_2^\pi$ as a function of $x_\pi$ for various $Q^2$ values.
    }
\end{figure}

The ratio of the uncertainty of the $F_2^\pi(x_\pi,Q^2)$ structure function resulting from a global fit with EIC projected data to that without it is displayed in Fig.~\ref{fig:F2pi}. We show various $Q^2$ values over a wide range between a few GeV$^2$ and a few hundred GeV$^2$, over the range $10^{-3}<x_\pi<1$, to investigate the $Q^2$ dependence of the impact. Strikingly, the $F_2^\pi$ structure function's uncertainties reduce by 80-90\% in the range of $x_\pi$ between $3\times 10^{-3}$ and $0.4$ in the presence of EIC data, independent of the values of $Q^2$. Within the whole range, the uncertainties reduce by 65\% or more. Below $x_\pi$ of 0.1, the error in the $F_2^\pi$ structure function reduces by a factor of 10 for the case when $Q^2=2$~GeV$^2$. The EIC provides a unique opportunity to improve our knowledge of the $F_2^\pi$ structure function over a large range in $Q^2$ and $x_B$.

\subsection{Neutron spin structure}
\label{subsec:double_tag}
%\subsection{Double spectator tagging measurement}
Polarized $^3$He plays an important role as an effective neutron target in many neutron spin structure experiments. For inclusive measurements, as often done with fixed targets, the two protons not only dilute the signal, but they also have a small net polarization which is not known, leading to rather large systematic uncertainties on the extracted quantities. The EIC has a unique capability to measure the two protons in the far forward region; this allows for the extraction of neutron information with reduced systematic, as well as an enhanced asymmetry, as compared to inclusive measurements, as will be shown in this section.  

\begin{figure}[t]
    \centering
    \includegraphics[scale=0.3]{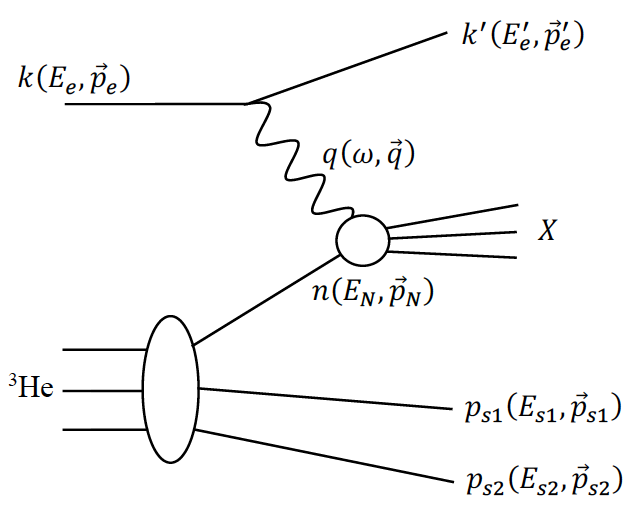}
    \caption{A diagram of Deep Inelastic $e\,^3$He scattering with double spectator tagging. The channel shown here is electron scattering off a neutron in $^3$He; the two spectator nucleons are the protons in the process $^3$He$(e,e'p_{s1}p_{s2})X$.}
    \label{fig:diagram}
\end{figure}

%==========Generator section==============%
\subsubsection{Event generation}

This study used the output of the DJANGOH 4.6.10~\cite{incl:djangoh1,incl:djangoh2} event generator to produce neutral-current DIS events from $^3$He, with a fully-calculated hadronic final-state from the leading nucleon. 
The event generation was performed using the CTEQ6.1 PDF set~\cite{Stump:2003yu}.
% Are more details about DJANGOH necessary given this citation? It uses LEPTO, HERACLES, JETSET, SOPHIA, etc.
As DJANGOH events already include the effects of QED radiation and final-state hadronization, it is only necessary to add the kinematics of the spectator system. 

The method used to determine the distributions of the spectators comes from the convolution approximation for nuclear structure functions in the Bjorken limit~\cite{Frankfurt:1981mk}:

\begin{equation}
    F_{1A}(x_B,Q^2)\approx\int \frac{d\alpha d\Gamma_s}{\alpha} \frac{A}{\alpha} \sum_{N=p,n} F_{1N}\left(\frac{x_B}{\alpha},Q^2\right) 
    \rho_N(\alpha,\Gamma_s)
\end{equation}
\begin{equation}
    F_{2A}(x_B,Q^2)\approx\int \frac{d\alpha d\Gamma_s}{\alpha} \sum_{N=p,n} F_{2N}\left(\frac{x_B}{\alpha},Q^2\right) 
    \rho_N(\alpha,\Gamma_s)
\end{equation}

Here, $\alpha=\frac{A}{m_A}p^+$ is the light-cone fraction of the struck nucleon, $\Gamma_s$ are the remaining kinematic degrees of freedom of spectator system, and $\rho_N(\alpha_s,\Gamma_s)$ is the light-front decay function of the $^3$He nucleus which gives the distribution of these kinematic variables (described in Ref.~\cite{friscic2021neutron}). 
Inserting these formulae into the DIS cross-section formula and removing the convolution, we arrive at the following cross-section differential in the spectator kinematics:

%\begin{equation}
 %\begin{split}
 \begin{multline}
    \frac{d\sigma}{dx_Bdy\frac{d\alpha d\Gamma_s}{\alpha}} = \sum_{N=p,n}
    \frac{4\pi\alpha_{em}^2}{x_ByQ^2} \rho_N(\alpha,\Gamma_s) \times \\
     \left[
   \left(1-y-\frac{x_B^2y^2m_N^2}{Q^2}\right)F_{2N}%\left(\frac{x_B}{\alpha},Q^2\right)
    +y^2\frac{x_B}{\alpha}F_{1N}%\left(\frac{x_B}{\alpha},Q^2\right)
   \right]
 %\end{split}
%\end{equation}
\end{multline}

\begin{figure}[t]
    \centering

    \includegraphics[width = 0.9\linewidth]{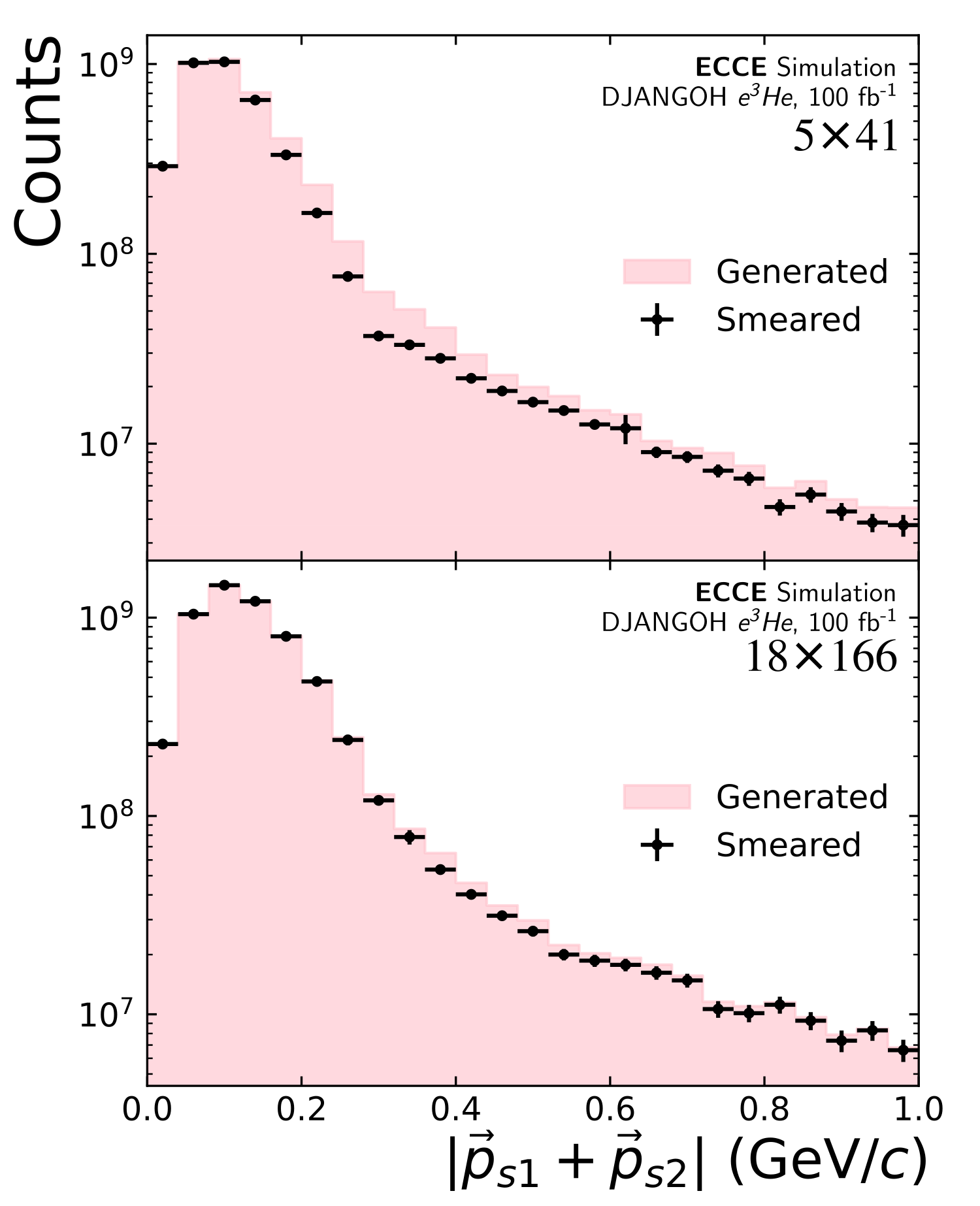}    
    \caption{Distribution of the momentum vector sum of two spectator protons, $\vec{p}_{s1}$ and $\vec{p}_{s2}$, in the ion rest frame, for $5 \times 41$ (top) and $18 \times 166$ (bottom).}
    \label{pmiss-doubletag}
\end{figure}

Here, $F_{iN}=F_{iN}\left(\frac{x_B}{\alpha},Q^2\right)$ , $i=1$, $2$.

%Dividing this differential cross-section by the cross-section differential only in inclusive variables $x_B$ and $y$ gives a distribution for the nuclear kinematic variables as a function of inclusive kinematics for a given nucleon:

For a given nucleon, the inclusive differential cross-section in terms of variables $x_B$ and $y$ gives a distribution 
\begin{multline}
    P(\alpha,\Gamma_s|x_B,y)=\\
    \frac{\left[
    \left(1-y-\frac{x_B^2y^2m_N^2}{Q^2}\right)F_{2N}%\left(\frac{x_B}{\alpha},Q^2\right)
    +y^2\frac{x_B}{\alpha}F_{1N}%\left(\frac{x_B}{\alpha},Q^2\right)
    \right]
    \rho_N(\alpha,\Gamma_s)}
    {\int \frac{d\alpha d\Gamma_s}{\alpha} \left[
    \left(1-y-\frac{x_B^2y^2m_N^2}{Q^2}\right)F_{2N}%\left(\frac{x_B}{\alpha},Q^2\right)
    +y^2\frac{x_B}{\alpha}F_{1N}%\left(\frac{x_B}{\alpha},Q^2\right)
    \right]
    \rho_N(\alpha,\Gamma_s)}
\end{multline}

This distribution was applied by sampling the light-front decay function and is included in the event-by-event weighting factors:

\begin{multline}
    w=
    \frac{\left[
    \left(1-y-\frac{x_B^2y^2m_N^2}{Q^2}\right)F_{2N}%\left(\frac{x_B}{\alpha},Q^2\right)
    +y^2\frac{x_B}{\alpha}F_{1N}%\left(\frac{x_B}{\alpha},Q^2\right)
    \right]}
    {\int \frac{d\alpha d\Gamma_s}{\alpha} \left[
    \left(1-y-\frac{x_B^2y^2m_N^2}{Q^2}\right)F_{2N}%\left(\frac{x_B}{\alpha},Q^2\right)
    +y^2\frac{x_B}{\alpha}F_{1N}%\left(\frac{x_B}{\alpha},Q^2\right)
    \right]
    \rho_N(\alpha,\Gamma_s)}
\end{multline}

The $F_{1N}$ and $F_{2N}$ models used in calculating these weights were provided by Ref.~\cite{Accardi:2016qay,Accardiprivate}.

%=====Kinematic plots and the occupancy plots=======%
\subsubsection{Event selection}
The full simulation framework, Fun4All, was used to process the generated event samples to account for the detector acceptance effects. For each EIC energy setting (eN: $5 \times 41$ and $18 \times 166$), a sample of 1M events was generated for each Bjorken-x region ($x_B<0.2$, $x_B>0.2$ and $x_B>0.5$). These samples are scaled by their corresponding normalization factors and combined to provide 3M events for the full $x_B$ range. The output of Fun4All is used as pseudo data for analysis. In this study we will select two different event samples, inclusive and tagging using selection cuts as below:

\subsubsection*{Inclusive sample e$^3$He(e,e')}
The event selection cuts were applied to the scattering electron leptonic reconstructed variables:

\begin{itemize}
    \item $E_{e}^{'} > 2$ GeV, $\eta_{e} > -3.5$
    \item $Q^2 > 2$ GeV$^2$
    \item $W^2 > 10$ GeV$^2$  
    \item $0.05 < y < 0.95$
\end{itemize}

\subsubsection*{Double tagging sample e$^3$He(e,e'$p_{1s},p_{2s}$)}

\begin{figure}
    \centering
    \includegraphics[width = 0.98\linewidth]{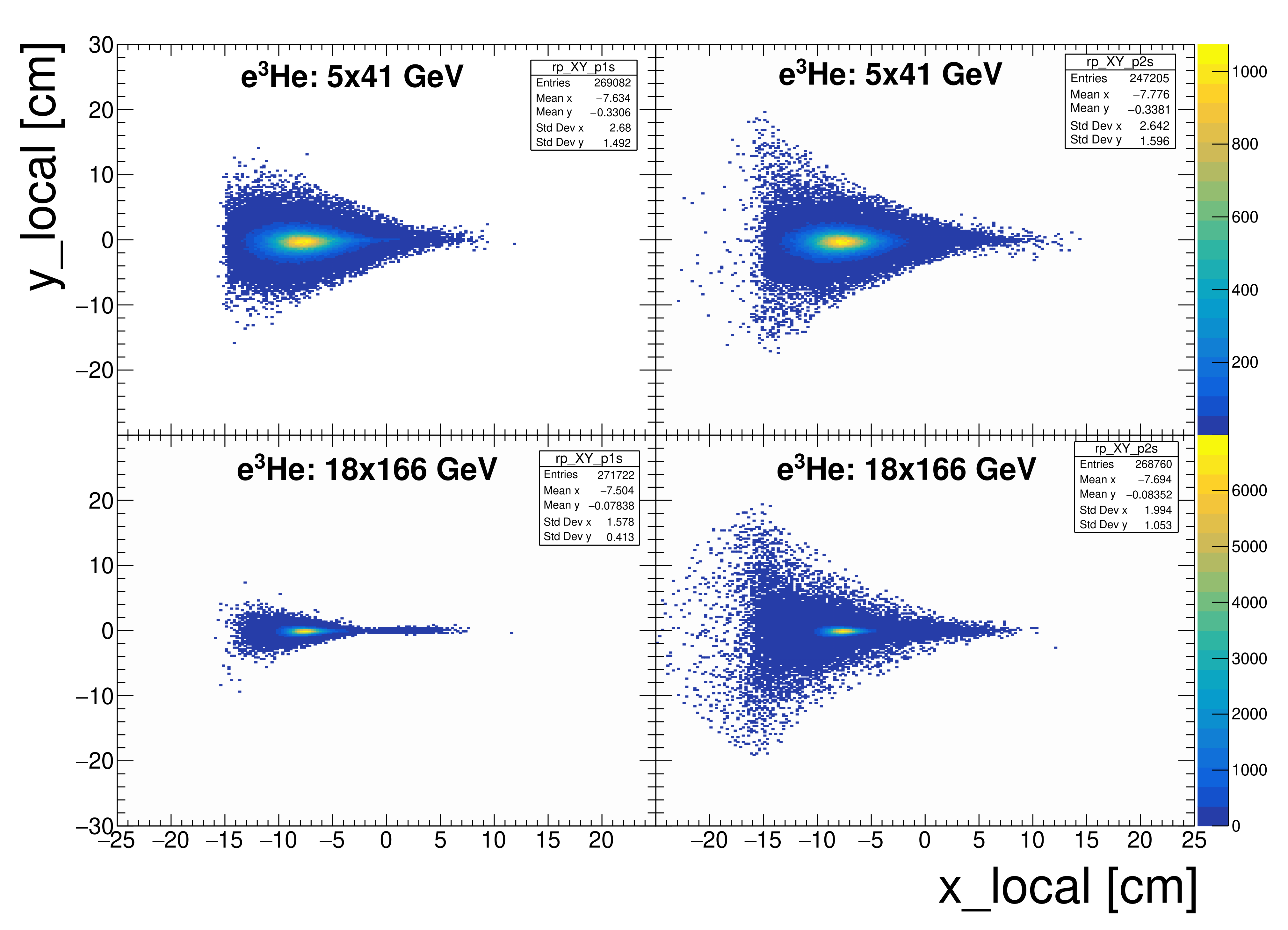}    
    \caption{The Roman Pot occupancy layer 1 for spectator proton 1 (left) and spectator proton 2 (right) for the double tagging events}
    \label{rp_doubletag}
\end{figure}

 In addition to inclusive cuts, the tagging sample requires two spectator protons to be detected. In order to identify the double tagging event, we use the hit information from the Roman Pot. Only the first layer was considered in the selection cuts. The occupancy plots for each spectator proton for two energy settings were shown in Fig.~\ref{rp_doubletag}. First, we require both spectator protons to have a hit on the first layer and the hit's local position to satisfy the condition: $-12.5 < x < 12.5$ cm and $-5 < y < 5$ cm. In addition, the beam contribution is excluded using the cut $-5 < x < 5$ cm and $-1 <y < 1$ cm.

 After the double tagging events are identified in the collider frame, the 4-vectors of two spectator protons are boosted to the ion rest frame, and their total momentum ($|\vec p_{s1}+\vec p_{s2}|$) as shown in the Fig.~\ref{pmiss-doubletag}. A cut of $|\vec p_{s1}+\vec p_{s2}|<0.1$ GeV was placed to ensure minimal nuclear effects, where $\vec p_{si}$ is the 3-momentum of spectator proton $i$. Due to the state of the far-forward reconstruction at the time of this study, we used only the truth information (directly from Geant4 simulation without smearing) of the far-forward protons.

%==========Extraction section =========%
%=======Uncertainty calculation and physics plots======% 
\subsubsection{Extracted $A_{1}^{n}$ vs double tagged $A^{n}_{1}$}

Uncertainties were calculated on $A^{n}$ given both extraction from $A_{1}^{^3\text{He}}$ and measurement via double-spectator tagging. The uncertainties were calculated given the estimated yields from the DJANGOH event samples, as well as systematic uncertainties in the case of inclusive extraction. Events were binned in $x_B$ and $Q^2$ and unfolded to Born-level from reconstructed values. The unfolding procedure was completed using a 4-iteration Bayesian unfolding algorithm using the RooUnfold \cite{roounfold} framework, trained with the Born-level and reconstructed values in the data; this means that the impact of unfolding was to increase the uncertainty, but to perfectly reconstruct the Born-level values. The effects of unfolding are reflected in the yield uncertainties in each kinematic bin.

We compare the uncertainty of extracted $A^{n}_{1}$ from $A_{1}^{^3\text{He}}$, and directly measured $A^{n}_{1}$, using the double spectator tagging measurements. In a simple approximation, the relation between $A_{1}^{n}$ and $A_1^{^3\text{He}}$ can be expressed as: 
\begin{equation}
    A_{1}^{^3\text{He}} = P_n\frac{F_2^n}{F_2^{^3\text{He}}}A_1^n + 2 P_{p} \frac{F_2^p}{F_2^{^3\text{He}}} A_1^p,
\label{a1he}   
\end{equation}
Eq.~\ref{a1he} is used to calculate the prediction value for inclusive $A_{1}^{^3\text{He}}$ where:

\begin{itemize}
\item The values of $A_1^n$ and $A_1^p$ are taken from the parameterization provided in~\cite{E99117-PRC}. The uncertainties and the correlation matrix associated with $A_1^n$ and $A_1^p$ parameterization have also been obtained from~\cite{E99117-PRC}. 

\item The structure functions $F_2^p$ and $F_2^\text{D}$ are taken from the world data fit NMC E155 \cite{Arneodo:1995cq}. The larger of the asymmetric uncertainties is chosen as the symmetric uncertainty for these structure functions.

\item Assuming no off-shell or nuclear-motion corrections, the value of $F_2^n$ is obtained using $F_2^n = F_2^\text{D} - F_2^p$. Similarly, $F_2^{^3\text{He}}$ is obtained by using $F_2^{^3\text{He}} = F_2^\text{D} + F_2^p$. The uncertainties of $F_2^n$ and $F_2^{^3\text{He}}$ are propagated from the uncertainties of $F_2^\text{D}$ and $F_2^p$.

\item The effective polarization of neutron and proton are $P_{n} = 0.86\pm 0.02$ and $P_{p} = -0.028\pm 0.004$ taken from~\cite{Bissey:2001cw}. 
\end{itemize}

Experimentally, the virtual photon asymmetry $A_1$ can be extracted from the measured longitudinal electron asymmetry $A_{||}$ and transverse electron asymmetry $A_{\perp}$, where 

\begin{equation*}
    A_{||}= \frac{\sigma_{\downarrow\Uparrow} 
    - \sigma_{\uparrow\Uparrow}}
    {\sigma_{\downarrow\Uparrow}
   + \sigma_{\uparrow\Uparrow}}
     \quad \text{and} \quad
    A_{\perp} = \frac{\sigma_{\downarrow\Rightarrow} - \sigma_{\uparrow\Rightarrow}}{\sigma_{\downarrow\Rightarrow} + \sigma_{\uparrow\Rightarrow}}.
\end{equation*}

Considering electromagnetic interaction only,
$\sigma_{\downarrow\Uparrow}(\sigma_{\uparrow\Uparrow})$ is the cross-section of the electron spin anti-parallel (parallel) to beam direction scatter off the longitudinally polarized target. $\sigma_{\downarrow\Rightarrow} (\sigma_{\uparrow\Rightarrow})$ is the cross-section of the electron spin anti-parallel (parallel) scatter off the transversely polarized target. The relation between $A_1$, $A_{||}$ and $A_{\perp}$ is 

\begin{equation}
A_{1}=\frac{A_{\|}}{D(1+\eta \xi)}-\frac{\eta A_{\perp}}{d(1+\eta \xi)},
\label{A1_formula}
\end{equation}
where $D = {y(2-y)(2+\gamma^2y)}/(2(1+\gamma^2)y^2 + (4(1-y) - \gamma^2y^2)(1+R))$,$\gamma = 4M^2x^2/Q^2$, $d= \sqrt{4(1-y)-\gamma^2y^2}D/(2-y)$, $\eta = \gamma(4(1-y) - \gamma^2y^2)/(2-y)/(2+\gamma^2y)$, $\xi = \gamma(2-y)/(2 + \gamma^2y)$,  \cite{Sato:2016tuz, E99117-PRC} and $R$ is the ratio of the longitudinal and transverse virtual photon absorption cross sections $\sigma_{L} / \sigma_{T}$ \cite{Feynman:1973xc}. The world fit parameters in Ref.\cite{E143:1998nvx} are used to calculate the value of $R$. 

\begin{figure*}[t]
    \centering
    \includegraphics[width = 0.49 \linewidth]{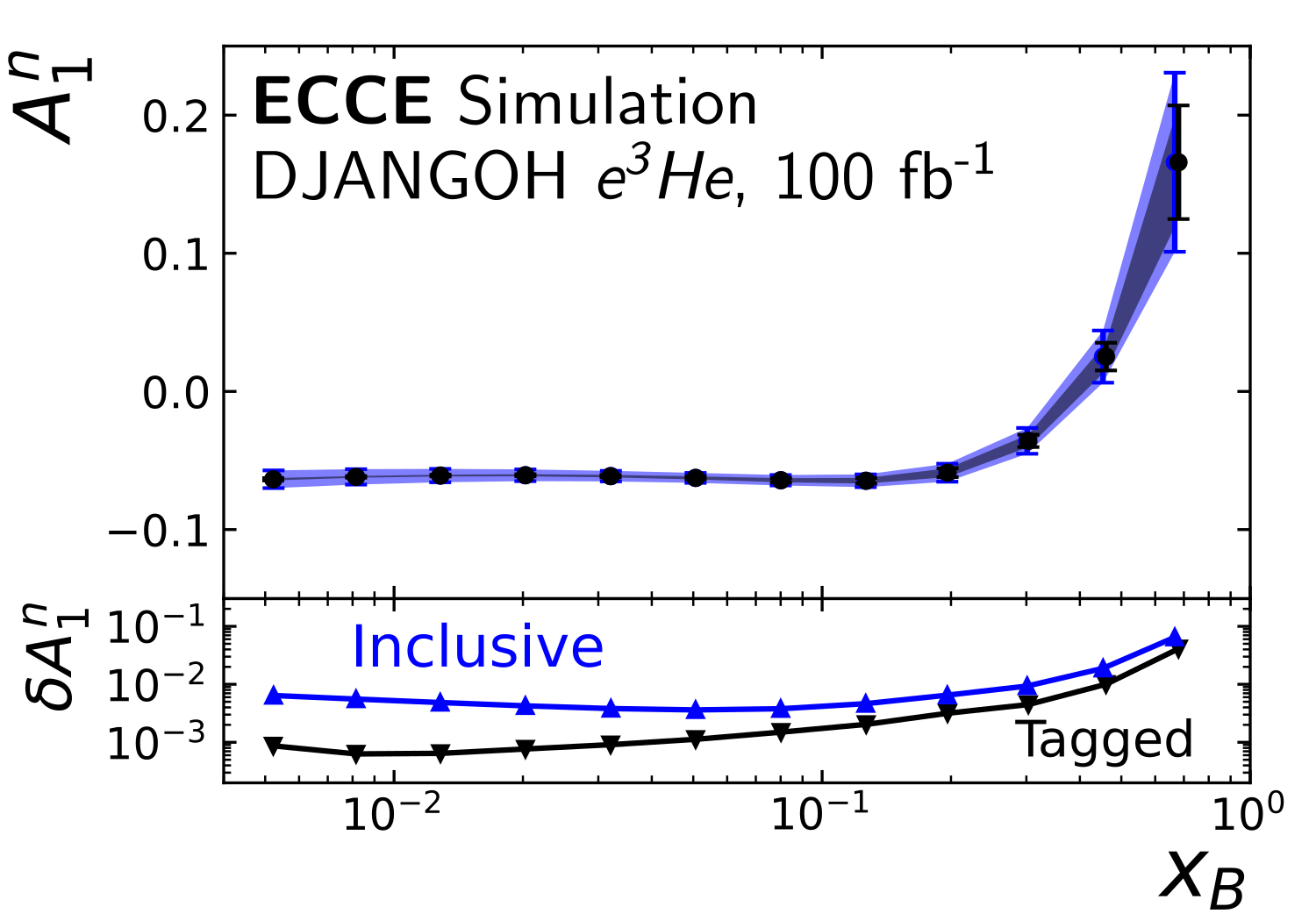}
    \includegraphics[width = 0.49 \linewidth]{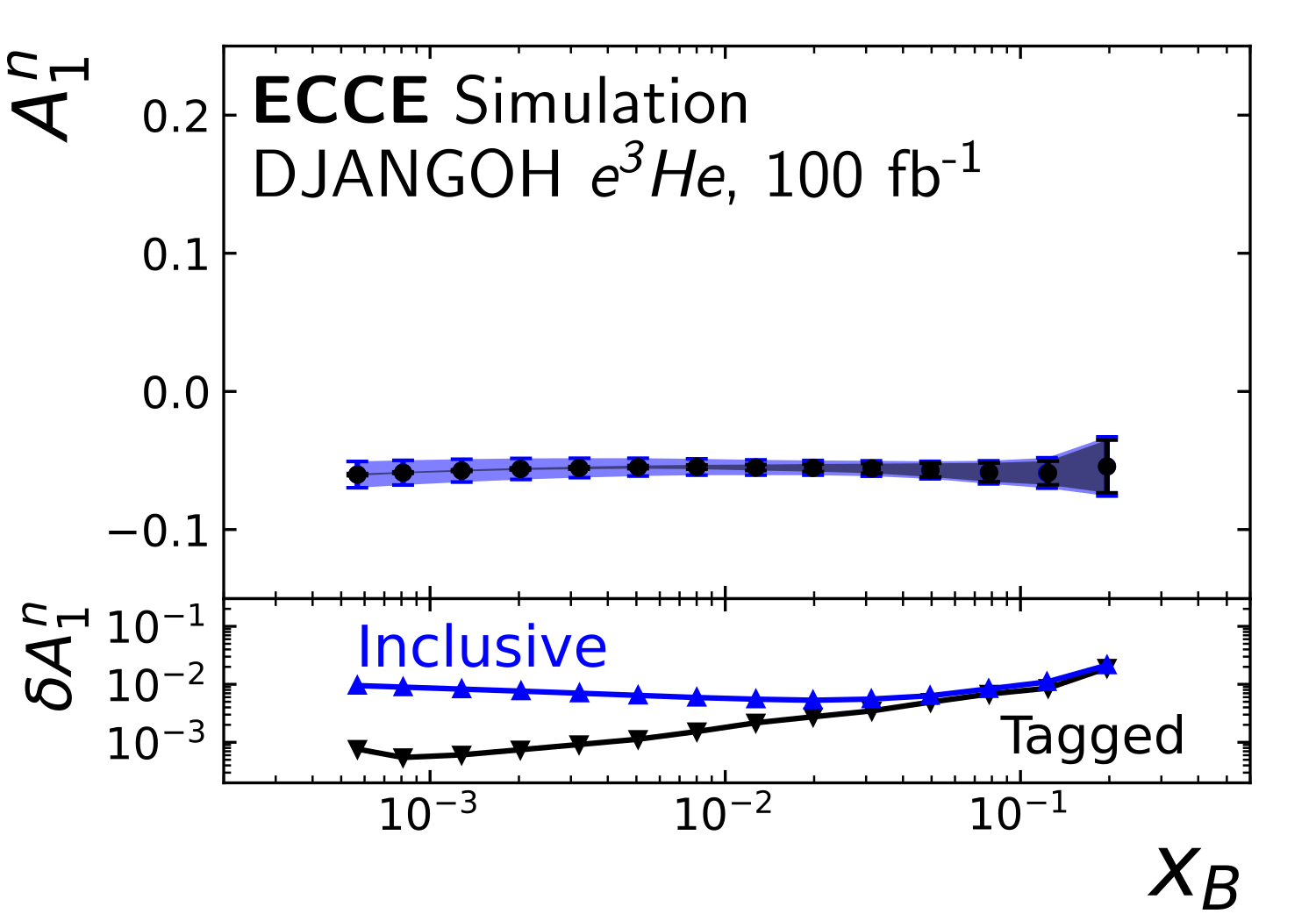}
    \caption{A direct comparison of extracted $A^n_{1}$ from inclusive measurements (blue band) and double tagging measurements (black square) which are superimposed on the blue band.  The left plot is for beam energy setting 5$\times$41 and the right plot is for 18$\times$166. The blue points are the $A_1^{^3\text{He}}$ measured values from inclusive measurements from which the blue band is extracted. The uncertainties for both techniques are compared in the bottom box where the blue (black inverted) triangles are the absolute uncertainties of inclusive (tagged) measurements. The data points were located at the average value for each $x_B$ bin. The asymmetry calculation for each data point corresponds to the average value of $Q^2$ for each $x_B$ bin.}
    \label{proj-doubletag}
\end{figure*}
%\fi

The $A_{1}^{n}$ extraction from $A_{1}^{^3\text{He}}$ follows the below procedure:

\subsubsection*{Inclusive $A_{1}^{^3\text{He}}$}

The number of DIS e$^3$He$(e,e')$ events passing the selection cuts were binned in $x_{B}$ and normalized up to the EIC total luminosity. Assuming that we will measure $A_{||}$ and $A_{\perp}$ using the same luminosity, 100 fb$^{-1}$, the statistical uncertainty can be defined as:

\begin{equation}
    \delta A_{||, \perp}^{^3\text{He}} = \frac{\delta N}{NP_{e}P_{N}},
\label{abs-unc}    
\end{equation}
where $N$ is the number of events for a given bin after normalization, and $\delta N$ is the uncertainty on the number of counts. and $P_{e}$ and $P_{N}$ are the polarization of the electron and ion beam respectively, both taken to be $70 \pm 1$\% as stated in the YR. $\delta N$ reflects the inflation of uncertainty related to the unfolding (during the reconstruction to the Born-level values). The $\delta A_{1}^{^3\text{He}}$ is the propagation uncertainty of $\delta A_{||, \perp}^{^3\text{He}}$ through Eq.~\ref{A1_formula}.
The prediction values for $A_{1}^{^3\text{He}}$ for each $x_{B}$ bin are calculated using Eq.~\ref{A1_formula} at the average values of $x_{B}$ and $Q^2$ for that given bin.

\subsubsection*{Inclusive extracted $A_{1}^{n}$}

For each $x_{B}$ bin, using the obtained value of inclusive $A_{1}^{^3\text{He}}$ from previous step, $A_1^{p}$ from~\cite{E99117-PRC}, $F_2^p$ and $F_2^\text{D}$ from fit NMC E155~\cite{Arneodo:1995cq}, and $P_{p(n)}$ from~\cite{Bissey:2001cw}, we extract $A_1^{n}$ using Eq.~\ref{A1_formula}. The total uncertainty of extracted $A_{1}^{n}$ is propagated from statistical uncertainty of $A_{1}^{^3\text{He}}$ and systematic uncertainty from $A_1^p$, $F_2^{n}$, $F_2^\text{D}$ and $P_{p(n)}$.

\subsubsection*{Double tagging $A_{1}^n$}

The double-tagging sample was binned in the same way as the inclusive sample and normalized to the same total luminosity. Also assuming $A_{||}^n$ and $A_{\perp}^n$ are measured with the same total luminosity, the statistical uncertainties $\delta A_{||, \perp}^n$ can be calculated similarly as Eq.~\ref{abs-unc}. The total uncertainty of double tagging $A_{1}^n$ is propagated from the $\delta A_{||, \perp}^n$ using Eq.~\ref{A1_formula}.
%\fi

\subsubsection{Projections and impacts}
We show the direct comparison of uncertainty from double tagging to the extracted $A_{1}^{n}$ for two energy settings (5$\times$41 and 18$\times$166) in Fig.~\ref{proj-doubletag}. The study shows that the double tagging method results in reduced uncertainties by a factor of $2$ on the extracted neutron spin asymmetries for overall kinematics, and by a factor of $10$ in the low-$x_B$ region for energy setting 5x41. 

\begin{figure}[t]
  \centering
  \includegraphics[width = 0.95 \linewidth]{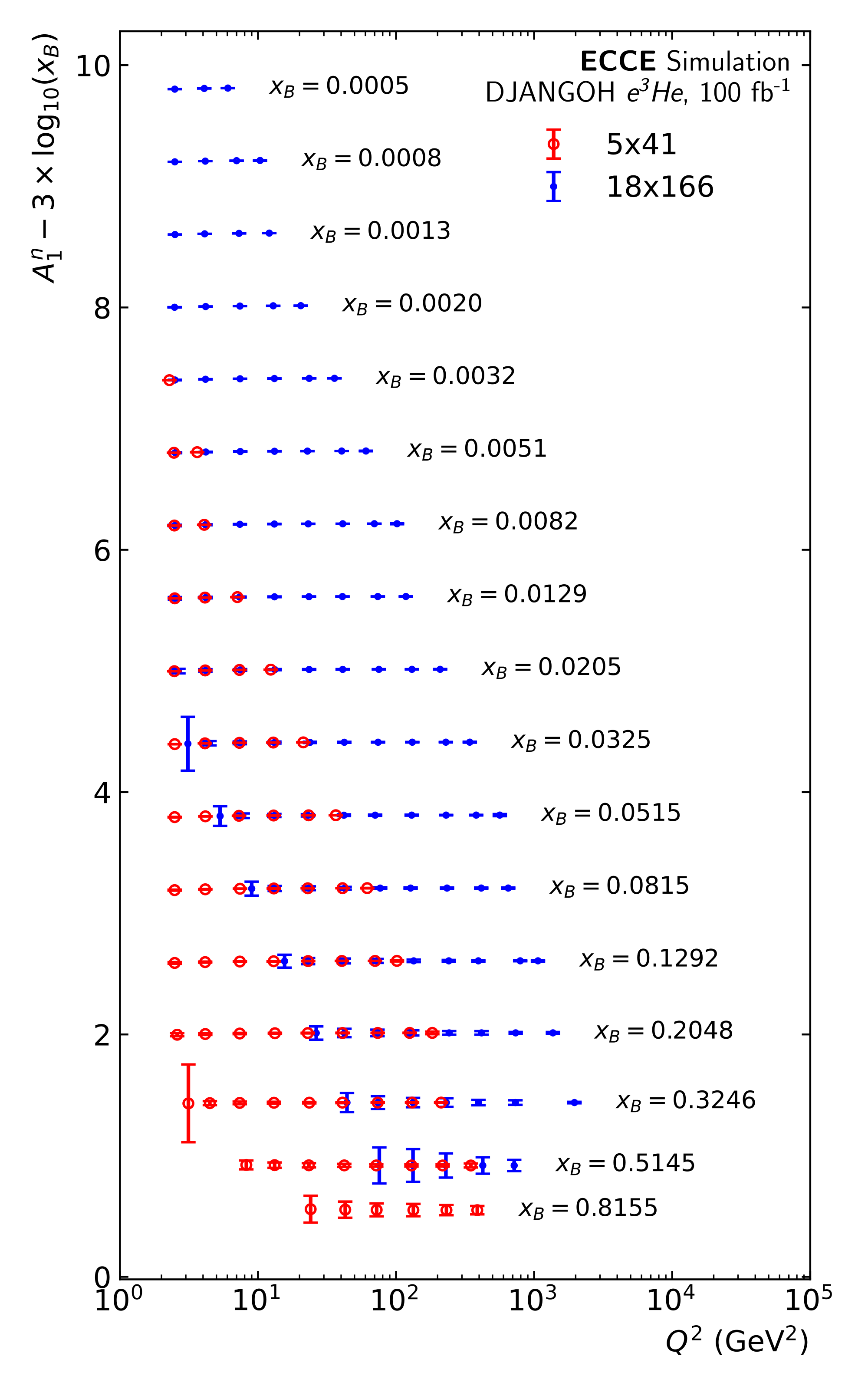}
  \caption{The EIC kinematic coverage of the neutron asymmetry $A_1^n$ as a functions of $x_B$ and $Q^{2}$ for two electron-nucleon energy settings 5$\times$41 and 18$\times$166.}
  \label{a1-global}
\end{figure}

The EIC coverage of $A_{1}^n$ as a function of $x_{B}$ and $Q^2$ is shown in Fig.~\ref{a1-global}. These new data points cover a previously unreachable kinematic region, especially for neutron spin structure function study. This provides valuable input for the polarized parton distribution global fit and the flavor separation. In addition, the overlap in the moderated $x_B$ region with much higher $Q^{2}$ compared to existing fixed target data will be the perfect place to test the nuclear correction that has been used to extract the neutron information.

\subsection{DVCS}
\label{subsec:dvcs}
\begin{figure}[ht]
\includegraphics[width=0.45\textwidth]{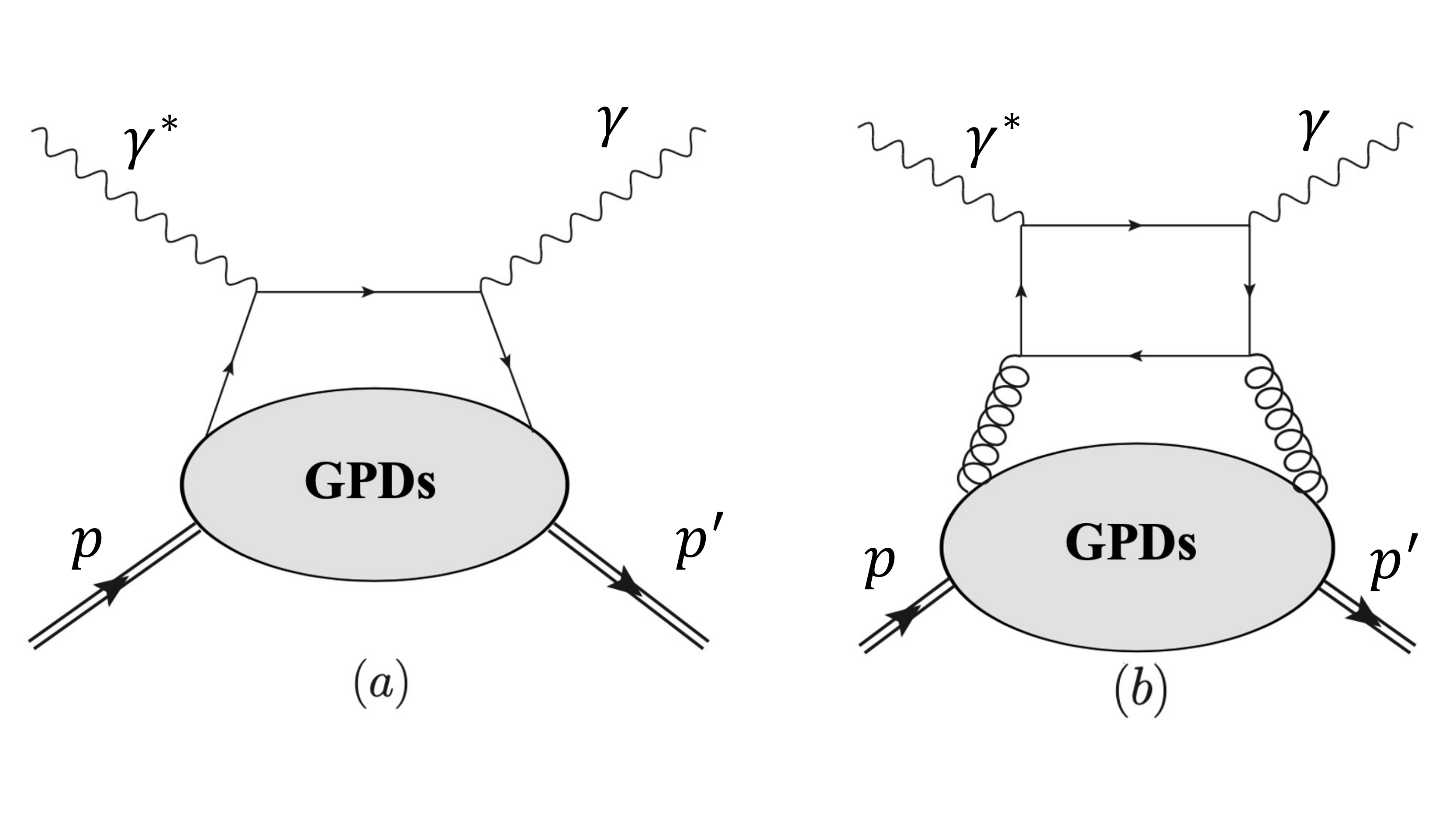}
\caption{\label{fig:FeimanDVCS} Feynman diagrams for the $ep$-DVCS process. (a) Quark and (b) Gluon contributions to GPDs.}
%\end{figure}
%\begin{figure}[ht]
\includegraphics[width=0.45\textwidth]{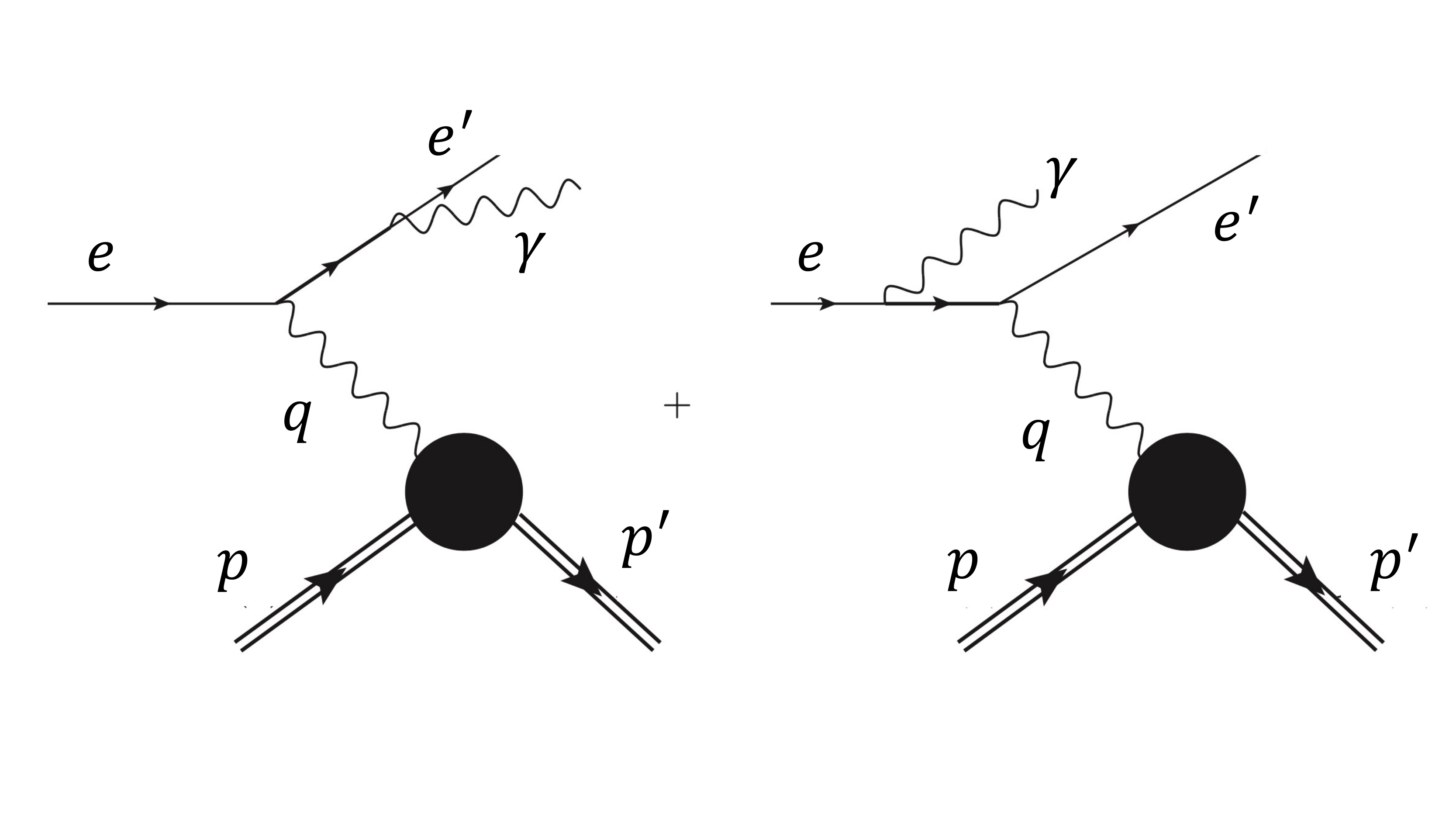}
\caption{\label{fig:FeimanBH} Feynman diagrams for Bethe-Heitler process, where the final state particles are identical to $ep$-DVCS.  }
\end{figure}

%\subsection{$ep$-DVCS}
Deeply Virtual Compton Scattering (DVCS), $(ep,e^{\prime}p\gamma)$, provides an excellent tool to study the Generalized Parton Distributions (GPDs) of the proton, Fig.~\ref{fig:FeimanDVCS}, and the three-dimensional structure of the nucleon. These non-perturbative quantities encode the correlated momentum and spatial distributions of the quark and gluons within the proton. In addition, these important quantities offer a unique opportunity to probe the energy-momentum tensor and thus open the door to deepening our understanding of the nucleon mass. 

Current knowledge of GPDs from DVCS is mainly based on data from fixed target experiments from JLab at high $x_B$, and the HERA collider at low $x$. EIC offers a unique opportunity in kinematics coverage which will create a linkage between the JLAB and HERA data, $ep$-DVCS was labeled as one of the future flagship measurements and was described extensively in the YR.

In this work, we estimated the feasibility of the ECCE detector for measuring $ep$ DVCS without addressing the separation between DVCS and pure Bethe-Heitler (BH) scattering. Both DVCS and BH have the same final states (see Fig.~\ref{fig:FeimanBH}), and both processes contribute to the total cross-section. This separation will be reported elsewhere, in the future. Additional background can originate from deeply virtual $\pi^{\circ}$ production. However, due to the high photon acceptance and resolution of the barrel and forward endcap calorimeters, this background is expected to play a minor role.

The $ep$-DVCS study used the MILOU3D generator~\cite{url:MILOU3D}~\cite{PEREZ04} at three beam energy configuration: 5$\times$41, 10$\times$100 and 18$\times$275. This generator was selected based on the existing YR studies and was selected in order to assess a comparison of the ECCE detector versus the original YR conceptual detector performance. However, this generator is not suitable for beam spin asymmetry studies, because it integrates the angular dependence between the leptonic and hadronic planes. Therefore, we have not reported on those studies here, but we are currently looking at other event generators for these purposes and will also report on that in the future.

Parameters input to the MILOU3D code for event generation were based on the settings used for the YR studies. We performed simulations for three beam energies 5x41, 10x100, and 18x275. For all three setups, we limited the kinematical range to:
\begin{itemize}
\item $1 < Q^{2} < 1000.0 $ (GeV)$^2$,
\item $10^{-5}<x_{B}<0.7$,
\item $-2.0 < t < -0.01 $  (GeV)$^{2}$.
\end{itemize}
For each setup, 500\,000 events were generated and used as an input to the ECCE Fun4All detector simulation. The particle kinematics generated by the MILOU3D software are described in the head-on (i.e. center of mass) frame. The Fun4All software took the head-on kinematics from the MILOU3D input files, applied the beam crossing effects (as described elsewhere), and then propagated each particle through the realistic ECCE detector.

In exclusive $ep$-DVCS measurements, the goal is to detect all three emerging particles, the electron, proton, and real photon. In all three kinematics studied, electrons and real photons are measured in the central detector (pseudorapidity $ -3.5<\eta<3.5$ ), while the scattered protons escape through the beam pipe opening in the hadronic endcap, and thus their detection is required in the far-forward region ($\eta>3.5$) described in Sec.~\ref{sec:ECCEDet}.

\subsubsection{Event Selection}
The starting point of this study was to reconstruct the detected electrons. This was done using the track reconstruction algorithm that was used in Fun4All software (SvtxTrackEval). To identify the scattered electron, we required a track multiplicity equal to 1 (almost 100\% of events) and the particle charge to be negative ($\sim$ 99.5\% of events). For most of the kinematical region, this selection technique yielded excellent performance, except at the geometric edges ($\sim$15\% of the events were lost). Most of this inefficiency can be recovered by the use of calorimeters for electron reconstruction (this was outside the scope of the work reported here, and again will be studied in the future).

In the second step of the DVCS event selection, the real photons were reconstructed. Their reconstruction was based on the identification of clusters in the electromagnetic calorimeters (EEMC, BECAL, and FEMC). The reconstructed photon energy was based on the energy deposition in the cluster. For this, the total momentum of the reconstructed photon was calibrated using the energy of the ``truth'' photon information, where the truth momentum is the known momentum of the particle from the MILOU3D generator output. The photon direction was reconstructed based on the electron vertex and the position of the cluster.

Finally, the scattered protons that elude the central barrel were detected using either the B0 detector or the Roman Pots in the far-forward region. As described in Sec.~\ref{sec:ECCEDet}, a realistic geometry of the B0 detector was encoded in the simulation, allowing for accurate modeling of the geometric acceptance directly in the Fun4All simulation. However, the Roman Pot beam pipe cutout was not included in the Fun4All software. Hits in the B0 were therefore selected directly, based on which layers were hit first per event. Geometric cuts of $\pm$5 cm in $x$ (detector horizontal plane) and $\pm$1 cm in $y$ (detector horizontal plane) were applied to the center of both Roman Pots in the analysis of the Fun4All output, to remove events which would have otherwise been lost down the beam pipe into the beam dump. For the results shown here, the analysis used ``truth'' momentum values, as currently there was no reconstruction of momentum from the far forward detectors. For each hit in the B0 detector planes or Roman Pots, the Geant4 particle ID was used to select the detected protons. To simulate the expected level of response of the detector, the ``truth'' momentum of the detected protons was smeared by 1\%. For these studies, position resolution effects were not studied and the proton directions were kept intact. This smearing level was selected as it is consistent with the proposed detector technology, AC-LGAD, and its expected segmentation.

\subsubsection{Results}
The results shown here present the acceptances of $ep$-DVCS photons and protons, which enabled us to assess the accessible $-t$ range with the ECCE detector, required for nucleon imaging purposes. The uncertainties shown in this study are only statistical for 10 fb$^{-1}$ integrated luminosity. The resulting projected differential cross-section measurements are also given. In the case of $ep$-DVCS, the $-t$ variable can be calculated using two different methods. The first one is based solely on reconstruction from $e^\prime + \gamma$, while the second corresponds to the more standard definition, which is $t = (p - p^{\prime})^2$. During the study, both methods gave comparable results. We chose to complete the study with the latter method because the former is subject to significant radiation correction which is poorly understood at the current stage (larger uncertainty at certain kinematics regions).

\begin{figure}[t]
\centering
\includegraphics[width=0.48\textwidth]{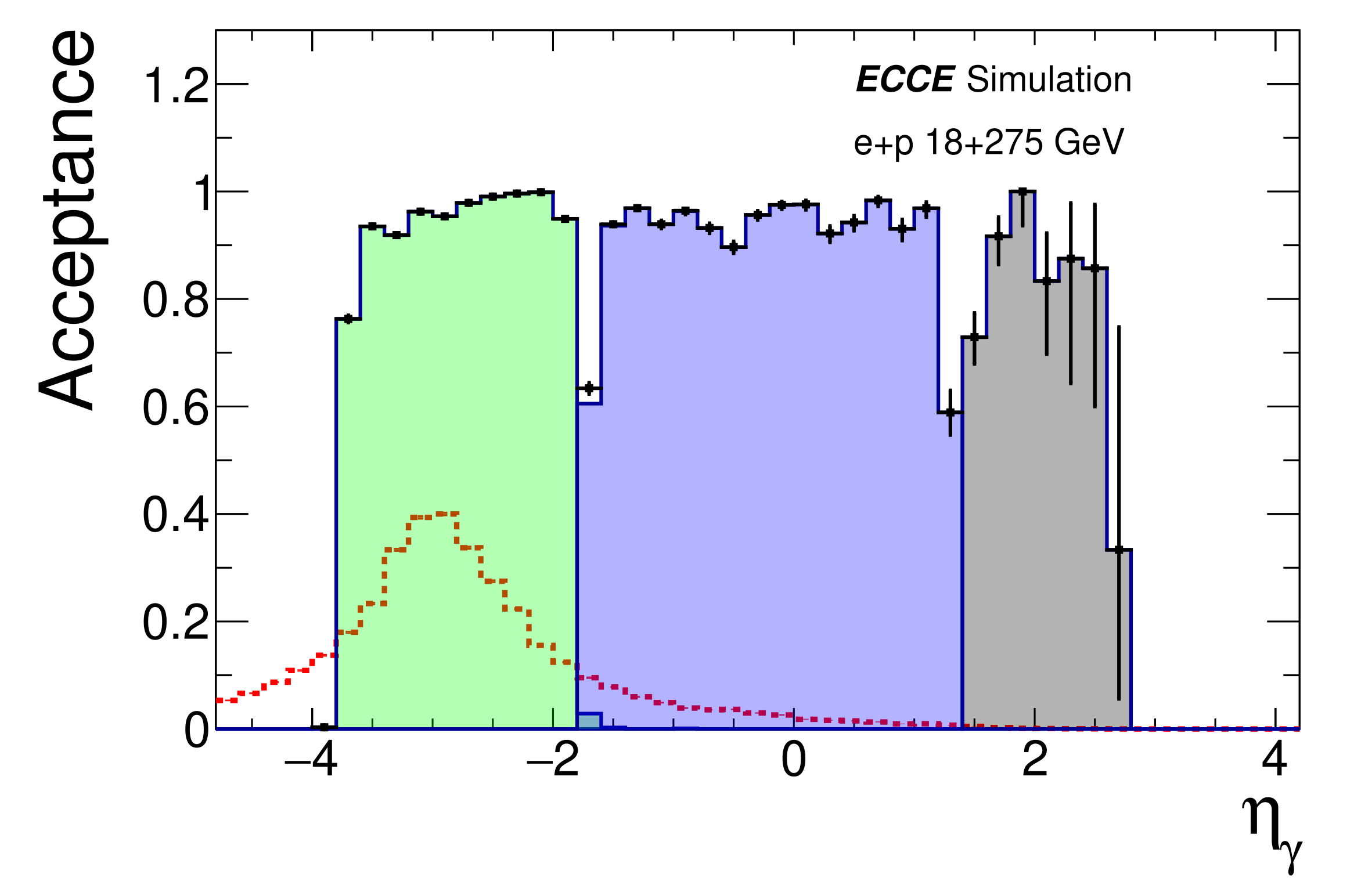}
\caption{  DVCS photon acceptance in the backward (green), barrel (blue), and forward (grey) ECAL's, as a function of pseudorapidity. The red dotted line shows the distribution of (generated) DVCS photons}
\label{fig:accptPhot}
\end{figure}

Simulation of the current detector configuration exhibits good performance for photon detection. Fig.~\ref{fig:accptPhot} presents the acceptance as a function of $\eta$ of the real photon for the highest beam setup of 18x275. The acceptance is defined as the ratio of reconstructed photons in the calorimeters to the number of generated photons in the MILOU3D generator.

\begin{figure}[ht]
\centering
\includegraphics[width=0.48\textwidth]{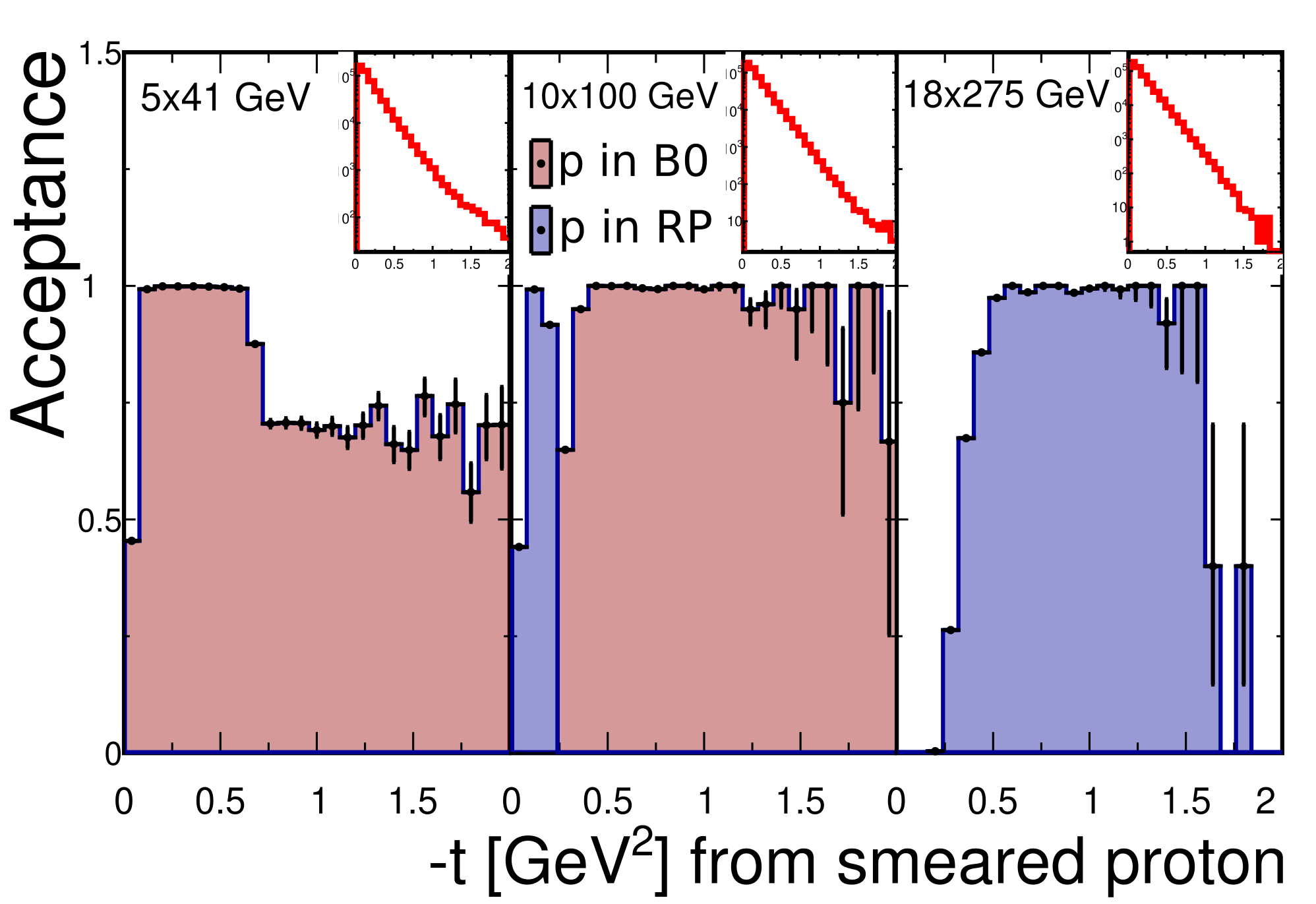}
\caption{\label{fig:accptProt}  Acceptance for DVCS protons as a function of $-t$ in the far-forward detectors for different beam energy configurations. The inserts show the $-t$ distributions of generated events.}
\end{figure}

Contrary to the photon acceptance, which exhibits similar behavior from the lowest to the highest beam configurations (the minimum energy of DVCS photons must be much higher than the detection limit of the calorimeters), in the proton case the acceptance is very sensitive to the beam energies. The recoil proton acceptances of the B0 spectrometer and Roman Pots for different energy configurations as a function of the momentum transfer to the proton $t = (p - p^{\prime})^2$, for each energy configuration studied, are shown in Fig.~\ref{fig:accptProt}. The resulting $-t$ acceptance is shown to be very wide, continuous, and extends to low-$t$. Such a wide coverage is essential for the precision extraction of the transverse position distributions of quarks and gluons inside the nucleon. It is also worth noting, that for the highest beam setup, the minimal $-t$ value is limited by the beam size and the mandatory gap between Roman Pots and the beam.

\begin{figure}[ht]
\centering
\includegraphics[width=0.48\textwidth]{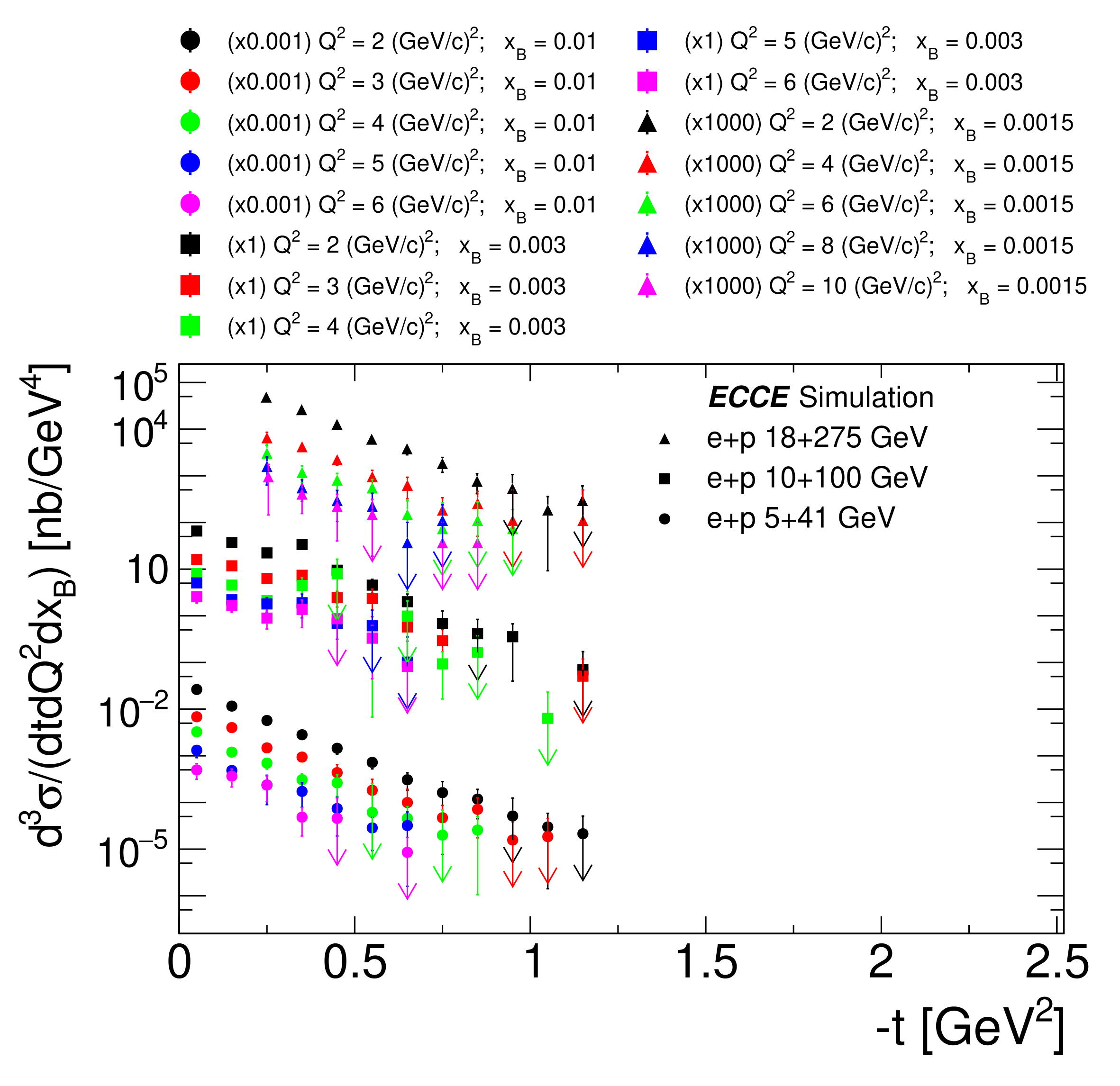}
\caption{\label{fig:crosSection}  Projected DVCS differential cross-section measurements as a function of the momentum transfer $-t$ for different bins in $Q^2$ and $x_{B}$. The assumed integrated luminosity is 10~fb $^{-1}$ for each beam energy configuration.}
\end{figure}

The full exploration of nucleon GPDs will require multi-dimensional measurements of the $ep$-DVCS differential cross-section in $Q^2$, $x_{B}$, $t$ and the azimuthal angle $\phi$ between the lepton and hadron planes in the initial hadron rest frame. Fig.~\ref{fig:crosSection} shows the projected precision and coverage of $ep$-DVCS differential cross-section measurements for several beam energy configurations and in multi-dimensional bins of $Q^{2}$, $x_{B}$ and $t$, whilst due to the aforementioned MILOU3D limitation the $\phi$ dependence is integrated. The uncertainties of the differential cross-section are based on the expected integrated luminosity of $\mathcal{L} =$10 fb$^{-1}$.

\subsubsection{Summary}
To summarize, our study shows that the ECCE detector is suitable to deliver a wide $Q^2$ and $x_{B}$ coverage for the $ep$-DVCS process with reasonable statistical uncertainties. Additional studies must be performed with a fully realistic implementation of  the far-forward region of the detector, due to the large proton acceptance sensitivity in this reaction. The Roman Pots must be sufficiently separated from the beam ($\sim10\sigma$ according to YR), in order to avoid saturation and radiation damage. This issue will be addressed in future work, through the study of different beam configurations, such as high-acceptance or high-divergence, and will be reported in future studies.

\subsection{eA DVCS}
\label{subsec:dvcsea}
%%%%%%%%%%%%%%%%%%%%%%%%%%%%%%%%%%%%%%%%%%%%%%%%%%%%%% general info stuff
This study aimed to investigate the feasibility of measuring coherent exclusive DVCS off $^{4}$He, i.e. $(e\,^{4}$He, $e^{\prime}\,^{4}$He$^{\prime}\,\gamma)$ with the ECCE detector.

Including measurements of exclusive DVCS off light nuclei at the EIC, in addition to DVCS off the proton discussed in Sec.~\ref{subsec:dvcs}, would provide access to several physics topics of interest. These topics are only named here, as further details may be found in Sec.~7.2.5 of the YR and the references therein. This reaction is thought to allow one to look in detail at the European Muon Collaboration effect in the transverse plane. As with the proton, coherent DVCS on light nuclei also allows one to extract GPDs which encode the spatial distributions of partons in the nucleon. Furthermore, with $^4$He, in particular, the separation of coherent and incoherent channels in DVCS is a recent theoretical milestone \cite{Hattawy_2017}. The EIC will be able to probe the required low x$_B$ values in wide ranges of momentum transfer $t$, offers a unique opportunity to make measurements of these topics.

\begin{figure}[h]
    \centering
    \includegraphics[width=0.85\linewidth]{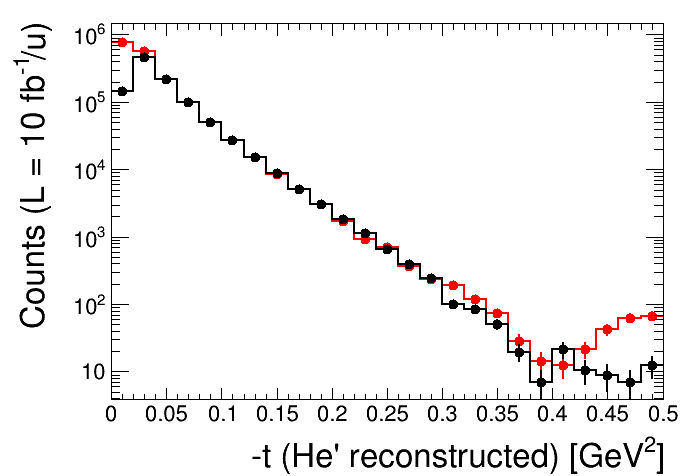}\\
    \includegraphics[width=0.85\linewidth]{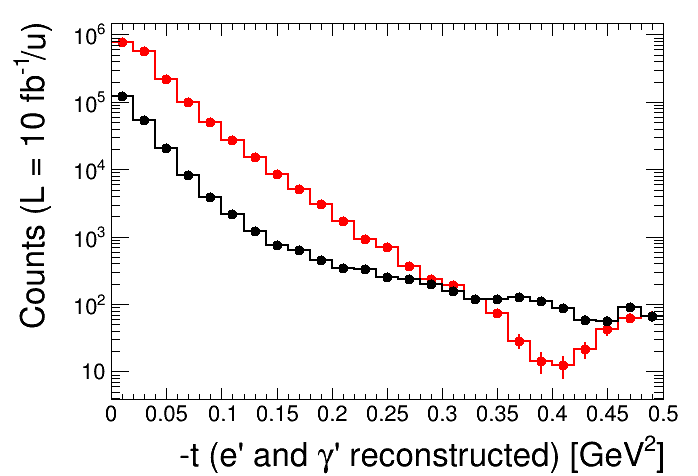}
\caption{Reconstructed (black) and generated (red) for $-t$ distributions for $(e\,^{4}$He, $e^{\prime}\,^{4}$He$^{\prime}\,\gamma)$, using different methods, as described in the body of text, and normalized to the EIC luminosity.}
\label{fig:dvcsea:tcuts}
\end{figure}

\subsubsection{Simulated Settings}
For this study, 1\,M events were generated with the TOPEG generator~\cite{TOPEG}. More information on our use of TOPEG is available in \ref{sec:dvcsea_appendix}. The nominal EIC beam configuration 5$\times$41 was used. For $^{4}$He, with four nucleons, this gives an ionic energy configuration of 5$\times$164.

The Fun4All~\cite{fun4all} simulation software was used to simulate the physics events in Geant4, using the generator data as an input. Initial results presented in the exclusive physics note of the ECCE detector proposal~\cite{ecce-note-phys-2021-03} were obtained using the prop.4 simulation build. However, the Fun4All software has developed since, as detailed in Sec.~\ref{sec:sim_framework}, and further results have been obtained. Specifically, using software build prop.7.1, this study was repeated for the high-acceptance and high-divergence beam parameterizations detailed in Sec.~\ref{sec:beam_parameter}. The most important differences in each of these simulation builds, and indeed in the feasibility of this measurement, manifests itself in the acceptance of scattered $^4$He ions in the far forward Roman Pot detectors. This is detailed in Table~\ref{tab:dvcsSIMS}.

\begin{table}[h]
    \centering
    \begin{tabular}{c|c|c|c}
        Simulation        & $x$ [cm] & $y$ [cm]  & RP Acceptance \\ \hline
        prop.4.0          & 5.0      & 1.0     & 14.4\%               \\
        prop.7.1 hi.-div. & 4.25     & 0.80    & 8.0\%                \\
        prop.7.1 hi.-acc. & 2.08     & 0.34    & 60.0\%               \\
    \end{tabular}
    \caption{10$\sigma$ cuts in x and y on RP hits (to reject the ion beam), and scattered $^{4}$He acceptance in the RP in each beam configuration (high divergence and high acceptance). The $x$ and $y$ boundary cuts are based on standard deviations of simulation beam spot widths and heights at different beam configurations at the maximum energy (18x275).} 
    \label{tab:dvcsSIMS}
\end{table}

Due to the unique charge and momentum considerations with an $^{4}$He beam, specific magnetic steering of 82 GeV\textsuperscript{2} was employed in the simulation.

%%%%%%%%%%%%%%%%%%%%%%%%%%%details of particle reconstructions and analysis
\subsubsection{Event Selection and Analysis}
\label{sec:dvcseAEventSelect}
Final event selection is exclusive with three particles: the scattered electron $e'$, the scattered $^{4}$He, $A'$, and the real photon produced by the DVCS process, $\gamma^{\prime}$. Due to obvious similarities in the channels, much of the final analysis methodology and selection is identical to the DVCS-$ep$ case described in Sec.~\ref{subsec:dvcs}. 

The electron selection is as in Sec.~\ref{subsec:dvcs}, solely using the SvtxTrackEval Fun4All container and choosing events with explicitly 1 track in this silicon tracker. All momentum information about the electron is available in the container, so reliance on truth information is the lowest for this particle. 

The photon selection is also very similar to the method described in Sec.~\ref{subsec:dvcs}. The highest energy photon in any of the three calorimeters is selected using its PID and energy in the container and assumed to match the DVCS $\gamma^{\prime}$. It is important to subsequently calibrate the response of the calorimeters. For this, the energy of the photons in each calorimeter is plotted against the true energy of the photons, and a straight line was fitted to the data. This is done separately for each calorimeter in the central detector (FEMC, EEMC, BECAL). The energy of the selected photon is then corrected using the coefficients extracted from the calibration fit. The momentum components of the photon are reconstructed using the calibrated energy and the available angular information of the track provided in the container. 

Finally, the scattered $^{4}$He is selected using hits in the B0 detector or Roman Pot detectors. For each, all hits in the container with correct Geant4 tracking IDs are selected. When the study was performed, an accurate B0 geometry had already been implemented in the simulation. However, the RP geometry was still preliminary. Hits in the B0 are therefore selected directly based on which layers are hit first per event. A geometric cut is applied to the center of both Roman Pots, to remove events that would have otherwise been lost down the beam pipe. The size of the cut is proportional to the size of the beam spot and as such is different in each simulation build. These are detailed in Table~\ref{tab:dvcsSIMS}.

Currently, the analysis uses truth momentum values for the $^{4}$He ions, as the reconstruction of momentum in the far forward detectors is limited. As in Sec.~\ref{subsec:dvcs}, a 1\% momentum smearing was applied to the scattered $^{4}$He ion to account for detector effects.
The IP8 detector configuration may offer further improved resolution in the far forward region, as well as a higher acceptance for the forward going $^{4}$He ions due to the secondary focus region. Because of this, it is planned to repeat this study for the IP8 setup in the near future.

\begin{figure*}[h]
\centering
        \includegraphics[width=0.49\linewidth]{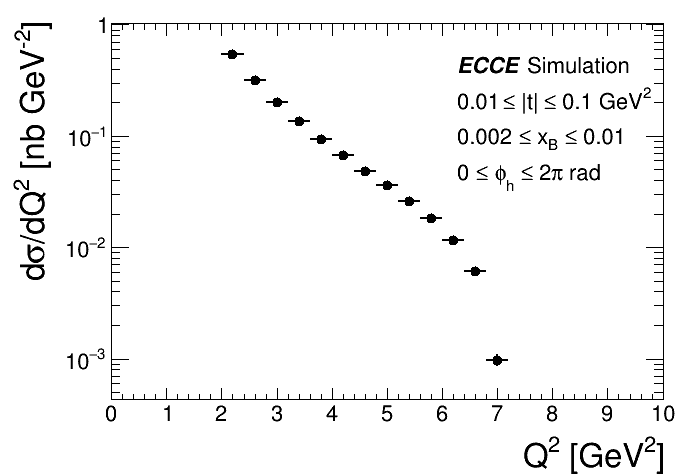}
        \includegraphics[width=0.49\linewidth]{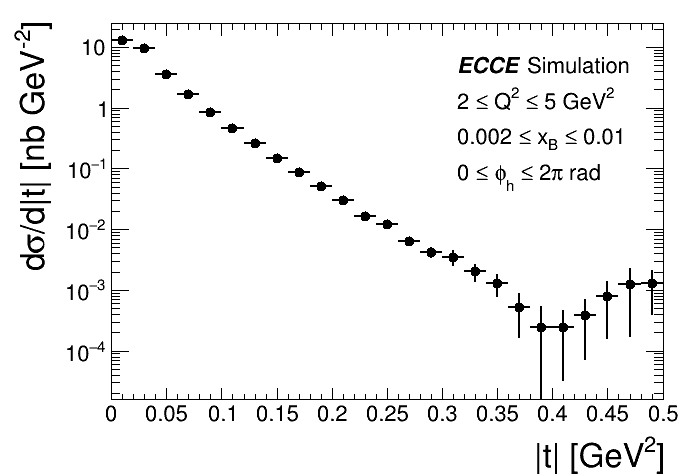} \\
        \includegraphics[width=0.49\linewidth]{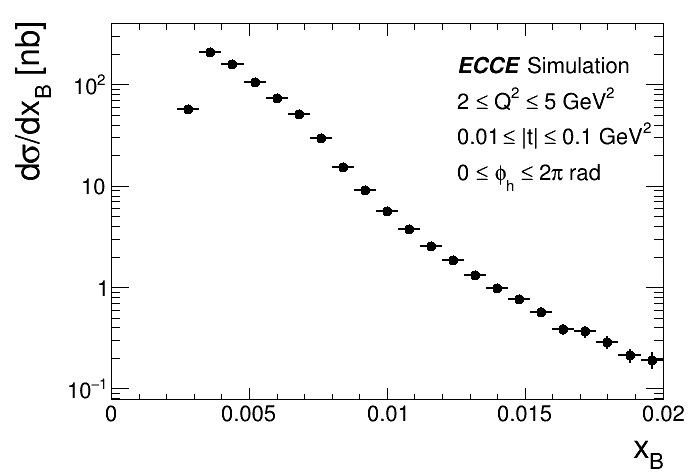} 
        \includegraphics[width=0.49\linewidth]{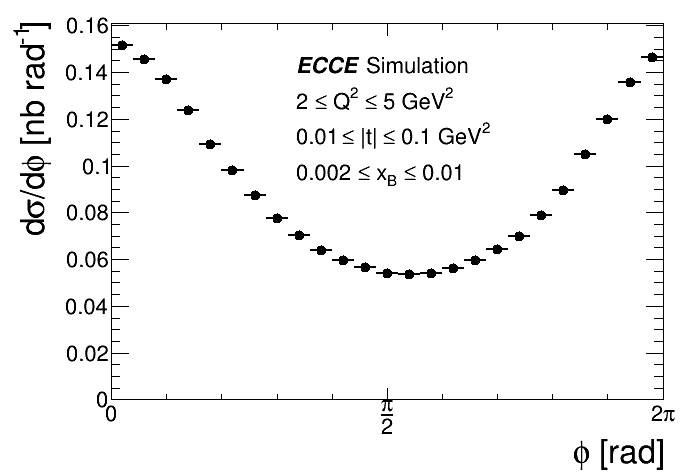}
        \caption{Projected differential cross-sections in ECCE as functions of physics variables $Q^2$, $x_B$, $-t$ and $\phi$ for DVCS-e$^{4}$He. Each plot is integrated over the phase space denoted in the legend.}
        \label{fig:dvcsea:xsecs}
\end{figure*}

The momentum transfer $t$ can be calculated using the reconstructed $^{4}$He or using only the reconstructed photon and electron as described in Eq.~\ref{eqn:dvcst}. Both reconstruction methods are shown in Fig.~\ref{fig:dvcsea:tcuts} for comparison's sake only. In the final analysis for this reaction, $t$ was reconstructed using only the scattered helium method, to avoid potential contamination of the channels cross-section with non-exclusive or incoherent events.  

\begin{equation}
t = -\frac{MQ^{2} + 2M\nu(\nu - \sqrt{\nu^{2} + Q^{2} }cos(\theta_{\gamma^{*}\gamma}))}{M + \nu - \sqrt{\nu^{2} + Q^{2} }cos(\theta_{\gamma^{*}\gamma})}
\label{eqn:dvcst}
\end{equation}

\subsubsection{Results}

%%%%%%%%%%%%%%%%%%%% Cross Section Plots %%%%%%%%%%%%%%%%%%%%

For analysis, the acceptance in each phase space bin is calculated using the reconstructed events in the bin divided by the number of events generated by TOPEG in the bin. The fourth differential of the cross-section is then given by the formula:
\begin{equation}
    \frac{d^4\sigma}{dQ^2dx_Bdtd\phi}=\frac{1}{\mathcal{L}\cdot Acc_{bin}\cdot\Delta\Omega}(N\pm\sqrt{N}),
    \label{xseccalc}
\end{equation}
where $\mathcal{L}$ is the integrated luminosity and in this case is equal to $\mathcal{L}$\textsubscript{eic}/u; Acc\textsubscript{bin} is the acceptance in the bin; $\Delta\Omega$ is the multidimensional bin width given as $\Delta\Omega = \Delta Q^{2} \Delta\ x_{B} \Delta t \Delta\phi_{h}$; and $N$ is the number of counts in the bin.

We then integrated over three-dimensional phase space, and the projected differential cross-sections are given as a function of $Q$\textsuperscript{2}, $t$ (using \textsuperscript{4}He and \textsuperscript{4}He$^\prime$ ions for $t$ reconstruction only), $x_B$ and $\phi$ (the angle between the leptonic and hadronic scattering planes). The results are shown in Fig.~\ref{fig:dvcsea:xsecs}.

\subsubsection{Analysis and Summary}
The simulation build which offered the best ion detection in the far forward region was chosen for the result presented here. From Table~\ref{tab:dvcsSIMS} it is clear that this is the high acceptance parameterization of the most recent prop.7.1 version. From a generated data sample of 10\textsuperscript{7} events, we find no hits in the B0 detector layers. A realistic beam pipe cut is implemented in the center of this detector. As such we conclude that due to the proximity to the interaction point and high pseudorapidity of deflected ions, almost all will pass through this central hole. Most simulated events can be reconstructed in the Roman Pot detectors, however, after photon and electron exclusivity cuts and the beam pipe cut on the Roman Pot geometry, we measure 607703 events. This yields a final acceptance for the IP6 high acceptance of 60.8\%.

$Q^{2}$ and $x_{B}$ acceptances are fairly high in the probed region of phase space and $t$ acceptance is non-zero across the entire region of generated space. Un-physical ($> 1$) acceptances in $t$ are within the statistical errors, and acceptance drops in the region of $<0.03$ are likely still due to forward ion acceptance

Overall, we can make the statement that these results are promising for realizing DVCS measurements on light ions at the EIC with ECCE if the design of the far forward region is fully considered. Initially, we were able to confirm the YR findings that the study would be heavily dependent on ion acceptance.
Our results with the latest Fun4All builds demonstrate that if this beam setting is realized at the EIC, large $\eta$ ions from a highly rigid beam will be detectable to an acceptable degree for the measurement. The kinematic region on which cross-sections are measured is most sensitive to the $Q^{2}$ versus $x_{B}$ space.
However, this is purely kinematics, and as such, probing a larger phase space becomes a task of generating data in a given region, rather than refining analysis or simulation. To conclude, these results lend much confidence that the ECCE detector matches extremely well with any stated possibilities and reaches of this measurement that were previously outlined in the YR.

\subsection{ep DVMP}
\label{subsec:dvmpep}
\label{sec:DVMPep}

Hard exclusive meson electroproduction processes, also known as deeply virtual meson production (DVMP), complement the DVCS reaction. In the DVMP case, the scattering reaction produces a meson instead of a photon. Heavy vector mesons, such as J/$\psi$, probe the gluon GPDs and ultimately provide information about saturation when studying the change of gluon spatial distribution from low to high  (see \cite{khalek2021science} page 114).

Following the lead of the YR (page 334), the detector performance and efficiency in the context of J$/\psi\rightarrow e^{+}e^{-}$ events from $ep$ collisions were evaluated with ECCE. The main goal of this study was to quantify the detector acceptance for this reaction in one of the kinematic regions. The final results are estimated for 10 fb$^{-1}$ luminosity. 

\subsubsection{Electroproduction of J$/\psi$ decaying in an electron-positron pair }

This section  summarizes the event selection and the simulation  details in the analysis of the J$/\psi \rightarrow e^{+}e^{-}$ reaction with the ECCE detector.  The event generator is summarized in \ref{sec:jpsi_generator}, and the kinematics studied in this analysis correspond to electron and proton beam energies of 18 GeV and 275 GeV, respectively. 

The generated events were selected with the requirements summarized in Table \ref{tab:Jpsi-kin-cuts}. 

The sample of reconstructed events was chosen such that only three tracks were detected. Two of the tracks were positive and the third one was negative. The J$/\psi$ selection had the negative track and the two possible combinations with the positive tracks. If the J$/\psi$ reconstructed mass was in the 2 to 5~GeV window for a single combination of tracks (1 negative and 1 positive), the event was processed, otherwise, the event was discarded. The proton was detected in the far forward region with the Roman Pots since B0 was out of the acceptance for this kinematic sample. 

\begin{table}[h!]
\centering
\begin{tabular}{|c|c|c|}
\hline
\multicolumn{1}{|l|}{\textbf{Variable}} & \multicolumn{1}{l|}{\textbf{Definition}}      & \multicolumn{1}{l|}{\textbf{Range}} \\ \hline
$Q^{2}$ {[GeV]}  & $Q^{2} = -q^{2}= - (k_{e} -  k{e'})$           & 0 - 50~GeV$^2$ \\ \hline
$x_{B}$          & $x_{B} = \frac{Q^{2}}{2\cdot k_{p} \cdot q} $  & 0 - 0.15 \\ \hline
\end{tabular}
\caption{\label{tab:Jpsi-kin-cuts} Kinematic limits in the  J$/\psi$ study. }
\end{table}

\begin{figure}[h]
 \centering
    \includegraphics[width=0.85\linewidth]{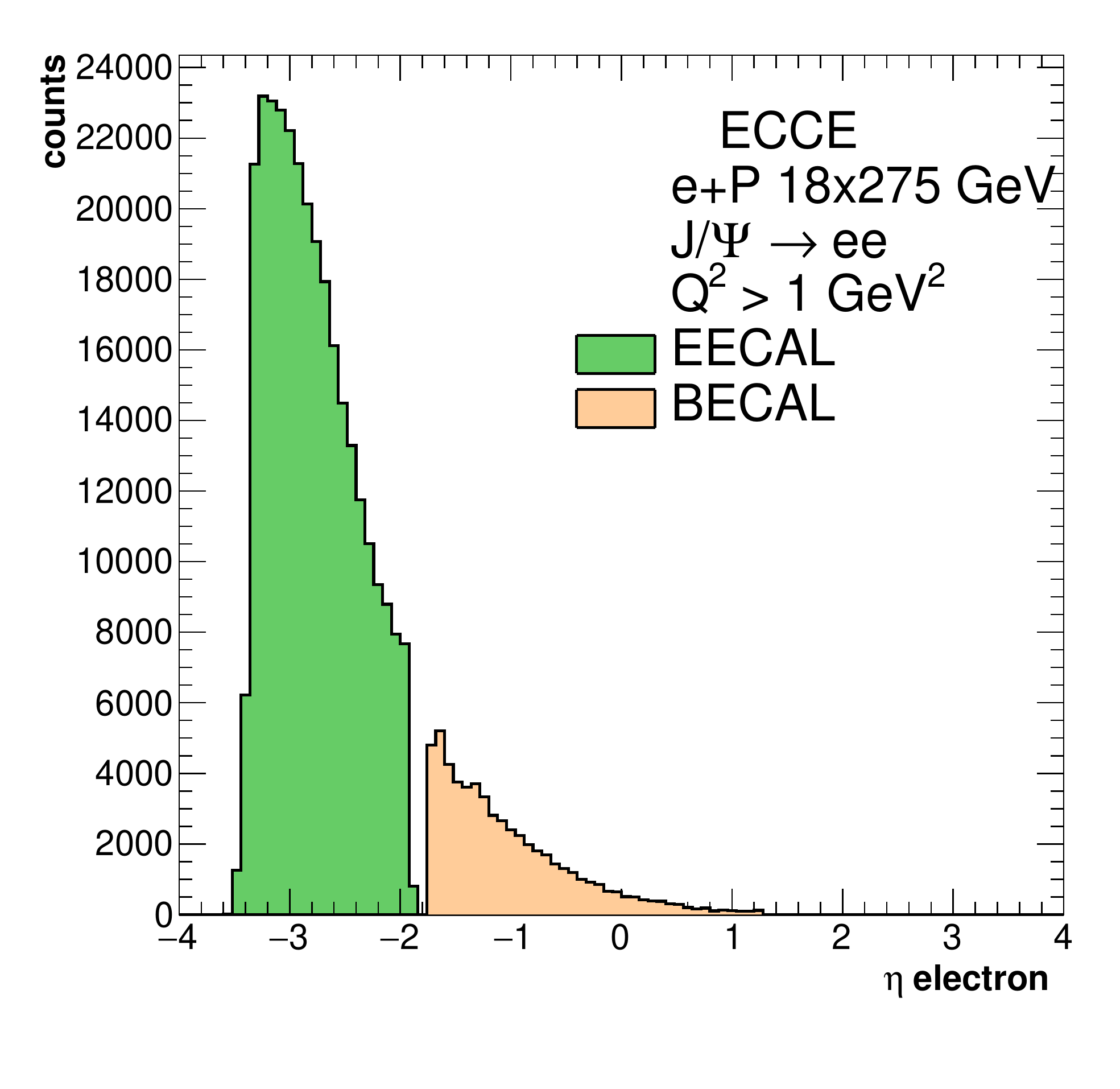}
  \caption{Scattered electron detection in the calorimeters. Most of the electrons go to the far backward region.}
  \label{fig:jpsi-scat-el-calo}
\end{figure}

\begin{figure}[h]
 \begin{center}
    \includegraphics[width=.75\linewidth]{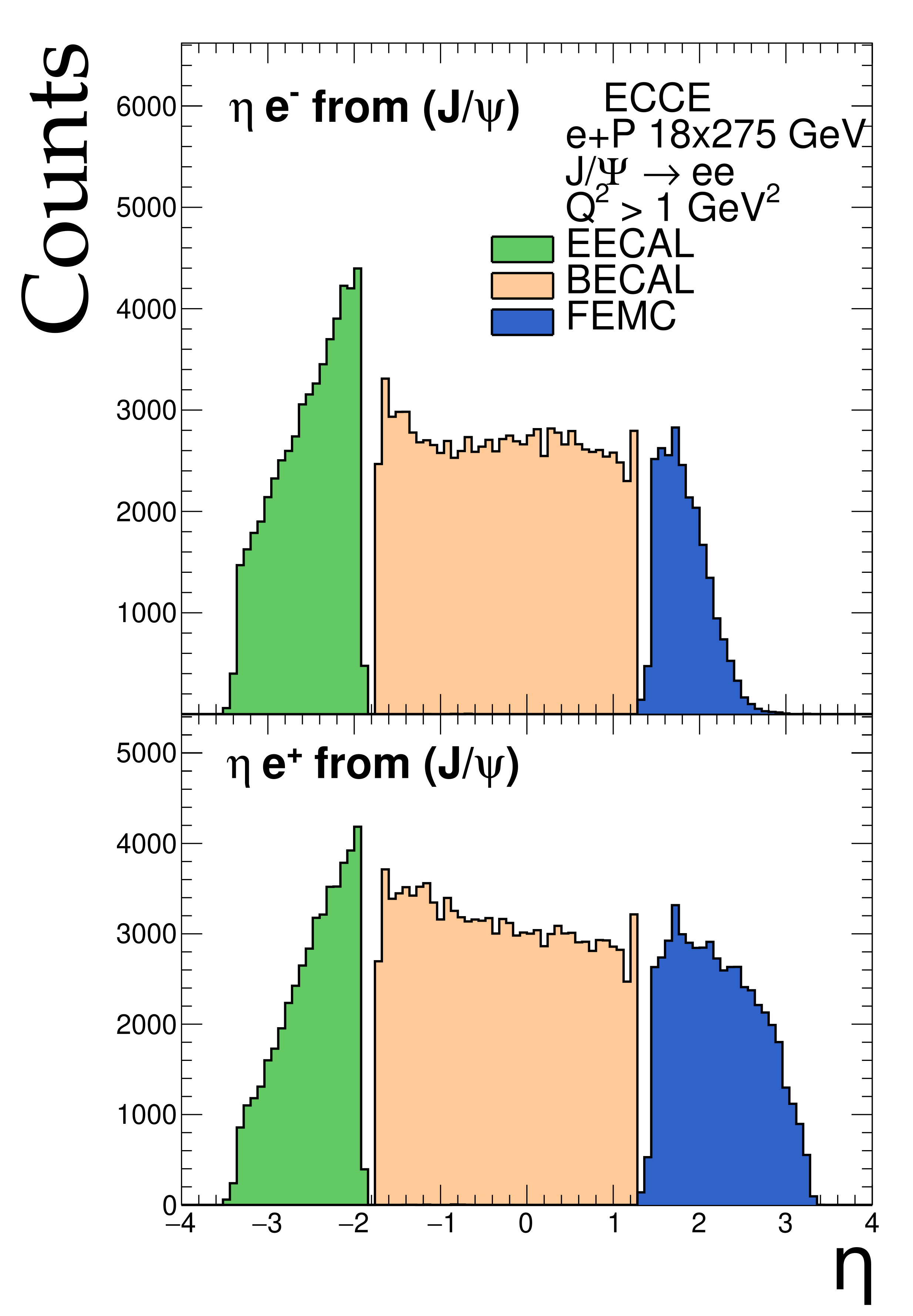}   
  \end{center}
  \caption{\label{fig:jpsi-scat-ep-calo} Electron (left) and positron (right) from J$/\psi$ detection in the calorimeters.}
\end{figure}

\begin{figure}[h]
 \centering
    \includegraphics[width=.49\linewidth]{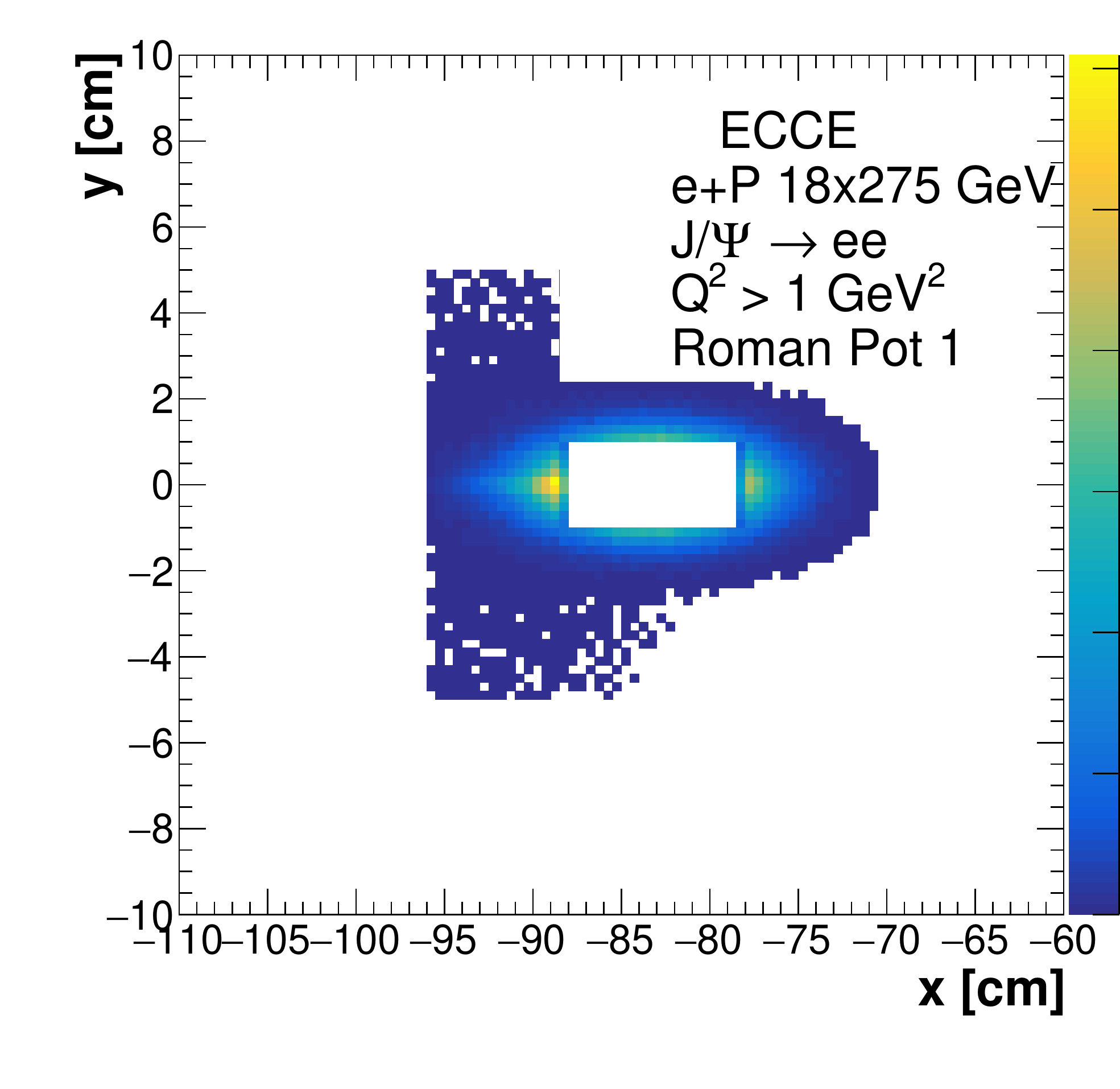}
    \includegraphics[width=.49\linewidth]{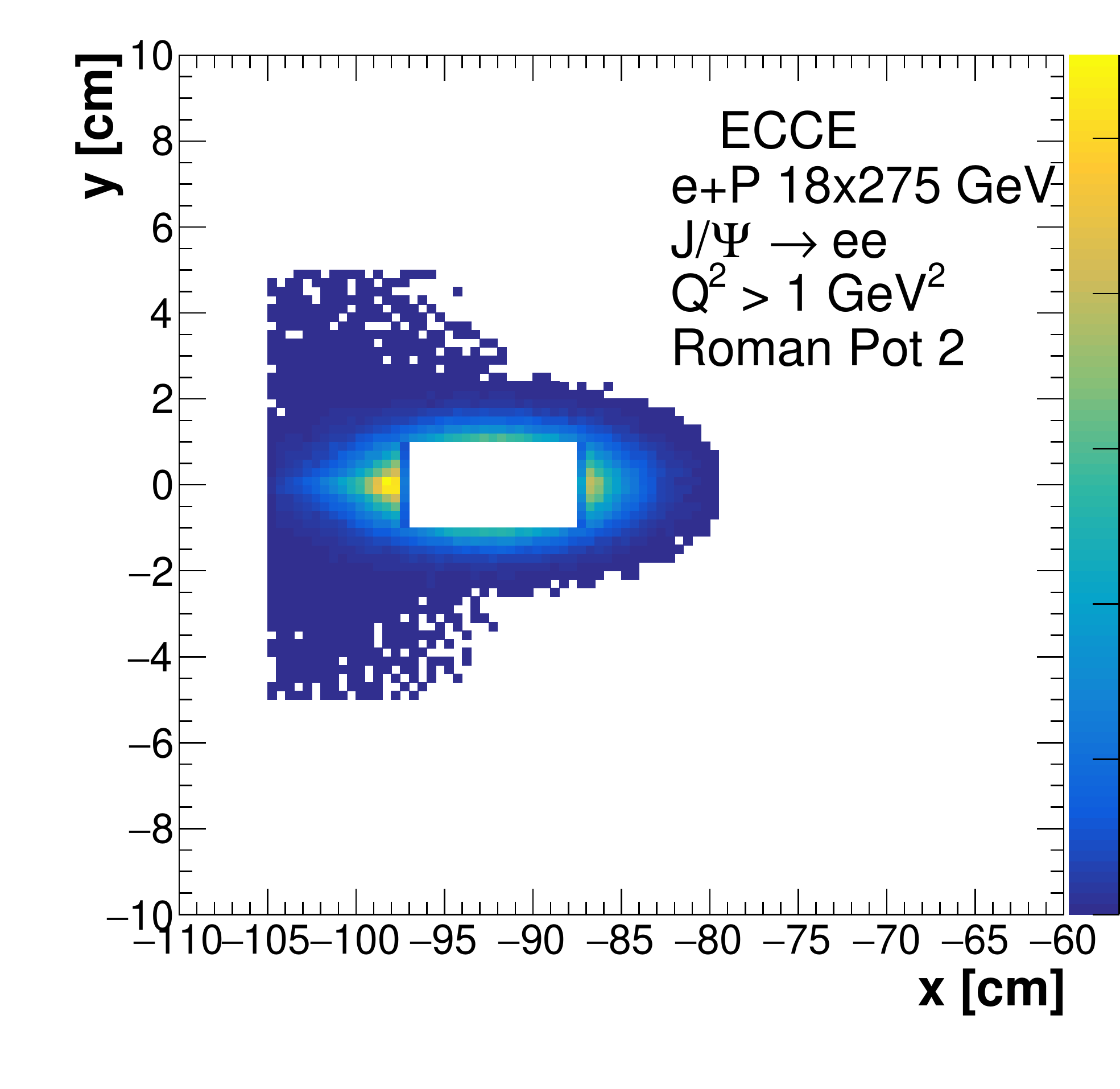}
  \caption{\label{fig:jpsi-Rp2} Proton detection in Roman Pot 1 (left) and Roman Pot 2 (right) for the kinematic setting studied in this work.}
\end{figure}

Fig.~\ref{fig:jpsi-scat-el-calo} shows that the scattered electron is detected mostly in the backward region, but also in the barrel. Fig.~\ref{fig:jpsi-scat-ep-calo} shows that the lepton pair daughter of the J$/\psi$ is detected in three regions (backward, central, and forward). In addition, Fig.~\ref{fig:jpsi-Rp2} shows the distribution of the protons detected in the Roman Pots, where the majority of the generated events that are not reconstructed are lost to the Far Forward region in an exclusive measurement. These studies have shown that, given the current IP6 design, a large number of protons go through the beam pipe and cannot be detected.

The acceptance of electrons and positrons is 80 $\%$ and appears to be independent of the kinematic setting. In the case of the protons, they are limited by the far forward region; protons with $\eta < 6$ in the head-on frame are not detected, and there is an average of 50$\%$ efficiency for the other $\eta $ regions. 

The acceptance of electrons and positrons does not seem to depend on the beam setting. The dips in those acceptances correspond to the transitions in the tracking system and/or calorimeters. 

\begin{figure}[h!]
\centering
\includegraphics[width=0.75\linewidth]{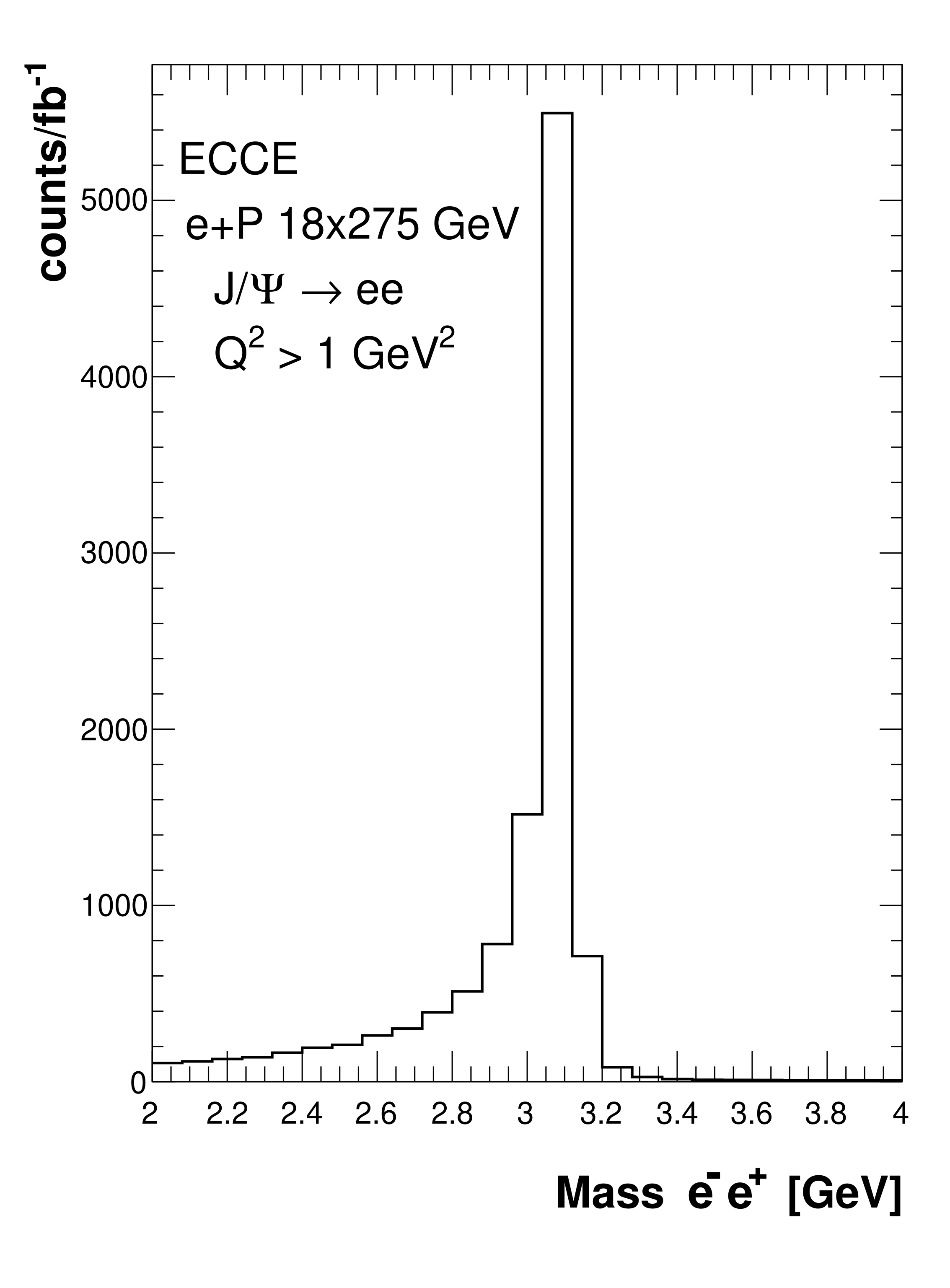}
\caption{Reconstructed J$/\psi$ mass, for the 18x275 GeV kinematic setting. 
}
\label{fig:jpsi_masses}
\end{figure}

The $e^{+}e^{- }$ invariant mass, and the missing mass reconstruction for the whole process, will be essential to check the exclusivity of the measurement. Fig.~\ref{fig:jpsi_masses} shows the  $e^{+}e^{-}$ reconstructed mass from this simulation, even if they are difficult to interpret in the absence of a background study. 

\begin{figure*}[h]
    \includegraphics[width=.33\linewidth]{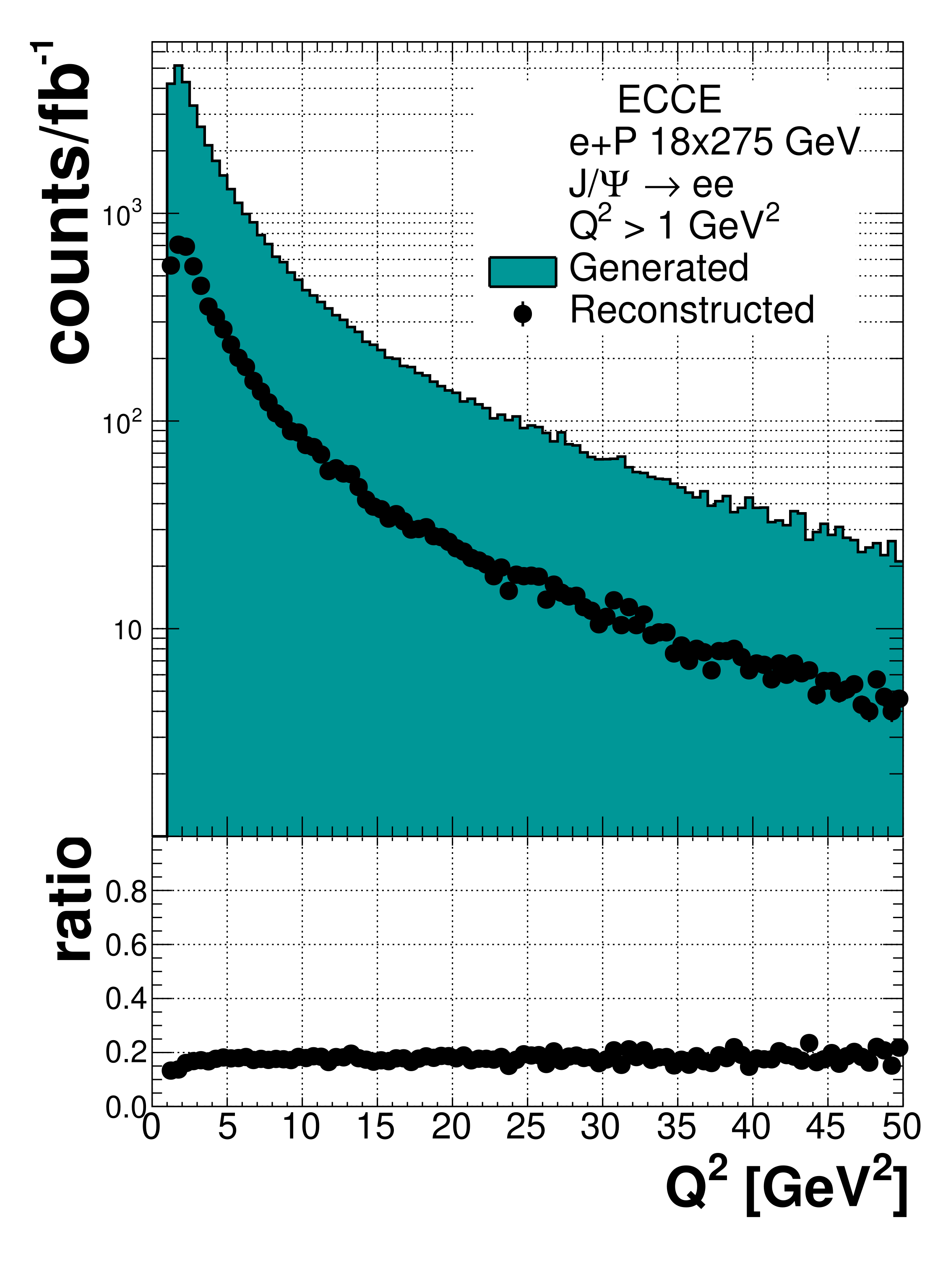} 
    \includegraphics[width=.33\linewidth]{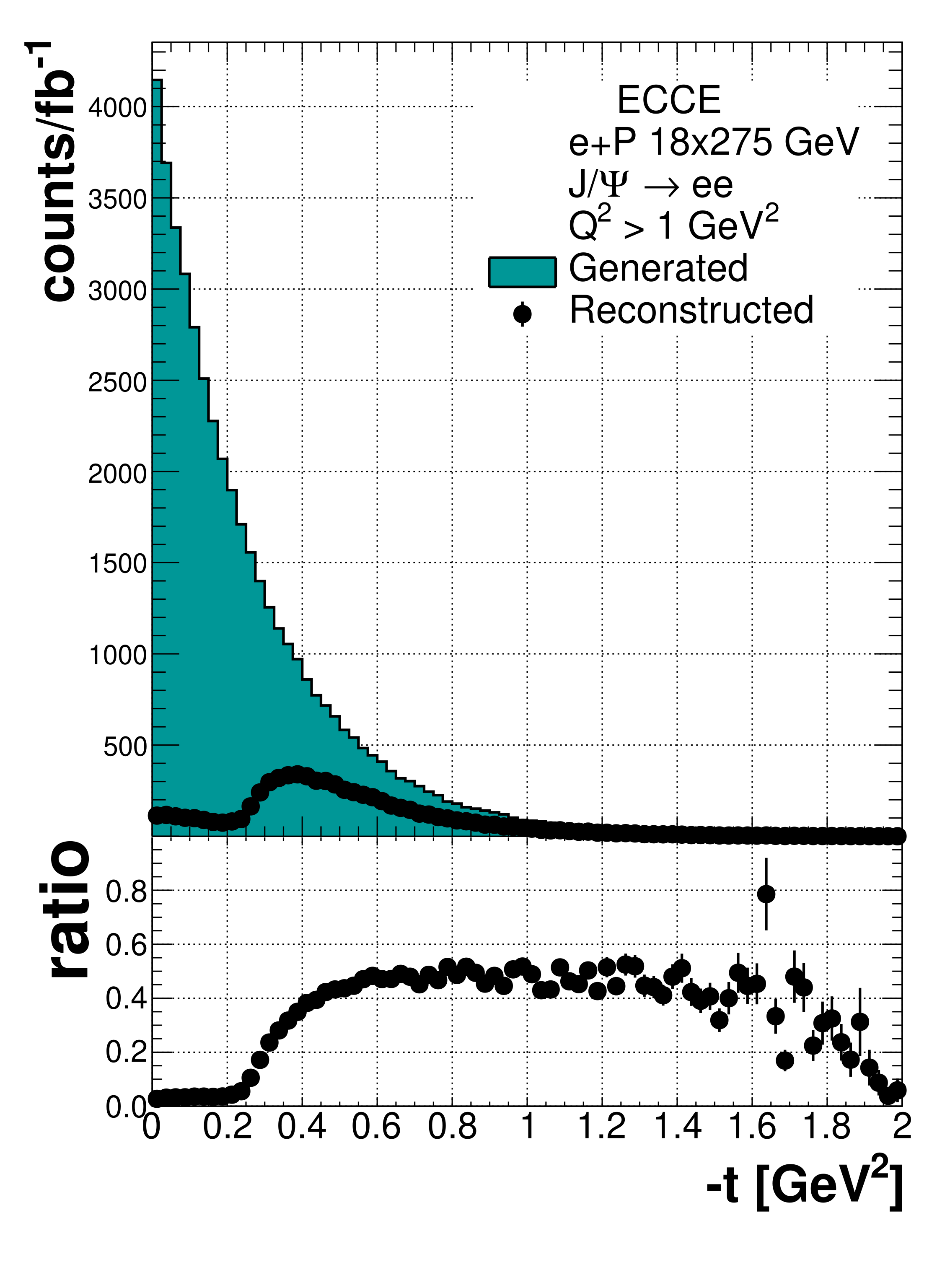}
    \includegraphics[width=.33\linewidth]{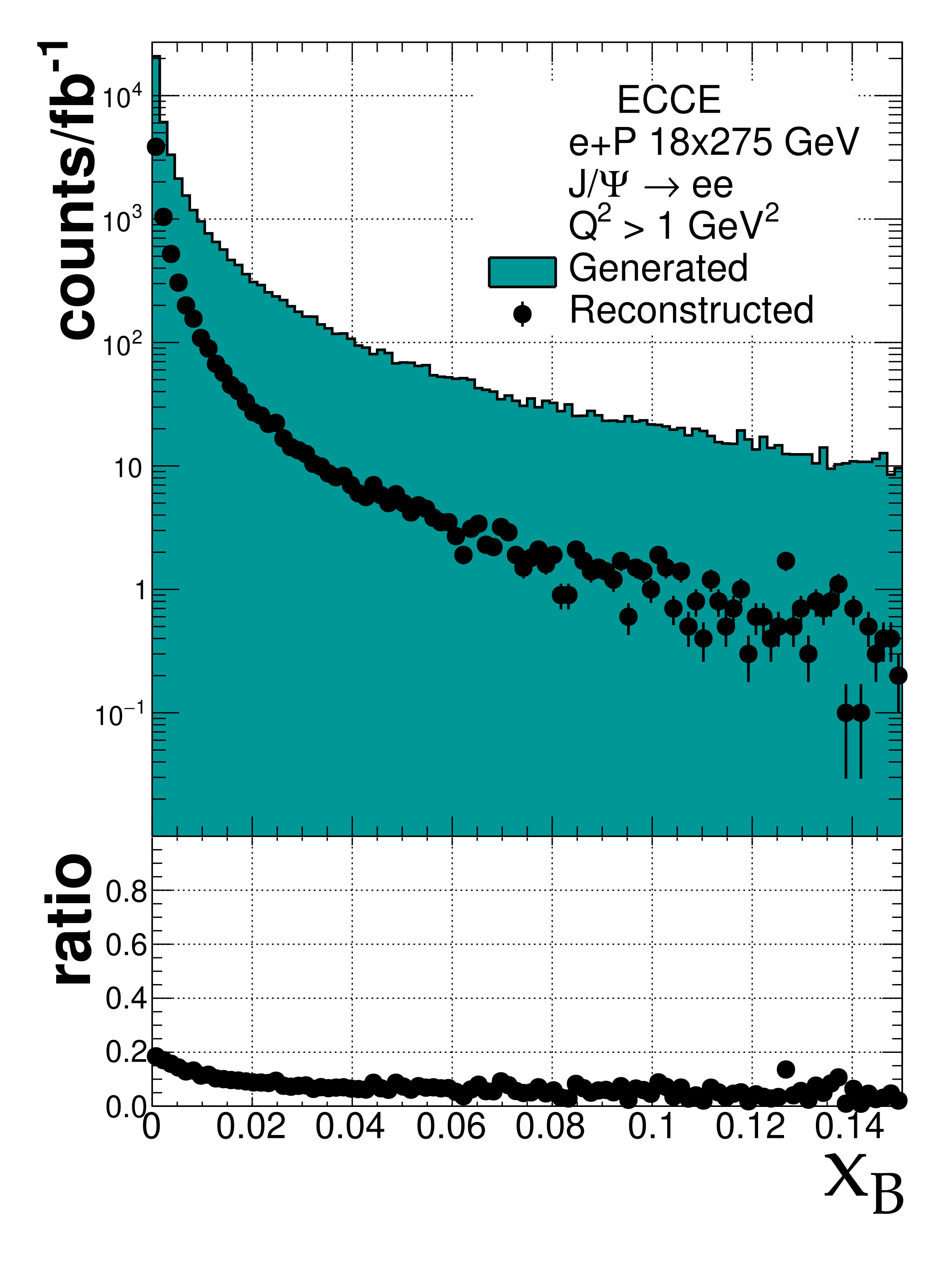}
  \caption{\label{fig:phys-kin1} Physics kinematicas variables and resolutions for $ep$ scattering of 18x275 GeV$^{2}$.}
\end{figure*}

\subsubsection{Physics Variables: distribution and resolutions}
The various quantities that are relevant to the physics at hand are $\phi$, $Q^{2}$, $-t$, $x_{b}$, $x_{v}$ and $x_{L}$. Fig.~\ref{fig:phys-kin1} shows the distributions of these variables.  As expected, the effective range of these physical variables is limited by the acceptance of the protons. This directly affects the range of the variable $-t$: events with $-t>1$~GeV$^2$ are not reconstructed due to the outer acceptance of the Roman Pots and the lack of statistics. For the case of 18$\times$275 in this study, events with small $-t$ ($<$ 0.2 GeV$^2$) are not detected because of the inside edge of the Roman Pots.

\begin{figure}[h]
\begin{center}
\includegraphics[width=0.85\linewidth]{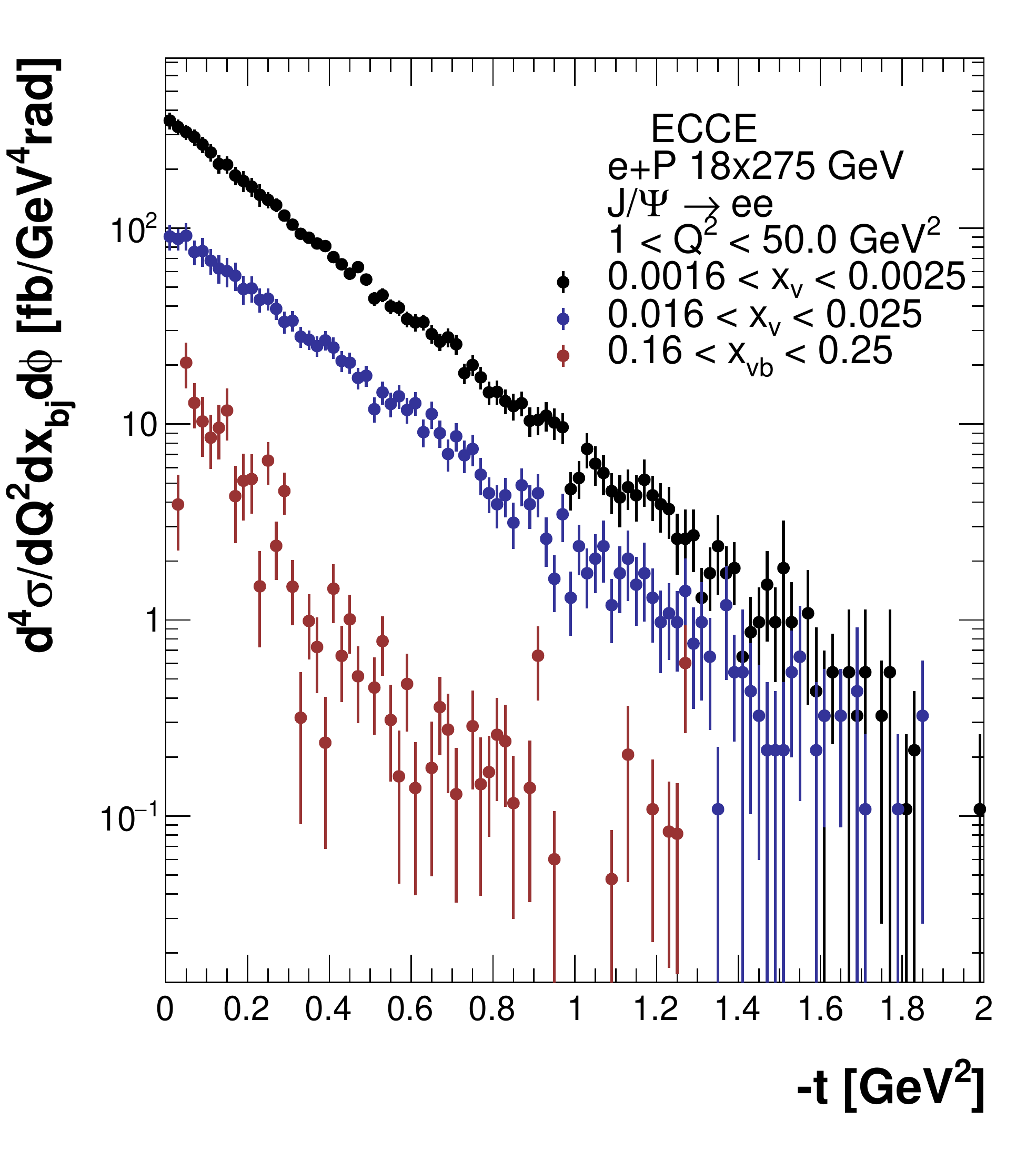}
\end{center}
\caption{Differential cross-section vs Momentum transfer t for the 18$\times$275 beam setting studied in x$_v$ slices, 0.0016 $<$ x$_v$ $<$ 0.0025 (black), 0.016 $<$ x$_v$ $<$ 0.025 (blue) and 0.16 $<$ x$_v$ $<$ 0.25 (red).}
\label{fig:cross_section_jpsi18x275}
\end{figure}

\subsubsection{Cross-section}

The cross-section, assuming a luminosity of 10~fb$^{-1}$, was extracted as a function of $-t$ and is displayed in Fig.~\ref{fig:cross_section_jpsi18x275}.  The acceptance of the ECCE detector was fully considered for the events generated, but as expected it is not the limiting factor in the measurement of these processes. The Far Forward detectors are the main limitation of this measurement.  

The statistical precision shown is for an integrated luminosity of 10 fb$^{-1}$, while the YR study (page 342) was performed for an integrated luminosity of 100 fb$^{-1}$. The physics interest resides in the evolution of the $-t$ dependence of the cross-section. The $Q^2$ dependence is also important to allow for multi-dimensional binning. To a large extent, the $Q^2$ accepted range is independent of the $-t$ range, and we have shown the evolution with $-t$ only here.

\subsection{Time-like Compton scattering}
\label{subsec:tcs}

The following study investigated the feasibility of measuring Time-like Compton Scattering (TCS) off the proton with the ECCE detector. The YR description on this topic (Sec.~8.4.4) was conducted with a toy Monte Carlo generator, eic-pi0-toy-MC~\cite{pi0}. The main aim for ECCE TCS activities was to conduct the same study performed for the YR, however, this time taking detector effects into account via the Geant4 simulation of the ECCE detector, available with the Fun4All software~\cite{fun4all}. Detector acceptance and reconstruction of the final state were key areas of study for these activities.

\begin{figure}[!tbp]
\centering
\includegraphics[width=9cm]{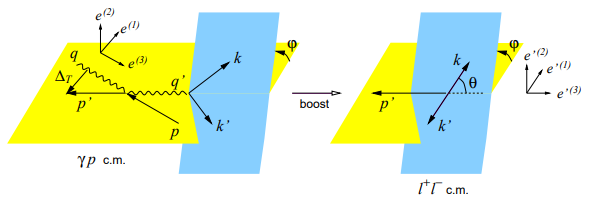}\caption{ Representation of TCS kinematics in the hadronic plane (yellow) and leptonic plane (blue). The planes are separated by an angle \begin{math} \phi \end{math}. Initial four momenta of the beam proton and the real photon are represented by convention as \begin{math} p, q \end{math}, and the final state four momenta (the scattered proton and produced virtual photon) are represented as \begin{math} p', q' \end{math}. The momenta of the decay lepton pair are represented as \begin{math} k, k' \end{math}. The angle between the decay lepton \begin{math} k \end{math} and the scattering axis of the proton is represented as \begin{math} \theta \end{math} \cite{BERGER}. For the study in this note, the decay lepton pair was $e^+e^-$.}
\label{fig:TCSFIG} 
\end{figure}
A sketch of the TCS process may be found in Fig.~\ref{fig:TCSFIG}.
TCS is an inverse process to DVCS. Both measurements provide access to the same Generalised Parton Distributions (GPDs), yet each have different experimental advantages over the other. Complimentary TCS and DVCS measurements at the EIC will be crucial for testing factorization, the transition between the space-like and time-like regimes, and the universality of GPDs. The physics accessible via TCS is further described in sections 7.2.2 and 8.4.4 of the YR. The dominating background channel for TCS is the BH process, in which an incoming or scattered electron radiates a photon, and scatters elastically off the proton, giving the same final state as TCS. Measuring the interference between TCS and BH allows access to the real part of the Compton Form Factor, and can therefore place constraints on the determination of GPDs.

%, developed by Pawel Sznajder (NCBJ) and Kemal Tezgin (UCONN)

\subsubsection{EpIC Generator Settings and Fun4All Version}
%For this study, the EpIC generator~\cite{EPIC} was used. 

The EpIC generator~\cite{EPIC}, used for this study, is an MC event generator that uses GPD models from the PARTONS framework~\cite{PARTONS}, plus mFOAM (a general purpose MC event simulator integrated with ROOT) to generate random events in phase space. EpIC takes in an input .xml file, within which parameters such as beam energy, kinematics, and the decay process are defined. Based on the input information, EpIC then generates the four vectors of all particles as the output. The EpIC generator is capable of generating pure TCS events, pure BH events, and events that combine TCS, BH, and the interference term (INT). The events simulated in this study were the combined set of TCS+BH+INT, extending on the result using only pure TCS in the YR. The EpIC generator has the capability to include radiative corrections, however, at the time of this study these were still in development and thus have not been included. 

EpIC was used to simulate TCS events at beam energy settings of 5\,$\times$\,41 and 18\,$\times$\,275, to study the anticipated two extremes in acceptances. The electron beam helicity was set to negative in the event generator, and a total of 1M events were generated for each energy setting.

The generated kinematics in EpIC was set to either match or slightly extend upon the original kinematics studied in the YR, and are detailed below.
\begin{itemize}
\item $0 < -t < 1$~GeV$^2$ to capture the physics region of interest.
\item $2 < Q'^{2} <20$~GeV$^2$ to ensure a hard scale for the scattering and to minimize background from the low resonance region. $Q'^{2}$ represents the virtuality of the produced virtual photon (see Sec.~\ref{sec:Kinematics} for the full definition).
\item $0 < \phi < 2\pi$ to obtain a full lab frame azimuthal (\begin{math} \phi \end{math}) angular coverage.
\item \begin{math} 0 < \phi_S < 2\pi \end{math}, (where \begin{math} \phi_S \end{math} represents the angle between the leptonic plane, see Fig.~\ref{fig:TCSFIG}, and the transverse component of the polarization of the target nucleon), to obtain a full \begin{math} \phi_S \end{math} angular coverage.
\item \begin{math} \frac{\pi}{6} < \theta < \frac{5\pi}{6} \end{math}, slightly widened from the range used in the YR study (please note that at this stage in the YR studies, BH singularities became apparent at extremes of theta. These have since been rectified via recent updates to EpIC, and the restricted range from the YR is thus widened here).
\item $0 < Q^{2} < 0.15$~GeV$^2$ to select a quasi-real photon.
\end{itemize}

The Fun4All~\cite{fun4all} simulation software prop 7.1 was used for the TCS studies. This more recent version was used due to an improved scattered proton acceptance compared to older Fun4All versions, as a result of the implementation of the high acceptance setting (detailed in Sec.~ \ref{sec:dvcseAEventSelect}).

\subsubsection{Event Selection, Reconstruction and Analysis} \label{sec:Kinematics}
Event selection of the final state particles centers around the scattered proton (\begin{math} p' \end{math}), the decay electron (\begin{math} e^{-} \end{math}), and the decay positron (\begin{math} e^{+} \end{math}).

For the $e^{-}e^{+}$ pair, the information from hits registered in the Fun4All EEMC, FEMC, and BECAL  was compared with momenta from the truth container and separated by PID, taken from calorimeter cluster information.
The virtual photon \begin{math} \gamma^{*} \end{math} produced by the interaction was then calculated using this decay $e^{-}e^{+}$ pair, via summation of four momenta (\begin{math}\gamma^{*} = k (e^+) + k'(e^-) \end{math}).

The energy determination for these particles could be improved by calibrating the calorimeters using a similar method as described in Sec.~\ref{subsec:dvcs}, however, due to time constraints the plots in this study have not been corrected for this calibration. We plan to implement this in the near future and from preliminary tests we anticipate the change to be relatively minor, around a \begin{math} \approx \end{math}3\% correction.
The scattered electron, \begin{math} e' \end{math}, in this study is calculated, as opposed to being detected. This is due to the original findings of the YR study, which indicated that the $e'$ would be difficult to directly detect, without implementation of a low \begin{math} Q^2 \end{math} tagger and that it is instead better to use the momenta of other final state particles compared to initial beam momenta to calculate it~\cite{khalek2021science}. 

In more recent versions of Fun4All, there has been an integration of a low \begin{math} Q^2 \end{math} tagger, which would mean that a missing mass study could be performed with the scattered proton as the 'missing' particle, as it is within detector resolution to calculate this at high energies, however, this has not been explored in this study. 

To reconstruct the scattered proton, $p'$, which is very forward-going, the Roman Pots and B0 detectors were essential. The geometrical acceptances for the Roman Pots and B0 detectors were handled in exactly the same way as previously described in Sec.~\ref{sec:dvcseAEventSelect}, i.e. the acceptance of the B0 layers was fully modeled in Fun4All directly and cuts to remove the contribution of the beam pipe in each of the Roman Pots were added to the analysis of the Fun4All output. The Roman Pots cuts for each simulation were as given in Table \ref{tab:dvcsSIMS}. In the analysis of the Fun4All output to mimic detector resolution effects (since these are not fully modeled in the forward region of Fun4All yet)  a $1\,\%$, smearing was applied to the truth proton. The track direction of the proton, however, was not smeared in this study. A cut on successfully detecting the scattered proton and the decay lepton pair in the final state was also included in the event selection stage of the analysis.

In the analysis stage, several physics quantities of the reaction were reconstructed, their definitions are below.
\begin{itemize}
\item \begin{math} Q^{2} = -q^{2} = -(e'-e)^{2} \end{math}, where \begin{math} e \end{math} and \begin{math} e' \end{math} represent the four momenta of the beam and scattered electron respectively.
\item \begin{math} Q'^{2} = -q'^{2} = -(k+k')^{2} \end{math}, where \begin{math} k \end{math} and \begin{math} k' \end{math} represent the four momenta of the decay positron and electron respectively.
\item \begin{math} \tau = \frac{Q'^{2}}{(s-M_{p}^{2})}\end{math}, where
%\begin{math} Q'^{2} \end{math} is the virtuality of the produced virtual photon, 
\begin{math} s \end{math} represents the center of mass energy calculated via \begin{math} (p+q)^2 \end{math} and \begin{math} M_{p}^{2} \end{math} represents the mass of the proton \cite{TCS}.
\item \begin{math} -t = -(p-p')^{2} \end{math}, where \begin{math} p \end{math} and \begin{math} p' \end{math} represent the four momenta of the beam and scattered proton respectively.
\end{itemize}
See Fig.~\ref{fig:TCSFIG} for visualization of the four momenta.
%ADD BACK IN WHEN CUT CHECKED AND APPLIED
Due to the background events caused by the J$/\psi$ channel, there should also be a windowed cut applied to \begin{math} Q' \end{math} around the J$/\psi$ mass, ($\approx3.1$~GeV), which would appear as a gap in the \begin{math} Q'^{2} \end{math} phase space between $\approx9-12$~GeV$^{2}$, however, a fuller analysis must be conducted to discern this range correctly, wherein a \begin{math} 3\sigma \end{math} cut would be taken around the J$/\psi$ mass peak in a set of J$/\psi$ generated data. This cut has not been included in this analysis due to time constraints, however, its effect would only be a slight reduction in statistics, and would not greatly affect the overall shape of the resulting distributions. 

\subsubsection{Results}
\begin{figure*}[h!]
\includegraphics[width=0.5\textwidth]{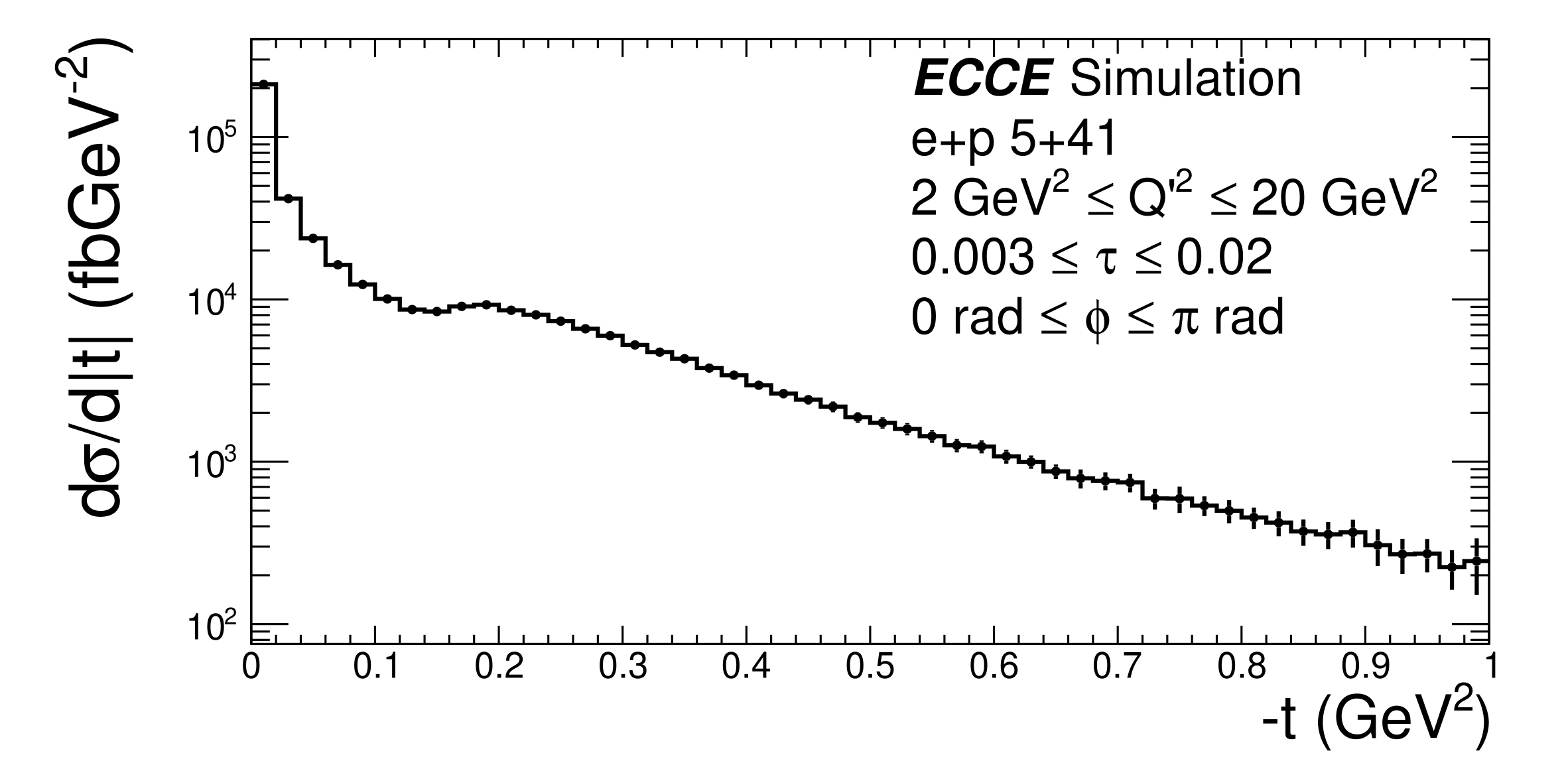}
\includegraphics[width=0.5\textwidth]{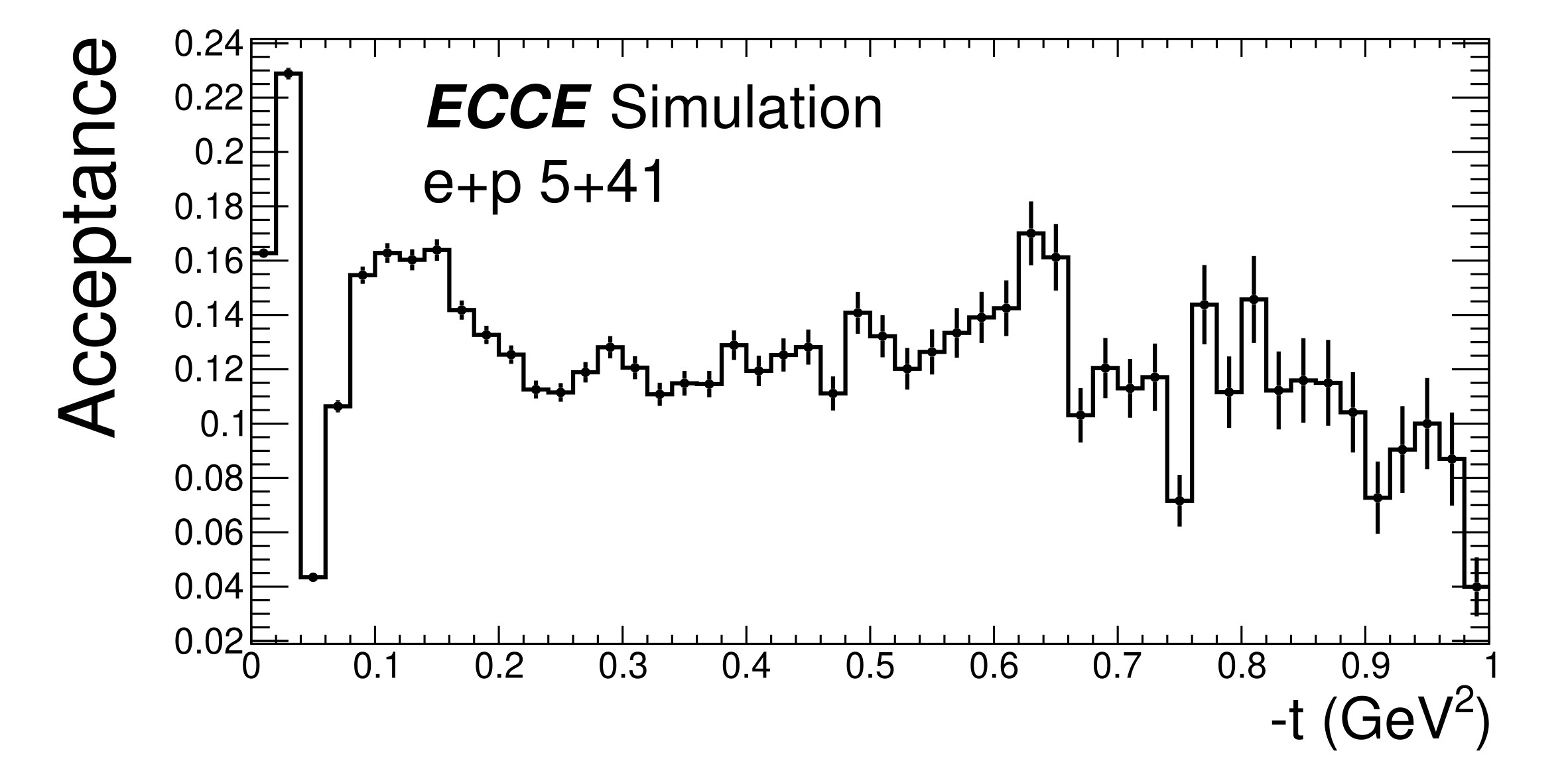}
\caption{ \label{fig:tcs5x41t} 5\,$\times$\,41 - TCS Differential cross-section versus the momentum transfer to the struck parton \begin{math} -t \end{math} reconstructed using the beam and scattered protons \begin{math} t=(p-p')^{2} \end{math} (left) and detector acceptance for \begin{math} -t \end{math} reconstructed using the beam and scattered protons (right).  Note acceptance is given as a value where 1 corresponds to 100\%} 
\includegraphics[width=0.5\textwidth]{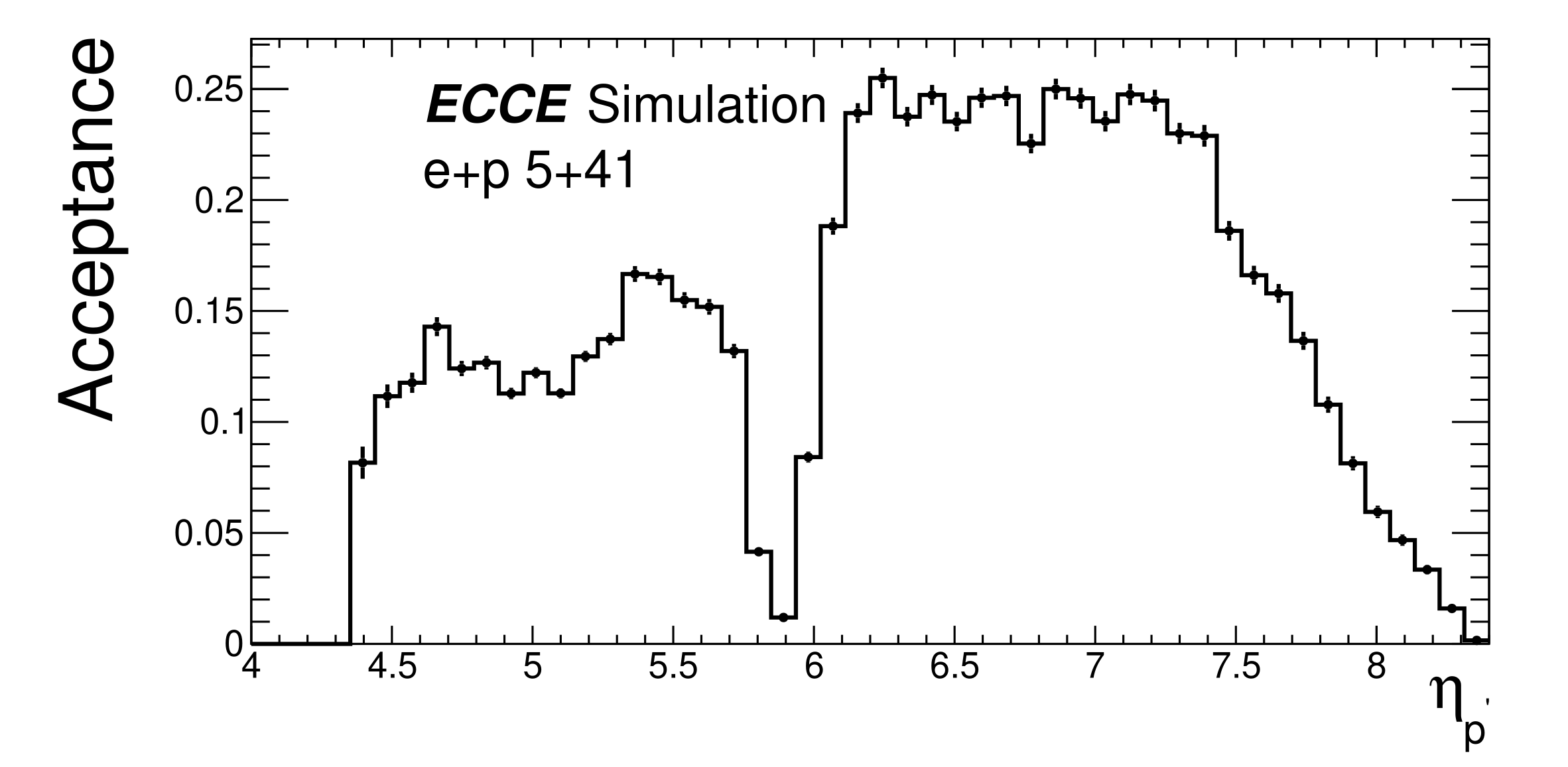}
%\label{fig:tcs5x41epTvseta_acc}
\includegraphics[width=0.5\textwidth]{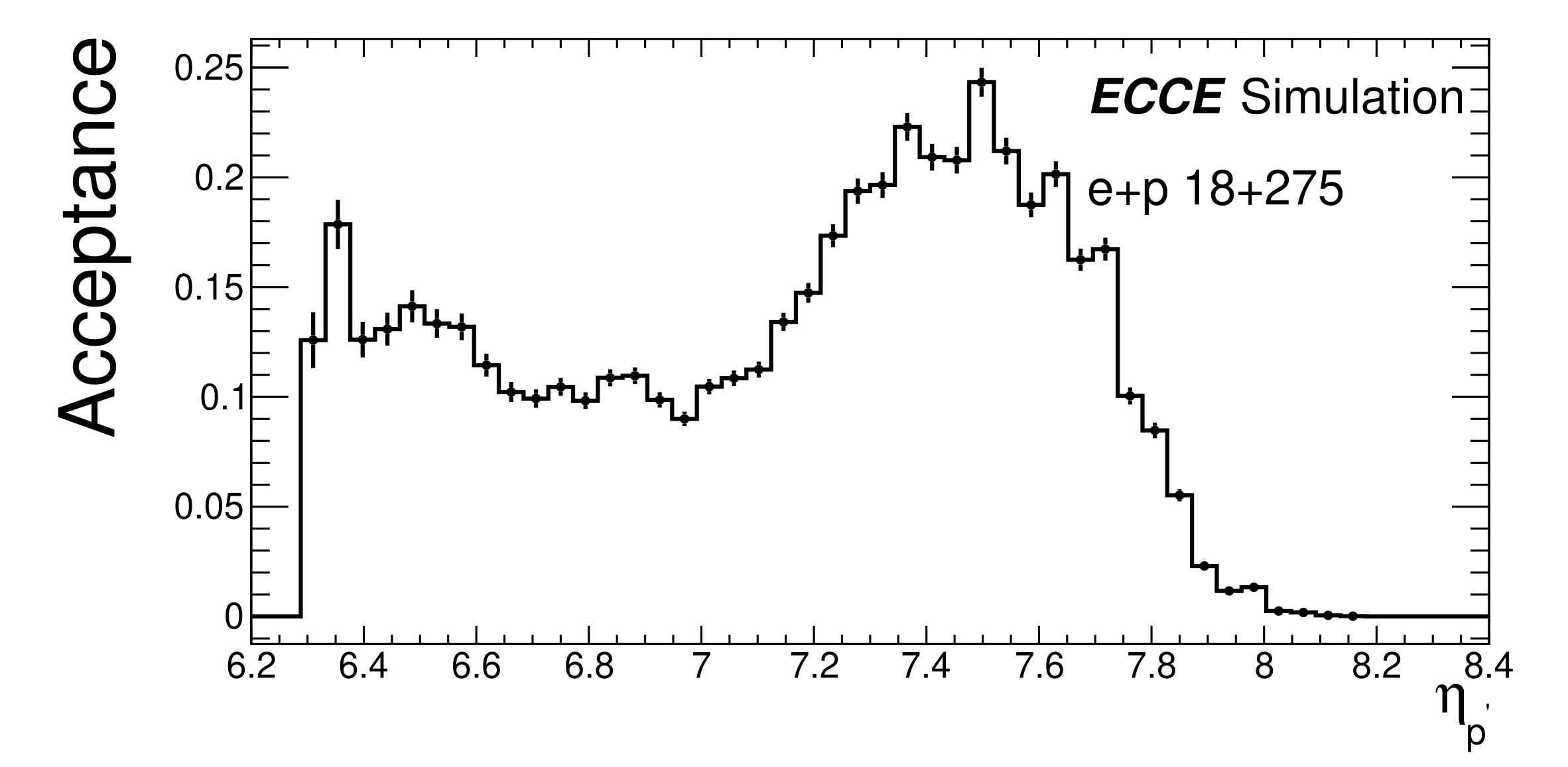}
%\label{fig:tcs18x275epTvseta_acc}
\caption{Left: 5\,$\times$\,41 acceptance vs pseudorapidity ($\eta$) of the scattered proton from TCS events. Right: 18\,$\times$\,275 acceptance vs pseudorapidity ($\eta$) of the scattered proton. Note acceptance is given as a value where 1 corresponds to 100\%}
\label{fig::5x41Pseudorapidity}
\includegraphics[width=0.5\textwidth]{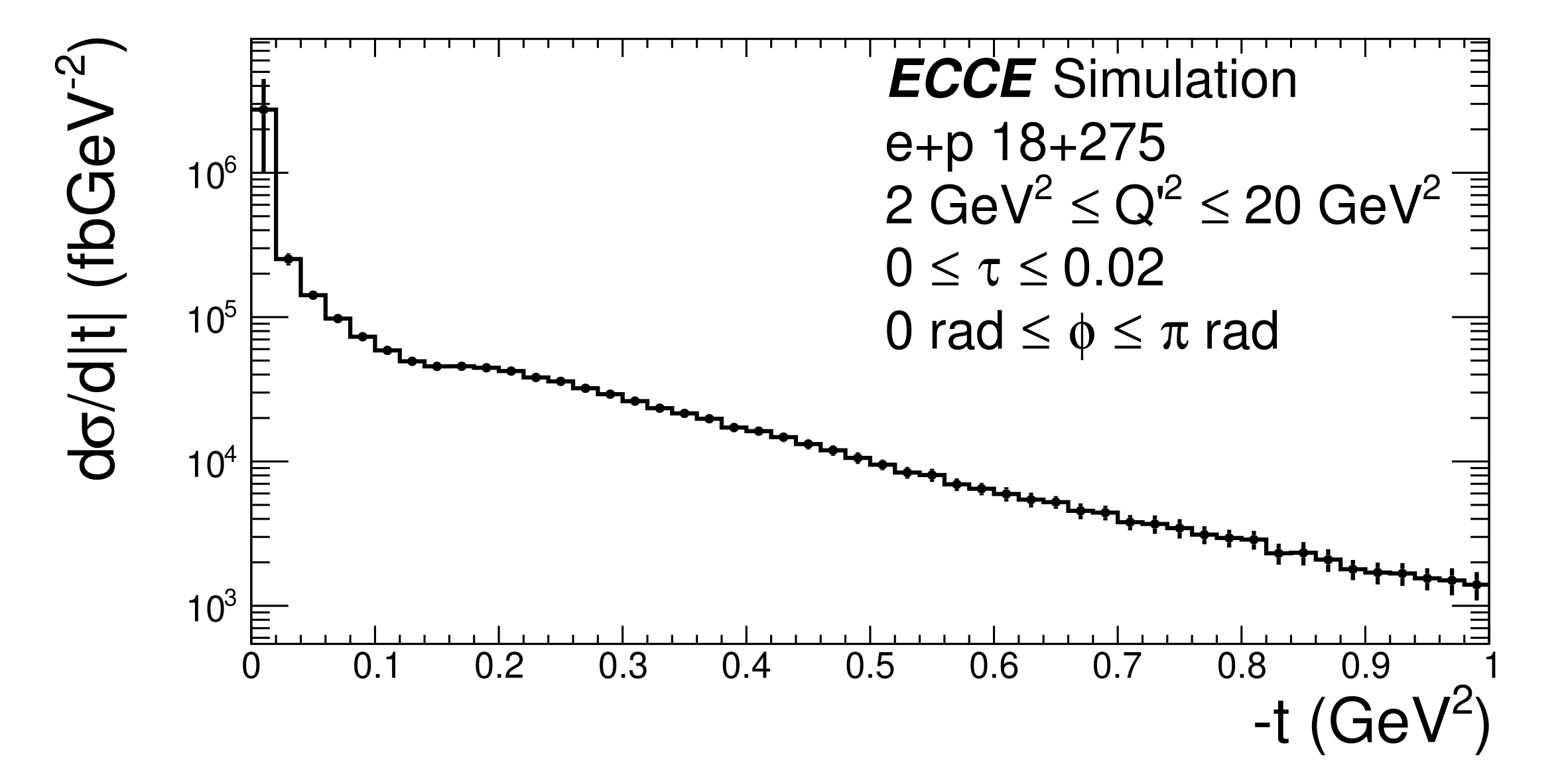}
\includegraphics[width=0.5\textwidth]{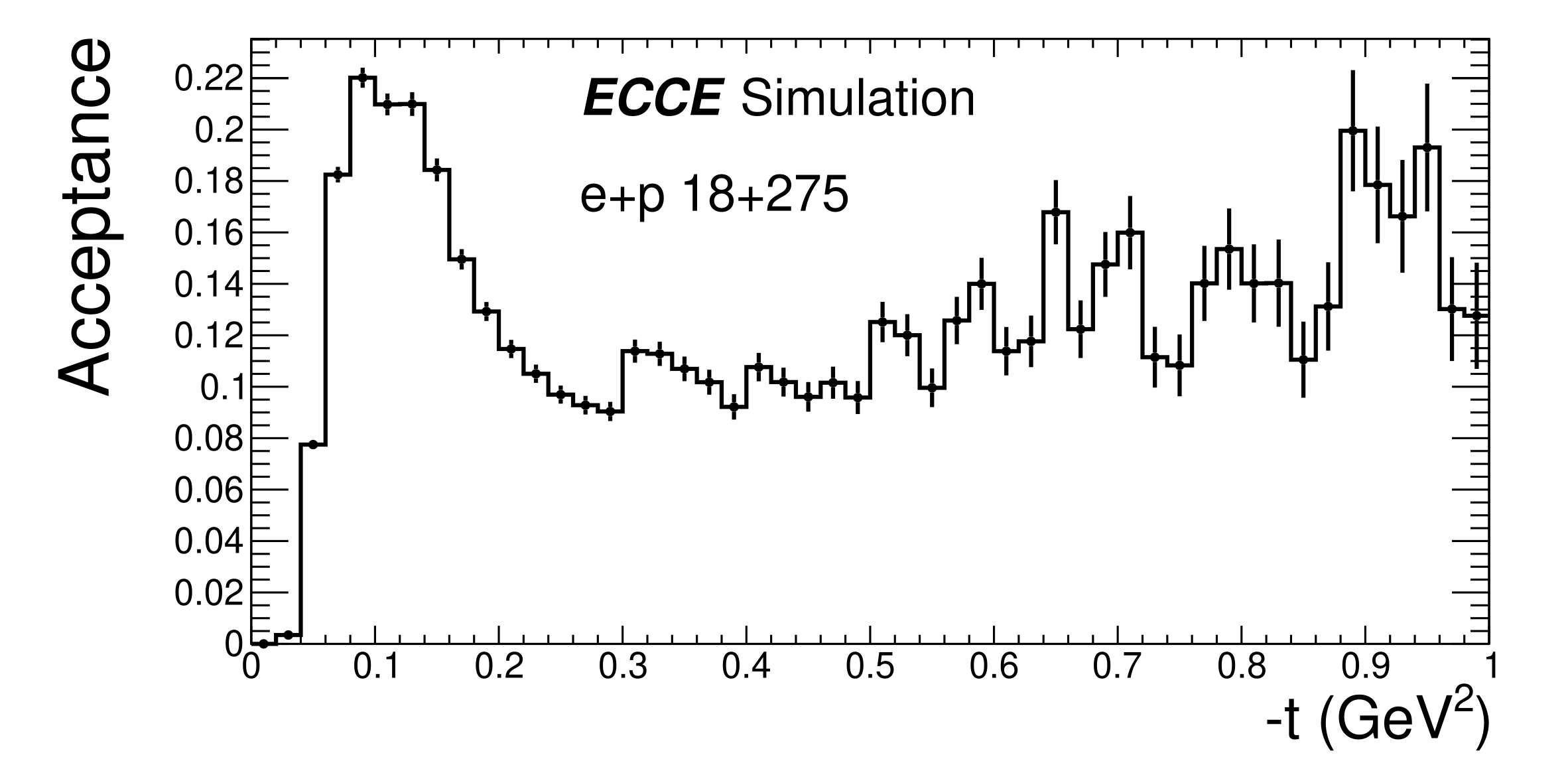}
\caption{\label{fig:tcs18x275t} 18\,$\times$\,275 - TCS Differential cross-section versus the momentum transfer to the struck parton \begin{math} -t \end{math} (left) and detector acceptance for \begin{math} -t \end{math} (right). Note acceptance is given as a value where 1 corresponds to 100\%}

\end{figure*}

Several of the results obtained from the analysis of the Fun4All output are given in Fig.~\ref{fig:tcs5x41t} -\ref{fig:tcs18x275t}. This includes TCS physics variables, the acceptance of the ECCE detector with regards to reconstructing these variables, and the kinematic phase space available for this reaction with the generated settings and the ECCE detector.

Please note that for any detector acceptance plots, the acceptance is calculated per bin and defined as the number of reconstructed events from the Fun4All output divided by the number of events outputted by the EpIC generator directly.

\subsubsection{Discussion and Summary}
As outlined in Sec.~8.4.4 of the YR, it is important to reconstruct the momentum transfer to the struck parton \begin{math} t \end{math} via \begin{math} -t=-(p-p')^{2}\end{math}, i.e. utilizing the four-momentum information from the target ($p$) and scattered proton (\begin{math} p' \end{math}) in TCS. This method provides a better resolution than using the reconstructed photon information. The YR also showed that the $p'$ is detected at very low transverse momenta ($p_{T}$), and very high pseudorapidities (\begin{math} \eta \end{math}), i.e. in the far forward direction. The far forward nature of the $p'$ is supported in the ECCE study by Fig.~\ref{fig::5x41Pseudorapidity}, which shows the detector acceptance for an \begin{math} \eta \end{math} range: $4.3 < \eta < 8.4$ at low energies, indicative of events captured with an acceptance of around 15-25\% in both the B0 and the RP. For high energies, we see an acceptance across an \begin{math} \eta \end{math} range: $6.3 < \eta <  8.4$, indicative of events captured only in the RP, not the B0, with an acceptance again of around 15-25\%. This result supports the need for both the B0 detector and Roman Pots in the detection of the scattered proton, as, similar to the $ep$ DVCS studies (Sec.~\ref{subsec:dvcs}), we see a high count of lower $\eta$ protons at the 5\,$\times$\,41 beam energies in the B0, and for higher energy settings a larger count is shown in the Roman Pots.

Utilizing the information from the Roman Pots, and for lower energies information from the B0, \begin{math} t \end{math} was successfully reconstructed across the full range with an acceptance of around 10-24\% for beam energy 5\,$\times$\,41 as shown in Fig.~\ref{fig:tcs5x41t}. For the 18\,$\times$\,275 beam energy setting in Fig.~\ref{fig:tcs18x275t}, we show again a full reconstruction of $t$, with an acceptance of around 10-22\%. The cross-section measurement for $t$ was calculated as in Eq.~\ref{xseccalc}, however with the variables $Q^{2} \rightarrow Q'^{2}$ and $ x_{B} \rightarrow \tau$. The cross-section for beam energy 5$\times$41 was averaged over $2 < Q'^{2} < 20$~GeV$^2$, $0.003 < \tau < 0.05$ and $0 < \phi < \pi$, determined by a phase space analysis of generated and reconstructed data, see \ref{sec:TCSep_appendix}. Beam energy 18$\times$275 simulations differed only in that \begin{math} 0 < \tau < 0.02 \end{math}. 

In summary, many of the main requirements observed for the TCS measurement outlined by the YR have been confirmed by the ECCE analysis. The forward acceptance is the main driving factor for the projected cross-section and statistics of this reaction.

It appeared initially that the detector performed better overall for the 5\,$\times$\,41 energy setting than for 18\,$\times$\,275, however with the new upgrade to the simulation software it has been shown that the two are much more comparable. An important next step would be to study an intermediate setting of 10\,$\times$\,100.

Another step to be taken in future studies is to calculate asymmetries rather than cross-sections, as in the leading order, the background contribution from pure BH can be removed. There is also a further background contribution stemming from measuring the final state \begin{math} e^{-} e^{+} \end{math} pair, where there is the potential that these are in fact misidentified pions, which could be dealt with by studying the \begin{math} \mu^{-} \mu^{+} \end{math} channel. This may also make for a simpler analysis procedure, due to issues with separating the decay electron \begin{math} e^{-} \end{math} from the scattered electron \begin{math} e' \end{math}. A final step would be to integrate analysis of the performance of the low \begin{math} Q^2 \end{math} tagger in detecting the scattered electron, and performing a missing mass study on the scattered proton, or the total reaction, to further discern how well each component of the detector system performs.

\subsection {XYZ spectroscopy}
\label{sec:spec_overview}

Spectroscopy of mesons with charmed quarks has provided some of the most surprising recent results and raised many interesting questions. These new states are commonly referred to as ``XYZ'' mesons and have unexpectedly small widths and masses inconsistent with quark model expectations. Instead many of these states are characterized by masses very close to two-meson decay thresholds. As a result, there are many possible means of describing the dynamics of these structures, for example, tetraquark states, di-meson molecular states, glueballs, hybrids, or kinematical effects due to thresholds and re-scattering interactions. For an overview of the subject, see Ref.~\cite{Brambilla:2019esw}.

In general, most of the new states have only been seen via single production mechanisms, such as B decays or $e^{+}e^{-}$ annihilation. This makes it difficult to resolve the dynamics contributing to the states. Photoproduction experiments offer an alternative production method with the advantage of, in principle, being able to produce all states within the center-of-mass range without the same potential for kinematic rescattering effects. This will be limited by the small production cross-sections for states with heavy quarks. However, it has been shown that production rates for many of these states are sufficiently high to be measurable with the EIC~\cite{Albaladejo:2020tzt}. In this section, we show that the proposed ECCE detector can deliver the event reconstruction required for investigating this exciting physics program. 

There are currently dozens of these potential new charmonium resonances. To make this study manageable, we limit the states under consideration to three: $\chi_{c1}(3872)$ or $X(3872)$, $Y(4260)$, and the well-established quark model state $\psi(2s)$. All of these states have decay branches to J$/\psi\pi^{+}\pi^{-}$, and so we focus on reconstruction of this final state with the J$/\psi$ decaying to $e^{+}e^{-}$. This allows us to compare the expected production of the exotic states to a regular quark-antiquark meson and check if we can distinguish the invariant mass peaks of these relatively close states.

As previously mentioned, with photoproduction we should produce many different mesons of exotic and non-exotic character. For example, the production of $Z_{c}$ isovector states will be of similar magnitude and decay to J$/\psi\pi$. We might also imagine production on deuteron producing both charge states of this manifestly exotic particle.

We may also search for poorly established or yet unknown states by looking through the many different final states accessible with the EIC, such as J$/\psi$ + vector mesons or kaons.
\subsubsection {Simulations}
\label{sec:sims}

To test the performance of the ECCE detector for XYZ production, an event generator was developed coupling realistic photoproduction amplitudes to low $Q^{2}$ virtual photons produced by electron scattering.  The photoproduction helicity amplitudes were calculated following the formalism and parameters given in \cite{Albaladejo:2020tzt}. The models therein are expected to give an order of magnitude estimates for meson production cross sections. Details of the generator are given in \ref{sec:elSpectro}.

For the J/$\psi \pi^{+} \pi^{-}$ final state events, we produced events via $\psi(2s)$, $\chi_{c1}(3872)$ and $Y$(4260) production. The number of events generated based on an integrated luminosity of  $10$ fb$^{-1}$ are summarised in Table \ref{tab:spec_numbers}. In each case the branching ratio of J$/\psi$ to $e^{+} e^{-}$ and of the meson to  J$/\psi\pi^{+}\pi^{-}$ were included. For the latter, the branching ratios assumed in \cite{Albaladejo:2020tzt} were used.

\begin{table}[hbt]
   \centering
        \caption{Generated event yields corresponding to 10 fb$^{-1}$ for the beam energy settings $E_e \times E_p$.}
        \label{tab:spec_numbers}
        \begin{tabular}{|c|c|c|c|c|}
            \hline
            $ep$ Setting &$\chi_{c1}(3872)$  &Y(4260)    &$\psi(2s)$     &Total \\
            \hline
            5x41        &96933               &9104      &71070          &177107 \\
            \hline
            5x100       &114906             &22384      &164942          &302232 \\
            \hline
            10x100      &125706             &37511      &270920          &434137 \\
            \hline
            18x275      &104291             &86199      &648881          &839371 \\
            \hline
         \end{tabular}
\end{table}

As all six final state particles were charged, we used the reconstructed tracks given
in the Fun4All DSTs SvtxTrackMap. Particle ID was taken from matched truth values. As we are primarily interested in high production rates, we included events with very low $Q^{2}$, and hence the electron was usually scattered below $2\degree$.  For these events, we investigated the benefits of a possible low $Q^{2}$ tagger. Similarly, the recoil proton was usually incident upon the far forward detector region
and so we investigated the acceptances given by the nominal Roman Pot and B0 detector systems. For the 
tracks in the central detector, the main uncertainty is from the $P_{t}$ threshold of the tracks, which can have quite a large effect on the decay pions we hope to detect.

Here, we focus on the results for the 5$\times$100 beam setting unless otherwise stated.

\subsubsection{Far Forward Models}

The far forward and far backward detectors were partially implemented in the simulations. The Roman Pot configuration used in the simulation is based on the high divergence $e+p$ scattering beam configuration, where its actual acceptance represents the $10\sigma$ beam boundary. Here, for the far forward detectors: Roman Pots and B0, a realistic physical coverage was used leading to reasonable estimates for proton acceptances. However, no realistic reconstruction was in place, so there are no genuine momentum components for deducing resolutions of variables requiring proton detection.

\begin{figure}[hbt!]
 \begin{center}
%   \raisebox{0.5mm}{\includegraphics[trim = 2 2 2 2, clip, width=0.95\linewidth]{figs/XYZ/RPHits.png}}
\includegraphics[width=0.95\linewidth]{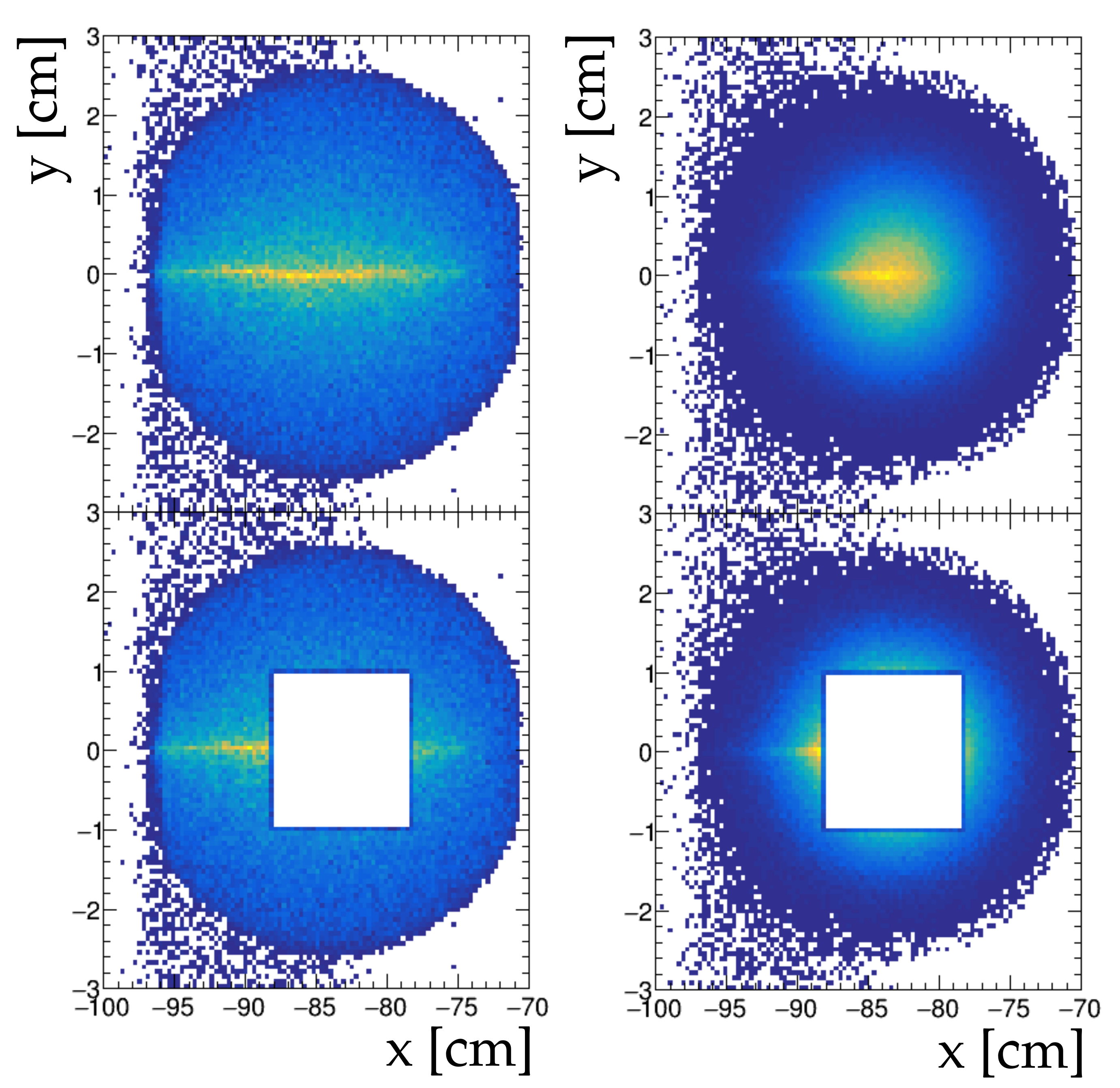}  
  \caption[]{\label{fig:spec_RP} Hit distributions on the first Roman Pot layer for beam settings, left-to-right 5$\times$100 and 18$\times$275. The top row plots show full event distribution; the bottom row plots show the RP acceptance cut applied to remove possible beam backgrounds/contributions.}
 \end{center}
\end{figure}

The hit distribution on the first layer of Roman Pots, at 26 m, is shown in Fig. \ref{fig:spec_RP}.
A further cut was applied on the position: $-1<y<1$, $ -88.22 < x < -78.22$ to remove the region where large backgrounds from beam divergence may occur, this cut is effectively around $10\sigma$ of beam divergence.

To summarize, for events for 5$\times$41, 13\% of events hit the first Roman Pot, and 9\% survive the cut; for 5$\times$100 it is 56 and 37\%; for 10$\times$100 58 and 39\%; and for 18$\times$275 99 and 28\%.

\begin{figure}[hbt!]
 \begin{center}
   \includegraphics[width=0.95\linewidth]{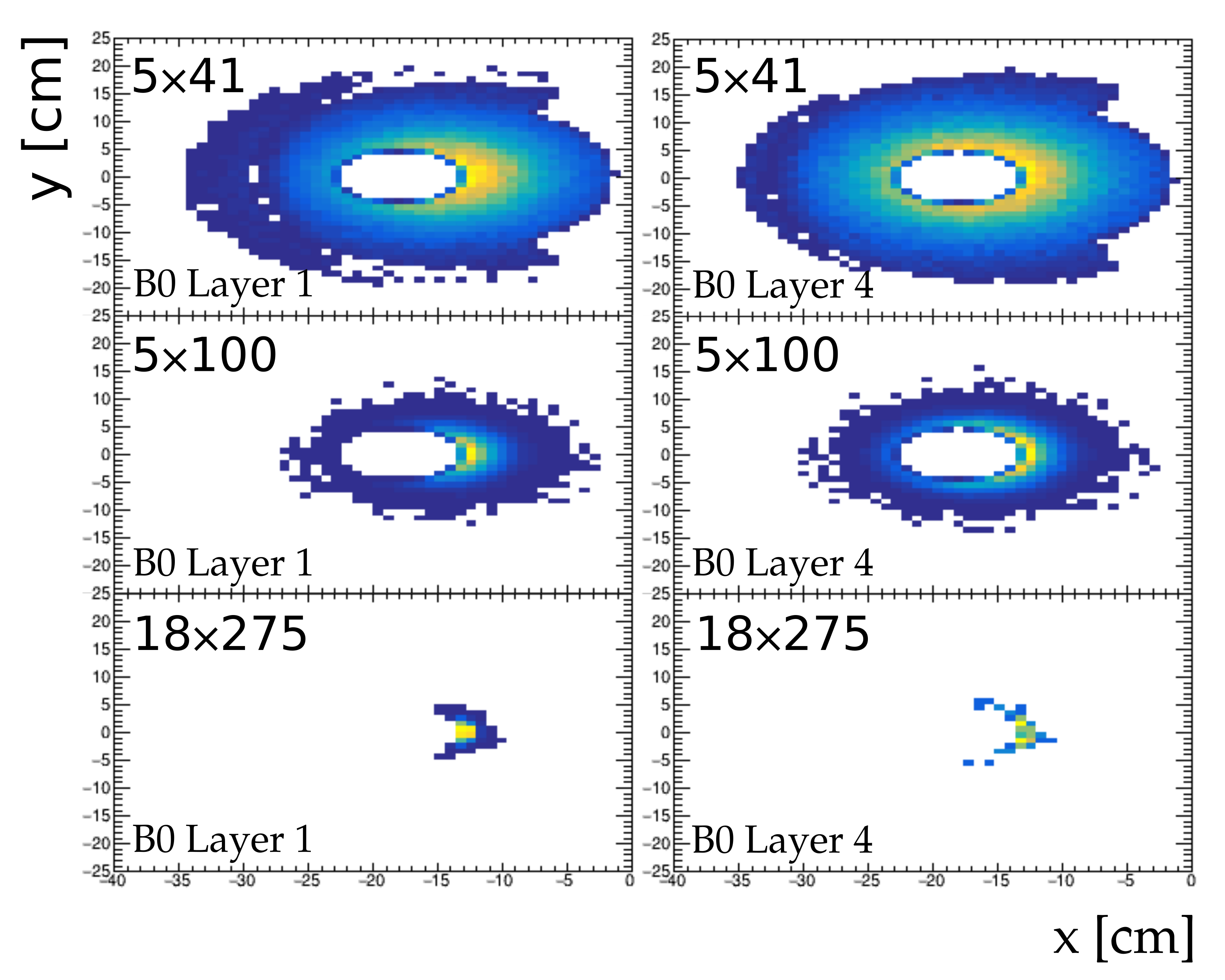}
  \caption[]{\label{fig:spec_B0} Hit distributions on the four B0 layers for beam settings, top-to-bottom 5$\times$100 and 18$\times$275; with left-to-right front-to-back.}
 \end{center}
\end{figure}

Fig. \ref{fig:spec_B0} shows the hit distributions for the 4 layers of the B0 detector for each beam momentum configuration. It is clear the B0 detectors play a far more important role at the lower CM energies.

The estimated far forward detection is also shown in Table \ref{tab:ff_numbers} as a percentage of the total number of events for Roman Pots and B0 detectors. Note, we take the number of B0 hits as the number in the highest occupancy layer.

\begin{table}[hbt]
   \centering
        \caption{Percentage of protons detected in the far forward detector systems.}
        \label{tab:ff_numbers}
        \begin{tabular}{|c|c|c|c|c|}
            \hline
            Setting     &5$\times$41   &5$\times$100  &10$\times$100     &18$\times$275 \\
            \hline
            Roman Pot   &9      &37     &39         &28 \\
            \hline
            B0          &66       &31   &30     &1 \\
            \hline
         \end{tabular}
\end{table}

\subsubsection{Particle Acceptances}

First, we show the event distributions and acceptances for the forward-going recoiling proton which low $Q^{2}$ $t$-channel exchange production process of concern here are all in the far-forward detectors. The overall average acceptance comes to around 63\%.

The majority of scattered electrons will also miss the main detectors and require detection in a low $Q^{2}$ tagger in the far backward region. Around 5\% of the higher $Q^{2}$ events do make it into the backward electron arm of the central detector. Overall, the two systems could detect around 52\% of the electrons, with 45\% potentially in the tagger. We observe unphysical acceptances greater than 1 when we plot in terms of pseudorapidity, this is presumably due to bin migration effects in this non-linear variable, perhaps due to beam divergence effects applied by the simulation afterburner.

To investigate the potential for a low $Q^{2}$ tagger to improve these spectroscopy measurements, we investigated scattered electrons reconstructed in the nominal tagger (cut $\eta<-6.5$) and the main detector. For the 5$\times$100 setting, the tagger supplies an order of magnitude more events with complete reconstruction and will be an important addition to the spectroscopy program particularly for measuring quantum numbers and spin density matrix elements.

\begin{figure}[t]
  \centering
  \includegraphics[width=1\linewidth]{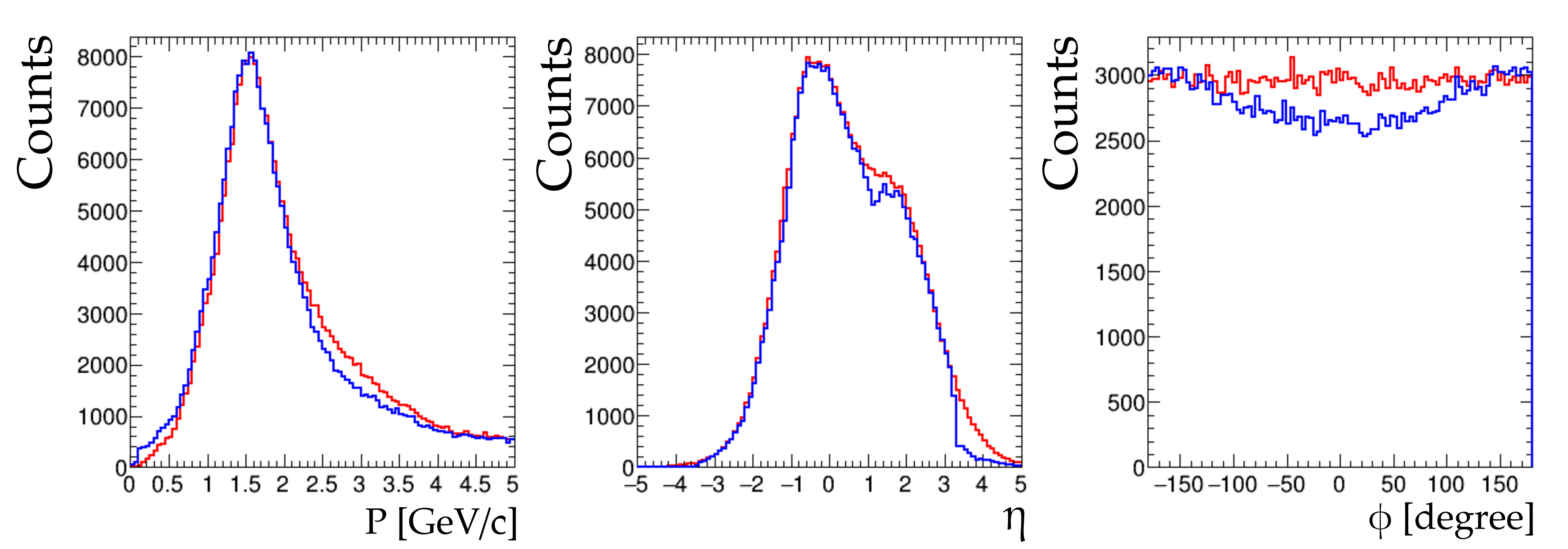}
  \caption[]{\label{fig:spec_JeAcc} 5$\times$100 generated (red) and reconstructed (blue) J$/\psi$ decay $e^{-}$ distributions of momentum (P), pseudorapidity ($\eta$), and angles ($\phi$). }

%\end{figure}
%\begin{figure}[hbt!]
   \centering
   \includegraphics[width=0.99\linewidth]{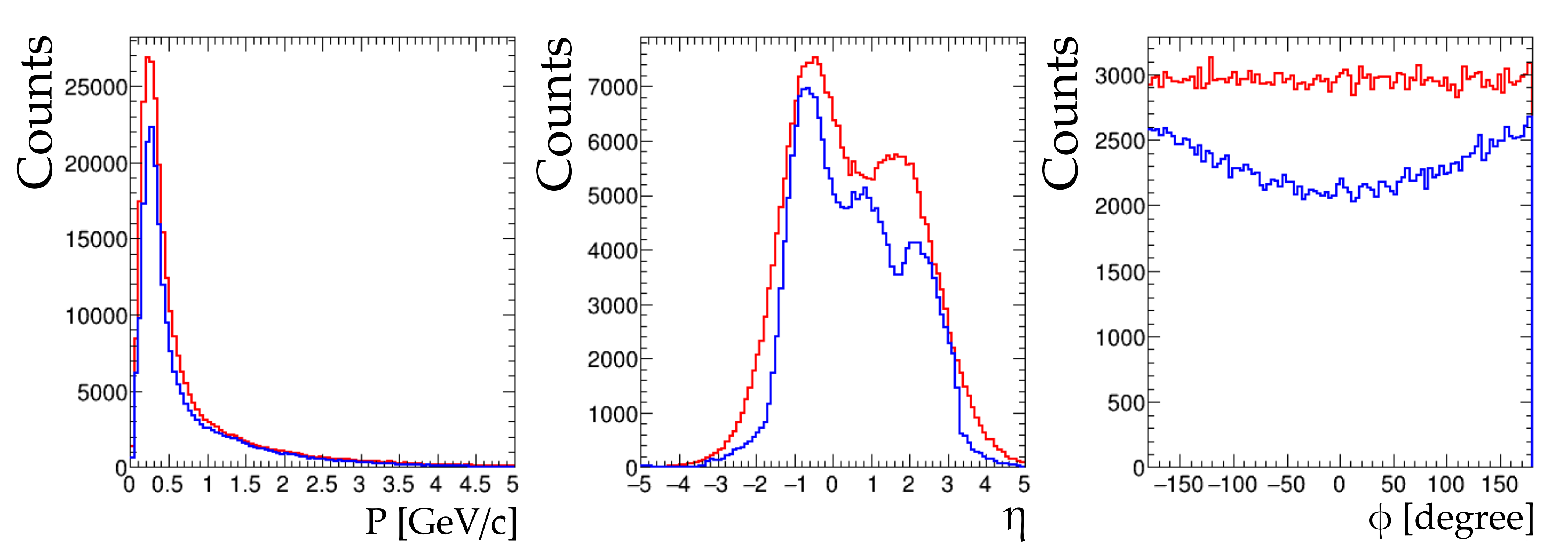}
   \caption[]{\label{fig:spec_pipAcc} 5$\times$100 generated (red) and reconstructed (blue) $\pi^{+}$ distributions of momentum (P), pseudorapidity ($\eta$), and angles ($\phi$).}
\end{figure}

One nice feature of these high-mass meson production processes at the lower CM energies is that the meson decay products populate the detector relatively uniformly allowing excellent acceptance for the states of interest. 
The $e^{+}e^{-}$ decay products from the J$/\psi$ are particularly well reconstructed and shown in Fig. \ref{fig:spec_JeAcc}. With very symmetric responses for both lepton charge states, the average acceptance is around 95\%.

Kinematically, the detection of the pions is more challenging due to their lower momentum, with significant numbers below 200 MeV, which is close to the tracking threshold. The distributions for the $\pi^{+}$ are shown in Fig.~\ref{fig:spec_pipAcc}. These show that the ECCE detector is capable of detecting pions with high efficiency above this threshold. Overall acceptance is higher at central angles, with some fall-off towards the forward detector systems.

\subsubsection{Particle Resolutions}

Good particle resolution is important to separate out background processes. The proton and electron reconstructed tracks were not available for the far forward/backward detectors, so we do not consider their effects here.

\begin{figure*}[htbp]
 \begin{center}
%  \raisebox{0.5mm}{\includegraphics[trim = 2 2 2 2, clip, width=0.6\textwidth]{figs/XYZ/ResolutionsPipJpsie.png}}

   \includegraphics[width=0.85\textwidth]{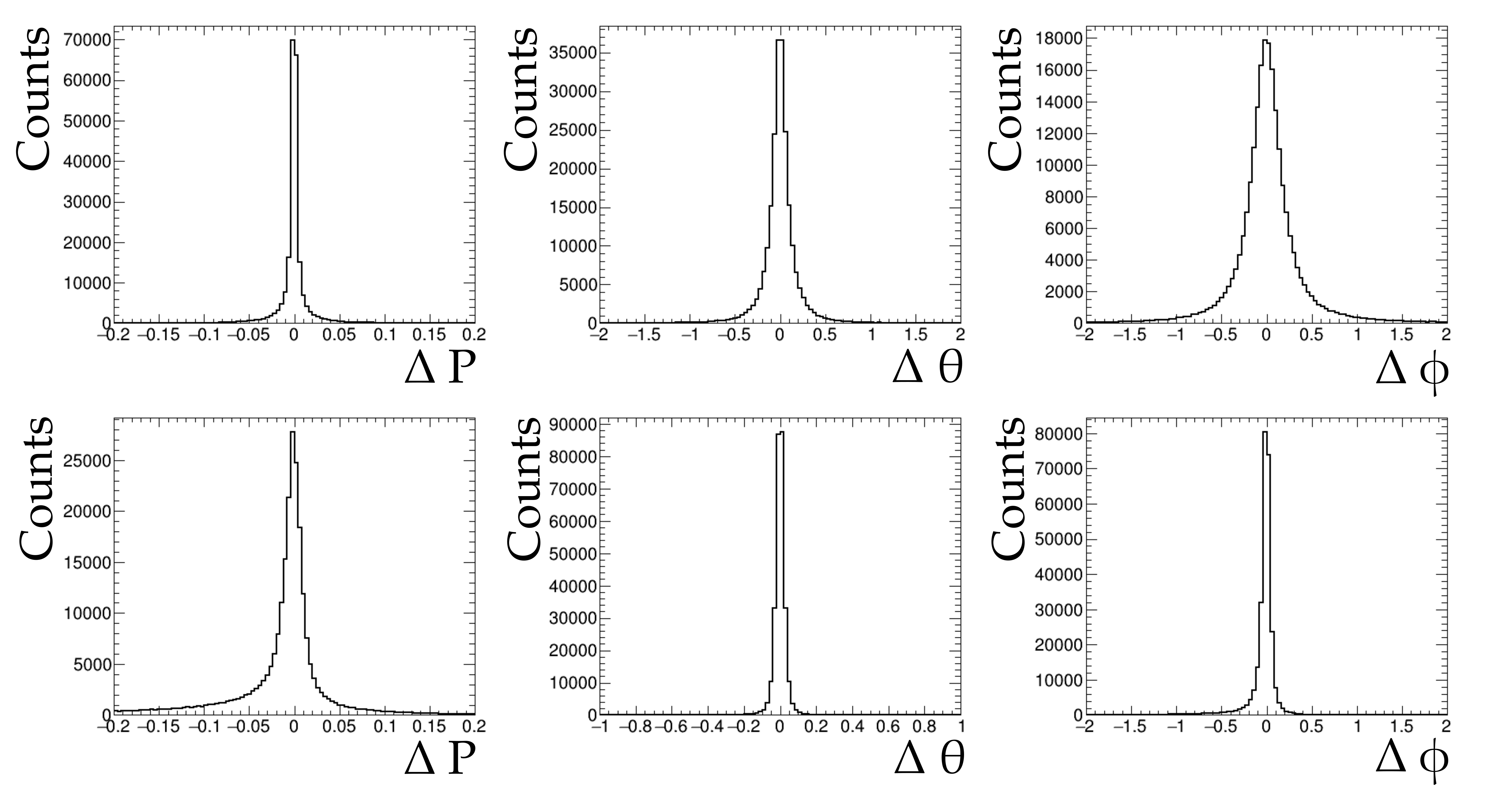}
  \caption[]{\label{fig:spec_pionRes} 5x100 $\pi^{+}$ (Top) and $e^{-}$ (Bottom), resolutions, $\Delta P$, $\Delta\theta$ and $\Delta\phi$ ($\degree$), calculated as the difference between reconstructed and truth values. }
 \end{center}
\end{figure*}

For the case of the pions, the difference in reconstructed to generated momenta is shown in Fig. \ref{fig:spec_pionRes} (Top). The widths of these distributions give an estimate for the resolutions and we find around 0.4\% for momentum; $1.1\degree$ for $\theta$ and $2\degree$ for $\phi$ averaged over all events.

%\begin{figure}[hbt!]
% \begin{center}
%   \raisebox{0.5mm}{\includegraphics[trim = 2 2 2 2, clip, width=0.95\linewidth]{figs/XYZ/ResolutionsPipJpsie.png}}
%  \caption[]{\label{fig:spec_pionRes} 5x100 $\pi^{+}$ (Top) and $e^{-}$ (Bottom), resolutions, $\Delta P$, $\Delta\theta$ and $\Delta\phi$ ($\degree$), calculated as difference between reconstructed and truth values. }
% \end{center}
%\end{figure}

For the J$/\psi$ decay leptons, the difference in reconstructed to generated momenta is shown in Fig. \ref{fig:spec_pionRes} (Bottom). Here, the estimated resolution is closer to 1\% for momentum with a slight radiative tail, and $0.25\degree$ for $\theta$ and $0.5\degree$ for $\phi$ averaged over all events. The significantly better angular resolution is probably related to the higher average momentum of the tracks.

\subsubsection{Event Acceptances and Resolutions}
\label{sec:xyz_AcceptRes}

The effect of the detector systems on the overall physics observables related to the meson photoproduction was considered. The study only involved the exclusive process where all final state particles were detected.

\begin{figure}[ht]
  \centering
  \includegraphics[width=1\linewidth]{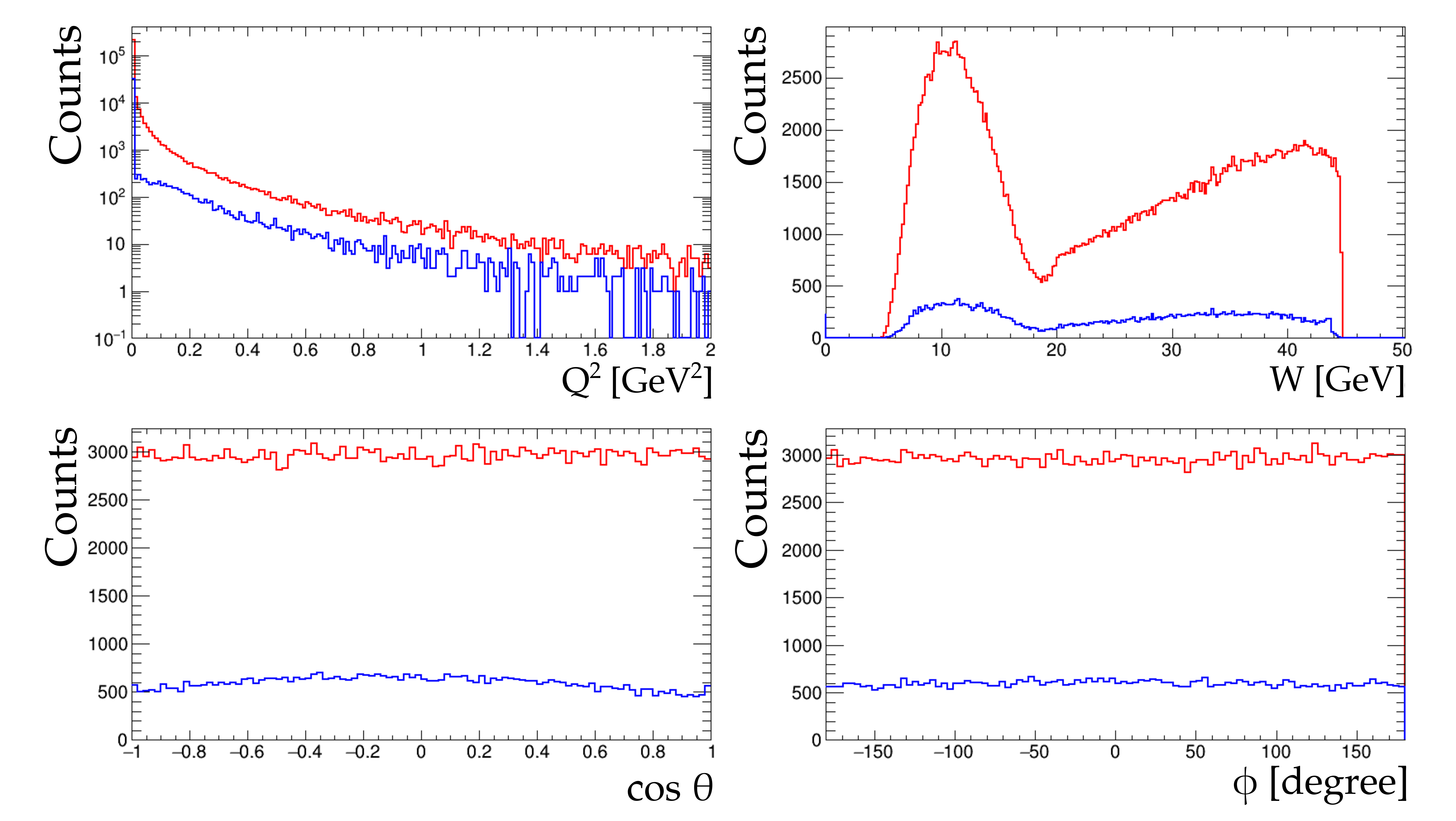}
  \caption[]{Top: 5x100 generated (red) and reconstructed (blue) distributions of $Q^{2}$ and $W$, for events where all particles were detected. Bottom: 5$\times$100 generated (red) and reconstructed (blue) distributions of the produced meson decay angles in the Gottfried-Jackson reference frame, $\cos{\theta}$ and $\phi$ (\degree), for events where all particles were detected. }
  \label{fig:spec_decayAcc}
\end{figure}

The $Q^{2}$, and $W$ distributions are shown in Fig. \ref{fig:spec_decayAcc} top row plots. The acceptances are shown in the bottom row. We do not consider the resolutions for these variables, as they depend on particles detected in the far forward/backward systems.

Overall, the full particle acceptance for this reaction at 5$\times$100 is found to be around 13\%.

Also shown in Fig. \ref{fig:spec_decayAcc} (bottom row plots), are the reconstructed decay angle distributions of the mesonic states. These are very uniform, suggesting ECCE is suitable for performing high-level analysis of the meson decay and therefore accessing quantum numbers and Spin Density Matrix Elements.

\begin{figure*}[ht]
 \centering
   \includegraphics[width=0.85\textwidth]{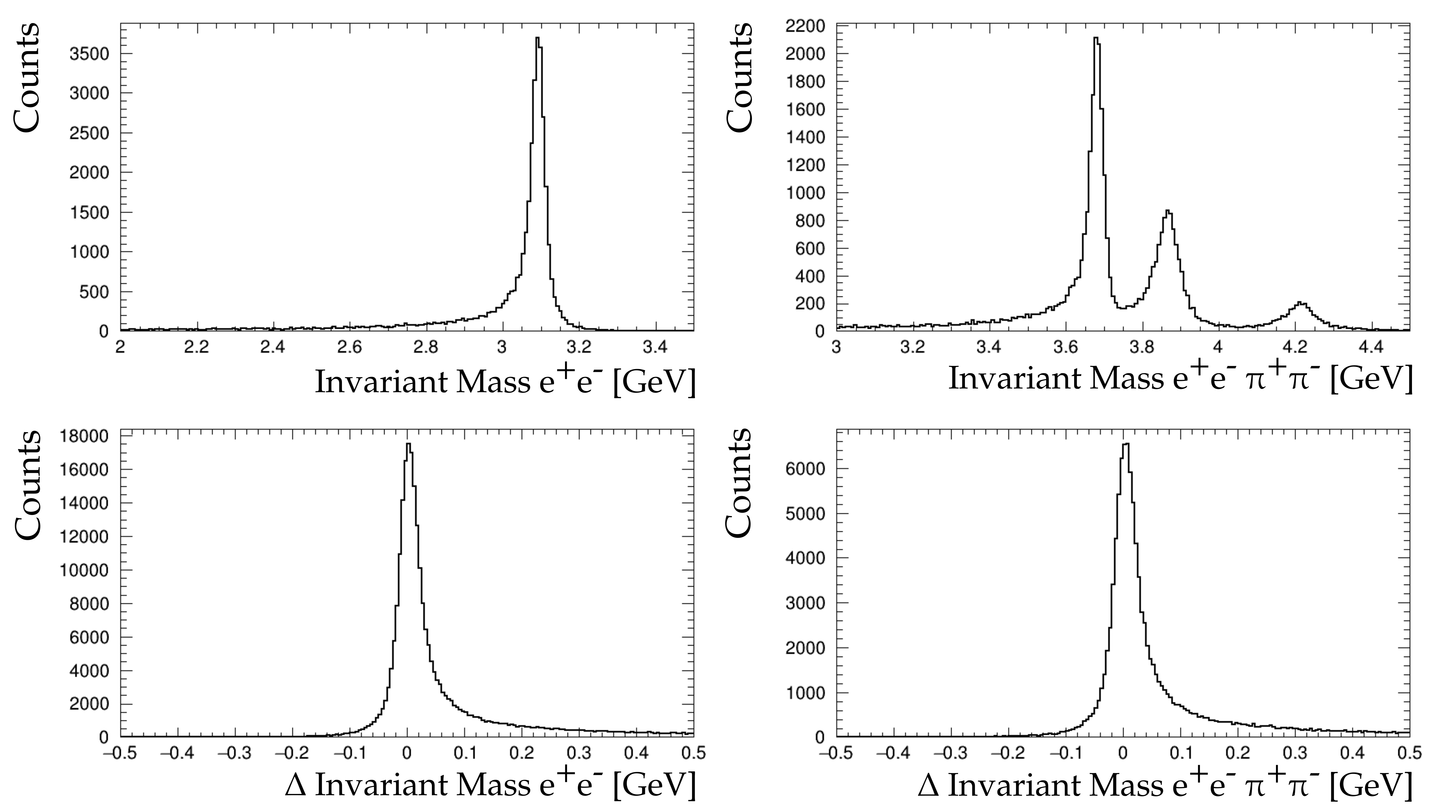}
  \caption[]{\label{fig:spec_masses} Top: reconstructed invariant masses for meson decay products, the three states of interest are clearly observed on the right plot. Bottom: shows the difference in reconstructed to truth masses.}
\end{figure*}

Finally, we show the reconstructed invariant mass distributions and resolutions in Fig. \ref{fig:spec_masses}. The resolutions for $M(e^{+}e^{-})$ and $M(e^{+}e^{-}\pi^{+}\pi^{-})$ are both around 30 MeV. This should be sufficiently narrow for distinguishing many of the final state mesons in the mass region. The good resolution for the J$/\psi$ mass also helps reduce the background from the signal without this meson.

\subsubsection{Summary}

The study presented shows the ECCE detector to be very promising for studies of exotic meson spectroscopy with the EIC. In particular, at mid-center-of-mass energies, the meson decay products are nicely distributed throughout the central detector. Four particle invariant mass resolutions provide sufficient separation to distinguish narrow states with mass differences greater than order 0.1 GeV, compared to typical masses of 4 GeV.

Excellent far-forward and backward detector systems will also be essential for reconstructing the overall reaction kinematics to allow for partial wave analysis and investigations of production mechanisms.

Given that, the fully reconstructed events yields were estimated to be around 10k for $X$, and 3k for $Y$ production for 10~fb$^{-1}$. This is very competitive with previously published experiments, as shown in Table \ref{tab:acc_numbers}.

\begin{table*}[hbt]
   \centering
        \caption{Expected event yields at the 5$\times$100 beam energy configuration for a luminosity of 10 fb$^{-1}$, compared to previous publications for the  J$/\psi\pi^{+}\pi^{-}$ final state. Note yields from published Y results are estimated from the publications rather than given explicitly.}
        \label{tab:acc_numbers}
        \begin{tabular}{|c|c|c|c|c|c|}
            \hline
            Lab.               &ECCE 5$\times$100 &CDF\cite{CDF_PhysRevLett.98.132002}    &LHCb\cite{LHCbX2_Aaij2020} &D$\Phi$\cite{DPhiX_PhysRevLett.93.162002}  &ATLAS\cite{ATLASX_Aaboud2017} \\
            \hline
            $\chi_{c1}(3872)$   &10000      &2292   &4230   &522    &30000 \\
            \hline
            Lab.         &ECCE 5$\times$100    &BABAR\cite{BABARY_PhysRevLett.95.142001}       &BABAR\cite{BABARY2_PhysRevD.86.051102}       &BELLE\cite{BELLEY_PhysRevLett.110.252002}       &BESIII\cite{BESIIIY_PhysRevLett.118.092001} \\
            \hline
            Y(4260)            &3000          &125  &200    &600    &7000\\
            \hline
         \end{tabular}
\end{table*}

\section{Insights for EIC detector 2 at IP8}
\label{sec:IP8}
The EIC accelerator site at BNL is capable of instrumenting a second interaction region at IP8 (Interaction Point 8, where the sPHENIX experiment is currently located) in addition to the primary interaction region planned for IP6 (where the STAR experiment is currently located). Although the current scope of the EIC project consists of only one detector at the IP6 location, the community is enthusiastic about the possibility of instrumenting IP8 with the second detector. The Exclusive, Diffractive, and Tagging working group performed physics impact studies to look for complementarities to the physics measurement of the first detector. Here, the general concept of these impact studies is to keep the design  of the central detector the same as the IP6 and modify the beamline components according to the official preliminary design of IP8 (documented in Ref.~\cite{IP6design}). It is worth noting that the design of the second detector at IP8 is still in the early conceptual stage. 

\begin{figure}[t]
\centering
\includegraphics[width=1.\linewidth]{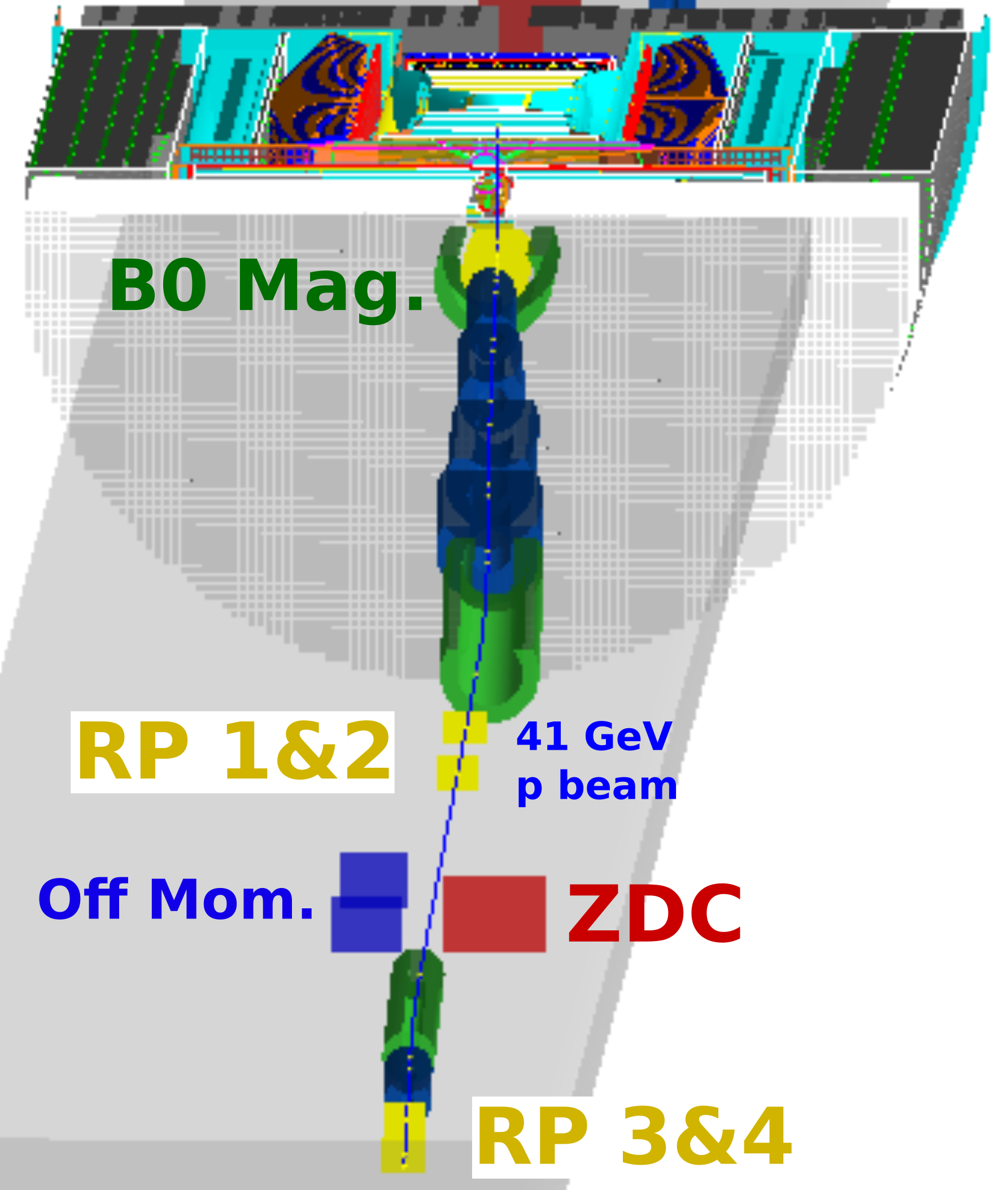}
\caption{41 GeV beam proton steering in IP8 configuration.}
\label{IP8_steering}
\end{figure}

Preliminary designs of the secondary IP feature a larger electron-ion beam crossing angle (35~mrad) and a region of high dispersion followed by a secondary focus.  Figure \ref{IP8_steering} shows a schematic diagram of IP8 with a 41~GeV proton beam being steered through the Far-forward detector stack.

The full ECCE simulation package, Fun4All, is capable of simulating physics processes using the IP6 and IP8 configurations. Currently, there are two main differences between the two configurations: 
\begin{enumerate}
    \item The addition of the secondary focus at IP8; 
    \item The crossing angle is 35~mrad at IP8, compared to 25~mrad at IP6. Consequently, the ZDC acceptance is larger than $\pm8$~mrad at IP8, compared to $\pm5$~mrad at IP6.
\end{enumerate}

It is important to note that the baseline magnet configuration only allows $\pm5$~mrad ZDC acceptance due to magnet aperture constraints at IP8. However, an alternative (improved) magnet design (with Nb3Sn) brings the possibility of an enlarged ZDC acceptance up to $\pm8$~mrad. The studies in this section are based on this optimized scenario~\cite{Gamage21}. 

Near the second focus, particles that are close in rigidity (momentum/charge) to the beam are separated, while the beam itself is focused on a small beam spot. This allows a set of Roman pot (RP) detectors (3\&4 as shown in Fig.~\ref{IP8_steering}) to be placed close to the beam where they can detect particles that are slightly off from the beam rigidity.  

Several physics processes can benefit significantly from the secondary focus capability. Examples include veto-tagging of incoherent diffractive vector meson production by detecting nuclear remnants, better kinematic acceptance for measuring the pion structure function (Sec.~\ref{subsec:pi_sf}), and better acceptance for $e$A-DVCS (Deeply Virtual Compton Scattering)~Sec.~\ref{subsec:dvcsea}.

In this section, we will discuss some studies using a very preliminary design and simulation of IP8. These include the basic Roman Pot acceptance in rigidity and angle, the impact on diffractive studies, and the impact on acceptance in exclusive physics processes. 

\subsection{Roman Pot Acceptance}
\label{ssec:RPOcc}

For this study, the BeAGLE~(Benchmark eA Generator for LEptoproduction) event generator~\cite{Chang:2022hkt} is used to simulate $e$Zr exclusive J$/\psi \rightarrow e^- + e^+$ events colliding with beam energies given by 18~GeV for the electrons and 122.22~GeV/nucleon for the Zr. 

The occupancy of hits registered in the RPs is studied in the first two RP layers (consisting of silicon trackers) in the IP6 configuration as well as all four RP layers in the IP8 configuration. Of particular interest are the RPs near the secondary focus in the IP8 configuration (third and fourth layers). Occupancies from layers two and four are similar to layers one and three, respectively.

\begin{figure}[htbp]
\centering
\includegraphics[width=0.8\linewidth]{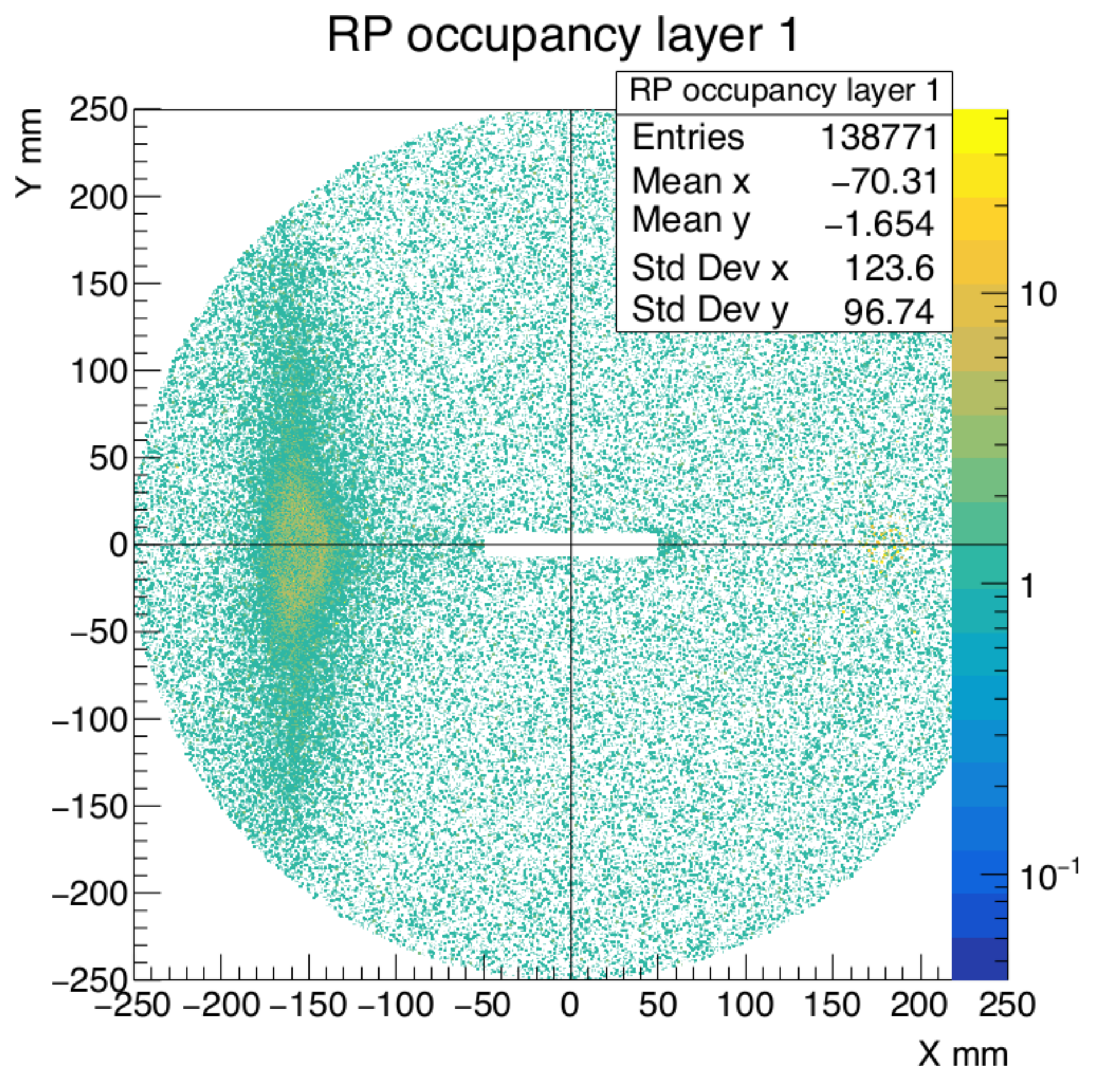}\\
\includegraphics[width=0.8\linewidth]{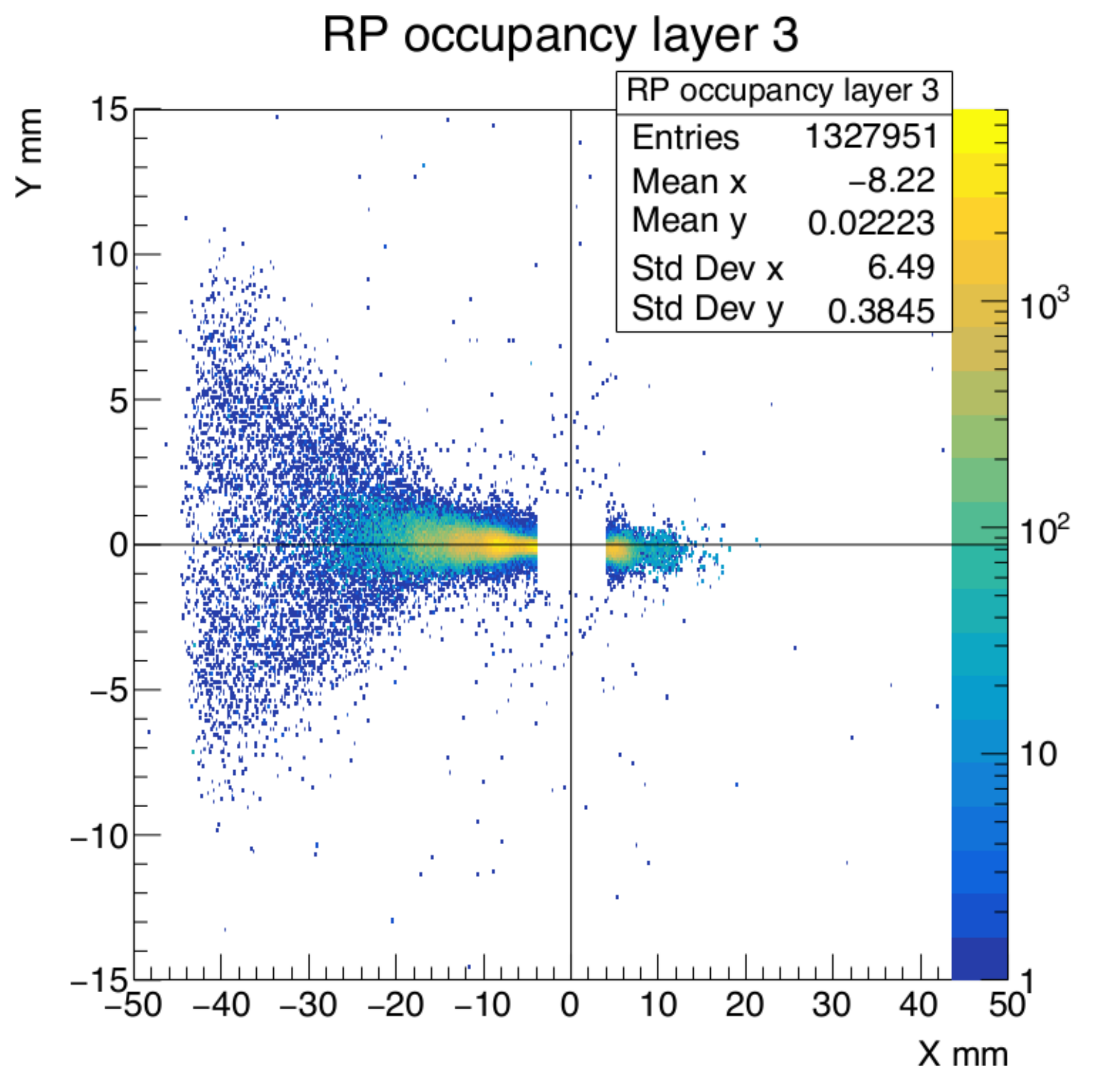}
\caption{Roman Pot occupancy Y versus X for layer 1 (top plot) and layer 3 (bottom plot) near the secondary focus.  The $10\sigma$ beam cut is visible at the center.  Note the different scales on the two plots.}
\label{fig:RP_XY_occup}
\end{figure}

The Geant4 truth hits are plotted versus the X and Y local coordinates of the particular layer in Fig.~\ref{fig:RP_XY_occup}. The rectangular $10\sigma$ beam cut is visible in the center and is much smaller for the RP near the secondary focus. A sharper focus of particles is evident for the third RP layer. For layer 1, $10\sigma_x = 5$ cm and $10\sigma_y = 0.7$ cm. For layer 3, $10\sigma_x = 0.4$ cm and $10\sigma_y = 0.16$ cm.

\begin{figure*}[htbp]
\centering
\includegraphics[width=0.45\textwidth]{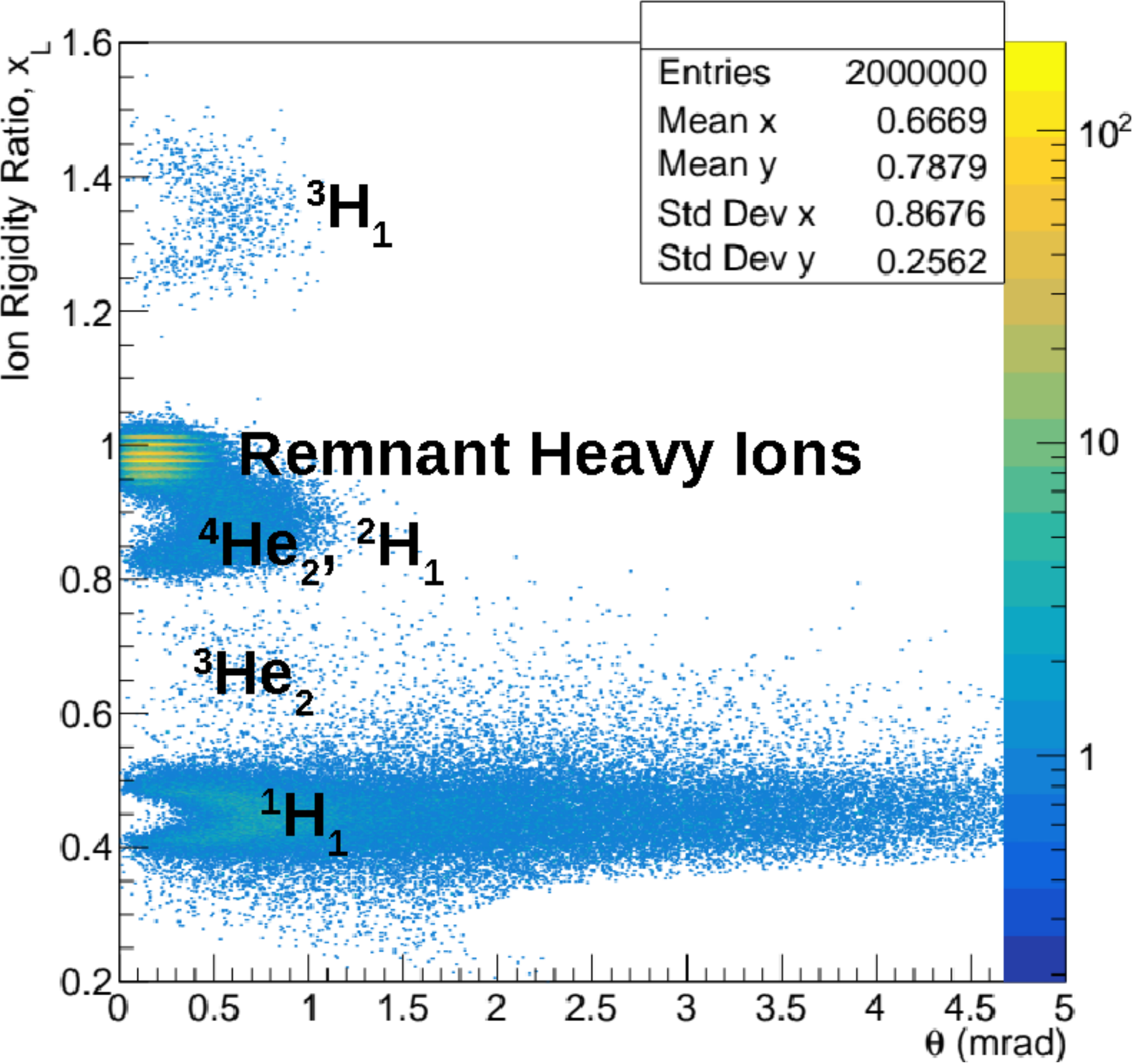} ~~
\includegraphics[width=0.45\textwidth]{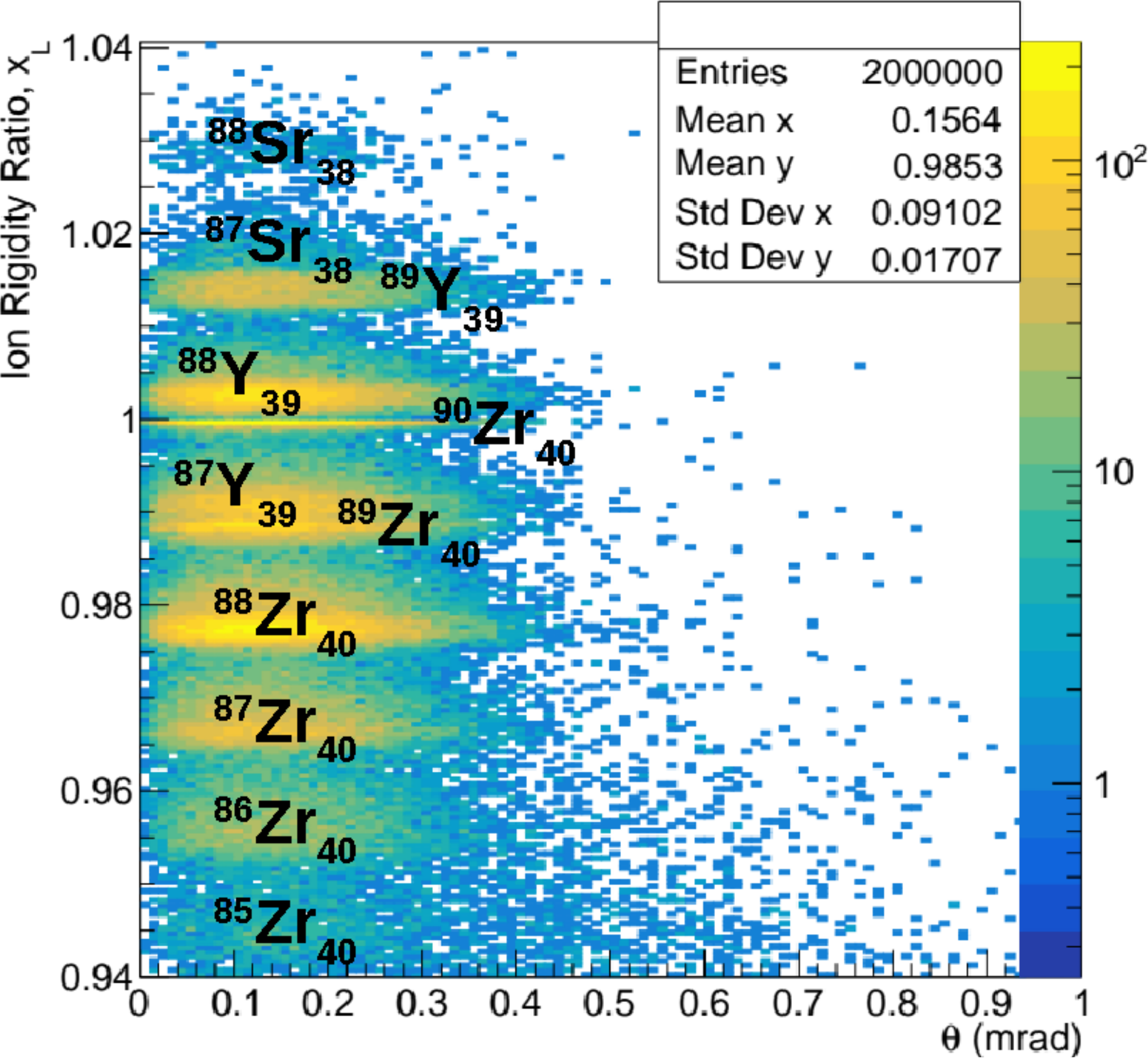}
\label{fig:xL_theta_noCutout}
\includegraphics[width=0.45\textwidth]{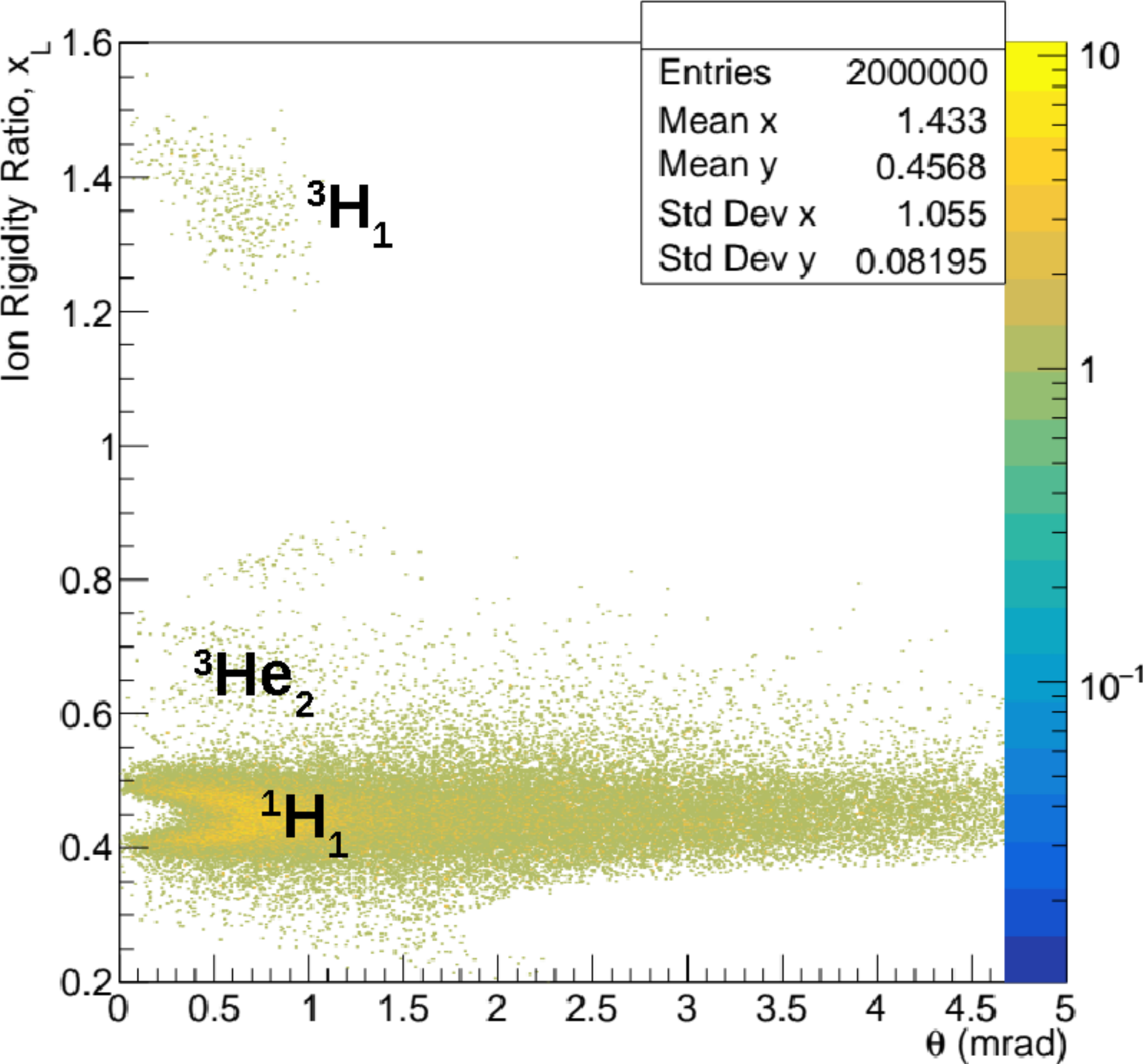} ~~
\includegraphics[width=0.45\textwidth]{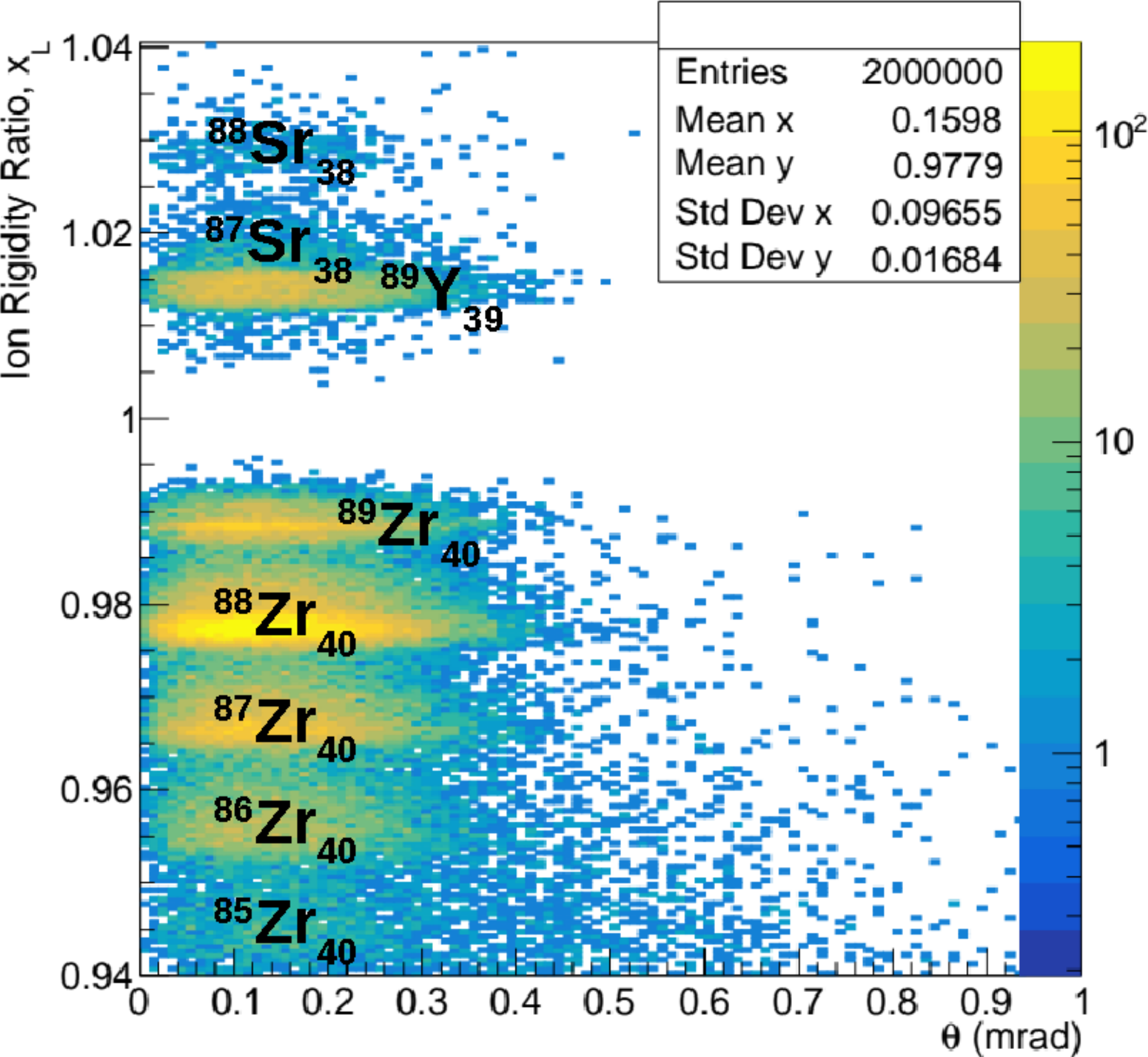}
\caption{Top left and right plots show RP occupancy $x_L$ versus $\theta$ in BeAGLE for layer 1 and layer 3 (near the secondary focus), while the $10\sigma$ beam cut is not applied; bottom left and right show the same for the case when the $10\sigma$ beam cut is applied. $x_L$ is defined as the rigidity fraction: $(p/Z) / (p/Z)_{\textrm{beam}}$. Note the different scales on the plots.}
\label{fig:xL_theta}
\end{figure*}

To better distinguish the identity of the ions detected in the RPs, the truth hits are matched to the generator-level particle, which is used to construct the rigidity ratio $x_L=(p/Z) / (p/Z)_{\textrm{beam}}$ and polar angle $\theta$. The RP occupancies in $x_L$ vs $\theta$ (after applying the $10\sigma$ cut) for all four layers of RP are shown  in Fig.~\ref{fig:xL_theta}. Note that the distribution in $x_L$ for layer 3 is much narrower than for layer 1.

%The bottom panel illustrates the RP occupancies after applying the $10\sigma$ beam cut. Each cluster is labeled by its associated ion. 

None of the remnant heavy ions are visible in layer 1, while most of them remain in layer 3, which is near the secondary focus.  
However, the lighter ions ($^{1}\textrm{H}_{1}$, $^{3}\textrm{H}_{1}$, $^{3}\textrm{He}_{2}$), with rigidities very different from the beam, are clearly visible in layer 1, while they are not detectable for the layers near the secondary focus. 

\begin{figure}[t!]
\centering
\includegraphics[width=0.99\linewidth]{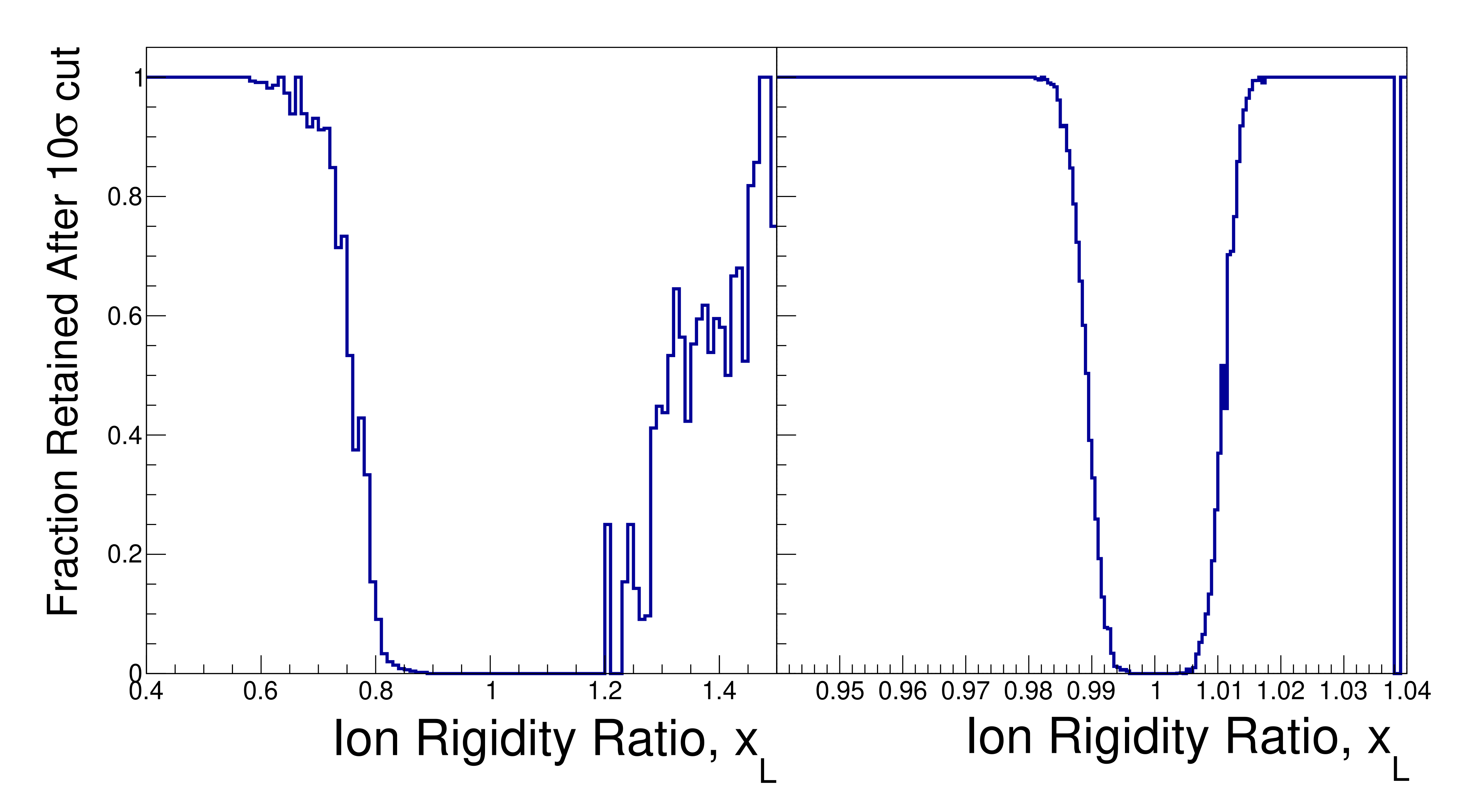}

\caption{Roman Pot $x_L$ acceptance (left) for layer 1 and (right) for layer 3 near the secondary focus for the range $\theta<1$~mrad in BeAGLE.  $x_L$ is defined as the rigidity fraction: $(p/Z) / (p/Z)_{\textrm{beam}}$. Note the different scales on the x-axis.}
\label{fig:xL_accep}
\end{figure}

The derived $x_L$ acceptances corresponding to $\theta<1$~mrad are shown in Fig.~\ref{fig:xL_accep}. It is clear that the Roman pots near the secondary focus greatly increase the $x_L$ acceptance to about 0.015 from the beam rigidity. 
%Note that $x_L$ is defined in caption of Fig.~\ref{fig:xL_accep}.
% and also in the text above. This line is not needed!

\begin{figure}[ht]
\centering
\includegraphics[width=0.99\linewidth]{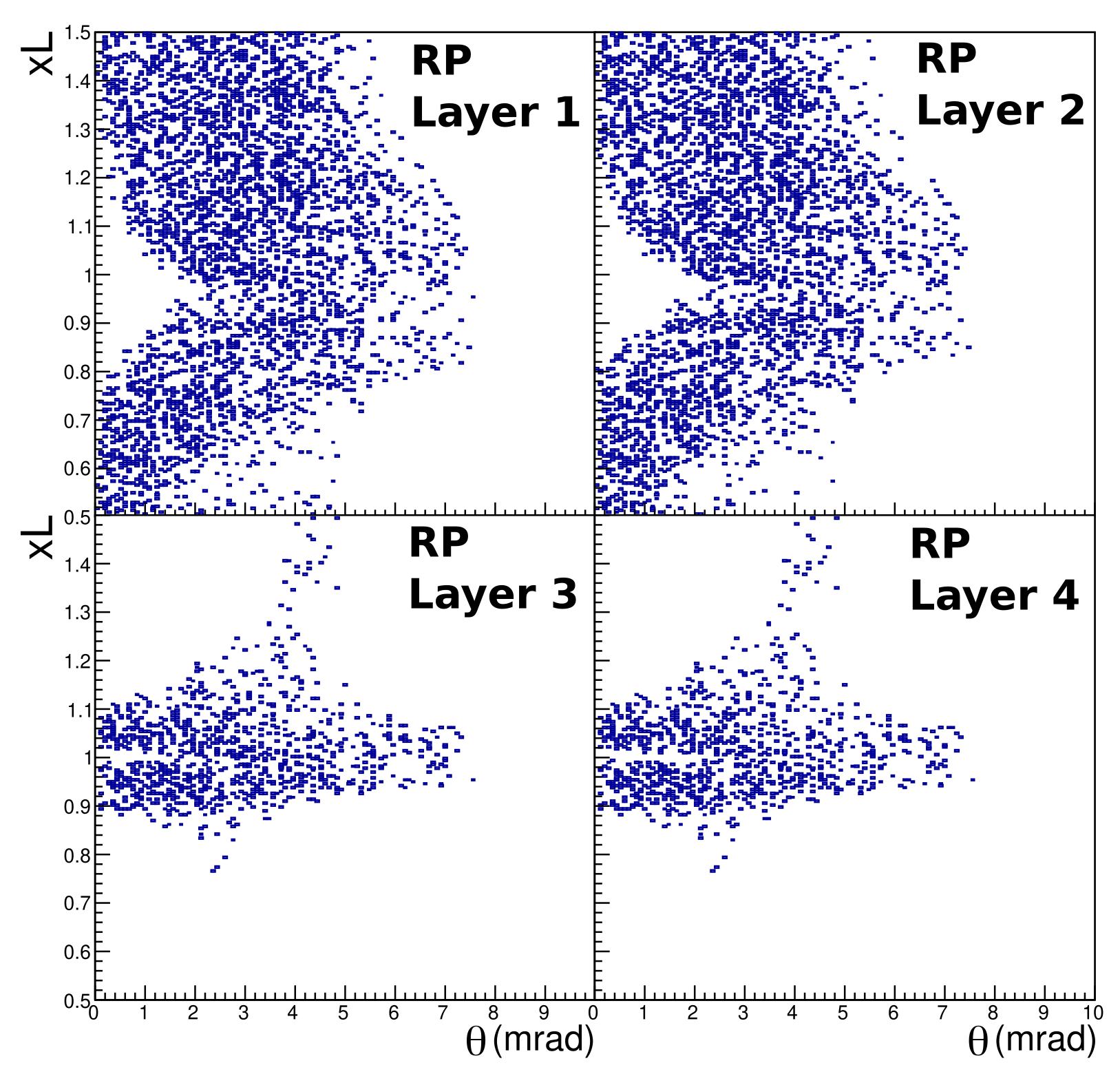}
\caption{Roman Pot $x_L$ vs.\ $\theta$ occupancy for all four layers (3 and 4 are near the secondary focus) in the proton Monte Carlo. $x_L$ is defined as the rigidity fraction: $(p/Z) / (p/Z)_{\textrm{beam}}$.}
\label{fig:accep_2d4lay}
\end{figure}

In order to get a more comprehensive view of this acceptance, a toy Monte Carlo was made with scattered protons simulated uniformly over $0.5<x_L<1.5$ and $\theta <10$~mrad. The occupancy for all four Roman Pot layers is shown in Fig.~\ref{fig:accep_2d4lay}. Recall that layers 3 and 4 are near the secondary focus in IP8, while 1 and 2 have a similar acceptance to that seen in IP6.

\begin{figure}[t!]
\centering
\includegraphics[width=0.99\linewidth]{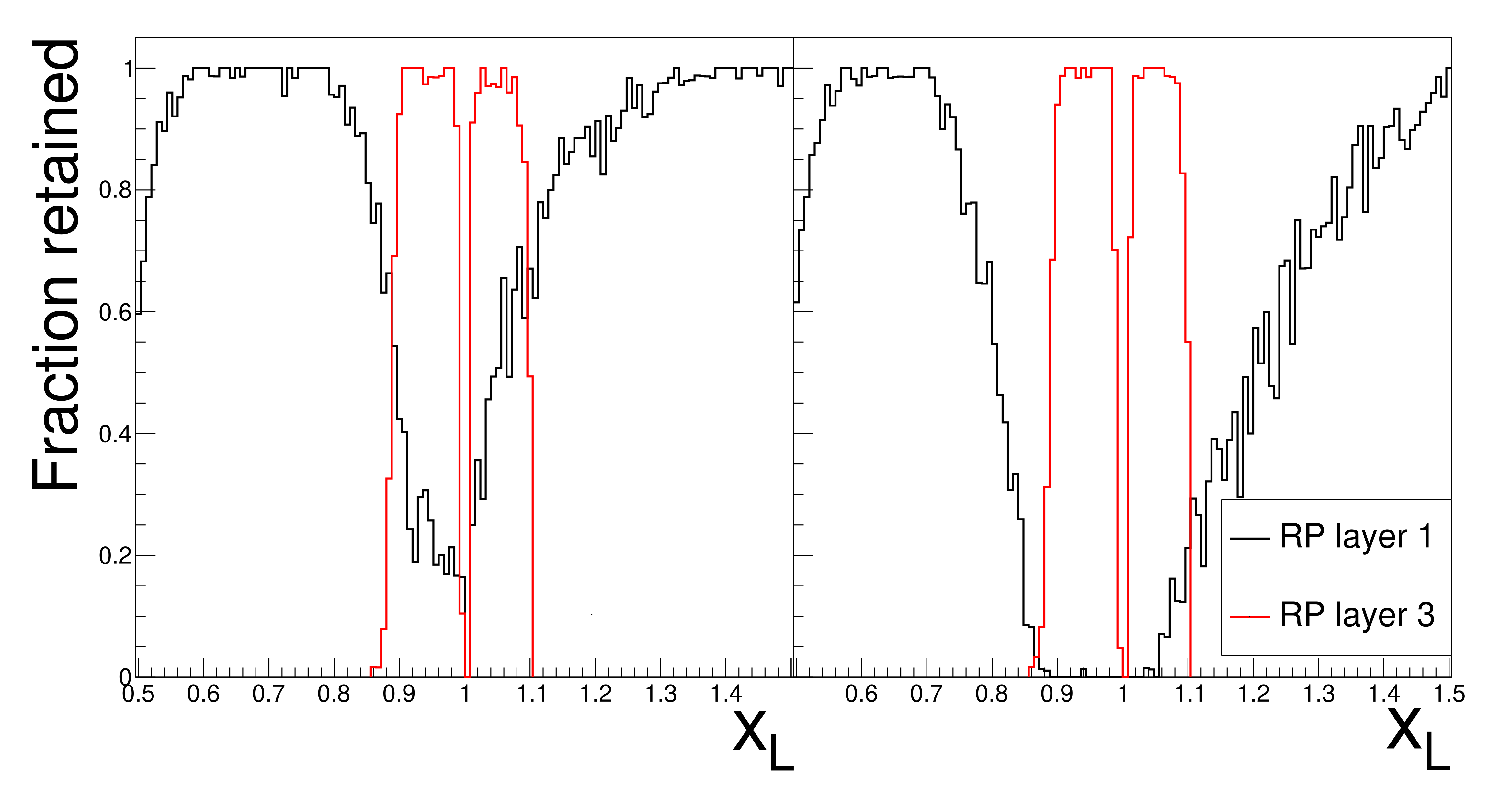}

\caption{Roman Pot $x_L$ acceptance overlaid for layers 1 (conventional) and 3 (secondary focus) for the range $\theta<1$~mrad for the high divergence (left) and high acceptance (right) settings of the beam configuration.  Proton Monte Carlo data was used.}
\label{fig:accep_1d2lay}
\end{figure}

The left panel of Fig.~\ref{fig:accep_1d2lay} shows the projected acceptance as a function of $x_L$ for the band $\theta<1$~mrad. Note that this is not the full range of coverage, but rather focuses on smaller polar angles.  It can be seen that the coverage is not complete.  The right panel of Fig.~\ref{fig:accep_1d2lay} shows the improved acceptance for the ``High Acceptance'' machine parameters, which allows the Roman pots to be moved closer to the beam. The coverage is significantly improved. It should be noted that these studies are preliminary and that the design of the IR and forward detectors for IP8 are expected to evolve. It should also be noted that these studies used the beam parameters and Roman plot placement appropriate to the proton beam, even for the Zr beam studies, as Zr beam parameters are not known at this time. Plots such as these will be useful for optimizing the detector placement and machine parameter decisions moving forward.

\begin{figure}[h!]
\centering
\includegraphics[width=0.9\linewidth]{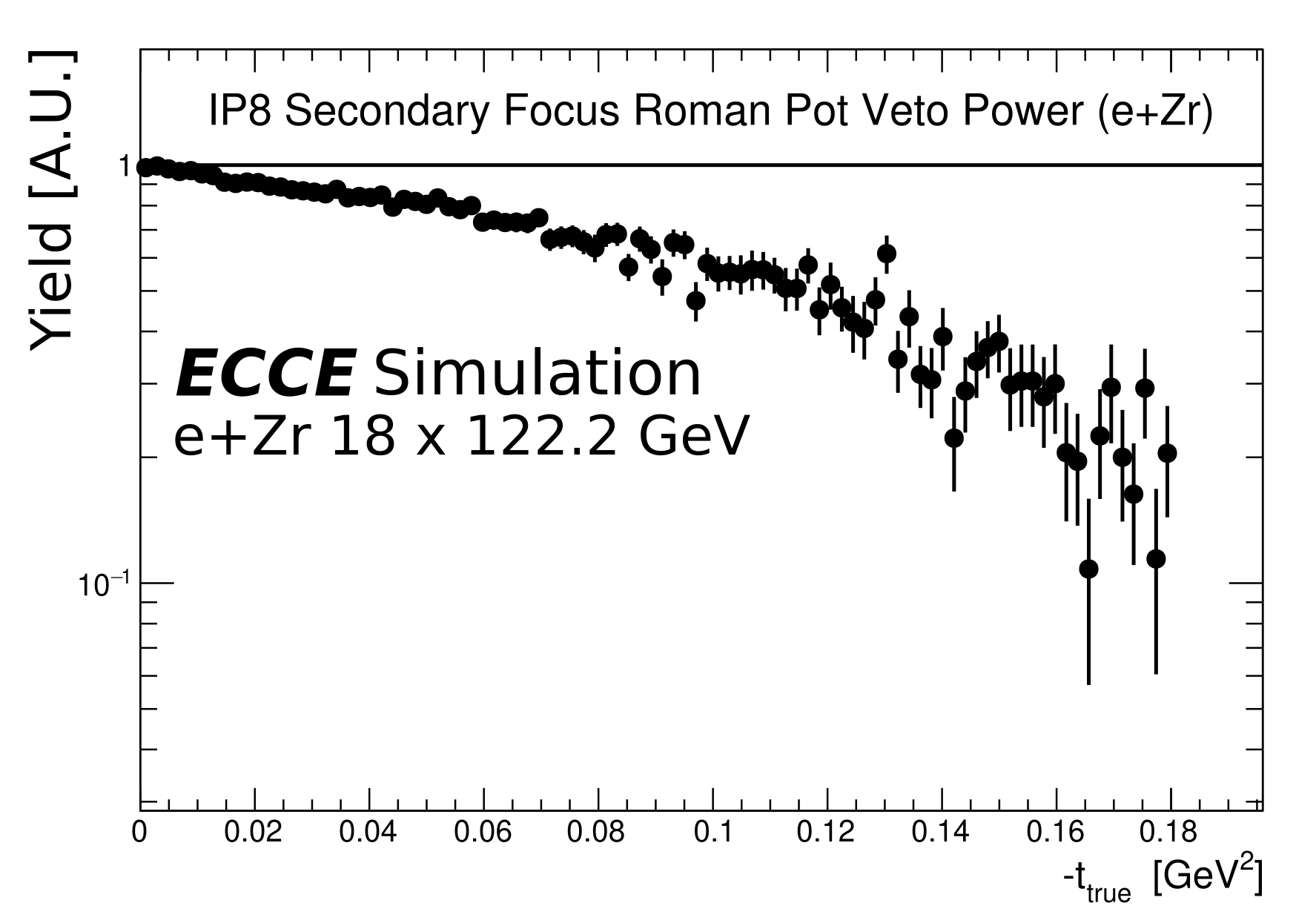}
\caption{Background veto efficiency effectiveness of the secondary focus Roman pots in IP8 as a function of $-t$ for the $e$Zr diffractive J/$\psi$ electroproduction study.
}
\label{diff_SFRP_veto_IP8_vs_t}
\end{figure}

\subsection{Effect of Secondary Focus on Veto Tagging in Diffraction}

Sec.~\ref{ssec:RPOcc} illustrates the ability of the secondary focus at IP8 to allow the detection of nuclear remnants close to beam rigidity (momentum/charge). We can use this ability to improve our efficiency in tagging incoherent diffractive events vs.\ coherent diffractive events.

Coherent $e$A diffractive J$/\psi$ production is an important measurement at the EIC, as it allows access to the spatial distribution of gluons in the nucleus~\cite{khalek2021science}. In order to make this measurement, incoherent diffractive events must be vetoed as they swamp the signal. The ability to veto incoherent $e$A diffractive J/$\psi$ production was studied using both the IP6 and IP8 configurations. Note: B0 photon detection and the beampipe were not yet implemented in this study. Fig.~\ref{diff_SFRP_veto_IP8_vs_t}  shows the impact of the secondary focus at IP8 in the case of $e$Zr diffractive events. The line at 1 corresponds to the amount of background remaining when all of the cuts are made except the secondary focus. The points show the relative effect of the cut using RP layers 3 and 4. In particular, they correspond to the ratio of the background after all cuts are made (including the secondary focus RP layer 3 and 4 cuts) to the background before that cut. The additional background rejection is significant, particularly at larger values of $|t|$.

\subsection{Pion SF IP8}

\begin{figure}
    \centering
    \includegraphics[trim={0 0 0 5cm}, clip, width=0.95\linewidth]{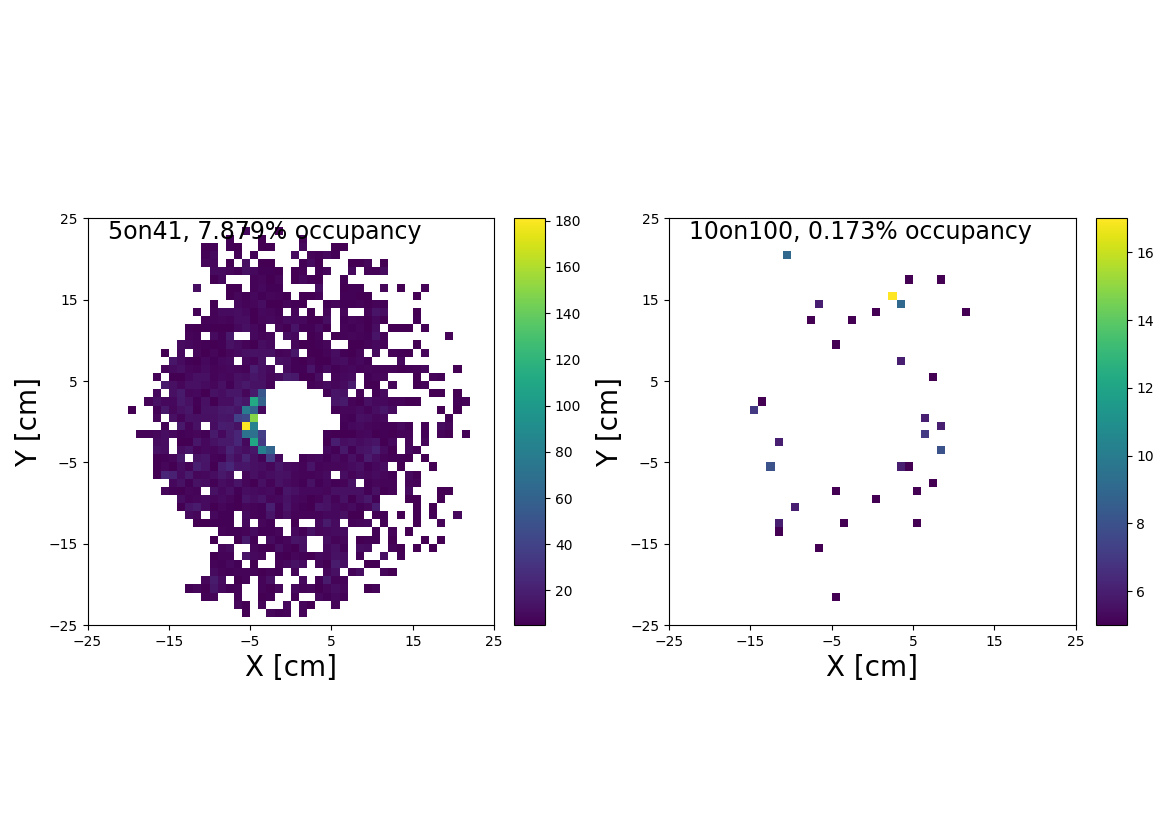}\\[-1cm]
    \includegraphics[trim={0 3.8cm 0 5cm}, clip, width=0.95\linewidth]{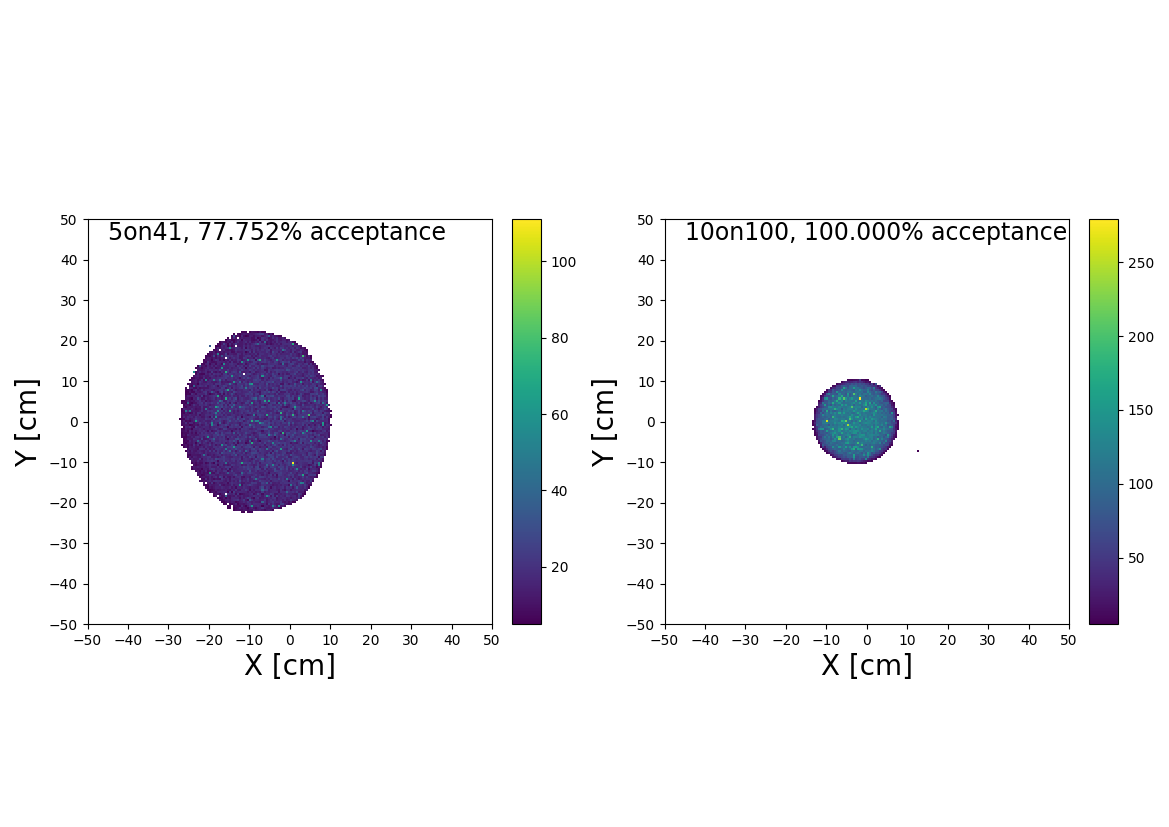}  
    \caption{ Top plots: B0 occupancy of the simulated leading neutron for a range of energies 5$\times$41~GeV (left) and 10$\times$100~GeV (right) at IP8. Bottom plots: ZDC acceptance of the simulated leading neutron for a range of energies 5$\times$41~GeV (left) and 10$\times$100~GeV (right) at IP8.}
    \label{fig:b0_zdc_ip8}
\end{figure}

 The B0 occupancy in Fig.~\ref{fig:b0_zdc_ip8} top row plots show a marginal decrease from IP6 to IP8. Similar to IP6, the ZDC acceptance for IP8 in Fig.~\ref{fig:b0_zdc_ip8} bottom row plots show a significant drop in neutron detection for the lowest energy setting (i.e. 5$\times$41) due to the increased occupancy in the B0. This drop is more prominent at IP6 than at IP8. 

\begin{table}
\begin{center}
\begin{tabular}{ |c|c|c|c|c |  }
 \hline
 \multicolumn{1}{|c|}{} &
 \multicolumn{2}{|c|}{IP6} &
 \multicolumn{2}{|c|}{IP8} \\
 \hline
  Energy [GeV] & Detector & $\Delta$t & Detector & $\Delta$t \\
               & Fraction &           & Fraction &           \\
 \hline
 5$\times$41  & 59$\%$ & 0.019 & 78$\%$ & 0.018\\
 5$\times$100 & 100$\%$  & 0.007 & 100$\%$ & 0007\\
 10$\times$100 & 100$\%$ & 0.007 & 100$\%$ & 0.007\\
 18$\times$275 & 100$\%$ & 0.005 & 100$\%$ & 0.008\\
 \hline
\end{tabular}
\caption{The neutron detection fractions in the ZDC from the above plots are laid out for a range of energies (5$\times$41~GeV, 10$\times$100~GeV) at IP6 and IP8 as well as the deviation of t from the detected value of t (i.e. $\Delta$t).}
\label{tab:det_frac_del_t}
\end{center}
\end{table}

\begin{figure}
    \centering
    \includegraphics[ width=0.98\linewidth]{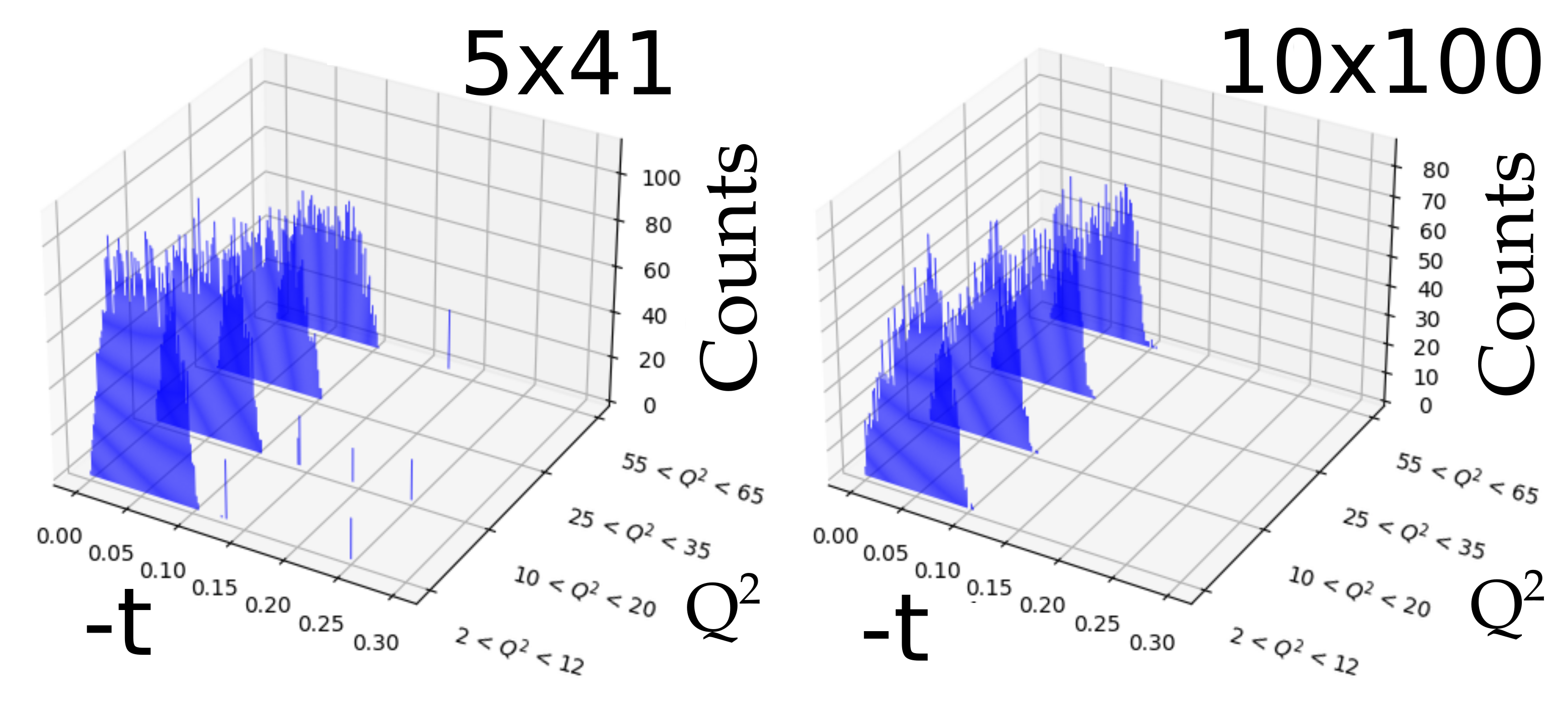} 
    \caption{\label{fig:t_Q2_3d_IP8}
    The $-t$ distribution for a range of energies at IP8 (5$\times$41~GeV, 10$\times$100~GeV) at IP8. There are four $Q^2$ bins presented (7, 15, 30, 60 GeV$^2$) of bin width $\pm$5 GeV$^2$. 
    }
\end{figure}

Fig.~\ref{fig:t_Q2_3d_IP8} shows the $t$-distribution for the two energies at IP8 for a range of $Q^2$ bins. The results are similar to those of IP6, with the drop in events at the higher $Q^2$ bins for the lower energy. This is best shown in Table \ref{tab:det_frac_del_t}, where the detection fraction of the ZDC and the deviation of $t$ from the detected $t$ (i.e. $\Delta{t}$) are broken down for four energies at IP6 and IP8.

\section {Summary}
\label{sec:summary}
This article presents a collection of simulation studies using the ECCE detector concept in the context of the EIC's exclusive, diffractive, and tagging physics program.  This program is a wide umbrella that covers a diverse set of reactions (as listed in Table~\ref{tab:physics}), and ultimately provides answers to the physics questions asked by the NAS white paper (in Sec.~\ref{sec:overview}).

The unifying theme of this program is the key role played by the far-forward and far-backward detector systems. These detector systems are used, either to ensure exclusivity, isolate diffractive reactions, or detect a particle that serves as a tag for a particular reaction of interest. 

Full simulation studies are the best way to study and establish the expected performance of the individual detector responses in far-forward systems. The preliminary results (Sec.~\ref{sec:physics}) demonstrate the design proposed by ECCE exceeds the requirements (in acceptance and resolution) underlined by the YR. Here, it is important to point out that the proposed technologies represent a snapshot in time in the development of the EIC, where further modification and improvements will be made in the near future.

For the physics impact studies, slightly different choices and assumptions about reconstruction are made and based on the general projected detector performances (based on full simulation results). The results represent the expected physics impact/significance with an integrated luminosity of 10$^{-1}$~fb (which corresponds to the first few years of EIC commissioning and operation). The results of these studies can be used for future comparison. As the EIC detector's design becomes more refined, and as inevitable trade-offs and compromises are made when turning an idea into reality, future simulations can be compared with these results to understand the impact on eventual physics the EIC can deliver. 

Similarly, the IP8 studies of Sec.~\ref{sec:IP8} can serve as a guidepost for the future development of a second EIC detector. The secondary focus at IP8 holds potential that may at some point be exploited.

The EIC will usher in a new era of exploration of the rich quark-gluon structure of nucleons and nuclei. As shown in these studies, the ECCE detector concept can deliver impactful results on a host of interesting questions through its exclusive, diffractive, and tagging physics program.

\section{Acknowledgements}
\label{acknowledgements}

We thank the EIC Silicon Consortium for cost estimate methodologies concerning silicon tracking systems, technical discussions, and comments.  We acknowledge the important prior work of projects eRD16, eRD18, and eRD25 concerning research and development of MAPS silicon tracking technologies.

We thank the EIC LGAD Consortium for technical discussions and acknowledge the prior work of project eRD112.

We acknowledge support from the Office of Nuclear Physics in the Office of Science in the Department of Energy, the National Science Foundation, the Los Alamos National Laboratory Laboratory Directed Research and Development (LDRD) 20200022DR, the Natural Sciences and Engineering Research Council of Canada (NSERC), and the UK Research and Innovation Science and Technology Facilities Council.

This research used resources from the Compute and Data Environment for Science (CADES) at the Oak Ridge National Laboratory, which is supported by the Office of Science of the U.S. Department of Energy under Contract No. DE-AC05-00OR22725.     

And we gratefully acknowledge the support of Brookhaven National Lab and the Thomas Jefferson National Accelerator Facility which are operated under contracts DE-SC0012704 and DE-AC05-06OR23177 respectively.
\appendix
\section{Detector layout inside B0 magnet}
\begin{figure*}[!ph]
    \centering
    \includegraphics[width=0.87\linewidth]{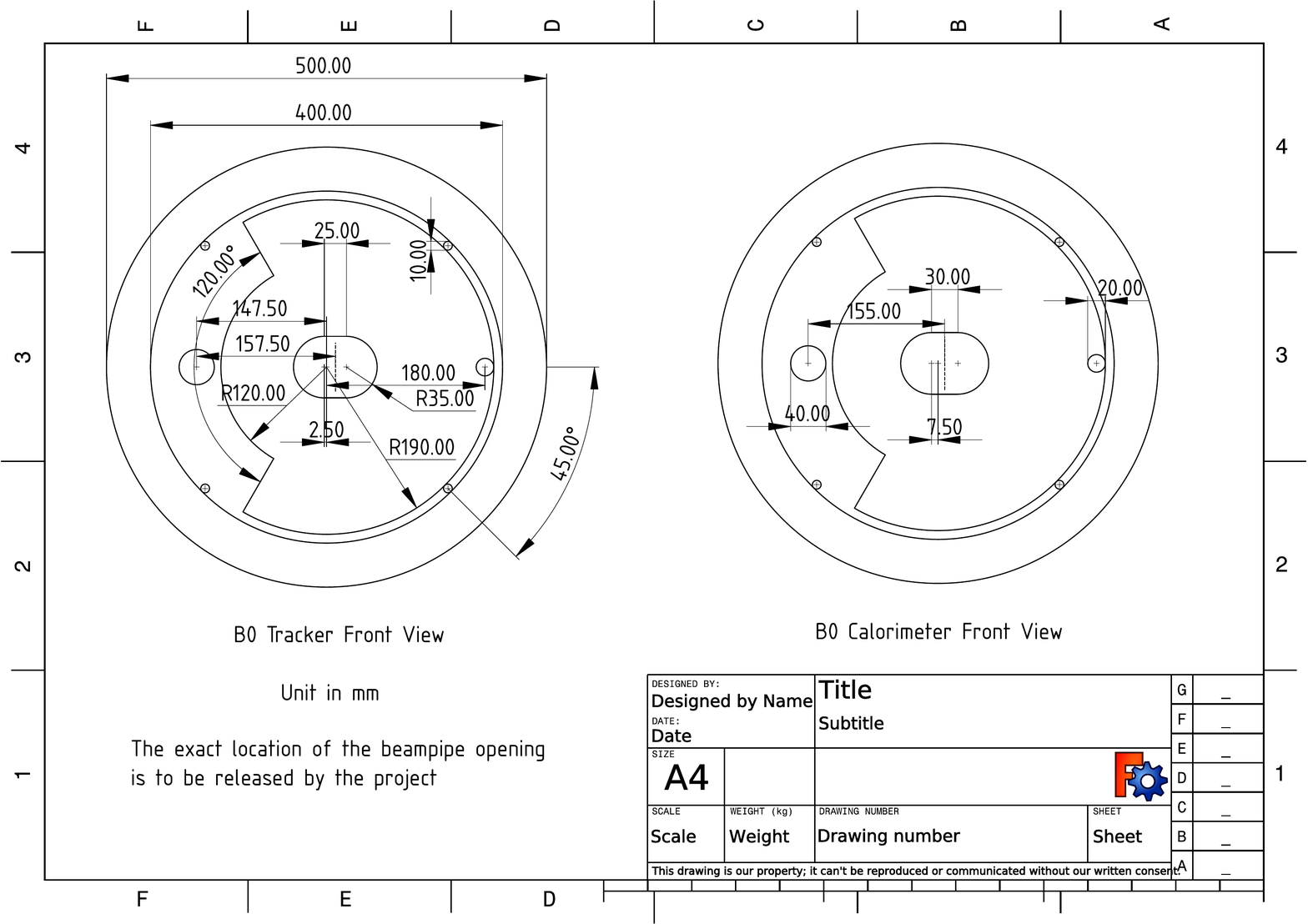}
    \\
    \includegraphics[width=0.87\linewidth]{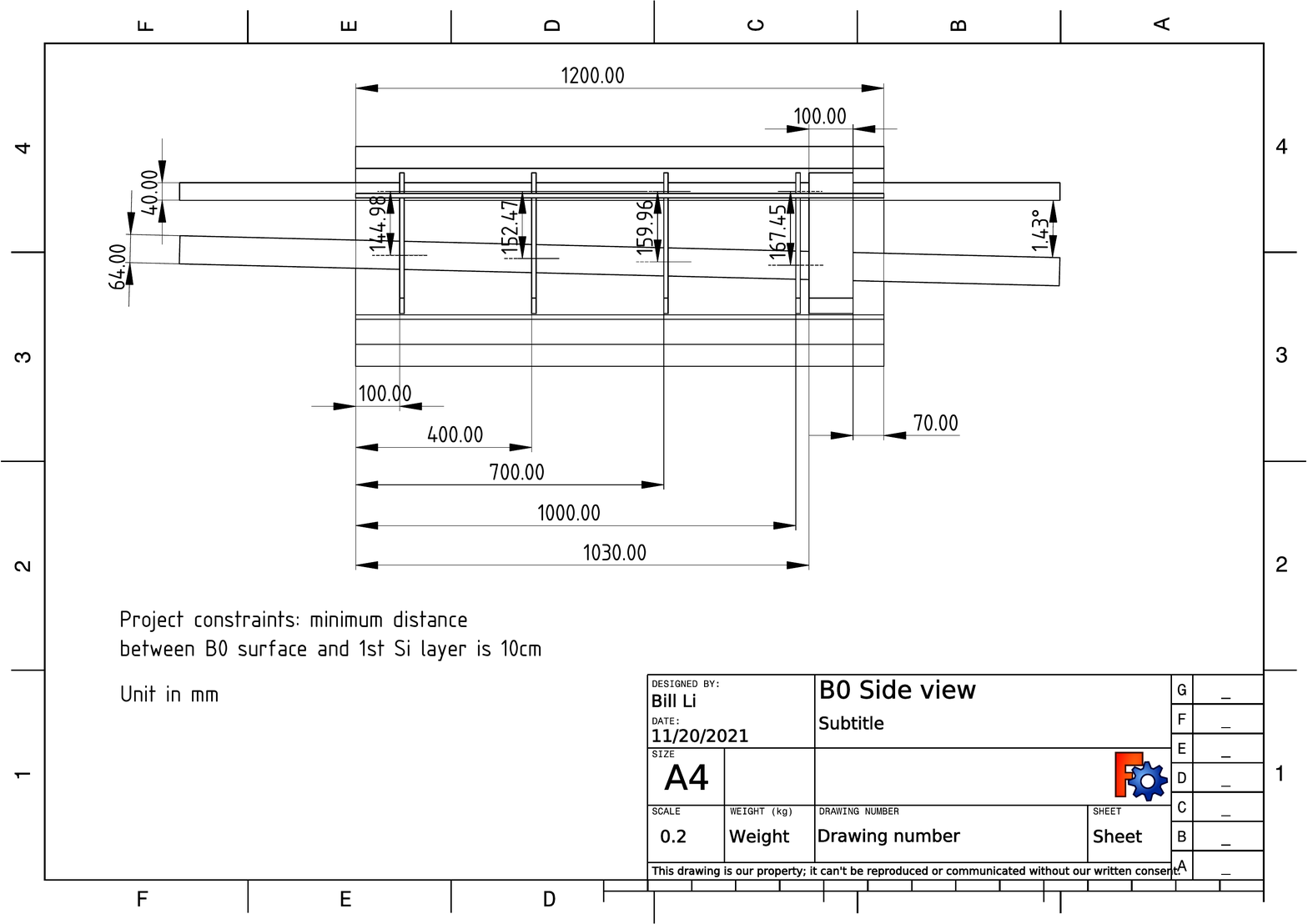}
    \caption{B0 design. Dimensions are given in mm.}
    \label{fig:B0_CAD}
\end{figure*}

To better visualize the layout of the trackers and calorimeter inside of the B0 magnet (see Fig.~\ref{fig:far-forward}), a CAD drawing (Fig.~\ref{fig:B0_CAD}) was created based on the official IP6 design specification from Ref.~\cite{IP6design}. The shape of the trackers and calorimeter resembles the `PAC-man', which is optical to accommodate the incoming (upstream) electron and outgoing (downstream) ion beam pipes. Note that the trackers and calorimeters are mounted on the guiding rods, which can be slid out for servicing. There is no access from the backside of the B0 magnet due to special constraints.

\label{app:B0}

%%%%%%%%%%%%%%%%%%%%%%%%%%%%%%%%%%%%%%%%%%%%%%%DVCS-eA, TOPEG, Gary Penman
\section{DVCS off Helium-4 and the TOPEG generator}
\label{sec:dvcsea_appendix}
The TOPEG event generator, which was originally used for the $^{4}$He DVCS studies in the YR, was used for our DVCS off-helium study. Full details on the TOPEG generator can be found in the YR and in~\cite{TOPEG} (as well as in the subsequent references provided within). For the ECCE studies, the model used neglects the real part of the $H$ generalized parton distribution in the full coherent DVCS $^{4}$He implementation (generator model 3, according to the TOPEG nomenclature). This model allows for a reasonable computation time, without sacrificing necessary physics precision. 

Complete generator card data corresponding to the settings used for the results shown in this document can be found in Table~\ref{tab:topegconfig}.
\begin{table}[ht]
    \centering
    \begin{tabular}{|l|l|}
        \hline
         Parameter & Values \\
            \hline
            \hline
         E\textsubscript{pz} (GeV\textsuperscript{2}) & 5 \\
         He\textsubscript{pz} (GeV\textsuperscript{2}) & 163.958 \\
         N TFoam Cells & 10\textsuperscript{4} \\
         N Cell Samples & 300 \\
         y & 0.05 $\le$ y $\le$ 0.85\\
         Q\textsuperscript{2} (GeV\textsuperscript{2})& 2.0 $\le$ Q\textsuperscript{2} $\le$ 30\\
         W\textsuperscript{2} (GeV\textsuperscript{2})& W\textsuperscript{2} $\ge$ 16\\
         $\theta^{e}_{max}$ (rad) & $\theta^{e}_{max} \ge$ 2.35\\
         t (GeV\textsuperscript{2}) & 0.01 $\le$ t $\le$ 0.5 \\
         eBeam helicity & $\pm$ 1\\
    \hline
    \end{tabular}
    \caption{TOPEG generator configuration used for these studies. Further value ranges are also under current study.}
    \label{tab:topegconfig}
\end{table}
In TOPEG, simulated data is constrained by kinematic limits set by the user in the generator input settings. Notably attempts to generate events at $t<$0.01 GeV\textsuperscript{2} and $Q$\textsuperscript{2} $\le$ 1 GeV\textsuperscript{2} often encountered issues. 

\begin{figure}[h]
\centering
    \centering
    \includegraphics[width=1\linewidth]{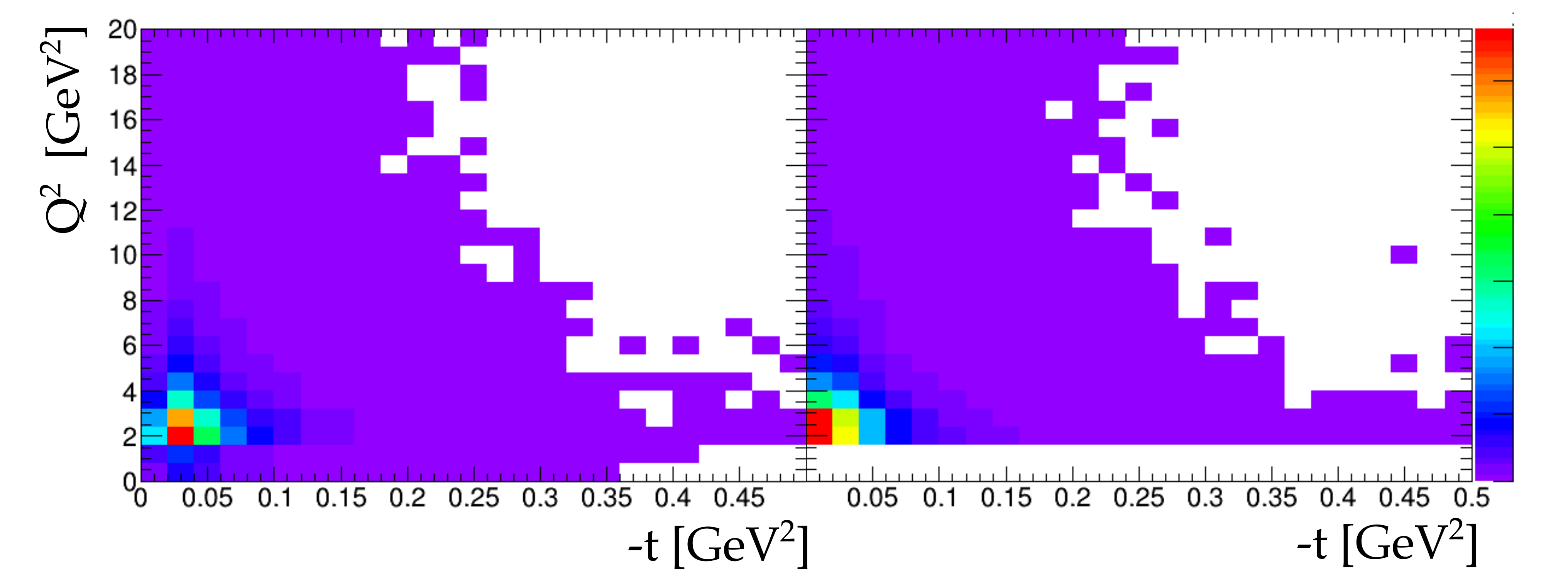}\\
    \includegraphics[width=1\linewidth]{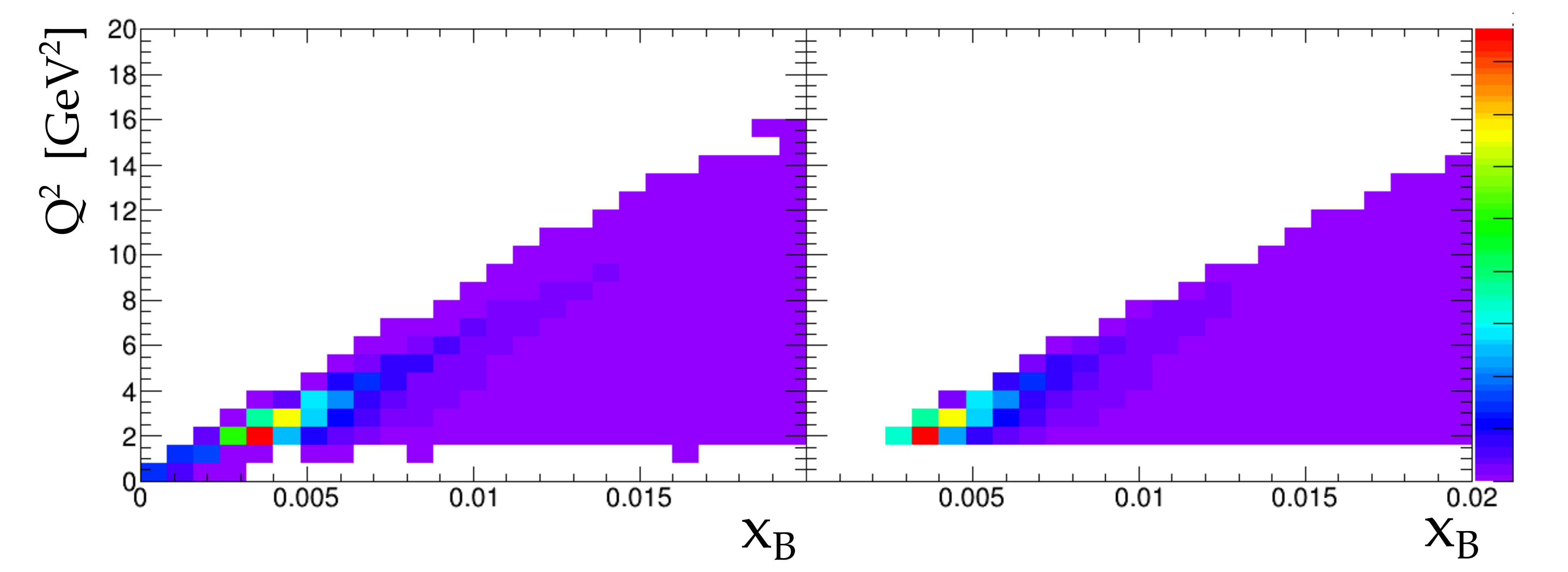}\\
    \includegraphics[width=1\linewidth]{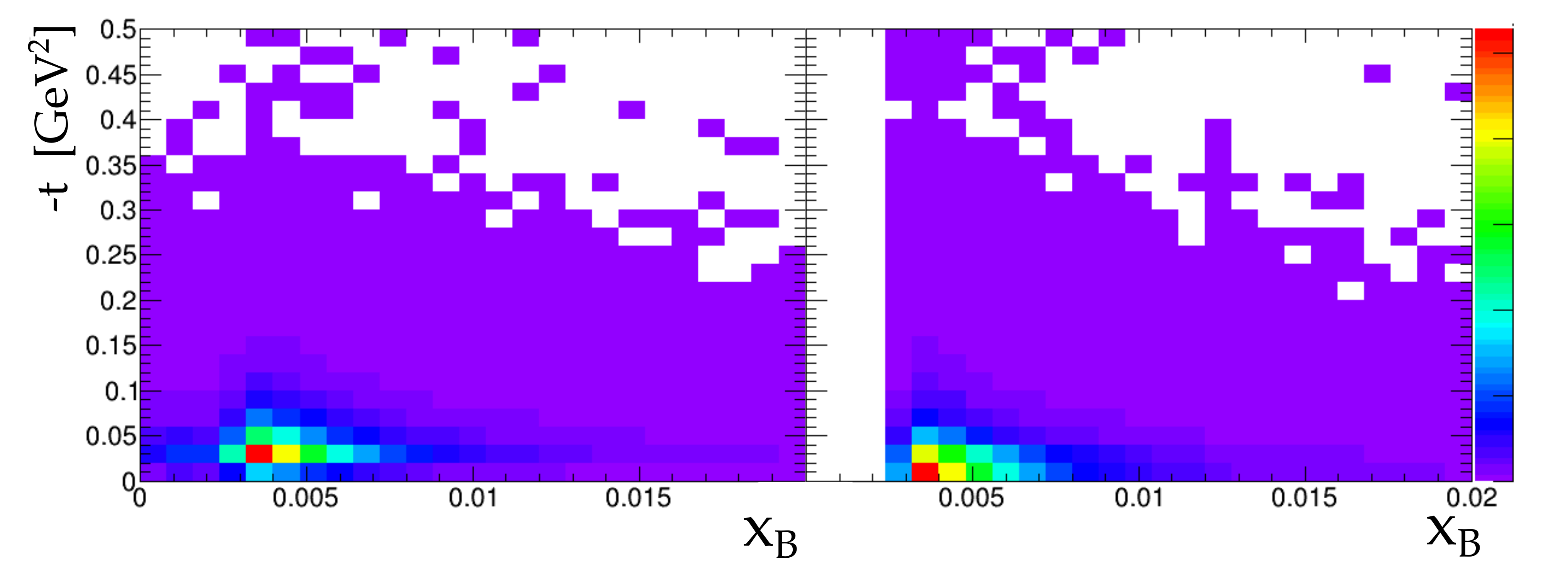}
        \caption{2D kinematic coverage plots for DVCS-e$^{4}$He. The left-hand side plots show the generated phase space from TOPEG directly. The right-hand side plots show the kinematic coverage as reconstructed from the Fun4All output for the ECCE detector. The color scales indicate raw counts and are not normalized to each other.}
        \label{fig:dvcsea:2dplots}
\end{figure}

%%%%%%%%%%%%%%%%%%%% 2D Phase Space Plots %%%%%%%%%%%%%%%%%%%%
%(for the physics quantities presented in Figures~\ref{fig:dvcsea:countplots} and~\ref{fig:dvcsea:acceptanceplots})
The matching kinematic phase space plots  are also given in Figure~\ref{fig:dvcsea:2dplots}. Cross sections are calculated by integrating over the largest possible phase space which is filled with events, as seen in these figures.

\section{XYZ production event generator}
\label{sec:elSpectro}

The event generator was custom developed for spectroscopy reactions at the EIC. 
It consists of two main parts: photoproduction helicity amplitudes; and virtual photon production. 

\subsection{JPAC Photoproduction Amplitudes}
The helicity amplitudes were calculated following the formalism and parameters given in Ref.\cite{Albaladejo:2020tzt}. The models therein are expected to give an order of magnitude estimates
for meson production cross sections.
For the J$/\psi \pi^{+} \pi^{-}$ events, we assumed only $\psi(2s)$, $\chi_{c1}(3872)$ and Y(4260)
states were produced. For the $\chi_{c1}(3872)$ production we used the pion exchange amplitudes 
and for $\psi(2s)$ and Y(4260) we assumed only pomeron exchange. As these models consist of high and low energy limits, for the current study we chose to combine the two via a simple linear interpolation of the helicity amplitudes between the high and low regions given.

\subsection{Virtual photoproduction}

To make estimates of exclusive electroproduction with low-$Q^2$ quasi-real virtual photons we first generate a beam of virtual photons that interacted with the proton beam producing the meson which we subsequently decayed to specific final states which were then run through the ECCE detector simulation,

\begin{equation}
    \frac{d^{4}\sigma}  {ds dQ^{2} d\phi dt} =  \frac{d^{2}\sigma_{e,\gamma* e'}} {dsdQ^{2}} \frac{d^{2}\sigma_{\gamma *+p \rightarrow V+p} (s,Q^{2})}  {d\phi dt}
    \label{eqn:spec_dsigma}
\end{equation}

the virtual photon flux factor was sampled from 

\begin{equation}
    \frac {d^{2}\sigma_{e,\gamma* e'}}  {dsdQ^{2}} = \frac{\alpha} {2\pi} \frac {K.L} { E } \frac { 1} { Q^{2}} \frac{ 1 } { (s-M^{2}+Q^{2})}
    \label{eqn:spec_photonflux}
\end{equation}
\begin{equation}
     K= \frac {W^{2}-M^{2}}{2M} \quad L = \frac{ 1+(1-y)^{2} } {y} - \frac{ 2m_{e}^{2}y } { Q^{2} } 
\end{equation}

and the two-body photoproduction cross-section was calculated as
\begin{equation}
    \frac{d^{2}\sigma_{\gamma^{*}+p \rightarrow V+p} (s,Q^{2})}  {d\phi dt} = \frac{ 1 }{ 128 \pi^{2} s } \frac{ 1 } { | \mathbf{p}_{\gamma^{*}_{CM}} | ^{2} } {| { M }(s,t)  |^{2}}
    \label{eqn:spec_twobodydsigma}
\end{equation}
with $M(s,t)$ the photoproduction amplitude. 
This cross-section was modified by an additional $Q^{2}$ dependence taken from \cite{H1:1999pji}.
Eqn. (\ref{eqn:spec_dsigma}) was integrated numerically to give the total cross section for determining event rates. 

The generation algorithm proceeded as
\begin{itemize}
    \item Generate the scattered electron by sampling from 2D distribution in Eq. \ref{eqn:spec_photonflux} in the rest frame of the proton.
    \item Sample the intermediate particle masses from Breit-Wigner distributions with parameters taken from PDG values.
    \item Given s  and the particle masses accept/reject the event based on the n-body mass phase space.
    \item Given s, $Q^{2}$ and final state masses accept/reject on the production t from Eq. \ref{eqn:spec_twobodydsigma}
    \item Sample random $\phi$ angle and complete the kinematics of the produced meson and recoiling proton.
    \item Decay produced meson to J$/\psi$ and 2 pions using flat decay angle distributions.
    \item Decay J$/\psi$ to $e^{+} e^{-}$ using flat decay angle distributions.
    \item Boost all stable particles to the lab system.
\end{itemize}

Prior to tracking in Geant4 the ECCE afterburner is applied to the 4-vectors to apply crossing angles and divergences.
\section{TCS $ep$ and the EpIC generator 2D phase space}
\label{sec:TCSep_appendix}
\begin{figure*}
    \centering
    \includegraphics[width=0.95\textwidth]{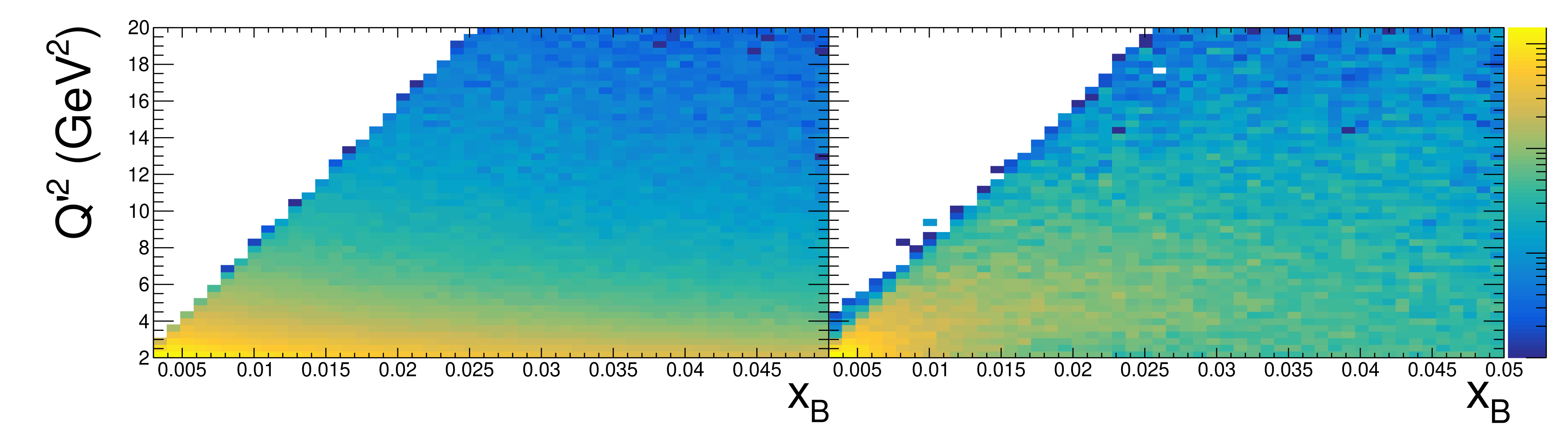}
    \caption{5\,$\times$\,41 -  \begin{math} Q'^{2} \end{math} versus TCS \begin{math} \tau \end{math} ($x_{B}$) for generated EpIC data (left) and reconstructed Fun4All data (right).}
    \label{fig::5x41_TCS_phase}
\end{figure*}

\begin{figure*}
    \centering
    \includegraphics[width=0.95\textwidth]{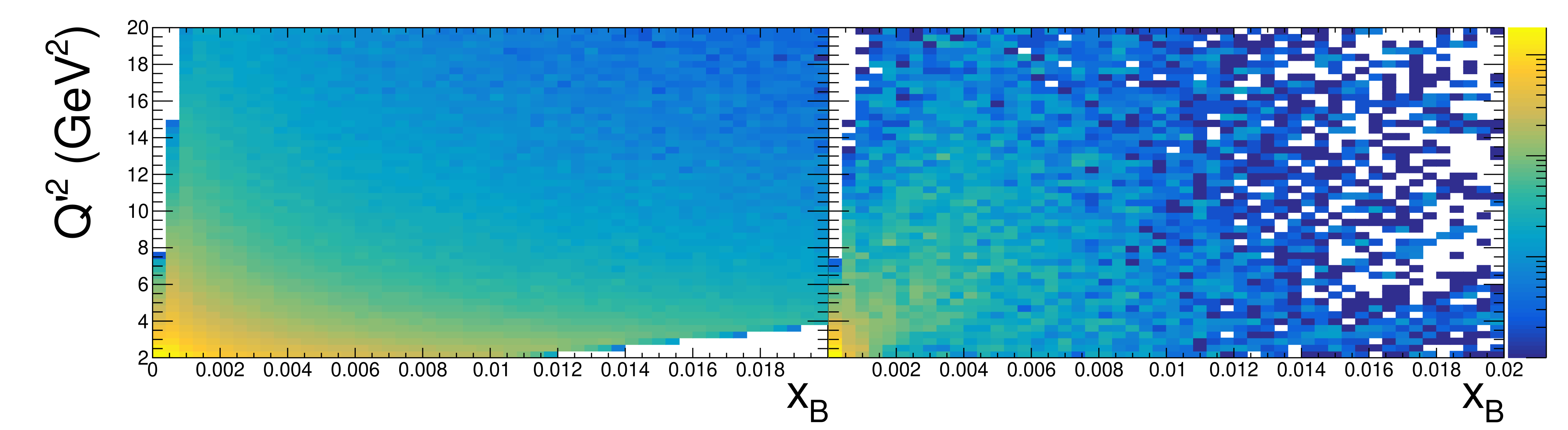}
    \caption{18\,$\times$\,275 - \begin{math} Q'^{2} \end{math} versus TCS \begin{math} \tau \end{math} ($x_{B}$) generated EpIC data (left) and reconstructed Fun4All data (right).}
    \label{fig::18x275_TCS_phase}
\end{figure*}

Figs. \ref{fig::5x41_TCS_phase} and \ref{fig::18x275_TCS_phase} are a representation of the phase space coverage of the ECCE detector, as compared with the generated data from EpIC.

%%%%%%%%%%%%%%%%%%%%%%%%%%% 2D 5X41 %%%%%%%%%%%%%%%%%%%%%%%%%%%%%%

%%%%%%%%%%%%%%%%%%%%%%%%%%  2D 18X275  %%%%%%%%%%%%%%%%%%%%%%

\section {LAGER generator for exclusive J$/\psi$ production}
\label{sec:jpsi_generator}

The LAGER generator \cite{lager} was used to produce event samples for the ECCE studies presented.
LAGER is described as a modular accept-reject generator, capable of simulating both fixed target and collider kinematics, and has previously been used for vector meson studies at EIC kinematics, with significant recent developmental effort in support of DVMP studies.

The event samples are processed through eic-smear and the resulting ROOT trees are provided to Fun4All, which simulates the full ECCE detector response in Geant4.
The final output of this process is the Fun4All DST files. %{\blue{JR: did you really use the DSTs? }}
All studies presented were performed at the IP6 detector location using Prop.4 (aka July detector designd). The kinematic presented in this study corresponds to  electron and proton beam energies of 18 GeV and 275 GeV, respectively.

\bibliographystyle{elsarticle-num} 
\bibliography{refs.bib,refs-ecce.bib,refs-xyz.bib,refs-dvcs.bib}

\end{document}